\documentclass[a4paper,11pt,twoside]{book}
\usepackage{caption}
\usepackage{comment}
\usepackage{setspace}
\usepackage{graphicx}
\usepackage{amssymb}
\usepackage{amsmath}
\usepackage{url}
\usepackage{natbib}
\usepackage{aastex_hack}
\usepackage{deluxetable}
\usepackage{longtable}
\usepackage{fancyhdr}
\usepackage[avantgarde]{quotchap}
\usepackage[calcwidth]{titlesec}
\usepackage{swinburnetitle}

\fancypagestyle{plain}{
\fancyhf{}
\fancyfoot[CO]{\thepage}
\renewcommand{\headrulewidth}{0pt}
}

\makeatletter
\def\cleardoublepage{\clearpage\if@twoside \ifodd\c@page\else
	\hbox{}
	\vspace*{\fill}
	\thispagestyle{empty}
	\newpage
	\if@twocolumn\hbox{}\newpage\fi\fi\fi}
\makeatother

\renewcommand\chapterheadstartvskip{\vspace*{-5\baselineskip}}

\titleformat{\section}[hang]{\sffamily\bfseries}
{\Large\thesection}{12pt}{\Large}[{\titlerule[0.5pt]}]

\newcommand{\degrees}{\ensuremath{^{\circ}}}
\newcommand{\Pdot}{\ensuremath{\dot{P}}}
\newcommand{\Gdot}{\ensuremath{\dot{G}}}
\newcommand{\Pb}{\ensuremath{P_{\mathrm b}}}
\newcommand{\Pbdot}{\ensuremath{\dot{P}_{\mathrm b}}}

\newcommand{\tauc}{\ensuremath{\tau_{c}}}
\newcommand{\taud}{\ensuremath{\tau_{d}}}
\newcommand{\BP}{\ensuremath{\overrightarrow{BP}}}
\newcommand{\vect}[1]{\ensuremath{\boldsymbol{\mathrm #1}}}
\newcommand{\unitvect}[1]{\ensuremath{\boldsymbol{\hat{\mathrm #1}}}}
\newcommand{\accb}{\ensuremath{\overrightarrow{a}_{\mathrm{bar}}}}
\newcommand{\accp}{\ensuremath{\overrightarrow{a}_{\mathrm{psr}}}}

\newcommand{\Vt}{\ensuremath{v_{\mathrm T}}}
\newcommand{\NH}{Northern Hemisphere}
\newcommand{\SH}{Southern Hemisphere}
\newcommand{\ns}{neutron star}
\newcommand{\msolar}{\ensuremath{\mathrm M_{\odot}}}
\newcommand{\kms}{\ensuremath{\mathrm {km\ s}^{-1}}}
\newcommand{\pone}{PSR J0108--1431}
\newcommand{\ptwo}{PSR J0437--4715}
\newcommand{\pthree}{PSR J0630--2834}
\newcommand{\pfour}{PSR J0737--3039A/B}
\newcommand{\pfive}{PSR J1559--4438}
\newcommand{\psix}{PSR J2048--1616}
\newcommand{\pseven}{PSR J2144--3933}
\newcommand{\peight}{PSR J2145--0750}
\newcommand{\esys}{\ensuremath{E_{\mathrm{sys}}}}
\newcommand{\phisys}{\ensuremath{\phi_{\mathrm{sys}}(t)}}
\newcommand{\etherm}{\ensuremath{E_{\mathrm{therm}}}}

\DeclareFontFamily{OT1}{pzc}{} 
\DeclareFontShape{OT1}{pzc}{m}{it}{<->[1.5] pzcmi8t}{}
\DeclareMathAlphabet{\mathpzc}{OT1}{pzc}{m}{it}

\begin{document}

\frontmatter
\author{Adam Travis Deller}
\title{Precision VLBI astrometry: Instrumentation, algorithms and pulsar parallax determination}
\date{January 2009}
\maketitle
\addcontentsline{toc}{chapter}{Abstract}
\chapter*{Abstract}                             
This thesis describes the development of DiFX, the first  
general--purpose software correlator for radio
interferometry, and its use with the Australian Long Baseline Array to complete the
largest Very Long Baseline Interferometry (VLBI)
pulsar astrometry program undertaken to date in the Southern Hemisphere.  
This two year astrometry program has resulted in the measurement
of seven new pulsar parallaxes, which has more than trebled the number of measured VLBI 
pulsar parallaxes in the Southern Hemisphere.  These measurements included a determination
of the distance and transverse velocity of \ptwo\ with better than 1\% accuracy, enabling
improved tests of General Relativity; the first significant measurement of parallax for the famous
double pulsar system \pfour, which will allow tests of General Relativity in this system to proceed to the
0.01\% level and also offers insights into its formation and high--energy emission; 
and a factor of four revision to the estimated distance of
\pthree, which had previously appeared to possess extremely unusual x--ray emission characteristics.
Additionally, the ensemble of refined distance and transverse velocity estimates have enabled a 
widely applicable improvement in knowledge of pulsar luminosities in several wavebands and
the Galactic electron distribution at southern latitudes.  Finally, the DiFX software correlator
developed to enable this science has been extensively tested and verified against three existing
hardware correlators, and is now an integral part of the upgraded Long Baseline Array 
Major National Research
Facility used by astronomers throughout Australia and the world; furthermore, it has been selected
to facilitate a major upgrade of the world's only full--time VLBI instrument, the Very Long 
Baseline Array operated by the National Radio Astronomy Observatory in the US.
\clearpage

\chapter*{Acknowledgements}

Like anyone who has navigated the hazards of completing a PhD,
I have many people to thank, personally and professionally -- and in many cases both --
for my progress to this point.  My supervisors -- Prof. Steven Tingay, Prof. Matthew Bailes and
Dr. John Reynolds -- have guided me from astronomy newcomer to where I am now, and set
me a high standard to aspire to for the remainder of my career.  Special thanks go to Steven, whose
door was always open as I crash--coursed my way through interferometry at the outset of my thesis.
There are countless other friends 
and colleagues from many institutes world--wide I would like to thank for their professional input, 
and to mention but a few, Emil, Joris, Chris, Michael, Walter and Craig have all
lent me sage advice, careful reading, 
a cool head in an apparent crisis and a good supply of jokes, puns and
general comic relief many times over the last three and a half years.

No PhD occurs in isolation and I would like to thank the many people who have enabled me to
remain a largely sane and well--adjusted individual.  To my friends from university, the Swinburne
R\&D course, triathlon, home, abroad and everywhere -- I'm lucky to know you all, and thanks for
everything.  Extra credit and thanks go to my flatmates past and present, all of whom I count amongst
my closest friends -- Gaz, Andy, Leeanne and Trish. 

I would also like to thank the observatory staff at the ATCA and Parkes telescopes where I undertook the 
majority of my observing for this thesis -- Phil, Robin, Dave, Euan and everyone; your helpfulness, 
professionalism and knowledge was extremely valuable to me and much appreciated. 

This thesis made extensive use of the PSRCAT tool made available by the Australia Telescope National Facility at http://www.atnf.csiro.au/research/pulsar/psrcat/.  The Long Baseline Array, without which
there would be no Southern Hemisphere astrometry, is part of the Australia Telescope which is 
funded by the Commonwealth of Australia for operation as a National Facility managed by CSIRO.  
The authors of the DIFMAP and ParselTongue software packages deserve universal acclaim for their 
excellent and freely available data reduction tools.

Finally, and most importantly, I would like to thank those closest to me for the love and support
that is bestowed regardless of the path I take in science or life.  To Anneke, who has patiently 
listened to innumerable tirades against LaTeX, AIPS, and any number of other
frustrations, and always had a gentle word of reassurance -- you have given me 
happiness and confidence during the long days of writing.
To my family -- Maxine, Rod, Tim, Loren, Norm, 
Craig, and others -- thank you for your unwavering support and encouragement in every endeavour 
I have chosen, not just throughout my undergraduate and postgraduate studies, but ever
since I was old enough to talk and question -- and talk back.  
\enlargethispage{2\baselineskip}
\clearpage
\chapter*{Declaration}

I hereby declare that all work within this thesis, with the exceptions noted below, is solely my own work and contains no material which has been accepted to the award of any other degree or diploma.

Much of the material in Chapters 4, 5 and 6 was drawn from the publications \citet{deller07a}, \citet{deller08b}, \citet{deller08c}, and \citet{deller08a}.  I gratefully acknowledge the useful discussions and critical analysis provided by my co--authors during the preparation of these publications.  Furthermore, I would like to acknowledge that the comparison between DiFX and the Bonn MkIV correlator in Chapter~\ref{difx:deploy:bonncomp} is based on the publication \citet{tingay08a} and I am grateful to Professor Steven Tingay for providing the figures.

\vspace{10ex}
\includegraphics{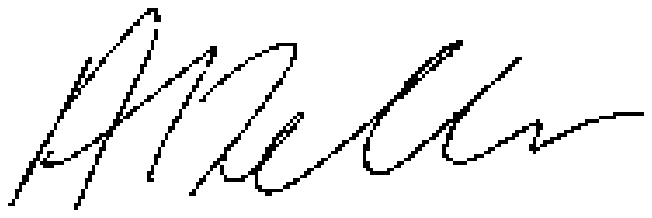}
\vspace{5ex}

Adam Travis Deller

January 2009
\clearpage

\tableofcontents
\addcontentsline{toc}{chapter}{List of Figures}
\listoffigures
\addcontentsline{toc}{chapter}{List of Tables}
\listoftables

\mainmatter

\setcounter{page}{1}    

\chapter{INTRODUCTION}
\label{intro}
\section[Thesis motivation]{Thesis motivation}
\label{intro:goals}
Radio astronomy, in particular radio interferometry and its high resolution sub--branch
Very Long Baseline Interferometry (VLBI, discussed in detail in Chapter~\ref{radio}), 
is a field in which advances in instrumentation --
driven in this case by developments in consumer and industrial electronics -- have 
enabled rapid, ongoing advances in science, by expanding the parameter space scientists
can explore. This has led to a dependency between engineer and scientist which is rarely
seen in other fields of astronomy -- most radio astronomers have at least a passing
knowledge of the systems they use, and many are themselves developers as well as users 
of instruments.  This is especially true for part-time VLBI arrays such as the Australian
Long Baseline Array (LBA).

Another field of study with a strong overlap between engineer and astronomer is
pulsar astronomy.  Radio pulsars (discussed in detail in Chapter~\ref{pulsars}) are 
rapidly rotating neutron stars that emit radiation from their magnetic
poles.  Neutron stars form from the collapsed cores of once--massive stars following 
a supernova explosion.  Due to the pulsar's very high moment of inertia, the pulsar spin period $P$ is
typically very stable.   The misalignment of the rotation and magnetic axes leads to the radiation being 
observed as a series of pulses (dispersed in frequency by intervening ionised matter)
at Earth. Analysis of pulsar data typically requires dedicated, high speed signal processing,
which has led to most pulsar groups developing and deploying their own digital electronic
systems on a telescope by telescope basis.

VLBI is an integral tool for the study of pulsars, allowing the determination of kinematic
parameters of individual pulsars in a (relatively) precise and model--independent fashion.
This use of high--resolution observations to accurately measure object positions
is known as astrometry.
As discussed in Section~\ref{pulsars:obs:vlbi}, the addition of independent kinematic information
allows the calculation of geometrical effects which alter the observed arrival time of pulses.
If the signature of the annual orbital parallax on pulsar 
position imposed by the Earth's motion around
the Sun can be detected, the resultant determination of pulsar distance can be used to accurately
calibrate the pulsar luminosity at all wavebands, as well as further refining the pulse
arrival time corrections.

\begin{deluxetable}{lcccc}
\tabletypesize{\tiny}
\tablewidth{0pt}
\tablecaption{Comparison of major VLBI arrays}
\tablehead{
\colhead{Array} & \colhead{Array} 	& \colhead{Maximum station}  	& \colhead{Maximum baseline} & \colhead{Active observing} 	\\
\colhead{name} 		 & \colhead{stations} 		& \colhead{data rate (Mbps)} 			& \colhead{length (km)} 				 & \colhead{(weeks/year)} 				
}
\startdata
VLBA\tablenotemark{1}				&10	& 512	& 8600	& 52		\\
EVN	\tablenotemark{2}				&18	& 1024	& 10000	& 10--15	\\
LBA (S2 -- pre--2005) \tablenotemark{3}	& 6	& 128	& 1700	& 3--4	\\
LBA (DiFX -- post--2005)  \tablenotemark{3} & 6 & 1024	& 1700	& 3--4	\\
\enddata
\tablenotetext{1}{http://www.vlba.nrao.edu/}
\tablenotetext{2}{http://www.evlbi.org/intro/intro.html}
\tablenotetext{3}{http://www.atnf.csiro.au/vlbi/}
\label{tab:vlbiarrays}
\end{deluxetable}

\begin{figure}
\begin{center}
\includegraphics[width=1.0\textwidth]{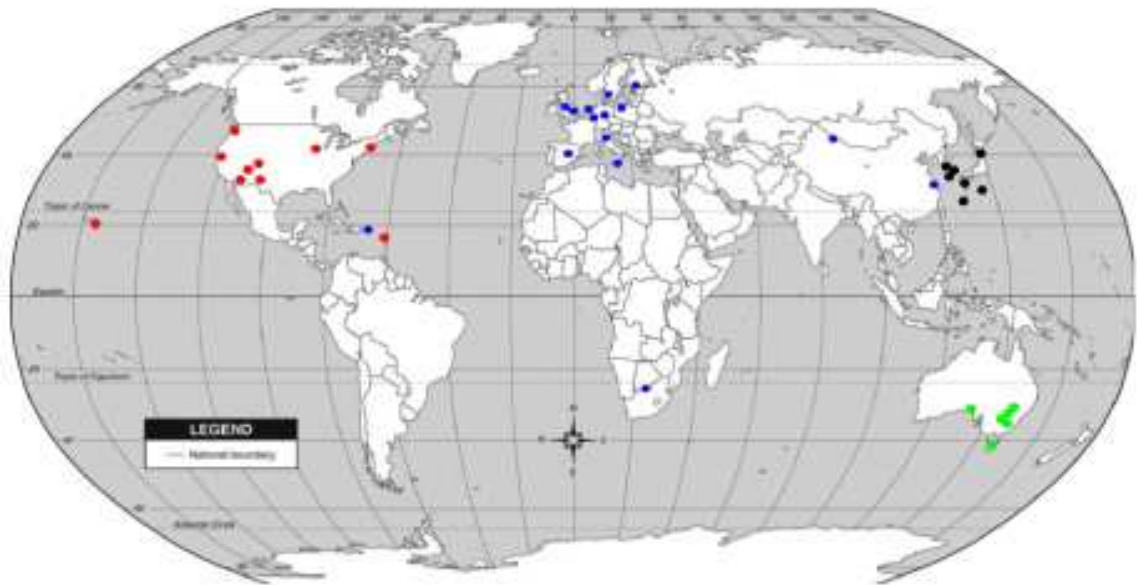}
\caption[Antennas that regularly participate in VLBI]
{The location of antennas regularly participating in VLBA (red), EVN (blue) and LBA
(green) observations.  Antennas which are sometimes added to one or more arrays on an
ad--hoc basis, or that belong to other arrays such as the Japanese VLBI Network (JVN) or the Korean VLBI Network (KVN) are shown in black.}
\end{center}
\label{fig:vlbiants}
\end{figure}

While \SH\ instrumentation has played a crucial role in the study of pulsars -- the
Parkes and Molonglo radiotelescopes in Australia have discovered over half of the known
radio pulsar population -- few pulsar VLBI observations have been made from the \SH.  Three
previous \SH\ surveys \citep{dodson03a,legge02a,bailes90a} have resulted in the measurement of
two pulsar parallaxes, whereas 16 \NH\ parallaxes were published at the time
of writing, with nine obtained in a single program \citep{brisken02a}.
This is primarily due to the capabilities of the American Very Long Baseline Array (VLBA) and
the European VLBI Network (EVN) instruments, both of which possessed advantages in 
recording bandwidth, support
and observation cadence compared to the LBA, which is the only \SH\ VLBI
array.  The LBA, VLBA and EVN antennas are shown in Figure~\ref{fig:vlbiants} -- full details of
these arrays are shown in Table~\ref{tab:vlbiarrays}. 

Despite the advantages of \NH\ arrays, 
there are many unique pulsars which lie too far south to be effectively observed from 
the \NH\, such as the unique double pulsar system \pfour, the longest period radio pulsar
\pseven, and the nearest and brightest millisecond pulsar \ptwo.  These objects and others
offer insights into pulsar formation, evolution and many other related fields of research,
but have yet to be the subjects of detailed study at the highest angular resolution.
The LBA offers the only means to undertake high angular resolution studies of these objects.

\clearpage

The key impediment faced by the LBA at the outset of this thesis (in early 2005), 
compared to the VLBA and EVN,
was a lack of sensitivity.  As discussed in Chapter~\ref{radio}, the sensitivity of an 
interferometer is proportional to the square root of the bandwidth of the signal it accepts,
which is limited by the digital sampling, recording and processing hardware employed by the array.
As shown in Table~\ref{tab:vlbiarrays}, in 2005 the LBA was significantly limited in the  bandwidth
it could record, compared to the VLBA and EVN.  This meant that targeting the most scientifically
desirable \SH\ pulsars, many of which are faint radio sources, would require impossibly 
large amounts of telescope time to obtain sufficient sensitivity.
An upgrade of the LBA was thus the only feasible alternative to 
obtain astrometric information on these objects.

This upgrade involved replacing the existing tape--based recorders and signal processing 
hardware (the "correlator", discussed in Chapters~\ref{radio} and \ref{difx}) with disk--based 
recorders and an alternate correlator capable of handling higher data rates.
At the commencement of this thesis,
new disk--based recorders were being tested with a preliminary correlator based on a 
software algorithm running on a small supercomputer \citep{west04a}.  Despite verifying the
functionality of the disk--based system, this initial "software correlator" was too slow (taking
weeks to correlate a day's observing) for production usage.  A refined, more efficient software
correlator was required, which was developed during this thesis and became known as 
DiFX (short for Distributed FX correlator -- the FX terminology is explained in Chapter~\ref{difx}).

Thus, from the outset, this thesis aimed to address the twin goals of a developing a flexible, 
powerful, and efficient software correlator to make use of the higher bandwidths available
with a disk--based system, and the integration of that correlator into the LBA 
to produce an instrument with the flexibility and sensitivity necessary to successfully undertake
an astrometric program encompassing the most scientifically fruitful \SH\ pulsars.  The 
undertaking of this astrometric program also necessitated the development of significant
new algorithms and tools for astrometric data reduction, and the characterisation and improvement
of many areas of LBA operations.

While the development of the DiFX software correlator was a necessary precursor to the desired
astrometric science, the improved sensitivity and flexibility of the new system extended
 the capabilities of the LBA for all science targets.  Indeed, DiFX has also been 
 adopted by several new or upgraded arrays external to Australia, most notably the VLBA.
 Section~\ref{difx:additional} briefly discusses some science highlights external to this thesis that 
 have been obtained using the DiFX software correlator on both the LBA and other
 arrays.
 
\section[Thesis outline]{Thesis outline}
\label{intro:outline}
An overview of the historical studies and present understanding of pulsars is given in 
Chapter~\ref{pulsars}, along with a discussion of the pulsar science which is possible
through the use of VLBI astrometry. Chapter~\ref{radio} presents a conceptual and 
mathematical overview of radio interferometry, including VLBI, and covers the 
application of VLBI to astrometric observations.   Chapter~\ref{difx} covers the development,
testing and verification of DiFX, the final version of the software correlator developed
to fulfil the first primary goal of this thesis.   Chapter~\ref{techniques} examines the post--correlation
data analysis undertaken on all astrometric datasets, showing the transformation from correlated
data to pulsar positions at a given epoch.  Chapter~\ref{results} highlights the results obtained
from the LBA astrometric program and shows the implications of the measured pulsar distances 
and kinematics, both for each pulsar individually and for population studies as a whole.  
Concluding remarks are made in Chapter~\ref{conclusions}.
\chapter{PULSARS}
\label{pulsars}

\section[Discovery and studies]{Discovery and studies}
\label{pulsars:discovery}
When the existence of highly compressed stellar objects consisting primarily of neutrons --
{\it neutron stars} -- was postulated as a possible result of a supernova explosion by
\citeauthor{baade34b} in \citeyear{baade34b}, the field of radio astronomy was 
barely taking its first tentative steps.  A third of a century later, however, it would be radio astronomy
that provided a remarkable confirmation of the existence of neutron stars, beginning
with PhD student Jocelyn Bell noticing a periodic ``little bit of scruff" while observing
at a Cambridge radiotelescope.  This discovery was published the following year 
\citep{hewish68a} and shortly after the rotating \ns\ origin of the signal was independently 
proposed by \citet{gold68a} and \citet{pacini68a}.

These periodic radio sources became known as ``pulsars"\!, and a flood of theoretical and 
observational results followed the initial discovery.  Before the end of the decade, pulsars 
were detected in the x--ray \citep{fritz69a} and optical \citep{cocke69a} wavebands,
with detection in gamma rays following shortly after \citep{fazio72a}.  By 1980 over 300
pulsars had been discovered and the neutron star origin, with the radio emission 
powered by the conversion of rotational energy, was well established.

Since that time, however, a series of observational surprises have shown that the neutron
stars can manifest themselves in a variety of guises:

\begin{itemize}
\item {\it recycled} or {\it millisecond} pulsars, with lower ($\sim10^{8}$\ gauss) magnetic
field strengths and spin
frequencies in the hundreds  of Hz, formed through accretion of matter in a binary system
\citep{alpar82a,van-den-heuvel75a};
\clearpage
\item Anomalous X--ray Pulsars (AXPs), which emit more energy than can be explained
by their spindown rate, and Soft Gamma Repeaters (SGRs), both of which are believed 
to be magnetars -- neutron stars  with extremely high ($\ge 10^{14}$\ gauss) magnetic 
fields, where the magnetic field decay powers repeated powerful outbursts in x--rays and 
gamma--rays \citep{thompson96a}; and
\item Nulling pulsars, intermittent pulsars and Rotating Radio Transients (RRATs), where the 
radio emission is intermittently suppressed \citep[see e.g.][]{wang07a,backer70a,mclaughlin06a}. 
In the case of RRATs, as few as  one pulse in thousands is emitted.
\end{itemize}

As the objects studied in this thesis are all rotation--powered, non--nulling radio pulsars, the 
remainder of this chapter will focus on these objects.

\section[Current understanding]{Current understanding}
\label{pulsars:understand}
\subsection[Formation]{Formation}
\label{pulsars:understand:form}
Ordinary radio pulsars are believed to form in the supernova explosions which result
when massive stars exhaust their supply of the light elements which had fueled nuclear fusion.  With the
abrupt removal of radiation pressure which had supported the star against gravity, a rapid 
contraction follows.  Depending on the stellar core mass, one of three compact objects is formed
in the final contraction -- a white dwarf, neutron star or black hole.  With a core mass of less than 
roughly 1.4 solar masses (\msolar), the stellar material becomes completely ionised during the core 
collapse, and the Fermi pressure of the resultant degenerate electron gas grows until, when the core is
several thousand km in radius, it balances the gravitational force and a stable, cooling white dwarf
remains.  However, for cores exceeding the Chandrasekhar limit of roughly 1.4 \msolar, the
rising degenerate electron pressure is not sufficient to halt further collapse into a denser state -- a 
process first recognised by \citet{chandrasekhar31a}.  This violent contraction and ejection of
stellar material is known as a core collapse supernova.

For core masses exceeding the Chandrasekhar limit, the collapsing material reaches densities
and temperatures sufficient to fuse electrons and protons into neutrons and electron 
neutrinos via the process of inverse beta decay:
\begin{equation}
e^{-} + p^{+} \rightarrow n + \nu_{e}
\end{equation}

The escape of these neutrinos from the collapsing core cools the collapsing material, and the 
simultaneous loss of thermal and Fermi pressure removes any means for the stellar material to 
resist gravitational collapse. It should be noted that the timescale of the neutrino emission remains 
somewhat uncertain, due to the difficulty of simulating the extreme environment of the supernova 
collapse \citep[see e.g.][and references therein]{fryer02a}.  The
collapsing matter largely conserves angular momentum and magnetic flux, and thus the initially
modest rotation speeds and magnetic fields of the progenitor star are amplified immensely
as the core compresses.

The production of a neutron star or black hole depends upon the core mass -- for masses below 
about 3 \msolar, the Fermi pressure of the degenerate neutron fluid grows until it balances the
gravitational pressure at a stellar radius of roughly 10 km.  The resultant neutron star has 
a core density of
$\ge10^{14}\ {\mathrm g\ \mathrm c \mathrm m}^{3}$, 
several times denser than
an atomic nucleus.  The inferred composition of neutron stars is discussed further in 
Section~\ref{pulsars:understand:compos}.  For collapsing core masses above about 3 \msolar, 
even neutron degeneracy pressure is insufficient to stop the collapse, and the core is
predicted to collapse completely to form a black hole in a {\it hypernova} explosion
\citep{iwamoto98a}.

Observations of pulsars have shown that the simple core collapse model described above
alone cannot completely explain typical pulsar characteristics.  One of the chief problems  is
the extremely high space velocity which many pulsars have been observed to possess.  Recent
estimates \citep{hobbs05a} put that average pulsar 3D birth velocity at 400\,\kms, with the
fastest known pulsar \citep[PSR B1508+55;][]{chatterjee05a} possessing an astonishing transverse 
velocity of 1100 \kms! These velocity values are much higher than those 
possessed by the massive stars which are neutron star progenitors 
\citep[$\sim\,20\ \kms$; see e.g.][]{feast65a}, which along with the small number of
pulsars in binary systems implies that some 
physical process imparts a large velocity on most neutron stars at birth
 \citep[e.g.][]{dewey87a, bailes89a} .  Whilst the 
disruption of binary systems may account for some pulsar velocities, it appears that
some kind of ``'kick" mechanism during the formation process is required to adequately
explain the full range of observed systems.  Counter--examples,
such as the \pfour\ system discussed in Section~\ref{results:binary:0737}, seem to imply that
kicks are not universal, further complicating interpretations of the physical mechanism.

\clearpage

While many theories have been advanced, generally 
requiring an asymmetry in the collapse and/or neutrino emission during the supernova formation,
or asymmetric electromagnetic radiation after the collapse \citep[see e.g.][]{fryer04a,lai01a},
the exact nature of the kick mechanism remains unclear.  It is important
to note that many theories of pulsar kicks predict that the kick is aligned with the
pulsar spin axis, which is tested observationally by measuring the polarisation
position angle of pulsars \citep[e.g.][]{johnston05a,rankin07a} or the position angle of an 
observed pulsar wind nebula and/or jet \citep[e.g.][]{gaensler02a, helfand01a} and 
comparing to the velocity position angle. Thus, studies 
of the space velocities of pulsars allow important insights into their formation processes.

\subsection[Composition]{Composition}
\label{pulsars:understand:compos}
The composition of the end state of matter in the interior of a neutron star is 
still the subject of controversy.  The
Equation of State (EoS) of the material at the core of a neutron star, which describes 
the relationship between density and pressure, is a much sought--after result which
could be obtained from a simultaneous measurement of a neutron star mass and radius
\citep[see eg.][]{lattimer07a}.  The extreme environment means that exotic forms of matter could
exist in the core, such as unconfined quarks \citep{prakash07a}.  The difficulty of such
measurements means that to date, a wide range of EoS's remain permitted.  The discussion
below assumes neutron stars do not contain exotic material.

\begin{figure}
\begin{center}
\includegraphics[width=0.85\textwidth]{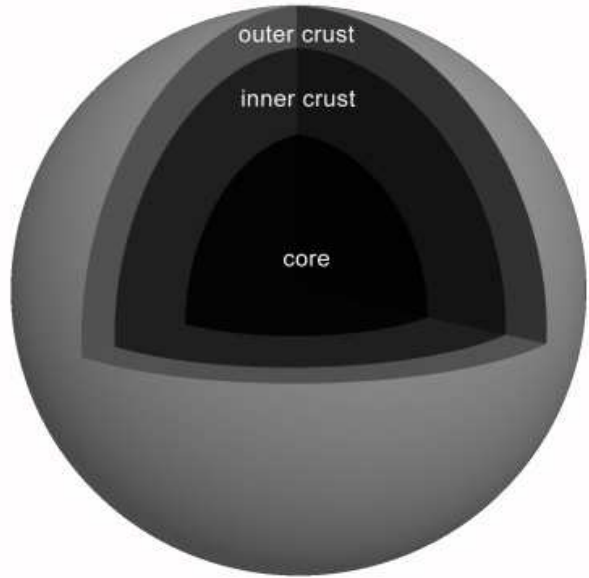}
\caption[The interior structure of a neutron star]
{The interior structure of a neutron star.  The (predominantly iron) nuclei in the crust form
a solid lattice, ``freezing" the magnetic field of the neutron star in place, 
while free neutrons in the inner crust and core are believed to exhibit superfluidity.}
\label{fig:pulsarcomp}
\end{center}
\end{figure}

Figure~\ref{fig:pulsarcomp} shows the generally accepted taxonomy of a ``normal" neutron star.
It can be broadly divided into the ``crust" and ``core" regions -- although the phase
transitions between the regions are poorly understood and intermediate layers could 
exist.  The crust consists of atomic nuclei and free electrons, since the pressure near the
neutron star surface is low enough to permit nuclei to remain intact.  The atomic nuclei
are locked in a solid lattice--like structure.  This rigid crust ``freezes" the magnetic
field configuration of the neutron star in place.  As the density increases further from the surface,
free neutrons (which are predicted to exhibit superfluidity) are present
along with nuclei and electrons \citep[e.g.][]{sandulescu04a} -- this is shown as the ``inner crust" in
Figure~\ref{fig:pulsarcomp}.  At a depth of several km, the pressure becomes too great for
atomic nuclei to exist and a transition to a neutron--only environment occurs.

\clearpage

The neutron core of a neutron star is expected to be superfluid, although the mathematical
treatment of the formation of Cooper pairs of neutrons at nuclear density is exceedingly
complex.  Some free protons are also expected to exist in the core, forming a superconducting
dynamo which is the source of the neutron star's magnetic field.  
Observational support for this model has come from pulsar glitches -- events where the 
rotational frequency undergoes a sudden increase (spin--up), before recovering over a longer
period to resume the steady spin--down caused by the loss of rotational energy.  The spin--up
can be explained by a transfer of angular momentum between the crust and core facilitated
by vortices in the superfluid core, changing the neutron star's moment of inertia \citep{larson02a}.

\clearpage
\subsection[Emission]{Emission}
\label{pulsars:understand:emission}
Despite decades of intense study, a complete picture of the pulsar emission mechanism has proved
elusive.  However, some aspects are well understood, and the generally agreed facts are presented
below.

\begin{figure}
\begin{center}
\includegraphics[width=0.85\textwidth]{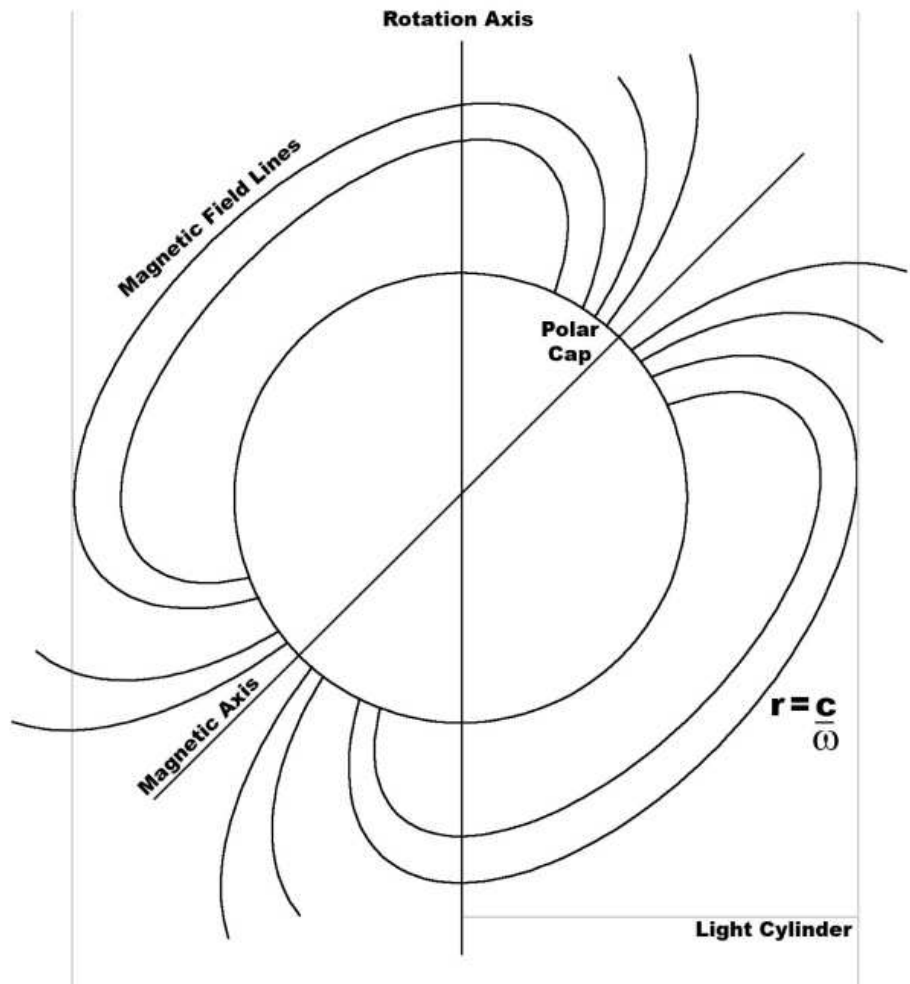}
\caption[Representation of the pulsar magnetosphere]
{Representation of the pulsar magnetosphere.  Magnetic field lines with an opening angle greater 
than a critical value are forced to remain open by the co--rotating charged plasma locked
to closed field lines within the light cyclinder \citep{michel74a}.  The resultant
rotating magnetic dipole emits electromagnetic radiation, and pair production in the intense
magnetic field provides a source of charged particles which are either accelerated away from the star --
the ``pulsar wind" -- or back onto the polar cap (the region defined by open field lines), 
heating it.  The heated polar cap is believed to be a source of thermal x--rays.}
\label{fig:pulsar_emission}
\end{center}
\end{figure}

A schematic diagram of an ordinary radio pulsar is shown in Figure~\ref{fig:pulsar_emission}.  
The ``light cylinder" reflects the distance from the pulsar at which a particle with the same angular
velocity as the pulsar would be required to travel at the speed of light -- thus, the region outside
the light cylinder is prevented from co--rotating with the pulsar.
As shown in Figure~\ref{fig:pulsarcomp}, the solid crust of the neutron star locks the 
magnetic field of the pulsar and forces it to co--rotate with the star.   

The pulsar magnetosphere (shown in Figure~\ref{fig:pulsar_emission}) is the relativistic
charged plasma which co--rotates with the pulsar, first postulated by \citet{goldreich69a}.  
This co--rotation of the magnetosphere means that magnetic field lines which originate 
close to the magnetic axis of the pulsar are forced to remain open, since their closure would 
require them to cross the light cylinder, meaning the plasma locked to these field lines would 
be traveling faster than the speed of light.  Thus, for some region around each
magnetic pole the field lines cannot close -- these regions are known as the ``polar caps".

The presence of charged plasma in the pulsar magnetosphere
can explain both the non--thermal and thermal emission of pulsars -- the non-thermal
emission by synchrotron emission from electrons and positrons 
spiralling away from the pulsar around the open magnetic field lines,
and the thermal emission from the polar cap regions, which would be heated by the impact of
infalling relativistic material.  The charged particles which form the pulsar
magnetosphere are continually replenished through a pair--production process
\citep{daugherty82a}, fed by high--energy $\gamma$--ray photons, themselves produced by 
curvature radiation from particles accelerated along the curved magnetic field lines 
in a very large electric potential produced in a vacuum gap somewhere in the magnetosphere.  
However, the process which produces the coherent radio emission is still not understood 
\citep[see e.g.][and references therein]{lyutikov99a}.

The precise site of the massive electric potential required to produce very high energy
$\gamma$--ray photons is not yet well understood.  Early models such as \citet{ruderman75a}
proposed the site of the acceleration was deep in the magnetosphere, near
the polar cap, while later models \citep[e.g.][]{cheng86a,chiang94a} 
proposed a location much higher in the
magnetosphere, in a region known as the outer gap.  
The difficulty of testing these models with the available observations means that one or both
could be correct, and the location of the gap could vary from pulsar to pulsar.

\clearpage

Another avenue for understanding the pulsar emission mechanism is higher--frequency observations.
Rotation--powered pulsars have been detected in the optical \citep[e.g.][]{zharikov04a},
ultraviolet \citep[e.g.][]{kargaltsev04a}, and x--ray \citep[e.g.][]{kargaltsev06a} wavebands, frequently
with contradictory results.  While some observations have suggested a non--thermal spectrum
for higher energy radiation, presumed to originate in the pulsar magnetosphere,
others have found emission consistent with a solely thermal source. 
Thermal emission will be 
generated over the entire neutron star surface, but as shown in Figure~\ref{fig:pulsar_emission}
the polar cap regions are expected to be heated relative to the remainder of the surface
and may dominate the high--energy thermal emission.  Models of pulsar surface temperatures
are presently poorly constrained, and hence accurate luminosities for rotation--powered pulsars
detected in the optical to x--ray wavebands are essential for ongoing attempts to understand the
pulsar emission mechanism.  The chief uncertainty in many pulsar luminosity estimates is
the large uncertainty in the pulsar distance.  Removing these uncertainties for specific objects
through direct measurement of pulsar distances is one of the applications of this thesis.

As shown in Figure~\ref{fig:pulsar_emission}, the misalignment of the rotational and magnetic axis
of the pulsar leads to the emission beam tracing a conical shape on the sky.   If the beam of radiation
intersects with Earth during its path, an observer on Earth can detect periodic pulsed emission.

\subsection[Isolated pulsar evolution]{Isolated pulsar evolution}
For ordinary radio pulsars, the emission of electromagnetic radiation due 
to the rotating magnetic dipole, governed by the classical electrodynamics, is the
primary mechanism for energy loss.  This ``magnetic braking" leads to a steady increase
in the pulsar's rotational period as rotational energy is lost.  Under the assumption that
other forms of energy loss are negligible, the following relation between $P$, \Pdot\ and surface
magnetic field strength $B$ can be obtained \citep[see e.g.][]{lorimer05a}:
\begin{equation}
B = \sqrt{\frac{3c^{3}}{8\pi^{2}}\frac{I}{R^{6}\sin^{2}{\alpha}} P \Pdot}
\label{eq:Bpdot}
\end{equation}

\noindent where $R$ is the neutron star radius (usually estimated as 10 km), 
$I$ is its moment of inertia (usually estimated as $10^{45}\,$g cm$^{2}$) and $\alpha$\ is the angle
between the magnetic moment of the neutron star and its spin axis.

\begin{figure}
\begin{center}
\includegraphics[angle=270,width=0.85\textwidth]{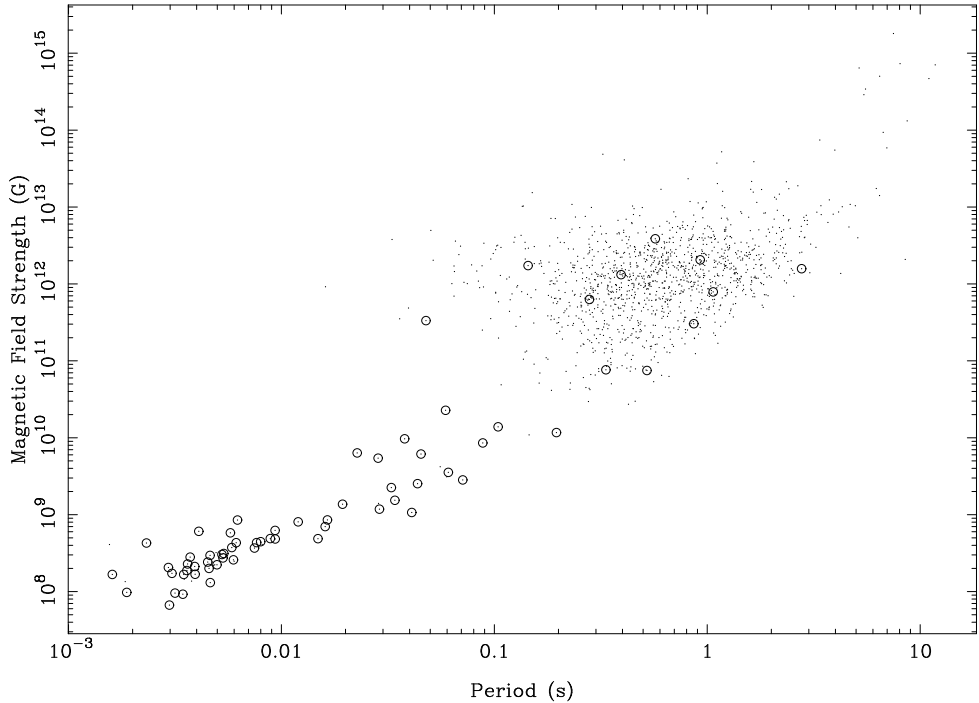}
\caption[The distribution of known pulsars in the $P$--$B$ diagram]
{Plot of known pulsars showing the distribution of spin period $P$ and surface magnetic field strength
$B$\ (as calculated from \Pdot\ using Equation~\ref{eq:Bpdot}).  
Binary systems are shown as semi-filled circles  and isolated
neutron stars with dots.  Magnetars lie in the upper right--hand corner of the plot, and recycled
pulsars in the lower left--hand corner.}
\label{fig:ppdot}
\end{center}
\end{figure}

A common way of visualising the known pulsar population is to plot $\log P$\ against 
$\log \Pdot$, or equivalently (under the assumption of pure magnetic dipole braking) 
$\log P$\ against $\log B$ using
equation \ref{eq:Bpdot}.  This is shown in Figure~\ref{fig:ppdot}.  This diagram
shows three distinct pulsar populations -- ``normal" radio pulsars in the centre, with 
$10^{11} < B < 10^{13}$\ G and $10^{-2} < P < 10^{1}$\ s, recycled pulsars with 
$B < 10^{10}$\ G and short rotational periods, and magnetars/AXPs with $B>10^{14}$\ G
and generally long rotational periods.  

While the evolution of the magnetic field strength of non-recycled, rotation--powered pulsars
over their lifetimes has been the subject of considerable debate 
\citep[see e.g.][]{goldreich92a,bhattacharya02a}, numerical simulations 
\citep[e.g.][]{bhattacharya92a} support the view that the field strength does not evolve with time, 
and so constant field strength is generally assumed.  Additionally, pulsars are often 
assumed to be born with initial spin periods of order 1--20 milliseconds, which is
reasonable based upon conservation of angular momentum of the progenitor, and is backed up
by simulations \citep{ott06a}.  However, various observational results suggest that pulsars can
also be born with considerably longer spin periods (e.g. PSR J1811--1925, 65\,ms, \citealt{torii99a};
PSR J0538+2817, 140\,ms, \citealt{kramer03a}).  Nevertheless, generally speaking, pulsars
are born close to the left of the $P$--\Pdot\ diagram and evolve towards the right along
lines of constant $B$.  The pulsar ``characteristic age" \tauc\ can be estimated by assuming the 
initial spin period is negligible compared to the current period using:
\begin{equation}
\tauc = \frac{P}{(\mathrm n - 1)\Pdot}
\label{eq:charage}
\end{equation}

\noindent where n is the braking index, which is equal to 3 for pure magnetic dipole braking
in a vacuum.

The pulsar's true age could vary considerably from its characteristic age if the initial spin period 
$P_{0}$\ was a significant fraction of the current spin period, or if the assumption that magnetic 
braking is the dominant form of energy loss is incorrect.  
An alternative way of estimating pulsar ages is to use the ``kinematic" age, which is 
calculated when the pulsar's position, proper motion and birth location are 
known.  This method can generally only be used for pulsars associated with a known supernova
remnant, or for whom the uncertainty in birth location (pulsars are generally assumed to be born
close to the Galactic plane, with a scale height of approximately 100--130 pc -- \citealt{cordes98a},
\citealt{faucher-giguere06a}) is 
unimportant, such as pulsars with a Galactic height many times the scale height, with a 
well--determined vertical proper motion.  Comparison of kinematic and characteristic ages
can lead to constraints on the birth location or initial spin periods of pulsars.

It is apparent from Figure~\ref{fig:ppdot} that a region of parameter space with low magnetic
field strength and long periods is completely depopulated of pulsars.  This pulsar ``death zone"
arises naturally from the pair--production cascade emission mechanism discussed in
Section~\ref{pulsars:understand:emission} -- once the curvature radiation no longer produces
photons with sufficient energy to initiate a pair--production cascade, the observable radio
emission ceases.  However, several observed pulsars (including most notably the $P=8.5$\,s 
\pseven) contradict this simple picture, being observable despite being past
the pulsar death line.  The existence
of observable long--period pulsars such as \pseven\ may be explained
by invoking inverse Compton scattering as the source of the pair--production cascade,
replacing curvature radiation \citep{zhang00a}.  Observations of pulsars in the death zone which
allow some insight into their fundamental properties such as luminosity may help to resolve
the questions as to how their radio emission is maintained.

\subsection[Binary pulsars]{Binary pulsars}
\label{pulsars:binary}
Pulsars exist in a wide range of binary systems, orbiting other neutron stars (including the
famous double pulsar system \pfour), white dwarfs, main--sequence and post
main--sequence stars.  Their presence in binary systems is a prerequisite for the
formation of recycled pulsars, where the accretion of matter from a companion star that
is overflowing its Roche lobe transfers angular momentum to the pulsar, spinning it
up to millisecond periods \citep[see e.g.][for a comprehensive discussion]{bhattacharya91a}.  
Although rarer than lone pulsars, binary pulsars offer a number of unique science opportunities,
discussed below.

The kinematics of binary systems including a pulsar offer an insight into the supernova
events which form neutron stars, since the binary system must survive the supernova
undisrupted\footnote{Globular cluster binaries, which have a much higher incidence of
interactions, do not necessarily offer the same insight}.  The orbital semi--major
axes and eccentricities of known binary pulsars allows some
constraints to be placed on the types of progenitor stars and supernovae that lead to these systems.

Pulsars in close binary orbits travel at relativistic speeds and offer the possibility to test
the predictions of General Relativity (GR) against alternate theories of gravity in
the strong--field gravity regime.  Examples of post-Keplerian 
effects\footnote{Those not predicted by classical mechanics} include decay in
orbital period \Pbdot\ due to gravitational wave emission, relativistic orbital precession and
Shapiro delay \citep{shapiro64a}.   The detection of these effects is dependent on
the precision timing of pulse arrival, which is discussed in Section~\ref{pulsars:obs:timing}.

Finally, mergers of compact objects in binary systems are expected to be one of the first 
sources of gravitational waves to be directly detected \citep[e.g.][]{belczynski02a}. 
The population size of relativistic binaries in our Galaxy is crucial when estimating the
frequency of merger events throughout the local Universe, and thus the probability of success
for the Laser Interferometer Gravitational Wave Observatory (LIGO).
Estimates of population size for relativistic binaries depend on the spatial density and luminosity
function of the systems, which require accurate distances. Thus, observations of existing binary 
systems can contribute to the expected frequency of gravitational wave events,
and hence event detection rates with LIGO.

\section[Observing pulsars in the radio waveband]{Observing pulsars in the radio waveband}
\label{pulsars:obs}
As shown in Section~\ref{pulsars:understand:emission}, pulsars generate beams of coherent,
broad--band radio
emission which is observed as a pulse train at Earth, due to the pulsar's rotation sweeping the
beam past Earth.  Pulsars are generally observed to have steep
spectra -- the mean spectral index for normal radio pulsars is $-1.8$ \citep{maron00a}.

In order for the pulsar signal to propagate to Earth, it must pass through the pulsar's local environment,
the interstellar medium (ISM) and Earth's own atmosphere.  Each of these environments is typically
composed, at least in part, by non--uniformly distributed ionised matter which interacts with the 
radio waves.  Essentially, the radiation traverses a path of continually varying refractive index,
which causes dispersion, scintillation (both refractive and diffractive), 
scattering and (for polarized radiation in the presence of magnetic fields) Faraday rotation.
These effects are summarised below in Figure~\ref{fig:propagation}.

\begin{figure}
\begin{center}
\includegraphics[width=0.95\textwidth]{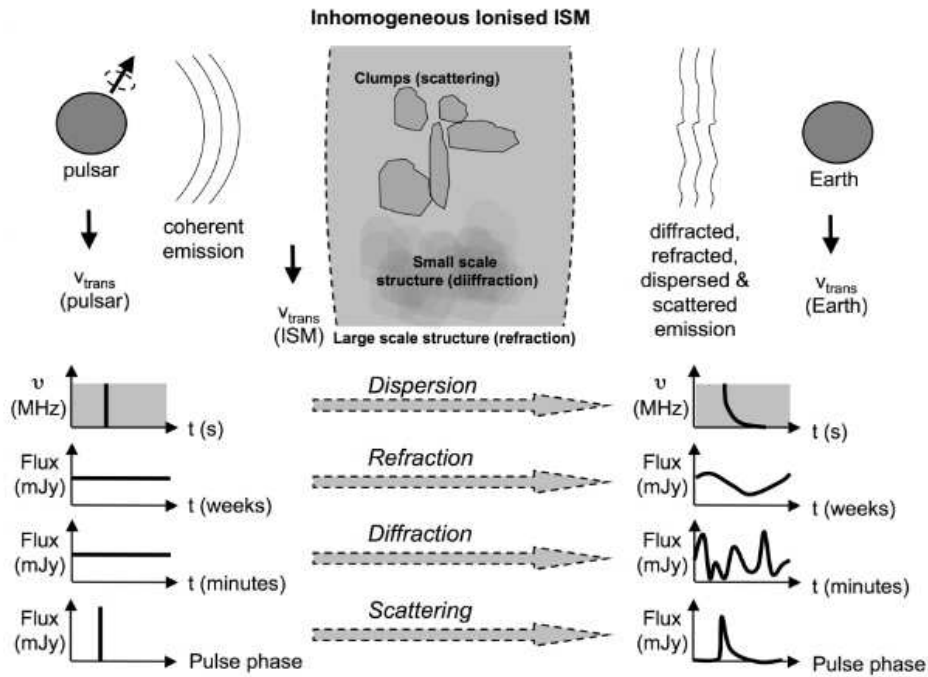}
\caption[Propagation effects on radio pulses in the ISM]
{The effect on pulsar radiation of travelling through a medium of non-zero density.  The 
broadband pulses, which are initially aligned in frequency, are dispersed by ionised
material along the line of sight.  Density variations cause refractive and diffractive scintillation 
and scattering.}
\label{fig:propagation}
\end{center}
\end{figure}

Much of the unique science made possible by pulsars depends upon their intrinsic rotational
stability, enabling their pulsed signals to be taken as accurate clock ticks.  For this approach to
be viable, the propagation effects  discussed above must be overcome, along with a host
of other error sources.  This discipline of {\em pulsar timing} is discussed below.

\subsection[Pulsar timing]{Pulsar timing}
\label{pulsars:obs:timing}
Pulsar timing determines a pulse time of arrival (TOA) by cross--correlating the observed 
pulse profile from an observation (obtained by averaging in time and frequency) and a template.
This is then compared with a timing model, and a ``timing residual" obtained.  A bootstrap
procedure follows, with the model undergoing refinement until an optimal model is obtained. 
Typical timing residuals can
be very small fractions of a pulse period \citep[e.g. 0.003\% for \ptwo;][]{verbiest08a}.
For most pulsars, the intrinsic average pulse profile is very stable over time \citep[see e.g.][]{hotan04a},
although this is not universal. Geodetic precision can cause secular changes in pulse profile
over long timescales, an effect which has been seen in PSR B1913+16 \citep{weisberg89a}
and PSR J1141--6545 \citep{hotan05a}.  On much shorter timescales, so--called 
``mode--changing" pulsars such as PSR B0329+54 \citep[e.g.][]{liu06a,lyne71a} 
switch between two or more modes, in which pulse components vary in relative and absolute strength.
Such complications are not relevent to this thesis and are not considered further.

The averaging of recorded data in time requires an accurate model for pulse arrival times -- 
the pulsar ephemeris.  The instantaneous position of the source (pulsar) and observer
(telescope on Earth) must be calculated to a high degree of precision.
The pulsar reference position, proper motion, binary motion (if  applicable), rotational period
and period derivative must be known, as well as the Earth's ephemeris and telescope
location.  This a priori model is summarised below in Figure~\ref{fig:geomodel}.

\begin{figure}
\begin{center}
\includegraphics[width=0.85\textwidth]{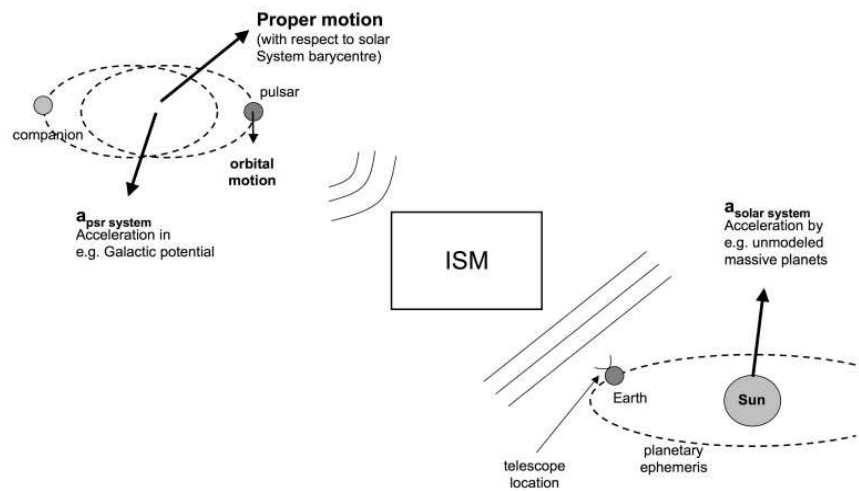}
\caption[Geometric delays in pulsar timing]
{Components of the geometric model used in recording pulse arrival times.  The orbital 
and transverse velocity and acceleration of both the pulsar and observer must be modeled,
requiring an ephemeris for both the pulsar and Earth.}
\label{fig:geomodel}
\end{center}
\end{figure}

\begin{figure}
\begin{center}
\includegraphics[angle=0,width=0.65\textwidth]{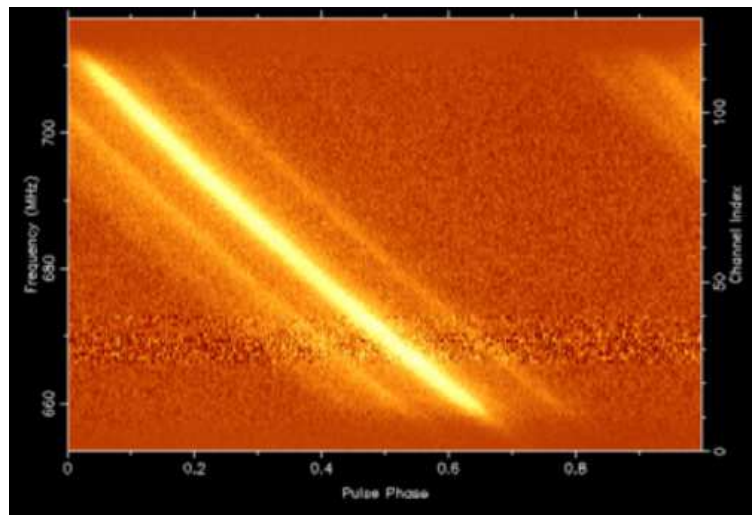}
\caption[The dispersed signal of \ptwo]{The intensity
of pulsar \ptwo, shown as a function of frequency and pulse phase (time modulo the pulse
period, expressed in units of fractional pulse period).  The increasing delay of the signal
with decreasing frequency is clearly apparent.}
\label{fig:dispersion}
\end{center}
\end{figure}

Averaging the received pulsar signal in frequency requires the removal  of the dispersive
effects of the ISM.  An example
of a dispersed pulsar signal is shown in Figure~\ref{fig:dispersion}.  The time delay $\taud$ 
in seconds experienced by a pulse at frequency $\nu$ GHz can be expressed as:
\begin{equation}
\taud = \frac{{DM}}{2.41 \times 10^{2} \nu^{2}}
\end{equation}
where $DM$ is the so--called ``Dispersion Measure" associated with a pulsar.  $DM$ is defined
as the integral of electron column density along the line of sight to the pulsar,
quoted here in pc cm$^{-3}$.  Observed values of $DM$ range from $<$ 5 pc cm$^{-3}$\ for 
very nearby pulsars, 
to $>$ 1000 pc cm$^{-3}$\ for distant pulsars in the Galactic plane.
While $DM$\ is often assumed to be constant, the relative motion of the pulsar, Earth, and 
ISM actually lead to continual small changes in $DM$\ due to the changing electron content
along the line of sight, and for precision timing the
time variation of $DM$\ must be measured and applied.

\clearpage

For pulsar timing, this frequency--dependent dispersion is simply an inconvenience to be characterised
and removed, as discussed below.  However, it also provides an accurate measure of the ionised
content of the ISM lying between the pulsar and observer, which can be translated into an estimate
of the pulsar distance, given an estimation of the density of the ionised ISM along the line of sight.
Widely used models of the Galactic electron distribution,
 which allow calculation of the ionised ISM content along arbitrary sightlines, 
have been constructed by \citet{taylor93a}, which is hereafter referred to
as the TC93 model, and \citet{cordes02a}, hereafter referred to as the NE2001 model.
Since pulsar luminosities vary over many orders of magnitude, this provides the most useful
estimate of distance which can be obtained for the entire pulsar population.  However, since
the density of the ionised ISM can also vary over many orders of magnitudes on small scales,
feedback into the electron distribution models in the form of model--independent distances 
is crucial to improve the quality of distance estimates for the bulk of known pulsars.  Methods of
obtaining such model--independent distances through VLBI are discussed in 
Section~\ref{pulsars:obs:vlbi} below, and demonstrated in Chapter~\ref{results}.

Removal of the frequency--dependent delay from the observed pulsar signal, to allow the
summation of data in frequency to improve the signal  to noise ration (SNR), can  be
accomplished in one  of two ways.  {\em Incoherent dedispersion} \citep{voute02a,large69a} 
makes use of a 
filterbank to divide the observed radio band into narrow frequency channels, and 
compensates the delay on a channel by channel basis, with delays appropriate
for the mean channel frequency.  Since the channels remain a finite width $B_{c}$\ MHz,
there is some small residual smearing which can be calculated for an observing frequency of 
$\nu$\ GHz by:

\begin{equation}
t_{\mathrm{smear}} = \frac{8.3 \times B_{c} \times {DM}}{\nu^{3}}\ \mu{\mathrm s}
\end{equation} 

Alternatively, {\em coherent dedispersion} may be employed. Essentially, this approach
applies a suitable filter to the baseband data (containing the inverse of the transfer function  of the ISM) before the channelisation process, minimising finite bandwidth effects (however, the 
fundamental limitation of the original sampling time remains).  This approach was first 
suggested by \citet{hankins75a}, and is becoming increasingly prevalent in
 high--precision timing campaigns \citep[eg][]{hotan06a,hessels06a}.

As shown in Figure~\ref{fig:propagation}, refractive scintillation is caused by large--scale structure
in the ISM, which acts as a large lens focussing or defocussing the pulsar radiation.  This naturally
leads to amplitude fluctuations in  the pulsar signal, and since it is caused by large--scale structure,
acts over long time periods and large observing bandwidths.  Diffractive scintillation, on the other 
hand, is caused by small--scale fluctuations in the ISM, with diffraction producing an
interference pattern on a plane which the Earth traverses.  Variations in amplitude are seen
as the Earth passes through the diffraction ``scintles" due to its transverse velocity, and the
diffraction pattern itself moves at the relative speed of the pulsar compared to the ISM
where the diffraction is occurring.  \citet{cordes86a} give a more detailed overview of the physics
of scintillation.

Since the pulsar velocity is usually much larger than that of the Earth or ISM, scintillation observations
of pulsars can be used to make an estimate of the magnitude of the pulsar transverse velocity.  
This requires an estimation of the
pulsar distance and a number of simplifying assumptions about the nature of the scattering
material.  A comprehensive discussion of the use of scintillation studies to make velocity estimates
can be found in \citet{cordes98b}.  For the commonly assumed case of a single, dominant thin
scattering screen, the scintillation speed $v_{\mathrm{ISS}}$ is given \citep[e.g.][]{gupta94a} by:

\begin{equation}
v_{\mathrm{ISS}} = A_{v}\frac{\sqrt{D\,\Delta\nu_{d} X}}{\nu\,\Delta{t_{d}}}
\end{equation}

\noindent where $A_{v}$\ is a constant related to the structure function of the ISM (equal to  
$3.85\times10^{4}$\,km s$^{-1}$\ for Kolmogorov turbulence in the thin--screen approximation; 
\citealt{gupta94a,cordes98b}), $D$ is 
the distance to the pulsar in kpc, $X$\ is the ratio of the Earth--screen distance
to the pulsar--screen distance, $\nu$ is the observing frequency in GHz and $\Delta\nu_{d}$
and $\Delta{t}$ are the decorrelation bandwidth in MHz and decorrelation time in seconds respectively.
The decorrelation bandwidth and time are determined observationally by averaging diffraction 
scintles and determining the mean bandwidth and time required for the pulsar intensity to fall to
$1/e$ of the peak value. The magnitude of the true pulsar velocity $|v_{\mathrm T}|$ is related to the
scintillation speed by $X\times|v_{\mathrm T}| = V_{\mathrm{ISS}}$ \citep{gupta94a, cordes98b}.
Comparison of predicted pulsar scintillation velocities to observed values 
from VLBI and pulsar timing are made in Chapter~\ref{results}.

Inhomogeneities in the ISM also lead to scattering, where reflected ``echoes" of 
the pulsed emission are seen after a time delay, as shown in Figure~\ref{fig:propagation}.
Scattering scales strongly with frequency, but the exact form of the scaling depends on the 
distribution of material in the intervening ISM -- for the commonly assumed Kolmogorov model
of turbulence in the ISM, the frequency dependence of scattering is 
$\nu^{-4.4}$\ \citep[e.g.][]{lee75a}.

All of these time--variable propagation effects lead to variations in the pulse arrival time estimates.
The dominant effect is that of $DM$\ variations, as shown by \citet{you07a}.  Scattering
variations are less noticeable, and variations due to refractive and diffractive scintillation have
traditionally been neglected, although simulations suggest that their effects are detectable
at low frequencies for well--timed pulsars \citep{foster90a}.

A final source of arrival time errors can be instrumental in nature.  The propagation of signals through
analog or digital filterbank and sampling system must invariably involve delays, which vary
from instrument to instrument, and telescope to telescope.  Changes to the signal path before
the pulsar hardware can also affect instrumental delays.  For long time series of pulsar
observations, which generally span multiple instruments, calibration of the unknown
relative instrumental delays introduces additional free parameters to the timing model.

A pulsar timing campaign requires regular observations to obtain a series of residual delays.
Whilst the arrival time errors due to finite signal to noise should be zero--mean, Gaussian
distributed random noise, incorrectly modeled or neglected effects manifest as
clear trends in the residual errors.  For example, an error in $P$\ will result in residuals which
linearly diverge from 0, while an unmodeled binary companion will lead to periodic
movement in residuals at the binary period  of the system.  Historically, analysis of
pulsar timing data and the fitting of pulsar parameters has used the software package
TEMPO\footnote{http://www.atnf.csiro.au/research/pulsar/tempo/}, 
although more recently a more advanced package known as 
TEMPO2\footnote{http://www.atnf.csiro.au/research/pulsar/psrtime/tempo2/} \citep{hobbs06a} 
is now available, and incorporates support for higher precision experiments
than TEMPO, as well as simultaneous timing of multiple pulsars.

A large proportion of the exciting science possible using pulsar timing involves binary
pulsars.  For example, exploration of GR effects is generally only possible in binary 
systems\footnote{A counter--example is microlensing of pulsars, although this is yet to be observed}.
The relevant equations in which a VLBI measurement of kinematics can contribute
to the precision of a timing result are presented below -- for an excellent review of all the 
equations relevant to pulsar timing, see the pulsar handbook of \citet{lorimer05a}.

For this thesis, the important timing equations are those dealing with orbital
motion in binary pulsars.  Equation~\ref{eq:pbdotobs} shows the factors which contribute 
to an observed change in binary period $\Pbdot^{\mathrm{obs}}$:
\begin{eqnarray}
\Pbdot^{\mathrm{obs}} & =  & \Pbdot^{\mathrm{int}} - \Pbdot^{\mathrm{kin}} \\
  & = & \left ( \Pbdot^{\mathrm{GR}} + \Pbdot^{\mathrm{drag}} + \Pbdot^{\mathrm{tid}} +  \Pbdot^{\mathrm{ml}} \right ) - \left ( \Pbdot^{\mathrm{acc}} + \Pbdot^{\mathrm{Shk}} \right )
\label{eq:pbdotobs}
\end{eqnarray}

\noindent where the intrinsic contributions to $\Pbdot^{\mathrm{obs}}$
due to energy loss from the system ($\Pbdot^{\mathrm{int}}$) consist of relativistic effects
such as the emission of gravitation radiation ($\Pbdot^{\mathrm{GR}}$), atmospheric drag
($\Pbdot^{\mathrm{drag}}$), mass loss ($\Pbdot^{\mathrm{ml}}$), 
and tidal dissipation ($\Pbdot^{\mathrm{tid}}$), and the
kinematic contribution $\Pbdot^{\mathrm{kin}}$\ consists of the relative acceleration of the pulsar
to the timing reference point (the solar system barycentre) $\Pbdot^{\mathrm{acc}}$\ and 
the apparent acceleration caused by the pulsar's proper motion $\Pbdot^{\mathrm{Shk}}$, which 
is known as the Shklovskii effect \citep{shklovskii70a}.   
The kinematic contributions to $\Pbdot^{\mathrm{obs}}$\ can be expressed as:
\begin{equation}
\frac{\Pbdot^{\mathrm{acc}}}{\Pb} = \frac{1}{c}\BP \cdot \left ( \accp - \accb \right )
\label{eq:pbdotacc}
\end{equation}
and
\begin{equation}
\frac{\Pbdot^{\mathrm{Shk}}}{\Pb} = \frac{\Vt^{2}}{c\, d}
\label{eq:pbdotshk}
\end{equation}

\noindent where \BP\ is a unit vector from the solar system barycentre to the pulsar, \accp\ is the 
acceleration of the pulsar system, \accb\ is the acceleration of the solar system barycentre,
\Vt\ is the transverse velocity of the pulsar, $d$ is the distance from the pulsar to Earth and
$c$ is the speed of light.  The acceleration terms \accb\  and \accp\ incorporate Galactic
rotation, the vertical potential of the Galaxy (and the parent cluster for globular cluster pulsars), 
and any unmodeled nearby perturbing massive bodies. These apparent and actual accelerations 
due to kinematic effects also affect the pulsar's spin period $P$\ in a similar manner, for both
isolated and binary pulsars.

Equations~\ref{eq:pbdotobs} -- \ref{eq:pbdotshk} show that in the presence of accurate
timing data for a binary pulsar and sufficiently accurate modeling of one set of parameters,
one of the contributing factors to \Pbdot\ can be calculated.  This was famously demonstrated
by \citet{taylor89a} for $\Pbdot^{\mathrm{GR}}$\ in the PSR B1913+16 system.  
Combining the measured value for \Pbdot\ with the observed and modeled components can
yield a limit on the anomalous acceleration of the pulsar with respect to the Solar System.
This anomalous acceleration can be interpreted as a change in the value of Newton's gravitational
constant $G$\ \citep{verbiest08a}, the presence of a distant, massive planetary companion in the Solar
System \citep{zakamska05a}, or an error in the estimate of the Galactic
gravitational potential \citep{bell96a}.

Often, the most uncertain contribution to $\Pbdot^{\mathrm{obs}}$ is the Shklovskii term
$\Pbdot^{\mathrm{Shk}}$, due to the large uncertainty which is generally associated with
most pulsar distance measurements, which contributes a large uncertainty to the transverse
velocity.  Even in the case of excellent timing data, such as the \citet{verbiest08a} measurement
of \ptwo, direct measurement of distance through detection of the annual parallax yields
an uncertainty of 8\%, resulting in an unacceptably large error in $\Pbdot^{\mathrm{Shk}}$, if
the aim is to constrain another contribution\footnote{
In this instance, the authors modeled all other contributions and used $\Pbdot^{\mathrm{obs}}$
to measure the distance to the pulsar} 
to $\Pbdot^{\mathrm{obs}}$.  If the transverse velocity of a pulsar can be supplied
through an independent measurement of parallax and proper motion, such as VLBI
astrometry, the uncertainty in $\Pbdot^{\mathrm{Shk}}$ can be reduced sufficiently to allow
significant constraints on other terms in $\Pbdot^{\mathrm{obs}}$.  An example of this
is shown, with limits on \Gdot\ and massive planetary companions, for \ptwo\ in 
Section~\ref{results:binary:0437}.  

\subsection[VLBI pulsar observations]{VLBI pulsar observations}
\label{pulsars:obs:vlbi}
For the purposes of VLBI, pulsars can be considered as unremarkable radio sources, except for 
the possibility of improving the sensitivity of observations through pulsar ``gating": blanking
the telescope data at times when the pulsar flux is low or zero. This has the effect of eliminating
noise which would otherwise be accumulated during these times, and hence improves
sensitivity by a factor which can be estimated by $\frac{1}{\sqrt{\mathrm{pulsar\ duty\ cycle}}}$.
To date, VLBI pulsar gating has always used incoherent dedispersion, as the small amount of 
smearing incurred compared to coherent dedispersion has a negligible impact on the recovered 
signal to noise ratio. Pulsar gating is discussed in more detail in Section~\ref{difx:difx:pulsar}.

Despite the impressive resolution obtainable with VLBI ($< 1$\,mas), the small physical 
size of the pulsar emission region means that even the nearest pulsars are completely
unresolved on VLBI baselines (assuming a pulsar emission height of 1000 km -- most likely a gross
overestimate -- yields an angular size of 0.07 $\mu$as at 100 pc).  Potential exceptions to this
rule are the interactions of pulsars which their surrounding environment, such as
pulsar wind nebulae (PWN), or interactions with companions, such
as PSR B1259--63 \citep{johnston99a}, although no such detections have yet been published.
It is also worth noting that VLBI observations of pulsars can be used to obtain very high resolution 
speckle images of scattering disks in the ISM.  The first example of such an observation
(which used the DiFX software correlator) is briefly discussed in Section~\ref{difx:additional:scint}.

Thus, the main application of VLBI observations of pulsars is to obtain astrometric information, 
either for the purpose of making associations with other observed structures such as 
supernova remnants \citep[e.g.][]{brisken05a}, or to improve the accuracy of kinematic 
and distance information for use in timing, luminosity, and Galactic electron distribution 
models. It is these latter applications 
which are the focus of this thesis, and they are explored in detail in Chapter~\ref{results}.
\chapter{RADIO INTERFEROMETRY}
\label{radio}

\section[Conceptual overview]{Conceptual overview}
\label{radio:concept}
\begin{figure}
\begin{center}
\includegraphics[width=0.85\textwidth]{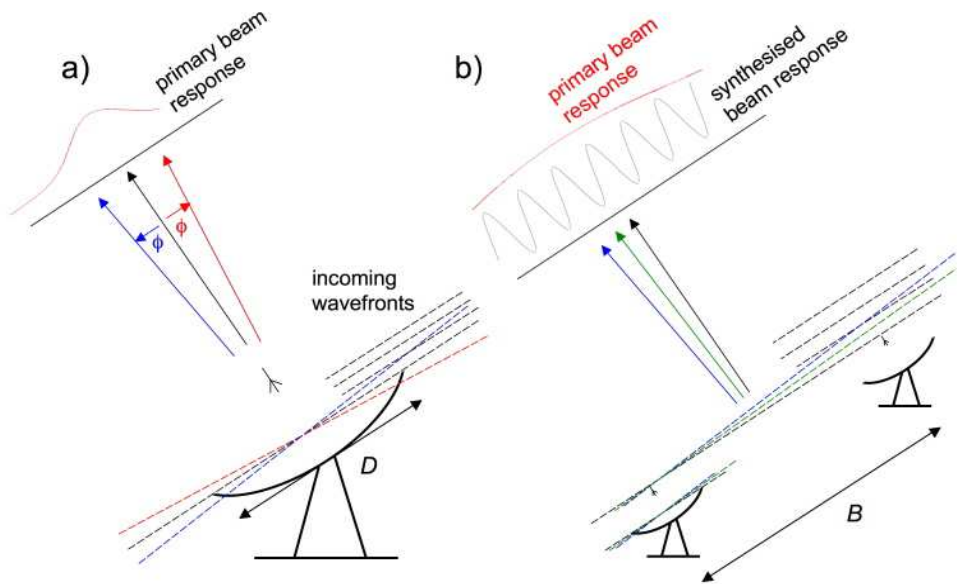}
\caption[Resolution of a single dish compared to a two--element interferometer]
{Resolution of a single dish compared to a two--element interferometer. (a) Response of a
single antenna element.  As diameter $D$ increases, the radiation collected at the edges
of the dish becomes further out of phase for a given angular offset $\phi$.  
As wavelength $\lambda$ decreases, a fixed time/distance offset at the edge of the dish
corresponds to a greater amount of phase.  
The full width half maximum (FWHM) of the antenna response --
the ``primary beam" -- is given by $1.22\lambda / D$.  (b) Response of a two--element
interferometer.  The FWHM of the ``synthesised beam" is given by the similar expression 
$1.22 \lambda / B$, where $B$ is the projected distance between the antennas, in a plane
perpendicular to the observation direction.}
\label{fig:beam-twoel}
\end{center}
\end{figure}

Radio interferometry makes use of the spatial separation of two or more antennas to 
obtain information about smaller angular structures in the radio sky than can be gleaned
from single--dish studies.  Whilst one can probe smaller angular scales with a single dish by
increasing the dish diameter, this has two undesirable side effects.  Firstly, the smaller beam 
means that the survey speed of the instrument is not improved, despite the improved sensitivity.
Secondly, the cost and technical difficulty imposed by the larger physical diameter of the antenna 
rapidly become prohibitive.  The inverse relationship between dish size and ``primary beam" -- the
full--width at which the antenna response has dropped to one half its peak -- is illustrated
in Figure~\ref{fig:beam-twoel}(a). Figure~\ref{fig:beam-twoel}(b) shows how a pair of antennas --
the classic two--element interferometer -- can discriminate between structure which lie within
the primary beams of the individual elements.  With a full mathematical treatment of 
interferometry deferred until Section~\ref{radio:math}, it is sufficient at this point to note
the potential degeneracies of a single measurement with the two--element interferometer, and 
to observe that different projected antenna spacings -- {\em baselines} -- would be required
to determine the structure of the source being observed.

\section[Historical development]{Historical development}
\label{radio:history}
Radio astronomy was founded in 1933, when Karl Jansky published the detection
of the Galactic background at low frequency \citep{jansky33a}.  Considerable progress in
the new wavelength regime, however, was delayed until the end of the second world war,
at which time large quantities of military radio equipment began to be used for radio astronomy.
The first interferometric observations were made around this time in Australia, using a single
receiving element mounted on a sea cliff \citep{pawsey46a}.  This arrangement made use of
the path delay provided by the reflection off the sea surface; however, arrays of separate
receiving elements soon appeared \citep{ryle46a}.

These early interferometers measured only the changing intensity of the summed interferometer
signal -- a direct analogue to the optical two--slit experiment.  A major advance came with 
the advent of phase--switching interferometers \citep{ryle52a}, which introduced a periodic 
phase inversion to one of the interferometer elements.  This allowed the measurement
of the multiplicative term between the elements without the addition of the individual
squared signals, considerably improving interferometer sensitivity.  This, in turn, was
made redundant through the improved stability of frequency standards which allowed
direct multiplication of the signals from interferometer elements.

The ongoing rapid improvements in the capabilities of digital electronic equipment has 
allowed correspondingly rapid improvements in interferometer capabilties.
The cost of new instruments is now generally dominated by the structural components
of the antennas and associated infrastructure, making the upgrade of existing instruments
with new electronic components an attractive proposition.  Such upgrades are underway or
recently completed for the Very Large Array \citep[the Expanded VLA:][]{perley04a}
and several VLBI arrays, which are discussed below.

\section[VLBI]{VLBI}
\label{radio:vlbi}
As radio interferometers developed, a natural tendency was to increase baseline length to 
achieve better angular resolution.  This trend quickly reached the limits at which 
information could be could be distributed to and received from antennas in real time
with existing technology.  To overcome this limitation, disconnected stations were equipped
with recording media to store baseband data until it could be brought to a common location,
and made use of independent frequency standards.  Early examples of science undertaken
with such arrays were published by \citet{clark67a} and \citet{moran67a}.
The technological limitations on VLBI were progressively lifted, and a number of ad--hoc
and part--time VLBI arrays functioned around the world from the 1970s onwards.
These included the Network Users Group in North America\footnote{discussed along with other early
North American efforts by \citet{kellerman88a}}, 
the European VLBI Network (EVN)\footnote{http://www.evlbi.org/}, 
the Asia--Pacific Telescope (APT)\footnote{http://www.vsop.isas.jaxa.jp/apt/},
and the Australian Long Baseline Array (LBA)\footnote{http://www.atnf.csiro.au/vlbi/}.
A purpose--built, full--time VLBI array was
commissioned in the US in the 1990s -- the Very Long Baseline Array 
(VLBA)\footnote{http://www.vlba.nrao.edu/}.
Throughout the development of VLBI, the use of independent frequency
standards and non--real time correlation has remained the defining distinction
between VLBI and regular ``connected element"  interferometry, although recently 
pseudo--realtime correlation has become possible on some VLBI arrays through the use of
high bandwidth fibre optic links (``eVLBI"), which is discussed further in 
Section~\ref{difx:additional:evlbi}.
For a comprehensive overview of the development of VLBI and
radio interferometry in general, see \citet{kellermann01a}.

VLBI allows the highest angular resolution available to imaging in astronomy at 
any waveband, which allows the detailed study of the most distant objects
in the Universe, as well as objects with small physical extent in the more nearby Universe.
Historically, targets of interest have included active galactic nuclei (AGN), radio galaxies,
masers and pulsars.  Studying the emission of these objects on the smallest scales has
given unique insight into the physics that power them, as well as throwing up surprising
discoveries such as apparent superluminal motion \citep{whitney71a}.  
One of the most important applications of VLBI has little to do with
the sources themselves, however, and is the creation and maintenance of a stable,
quasi--inertial reference frame upon which astronomical positions can be based.  Historically
the domain of optical astronomy 
\citep[and defined most recently by the FK5 reference frame;][]{fricke88a}, 
this responsibility was assumed by VLBI when the International Celestial Reference Frame
(ICRF) superseded the FK5 reference frame in 1998.  Defined by 212
distant radio sources \citep{ma98a}, it has been extended to include several hundred 
additional ``candidate" and ``new" sources \citep{fey04a}.  VLBI measurements
also form an important contribution to the International Terrestrial Reference Frame
(ITRF), which along along with the ICRF is discussed further in Section~\ref{radio:astrogeo}.

\section[Instrumentation and hardware]{Instrumentation and hardware}
\label{radio:hardware}
\begin{figure}[t!]
\begin{center}
\includegraphics[width=0.85\textwidth]{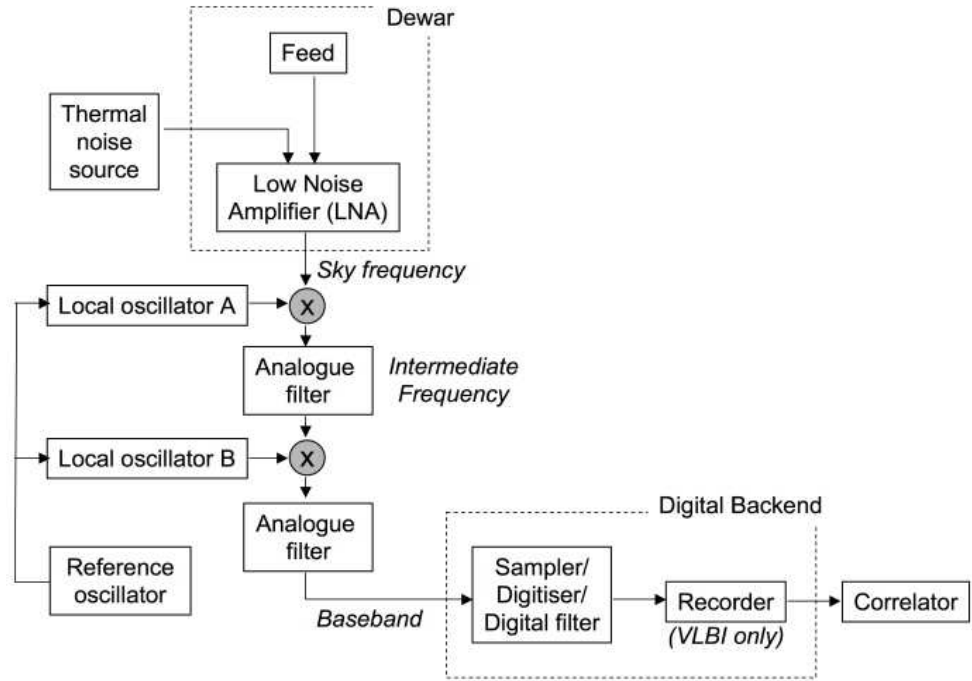}
\caption[Hardware components of an interferometer]
{Hardware components of an interferometer.}
\label{fig:hardware}
\end{center}
\end{figure}

The purpose of an element in a modern radio interferometer is to obtain a digitised time--domain
representation of the radio energy within a desired frequency band incident upon the element, 
adding as little noise as possible in the process.  Once this has been achieved,
digital signal processing can be used to compute the correlation between array elements,
which along with a knowledge of the array geometry can be used to generate a model
of the observed sky in the first the spatial frequency domain, then the image domain. 
This section describes the hardware required -- a mathematical description of the foundations of
interferometry is presented is Section~\ref{radio:math}.  The description below assumes 
a typical reflecting concentrator (dish) type radio telescope, referred to throughout as an antenna.
Figure~\ref{fig:hardware} shows the components discussed below.

The electric field present at the focal point of an antenna is converted to an electrical voltage
by a sensor referred to as a feed.  Typical feeds consist of two components sensitive to two
(usually linear, although circular is possible) orthogonal polarisations.  
The feed itself typically resides within a {\em feed
horn}, whose purpose is to shape the illumination of the feed to the antenna surface and
hence optimise the antenna primary beam response.  The time--varying voltage signal
$v(t)$ at the feed is linearly proportional to the electric field $E(t)$ present at the feed.

The voltage signal induced in the feed is extremely small, as the radio power influencing
the voltage is itself extremely small. Thus, the first stage of signal manipulation is to amplify
the signal using a Low Noise Amplifier (LNA).  This first stage of amplification dominates the
noise budget of the receiver system, and so considerable effort is made to minimise the noise
contribution of the LNA.  The dominant noise contribution comes from thermal electron
motion, and so feeds and LNAs are often housed in a cooled chamber -- a dewar -- as
shown in Figure~\ref{fig:hardware}.

\clearpage

The calibration of received power is generally undertaken through the injection
of a known noise source prior at the beginning of the receiver system, prior to the LNA.  
By modulating the noise source on and off, the rise in measured power can be used
to calculate the total power.  This procedure is particularly important for VLBI, where there
are no unresolved, constant flux sources that can be used for absolute flux calibration, as
discussed in Section~\ref{radio:calib}.  The noise source is usually a thermally controlled
resistive load, which is coupled to the input of the LNA.

Once the received signal has been amplified, it must generally be downconverted from the
sky frequency to a lower frequency where digital electronics can be efficiently used.  This
is achieved by a series of mixing operations, often with filtering applied at each stage to 
prevent aliasing of the signal.  The reference oscillator for the mixing stages is generally
based on an atomic oscillator, such as rubidium or hydrogen (commonly used for VLBI
due to the excellent frequency stability).  This ``local oscillator" is distributed to all 
antennas in a connected element array, but this is not feasible for VLBI arrays.

Once the data is at or close to ``baseband" (frequency range starting at 0 Hz), it can
be sampled and digitised, at which point digital signal processing can be applied.  
The sampler/digitiser (and recorder, for VLBI) is collectively referred to as a 
``digital backend", and modern examples incorporate features such as digital
filtering \citep[e.g.][]{iguchi05a} to improve the data quality.  The digitised data is then transported
to the correlator, which produces the sampled visibilities (scaled fractional correlation
between antennas) as a function of frequency and time.  The correlator is the focal point of
the hardware chain in terms of this thesis, and correlators in general and the DiFX software
correlator in particular are discussed in more detail in Chapter~\ref{difx}.

\section[Mathematical description]{Mathematical description}
\label{radio:math}
\begin{figure}
\begin{center}
\includegraphics[angle=270,width=0.65\textwidth,clip]{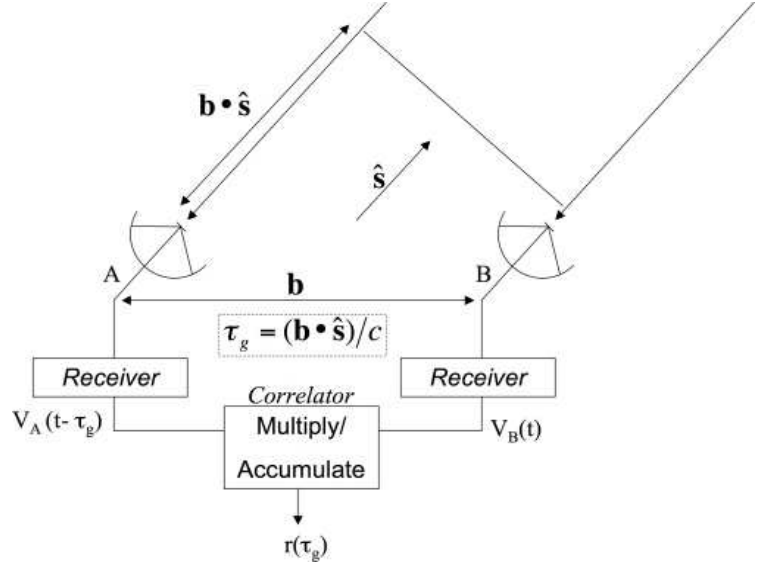}
\caption{Response of a two--element interferometer}
\label{fig:twoelement_response}
\end{center}
\end{figure}

All of the salient concepts of interferometric Fourier synthesis imaging can be succinctly illustrated
using a two--element interferometer, so this simplified example will  be presented here.  
It should be noted that this section draws heavily on the explanation of interferometer 
theory presented in \citet{thompson99a}.  Throughout, the standard interferometric assumptions
of a far--field, spatially incoherent source are made. A diagram
of a two--element interferometer is shown in Figure~\ref{fig:twoelement_response}.  The elements are
separated by a baseline \vect{b}, and are both pointing in the direction of the unit vector
\unitvect{s}.  The geometric time delay between the signal arriving at antenna A and antenna B
is given by $\tau_{g}(t) = \vect{b} \cdot \unitvect{s} / c$, where $c$ is the speed of light.  It is shown as
a function of time since, depending on the choice of reference axis, either \vect{b}\ or 
\unitvect{s}\ will change with time as the Earth rotates.

Considering the astronomical signal initially to be a monochromatic wave of frequency
$\nu$ considerably simplifies the analysis, reducing the correlator function to a simple 
multiply and accumulate\footnote{the non monochromatic case will be considered in the
following chapter}, while still demonstrating the Fourier relation between interferometer
observables and the actual sky brightness distribution.
In this case, the response of a single element will be of the form
$V = v \cos{2\pi \nu t}$, and so the correlator output can (in the absence of any delay tracking)
be written as:

\begin{eqnarray}
r(t) &=& <V_{A}(t)V_{B}(t)> \\ \nonumber
& = & v_{A}v_{B}\cos{2\pi \nu (t - \tau_{g})}\cos{2\pi \nu} \\ \nonumber
& = & v_{A}v_{B}\cos{4\pi \nu t - 2\pi \nu \tau_{g}}+ v_{A}v_{B}\cos{2\pi \nu \tau_{g}} \\ \nonumber
& \simeq & v_{A}v_{B}\cos{2\pi \nu \tau_{g}}
\label{eq:twoeloutput}
\end{eqnarray}

Thus, assuming the averaging time is long compared to the term at twice sky frequency, but short compared to the changing $\tau_{g}$, the correlator output will be a sinusoid--like 
function (since $\tau_{g}$ is not varying
linearly, it is not a true sinusoid) with amplitude $V_{A}V_{B}$, which
is proportional to the received power.

If radio brightness is represented by $I$, then the radio brightness in the direction of \unitvect{s}
can be written as $I(\unitvect{s})$.  Brightness, which is the desired quantity when mapping
the radio sky, is measured in power per unit area, per unit bandwidth, per unit solid angle, and
can be placed in units of W m$^{-2}$\ Hz$^{-1}$\ sr$^{-1}$.  From this it follows that
from a source element $d\Omega$ in the direction \unitvect{s}, an antenna with effective area 
$A(\unitvect{s})$\ accepting bandwidth $\Delta \nu$\ will receive power equal to 
$A(\unitvect{s}) I(\unitvect{s}) \Delta \nu\, d\Omega$.  Since the correlator output was proportional
to received power, by neglecting constant gain factors the correlator output from
solid angle $d\Omega$\ can be written as:
\begin{equation}
dr = A(\unitvect{s}) I(\unitvect{s}) \Delta \nu\, d\Omega \cos{2\pi \nu \tau_{g}}
\end{equation}

The correlator output can then be written as integration over the whole sky 
($S\equiv4\pi$\ steradians)\footnote{In practice, however, the limitations of the primary 
beam of the antenna elements means that
$A(\unitvect{s})$ is only non--zero for a small solid angle}:
\begin{equation}
r = \Delta \nu \int_{S} A(\unitvect{s}) I(\unitvect{s}) 
\cos{\frac{2\pi\nu \vect{b} \cdot \unitvect{s}}{c}} d\Omega
\label{eq:corrout}
\end{equation}

In general, interferometric images are made in a relatively small solid angle, as constrained by the
antenna primary beams.  Thus, it is convenient to rewrite \unitvect{s} as 
$\unitvect{s} = \vect{s_{0}} + \vect{\sigma}$, where the image centre \vect{s_{0}} is fixed and 
is referred to as the {\em phase centre}.  Substituting this into equation~\ref{eq:corrout} yields:
\begin{eqnarray}
r &=& \Delta \nu \cos{\frac{2\pi\nu \vect{b} \cdot \vect{s_{0}}}{c}} \int_{S} A(\vect{\sigma}) 
I(\vect{\sigma}) \cos{\frac{2\pi\nu \vect{b} \cdot \vect{\sigma}}{c}} d\Omega \\ \nonumber
 & - & \Delta \nu \sin{\frac{2\pi\nu \vect{b} \cdot \vect{s_{0}}}{c}} \int_{S} A(\vect{\sigma}) 
I(\vect{\sigma}) \sin{\frac{2\pi\nu \vect{b} \cdot \vect{\sigma}}{c}} d\Omega
\label{eq:corroutsigma}
\end{eqnarray}

At this point, it is necessary to introduce the term {\em visibility}.  The visibility is a measure of the 
coherence of the radio sky brightness distribution, modified by the antenna response, between an
antenna pair.  It is a complex quantity defined as:
\begin{equation}
V \equiv |V|e^{i\phi_{V}} = \frac{1}{A_{0}} \int_{S} A(\vect{\sigma}) I(\vect{\sigma}) 
e^{2\pi i \nu \vect{b} \cdot \vect{\sigma} / c} d\Omega
\label{eq:visdefinition}
\end{equation}

where $A_{0}$ is the antenna response at the phase centre.  Subsequently, it will be shown
that the visibility is, under certain assumptions, actually the Fourier transform of the
radio sky brightness $I(\vect{\sigma})$.  First, however, the relationship of correlator
output to visibility will be shown.  By separating the real and imaginary components of 
Equation~\ref{eq:visdefinition}, the following expressions are obtained:
\begin{equation}
A_{0}|V|\cos{\phi_{V}} =  \int_{S} A(\vect{\sigma}) I(\vect{\sigma}) 
\cos{\frac{2\pi i \nu \vect{b} \cdot \vect{\sigma}}{c}} d\Omega
\end{equation}
\noindent and
\begin{equation}
A_{0}|V|\sin{\phi_{V}} =  - \int_{S} A(\vect{\sigma}) I(\vect{\sigma}) 
\sin{\frac{2\pi i \nu \vect{b} \cdot \vect{\sigma}}{c}} d\Omega
\end{equation}
\noindent which can then be substituted into Equation~\ref{eq:corroutsigma} to obtain:
\begin{equation}
r = A_{0}\Delta\nu|V|\cos{\left(\frac{2\pi\nu \vect{b} \cdot \vect{s_{0}}}{c} - \phi_{V}\right)}
\label{eq:corroutvis}
\end{equation}

Thus, a correlator with no delay tracking measures the visibility, modulated by 
a fringe pattern with frequency dependent on frequency and baseline, and an amplitude
dependent on the antenna response. The removal of this
fringe pattern by a realistic correlator to enable direct measurement of visibility,
as well as dealing with the effects of finite bandwidth and frequency down--conversion, 
are discussed in Chapter~\ref{difx}.

\begin{figure}
\begin{center}
\includegraphics[width=0.85\textwidth]{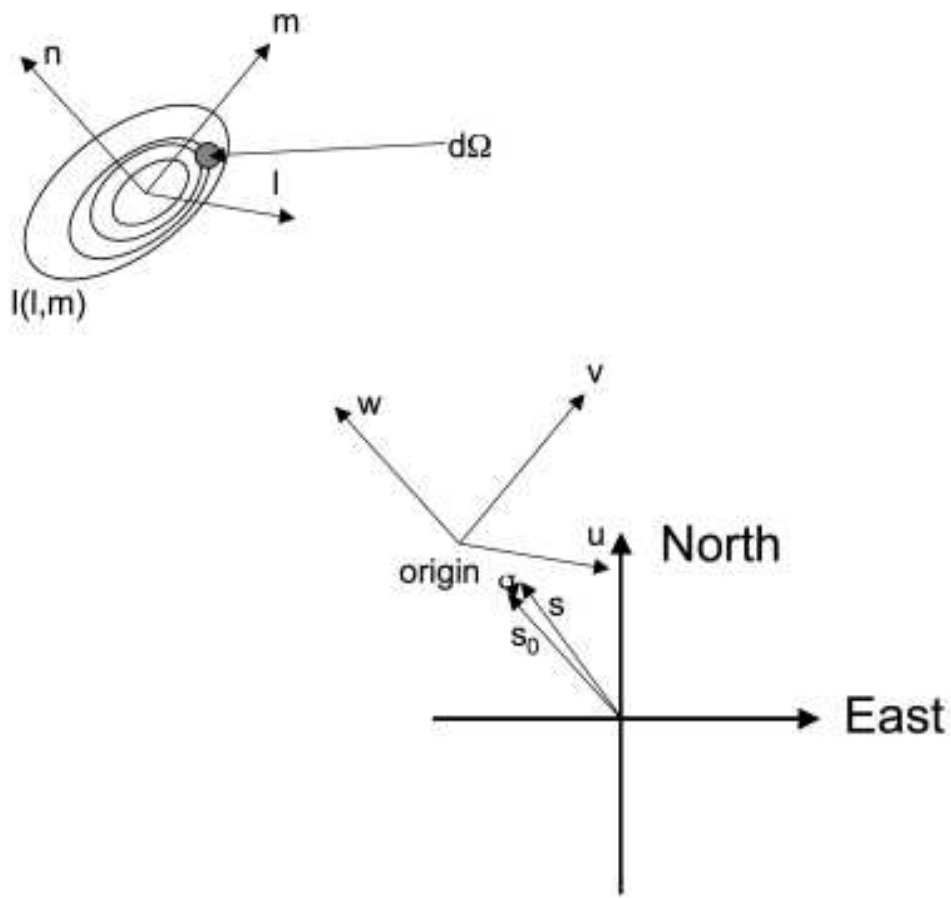}
\caption[Definition of interferometer coordinate systems]
{Definition of interferometer coordinate systems used throughout this thesis.}
\label{fig:coords}
\end{center}
\end{figure}

Once the complex visibility has been sampled by an interferometer, the desired result is 
generally an image of the radio sky brightness distribution.  In order to show the validity
of synthesis imaging, it is necessary to define the coordinate systems for both visibility and 
sky brightness.  These are illustrated in Figure~\ref{fig:coords}.  The three orthogonal
coordinate axes for
a visibility point are $(\unitvect{u},\unitvect{v},\unitvect{w})$, 
with the corresponding axes in the brightness distribution
being $(\unitvect{l},\unitvect{m},\unitvect{n})$.  
The natural units for the visibility axes are wavelengths corresponding to
the observed frequency.  This convention will be followed for the remainder of this thesis.
In general radio astronomy convention, the $\unitvect{w}$ axis points in the
direction \vect{s_{0}}, while \unitvect{u} and 
\unitvect{v} are orthogonal to each other and the \unitvect{w} axis, and are
on the plane containing the \unitvect{w} axis and celestial East and North
respectively.  The synthesised image is assumed to be two--dimensional, and
so positions on the sky are described in terms of $l$ and $m$.  The vector quantity \unitvect{s}
can be expressed as $l\unitvect{l} + m\unitvect{m} + n\unitvect{n}$.

The quantities \vect{b}, \unitvect{s}, \vect{s_{0}} and $\Omega$ referred to in the above
derivation can then be expressed as:
\begin{equation}
\frac{\nu \vect{b}\cdot \unitvect{s}}{c} = ul + vm + wn
\label{eq:coord1}
\end{equation}
\begin{equation}
\frac{\nu \vect{b}\cdot \vect{s_{0}}}{c} = w
\label{eq:coord2}
\end{equation}
\begin{equation}
d\Omega = \frac{dl\,dm}{n} = \frac{dl\,dm}{\sqrt{1-l^{2} - m^{2}}}
\label{eq:coord3}
\end{equation}

Substituting Equations~\ref{eq:coord1}--\ref{eq:coord3} into 
Equation~\ref{eq:visdefinition} yields:
\begin{equation}
V(u,v,w) = \frac{1}{A_{0}} \int_{-\infty}^{\infty} \int_{-\infty}^{\infty} A(l,m) I(l,m) e^{-2\pi i 
[ul + vm + w(\sqrt{1-l^{2} - m^{2}}-1)]} \frac{dl\,dm}{\sqrt{1-l^{2} - m^{2}}}
\label{eq:vis_coord}
\end{equation}

Since, as components of a unit vector, $l^{2} + m^{2} + n^{2} = 1$, the integrand is necessarily 0
where $l^{2} + m^{2} \geq 1$.

To obtain the radio sky brightness distribution $I(l,m)$, it is necessary to invert 
Equation~\ref{eq:vis_coord}.  Whilst possible in the generalised form given above, it is
conceptually difficult and computationally expensive.  If the visibility equation
given above can be reduced to the form of a two--dimensional Fourier transform (necessitating
the removal of the $w$ term), the inversion can be carried out using a Fast Fourier Transform
\citep{cooley65a}, which is much more computationally tractable.  If $l$ and $m$ are constrained
to be small -- restricting the field of view to a small solid angle -- then the term 
$\sqrt{1 -l^{2} -m^{2}}$ reduces to 1, eliminating $w$ from Equation~\ref{eq:vis_coord}
and leaving:
\begin{equation}
V(u,v) = \frac{1}{A_{0}} \int_{-\infty}^{\infty} \int_{-\infty}^{\infty} A(l,m) I(l,m) e^{-2\pi i [ul + vm]} dl dm
\label{eq:vis2d}
\end{equation}

Equation \ref{eq:vis2d} can be inverted via the Fourier relation to yield:
\begin{equation}
\frac{1}{A_{0}} A(l,m) I(l,m) = \int_{-\infty}^{\infty} \int_{-\infty}^{\infty} V(u,v) e^{2\pi i [ul + vm]} dl dm
\label{eq:skybright2d}
\end{equation}

Thus, the correlator output can be transformed, under certain assumptions, to yield the 
original sky brightness.  For a more detailed description of the limitations of the small--field
assumption and methods of circumventing it, see \citet{thompson99a}.

\section[Calibration and editing]{Calibration and editing}
\label{radio:calib}
Before the transformation of visibilities into a sky brightness distribution can take place,
incorrect and uncorrectable visibilities must be removed (a process known as flagging), 
and the remaining
visibilities correctly calibrated.  Visibilities which must be flagged are typically those which
have been affected by equipment failure, radio frequency interference (RFI) or
severe propagation problems (rapid ionospheric visibility at low frequencies, water vapour
variations at high frequencies, and physical problems such as antenna shadowing).
Flagging can be automated (based on logs of telescope
and electronic failures, for example, or on the visibility values themselves) or manual, 
where visibilities are inspected by eye and questionable samples removed.  Flagging is an
inexact procedure which has been the subject of numerous analyses 
\citep[e.g.][]{middelberg06a, ekers99a}.

Calibrating the interferometer output amplitude to obtain the true visibility amplitude
requires, as shown in Equation~\ref{eq:corroutvis}, a knowledge of the peak
antenna response.  As discussed in Section~\ref{radio:hardware}, this can be achieved
through the use of noise calibration at the individual antennas.  For short--baseline arrays,
this calibration is typically refined by including an observation of a known ``flux calibrator",
a compact, unconfused source known to possess stable flux.  Antenna--based correction
factors can be derived by comparing the measured visibilities with the known values.
This is a luxury not available to VLBI arrays, since there are no constant flux sources
which are still compact on VLBI baselines \citep{walker99a}.  Thus, accurate logging
of antenna noise calibration is especially crucial for the absolute, as well as relative,
calibration of VLBI observations.

At this stage, it is also possible to make corrections to the visibility phases, to compensate
for dispersive and non-dispersive delays which were not correctly accounted for in the 
correlator model -- that is, the first term in the cosine of Equation~\ref{eq:corroutvis} was 
not correctly subtracted.  Typically, this involves measured atmospheric and ionospheric 
quantities that were not available at the time of correlation, and are most commonly required
in VLBI observations.  Other potentially correctable errors which can occur during VLBI
correlation include geometric model errors due to incorrect Earth Orientation Parameters (EOPs).
These values, which collectively describe the orientation and rotational phase of the Earth
to very high precision, are made available by the International Earth Rotation 
Service (IERS\footnote{http://www.iers.org/}), and are calculated using intensive geodetic VLBI 
observations (see Section~\ref{radio:astrogeo} below).

Once calculated,  the application of corrections is simple, 
amounting to a simple scaling of visibility amplitudes
or rotation of phase.  Typically, the rate of change 
of corrections is small, and so the error caused by the finite visibility sampling duration
is not a concern.  For very rapidly varying corrections, or for errors which cannot be
modeled and subtracted, however, the sampled visibilities remain in error.
Residual visibility errors add noise to an interferometric image, but may be corrected 
(in some circumstances) in the imaging stage as discussed below.

\section[Imaging]{Imaging}
\label{radio:imaging}
This section is meant as an overview of the main techniques used in VLBI to produce radio 
images.  It does not cover complicating factors such as the (potentially inhomogeneous)
antenna primary beam responses, wide--field effects, the effect of spectral and
temporal averaging, or gridding techniques.  For a complete
reviews of radio synthesis imaging, see (for example)
\citet{thompson94a} and \citet{pearson84a}.

Once an edited, calibrated visibility set has been produced, it may be transformed into
an image using the Fourier relation shown in equation~\ref{eq:skybright2d}.  Generally
speaking, however, the sampling of visibilities in the $uv$ plane will be incomplete (especially
for VLBI, where small numbers of antennnas often mean extremely sparse sampling), which
introduces considerable distortion in the transformed image.  Essentially, the measured visibilities
can be considered to be the product of the true Fourier transform of sky brightness with a
sampling function $S(u, v)$, which is 1 (or indeed any non--zero weight, if weighting is applied
in the $uv$ plane) where visibilities were measured and 0 elsewhere.  
By virtue of the Fourier transform duality between multiplication and convolution, this leads
to the transformed image effectively becoming the convolution of the true sky brightness with 
the Fourier transform of the sampling function $\mathpzc{F} \!\left \{ S(u,v) \right \} = s(l,m)$.  This
directly generated map is generally referred to as the ``dirty" map, and the Fourier
transform of the sampling function, $s(l,m)$ is known as the synthesized beam.

Thus, to obtain the true sky brightness distribution, it is necessary to ``deconvolve" the effects of
limited visibility sampling.   In general, this takes the following iterative form:

\begin{enumerate}
\item Adjust sky brightness model (add/subtract/move components).
\item Compute visibilities expected due to this model.
\item Subtract expected visibilities from observed visibilities to obtain a ``residual" brightness
distribution.
\end{enumerate}

Once the best source model has been obtained, it is convolved with a ``restoring" beam, which
is typically taken as an elliptical gaussian truncated at the first null of the synthesized beam,
and the remaining residual map added.  This generates a ``clean" map with similar resolution
to the original dirty map, but with the sidelobes generated by the synthesised beam 
minimised.

Various procedures have been developed to make use of a priori information to help guide
step 1 above, and converge towards the best (and simplest) global solution for the true sky
brightness, which is complicated by the fact the the measured visibilities are corrupted by noise.
Popular examples of imaging algorithms include CLEAN \citep{hogbom74a}, least squares 
model fitting \citep{pearson99a}, maximum entropy \citep[MEM;][]{ables74a}
and various derivatives and combinations of the above.  In this thesis, both model
fitting and CLEAN were used.

During the imaging process, it is possible to make use of additional relationships between 
groups of visibilities to improve the a priori calibration and remove errors in the visibilities.
These ``closure" quantities result from the relationship between the true and observed visibilities,
which can be written as shown in Section~\ref{radio:math}:
\begin{eqnarray}
V^{\mathrm{obs}}_{mn} & = & g_{m}g_{n}^{*}V_{mn} \\
|V^{\mathrm{obs}}_{mn}| e^{i\,\phi^{\mathrm{obs}}_{mn}} & = & 
|g_{m}||{g_{n}||V_{mn}| e^{i\,(\phi_{g_{m}} - \phi_{g_{n}} + \phi_{mn}})}
\label{eq:visgains}
\end{eqnarray}

\noindent where $g_{m} = |g_{m}|\exp{i\,\phi_{g_{m}}}$ is  the complex gain associated with 
antenna $m$.  Considering only terms in the exponent of Equation~\ref{eq:visgains} yields:
\begin{equation}
\phi^{\mathrm{obs}_{mn}} = \phi_{g_{m}} - \phi_{g_{n}} + \phi_{mn}
\end{equation}

\noindent and considering the summation of the phase terms from the common baselines of 
three antennas $m$, $n$\ and $p$\ gives the phase closure relationship for the ``closure phase" 
$\phi^{\mathrm{close}}_{mnp}$:
\begin{equation}
\phi^{\mathrm{close}}_{mnp} = 
\phi^{\mathrm{obs}}_{mn} +  \phi^{\mathrm{obs}}_{np}+ \phi^{\mathrm{obs}}_{pm}
= \phi_{mn} +  \phi_{np} +  \phi_{pm}
\label{eq:phaseclos}
\end{equation}

Equation~\ref{eq:phaseclos} shows that the closure phase is a quantity which is independent of the 
antenna--based gains, which include the effect of varying atmospheric propagation and other
antenna--based errors, and as solely determined by the true visibility phase.  Thus, if a sufficiently
accurate model of the brightness distribution is available, deviations in phase from the model phase
can be attributed to antenna-based errors and adjustments made to the value of $\phi_{g_{m}}$ for
each antenna to minimise the difference between model and observed visibility phases.  This
procedure is known as ``phase self--calibration".

By considering the magnitude terms of Equation~\ref{eq:visgains}, the following equation for
closure amplitude $|V^{\mathrm{close}}_{mnpq}|$ can be constructed from baselines of
four antennas $m$, $n$, $p$ and $q$:
\begin{equation}
|V^{\mathrm{close}}_{mnpq}| = \frac{|V^{\mathrm{obs}}_{mn}| |V^{\mathrm{obs}}_{pq}|}{|V^{\mathrm{obs}}_{mp}| |V^{\mathrm{obs}}_{nq}|} = 
\frac{|V_{mn}| |V_{pq}|}{|V_{mp}| |V_{nq}|}
\label{eq:ampclos}
\end{equation}

Equation~\ref{eq:ampclos} shows that the closure amplitude, like closure phase, is independent
of the antenna--based complex gains, and similarly can be used to compute a correction to the
gains to minimise the difference between model and observed visibilities.  Unsurprisingly,
this procedure is known as amplitude self--calibration.

While self--calibration is a useful tool for removing antenna--based errors and improving the
fidelity of radio synthesis images (usually measured by {\em dynamic range} -- the ratio of
peak image flux to image noise), its principal drawback is that it can only attempt to reduce the 
difference between the observed visibilities and the model visibilities.  If the input model
does not resemble the actual sky brightness distribution, application of self calibration
will not improve image quality, and may limit the potential to improve the model
in the future.  Furthermore, if the closure quantities are extremely noisy, self calibration
may act to fit noise to the model.  For more details on the use of self--calibration
in interferometric imaging, see \citet{cornwell99a}.

\section[Astrometry and geodesy]{Astrometry and geodesy}
\label{radio:astrogeo}
As shown in Equation~\ref{eq:corroutvis}, the uncorrected phase term of correlator output
contains the true visibility phase for a source, corrupted by a geometric term.  Actual
correlator implementations, as discussed in the following chapter, attempt to remove
this non--intrinsic term as precisely as possible.  However,  the ability to do this depends on
the exact knowledge of both the \vect{b} and \unitvect{s} vectors, and thus on the exact
location of the source and antennas.  Errors in the assumed values for  \vect{b} or \unitvect{s}
will cause the correlator output phase to differ from the true visibility phase.

The use of the visibility phase error to calculate error in assumed source position (holding
the baseline vector fixed) is known as {\em astrometry}, while calculating errors
in assumed baseline vectors is known as {\em geodesy}.  An excellent overview of the theory
and practical difficulties of VLBI astrometry and geodesy is given in \citet{fomalont99a}.  
It should be immediately apparent that disentangling the phase contributions due to
unknown baseline and source vector errors {\em simultaneously} is a challenging
task, and multiple baselines and sources are required to make such a global solution.
Further complicating the problem are the residual visibility phase errors due to the
unmodeled atmosphere and ionosphere.  The contribution from the ionosphere is
frequency--dependent and can be estimated and subtracted if observing frequencies span
a wide fractional bandwidth and sufficiently high signal to noise
ratios can be obtained, and astrometric/geodetic observations often use a dual--band 
``S/X" (2.3 GHz and 8.4 GHz) receiver for the purpose of obtaining widely separated 
bands \citep[see e.g.][]{petrov08a}.  The contribution from the atmosphere 
(predominantly the wet and dry troposphere) is independent of frequency, however,
and so improvements to the a priori model can only be made through the use of
of station weather information, such as water vapour radiometers \citep{roy06a}.

The principal difficulty confronting astrometric and geodetic observations is that the
powerful self--calibration technique cannot be used freely -- 
since it acts to converge visibilities with a pre--existing model, incorrect usage will corrupt the 
measurements of source and antenna positions which are being sought.  Sources
used for astrometry and geodesy should ideally have no detectable structure, simplifying the 
problem, but suitably bright examples of such sources are rare \citep[see e.g.][]{gontier01a}.  
Thus, correct calibration of visibility phases through the careful modeling of propagation effects 
and source structure is of the utmost importance in astrometric and geodetic observations.

The fundamental reference frames which are the product of absolute astrometry and geodesy
are the ICRF and ITRF, introduced in Section~\ref{radio:vlbi}.  They are maintained and improved
by the IERS.  The ongoing maintenance of these frames allows other users to
undertake relative astrometry and (less commonly) geodesy, under the assumption that positions
of the defining members of these frames are correct.  This procedure of relative VLBI astrometry forms
the basis of the measurement of pulsar position measurements made in this thesis, and as such is
discussed in further detail in Chapter~\ref{techniques}.  Geodesy is not a fundamental component 
of this thesis, but will be mentioned where appropriate as it impacts on the astrometric measurements.

When undertaking relative, as opposed to absolute astrometry, one of the major practical differences
is the use of single--band phase, as opposed to multi--band delay which is commonly used for
absolute astrometry and geodesy.  Even the best available a priori delay modeling is uncertain
at the level of nanoseconds, which corresponds to multiple turns of phase at typical observing 
frequencies.  Thus, for absolute astrometry the delay is generally estimated by the phase gradient
between observing bands -- the so--called ``multi--band delay".  The concept of multi--band
delay is explained in detail in \citet{fomalont99a}.  When conducting relative astrometry, however,
the delays are calibrated to higher precision through observations of a known source, meaning
the single--band phases no longer suffer from wrap ambiguities.  This allows a higher precision 
estimation of relative position.  However, the accuracy of the delay solution transfer from calibrator
to target in relative astrometry depends on the angular separation of the two sources, and the gradient
in residual delay between them.  Clearly, it is desirable to minimise this angular separation, in 
addition to attempting to obtain the best possible modeling to minimise the residual error gradient.
\citet{pradel06a} present an comprehensive simulation--based analysis of errors in relative astrometry,
including the effect of different error sources such as the wet and dry troposphere at different
angular separations.

\chapter{DIFX: AN FX STYLE SOFTWARE CORRELATOR}
\label{difx}

An in--depth description and analysis of the DiFX code was published in
\citet{deller07a}, and the remainder of this chapter draws upon the material 
presented there, as well as the description of correlator functionality presented
in \citet{romney99a}.

\section[Mathematical description of a correlator]{Mathematical description of a correlator}
\label{difx:math}
Chapter~\ref{radio} showed that a radio interferometer measures the spatial frequency components of
the observed sky brightness distribution, and that this information, under certain restrictions, could
be transformed into an estimation of the actual sky brightness distribution.  However, this treatment
was valid only for the unphysical case of a monochromatic radio source, and neglected the effect
of data sampling, which is crucial to the operation of modern interferometers.
In this section, I show the effects of relaxing these restrictions, and the functionality which
must be incorporated into a realistic correlator to compensate.  The description will be appropriate for
an unconnected interferometer, and simplifications that can be adopted for connected interferometers
will be noted where applicable.  The mathematical derivation will be appropriate for an FX--type
correlator (see Section~\ref{difx:impl:fxxf} for the definition of an FX--type correlator), 
and the advantages and limitations compared to an XF--type correlator will be highlighted.

\subsection[Quasi--monochromatic formalism]{Quasi--monochromatic formalism}
\label{difx:math:formalism}
It is convenient to represent the electromagnetic radiation from the source as a quasi--monochromatic
plane wave incident on the antennas, described (for a given antenna $p$) by:
\begin{equation}
E_{p}(t+\tau_{p}) = A(t+\tau_{p}) e^{i2\pi \nu_{0} (t+\tau_{p})}
\label{eq:qmepw}
\end{equation}

The electric field component is given by the real component of Equation~\ref{eq:qmepw}.  It is clear
that for a realistic, band--limited process $A(t+\tau_{p})$, 
the electric field component will be a band--limited
signal, centred on $\nu_{0}$.  $A(t+\tau_{p})$ is a 
complex signal which must also be covariance--stationary,
ergodic and stochastic.  The term $\tau_{p}$ shown in Equation~\ref{eq:qmepw} above represents
the delay between antenna $p$ and a reference point\footnote{Usually the geocentre for VLBI, or
some point near the centre of the array for connected element interferometers}.

Since $A(t+\tau_{p})$ has been asserted to be band--limited, it can be written as:
\begin{equation}
A(t+\tau_{p}) = \int_{-\infty}^{\infty} s(\nu)e^{i2\pi \nu (t+\tau_{p})}d\nu
\label{eq:specrep}
\end{equation}
\noindent with the proviso that $s(\nu) = 0$ outside some limiting band $|\nu| < B$.

As shown in Section~\ref{radio:math}, the correlator's function is to compute the expectation of
the cross correlation of the two received signals from antennas $p$ and $q$:
\begin{eqnarray}
r_{pq}(\tau) 	&=& \langle E_{p}(t + \tau_{p})E_{q}^{\ast}(t+\tau_{q}) \rangle \\ \nonumber
		 	&=& \langle E_{p}(t + \tau_{g})E_{q}^{\ast}(t) \rangle \rangle \\ \nonumber
		 	&=& \langle A(t+\tau_{g})A^{\ast}(t) \rangle e^{i 2 \pi  \nu_{0} \tau_{g}} \\ \nonumber
			&=& \gamma_{pq}(\tau) e^{i 2 \pi  \nu_{0} \tau_{g}}
\label{eq:eestar}
\end{eqnarray}

Throughout this thesis, the superscript $^{\ast}$ will be taken to mean complex 
conjugation\footnote{The real (non--complex) correlator described in Chapter~\ref{radio} did not
require the second signal to be conjugated in order to calculate the correlation function, for 
the obvious reason that all signals were real}.  Since $A(t)$ is ergodic, the expectation
can be approximated  by a time average.  As in
Chapter~\ref{radio}, $\tau_{g}$ represents the geometric delay $\tau_{p} - \tau_{q}$ 
between antennas $p$\ and $q$.  The term $\gamma_{pq}(\tau)$\ is the covariance function
of the stochastic process $A(t)$, and represents the unmodulated
correlation between the electric fields on this baseline -- the visibility, as shown in the 
previous chapter.  This visibility is modulated by the time--varying phasor $e^{i 2 \pi  \nu_{0} \tau_{g}}$.

Ultimately, the visibility is required as a function of frequency, which can be obtained by
substituting the spectral representation for $A(t)$ into the expression for $\gamma_{pq}(\tau)$ as 
shown:
\begin{eqnarray}
\gamma_{pq}(\tau) 	&=& 	\langle A(t+\tau_{g})A^{\ast}(t) \rangle \\ \nonumber
				&=& 	\langle \int_{-\infty}^{\infty} \int_{-\infty}^{\infty} s_{p}(\nu)s^{\ast}_{q}(\nu')
					e^{i2\pi [(\nu-\nu')t +\nu\tau_{g}]}d\nu d\nu' \rangle \\ \nonumber
				&=&	 \int_{-\infty}^{\infty} S_{pq}(\nu) e^{i2\pi \nu \tau_{g}} d\nu
\label{eq:Spq}
\end{eqnarray}
\noindent where $S_{pq}(\nu) = s_{p}(\nu)s^{\ast}_{q}(\nu)$\ is the cross--power spectrum of $A(t)$, and 
can now be shown to be simply the Fourier transform of the covariance function:
\begin{equation}
S_{pq}(\nu) = \int_{-\infty}^{\infty} \gamma_{pq}(\tau) e^{-i2\pi \nu \tau} d\tau
\end{equation}

Thus, in order to obtain the term $S_{pq}(\nu)$, which is the desired visibility term,
a realistic correlator must convert the time domain signal to the frequency domain and
account for the changing geometric delay $\tau_{g}$ 
and other practicalities such as frequency downconversion and sampling -- in effect, 
compensating for the phasor term $e^{i 2 \pi  \nu_{0} \tau_{g}}$.   These steps are detailed 
in the following sections, assuming station--based corrections.  Station--based
removal of the phasor fringe term is generally only computationally feasible in an FX--style 
correlator architecture, as explained in Section~\ref{difx:impl:fxxf}.

\subsection[Baseband conversion and sampling]{Baseband conversion and sampling}
\label{difx:math:baseconv}
Despite continual improvements in digital electronics which have produced cheaply available,
high--speed (multi--Gsample s$^{-1}$) digital samplers, some form of frequency downconversion is
ubiquitous in all arrays, with the exception of dedicated low--frequency instruments.  For observations
at wavelengths of several cm and shorter, this situation is unlikely to change in the near future.
Typically, frequency conversion is a multi--stage process, utilising broad front--end filters,
sharp intermediate--frequency (IF) filters and potentially digital filtering after sampling.  The
mathematical treatment below assumes a single conversion step directly to baseband, which
is unrealistic but mathematically convenient.  For an in--depth discussion of frequency conversion
systems, see \citet{thompson94a}.  

The application of a real, single sideband mixer with frequency $\nu_{0}$\ to the 
received signal $E(t)$\ from Equation~\ref{eq:qmepw} yields the following signal:
\begin{equation}
V'_{p}(t + \tau_{p}) = E_{p}(t+\tau_{p})\times \cos{2\pi \nu_{0}} = \frac{A(t+\tau_{p})}{2} \left [ e^{i2\pi \nu_{0} \tau_{p}} + e^{i2\pi \nu_{0}(2t + \tau_{p})} \right]
\label{eq:baseband}
\end{equation}

The application of a low--pass filter removes the high--frequency term from Equation~\ref{eq:baseband}.
At this point, it should be noted that in a connected element interferometer with a common clock,
the phase of the downconversion signal can be varied continuously to compensate for the changing
geometric phase (effectively replacing the oscillator signal $e^{i 2 \pi\nu_{0}t}$ with 
$e^{i 2 \pi\nu_{0}\left(t - \tau_{p} \right)}$).  This renders the fringe--rotation and fractional sample
corrections detailed in Section~\ref{difx:math:geocomp} unnecessary.  However, as this approach
is not presently implemented on major VLBI arrays, the following analysis will 
assume that no phase compensation has been performed at the downconversion stage.

After frequency downconversion, the resultant bandlimited, baseband 
signal $V_{p}(t)$\ is sampled.  Since the 
process $A(t)$\ is bandlimited with bandwidth B, the sampling interval must be no longer
than $\Delta t = 1/2B$\ in order to prevent aliasing of the sampled spectrum.  Quantisation
can be as coarse as 1 bit, although 2 bits has been widely used and newer instruments have tended
to push towards higher precision in order to mitigate the effects of RFI, and obtain higher spectral
dynamic ranges for spectral line observations.
At the low bit precision typically used, the sampling function $\mathpzc{S}$\ is typically 
non--linear, allowing an optimum placement of levels given the Gaussian distribution of $V$.
The sampled signal $R_{p}(n)$\ can then be written as:
\begin{equation}
R_{p}(n) = \mathpzc{S}\left[ V_{p}(t=n\Delta t)\right]
\label{eq:sampled}
\end{equation}

For the sake of clarity, the following discussion of geometric compensation will use the continuous 
notation for the baseband time series $V_{t}$, and the limitations placed by the 
sampling process will be highlighted where appropriate.

\subsection[Geometric compensation and channelisation]{Geometric compensation and channelisation}
\label{difx:math:geocomp}
\begin{figure}
\begin{center}
\includegraphics[width=0.8\textwidth]{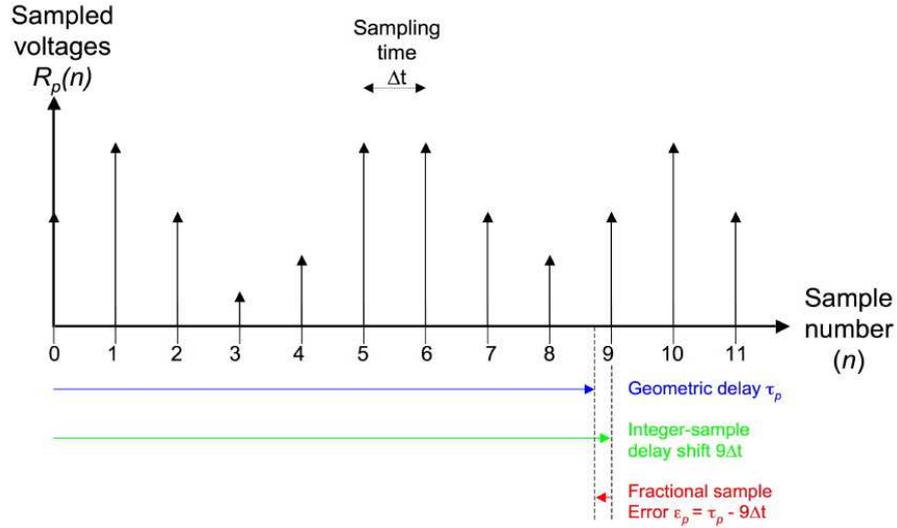}
\caption[Illustration of the geometric delay in integer--sample and fractional--sample components]
{Geometric delay for a sampled signal $R_{p}(n)$\ for an antenna $p$ at a given instant of time.  
The exact geometric 
delay $\tau_{p}$ is shown in blue.  The nearest integer--sample delay (shown in green) can be 
compensated by shifting the sampled datastream, leaving a fractional sample error $\epsilon_{p}$
(shown in red).}
\label{fig:delays}
\end{center}
\end{figure}

In order to form the correct correlation between the array antennas, the geometric delay $\tau_{p}$ for
each antenna must be removed.  Initially, this is implemented as an integer--sample shift of the
sampled signal $R_{p}(n)$ by N samples, where N is the rounded integer value of
$\tau_{p}/\Delta t$. The shifted baseband signal is thus 
$V_{p}(t+\tau_{p} - n\Delta t) = A(t+\tau_{p} - n\Delta t)e^{i 2 \pi \nu_{0} \tau_{p}}$.  Since the 
integer--sample shift cannot exactly compensate for the geometric delay $\tau_{p}$, a 
``fractional--sample" error $\epsilon_{p} = \tau_{p} - n\Delta t$ remains at this point.  
A diagram of the sampled signal $R_{p}(n)$ which illustrates the geometric delay $\tau_{p}$, the 
integer--sample delay N$\Delta t$\ and the fractional--sample error $\epsilon_{p}$\ for an 
antenna $p$\ at a given time is shown in Figure~\ref{fig:delays}.

The compensation of the fringe phasor $e^{i 2 \pi  \nu_{0} \tau_{p}}$\ can now performed.  This 
step is generally referred to as {\em fringe rotation}.  Conceptually this is simply applied
as a complex multiplication to $V_{p}$\ by the function $e^{-i 2 \pi  \nu_{0} \tau_{p}}$, yielding
the fringe--rotated, shifted baseband signal 
$V_{p}(t+\epsilon_{p})e^{-i 2 \pi  \nu_{0} \tau_{p}} = A(t + \epsilon_{p})$, which is complex--valued.
An alternative to a complex multiplication is to implement the multiplication with separate real--valued
cosine and sine components.  This ensures that the signal remains real--valued and allows the
use of real--valued Fourier transformations.  The ``cosine" and ``sine" arms of the correlator 
are then combined after Fourier transformation by applying a Hilbert transform to the sine
arm before addition.  For an example of this form of implementation of a complex correlator,
see \citet{thompson94a}.

The shifted, fringe--rotated signal may now be transformed into the frequency domain as shown:
\begin{equation}
\mathpzc{F}\!\left\{ A(t+\epsilon_{p} \right\} = s_{\nu}e^{i 2 \pi \nu \epsilon_{p}}
\end{equation}

In practice, this step is implemented using a Fast Frequency Transform \citep[FFT;][]{cooley65a}.
This imposes a segmentation upon the data, since the FFT windows the time domain data with
a window size equal to twice the desired number of spectral points.

The final geometric compensation step is the removal of the phase gradient generated by the
presence of fractional sample error.  Since the maximum magnitude of $\epsilon$\ is half the 
sampling time ($1/(4B)$\ seconds),
the phase induced at the upper edge of the band ($\nu = B$) can range between $-\pi$\ to $\pi$.
Fractional sample correction is implemented as a complex multiplication of the frequency--domain
data. After fractional sample correction, the spectral representation of the original signal $s(\nu)$
has been recovered exactly.

Finally, it is important to note at this point that realistic correlators approximate the true geometric
delay function $\tau_{p}$\ over short time intervals by a polynomial expansion.  Over a the time duration
of a single FFT window (typically microseconds, for bandwidths $B\sim$MHz and spectral
points $\mathrm{n}\sim$hundreds), 
this can be (and usually is) approximated by a linear expansion.  If the difference in fringe
phase across an FFT window 
$\delta\phi =2\pi  \nu_{0} \left( \tau_{p}(t_{1}) - \tau_{p}(t_{1}+2\mathrm n \Delta t) \right)$ 
is small, the fringe term may be approximated by a constant over the entire FFT window.  
Since the Fourier transform is a linear transformation, this single complex factor then may 
equivalently be applied after the FFT, rather than to every individual sample before the FFT. 
In practice, this allows the fringe rotation operation to be merged with the fractional sample correction.
``Post--F" fringe rotation allows a considerable saving in computational load, since for every 
FFT window it removes the need
for 2n multiplications and 2n trigonometric operations, at the cost of n complex additions and
one trigonometric operation.  However, if the FFT window is long (high spectral resolution) and/or the fringe rate is very high (high frequencies and/or long baselines), the change in fringe phase over an FFT 
window may incur an unacceptable amount of decorrelation.  The decorrelation can be calculated
as $\rm sinc \left(\delta\phi/2 \right)$. Table~\ref{tab:decor} shows the
maximum decorrelation which would be incurred through the use of post--F fringe rotation
for some typical observations.

\begin{deluxetable}{lrrrr}
\tabletypesize{\tiny}
\tablewidth{0pt}
\tablecaption{Maximum decorrelation incurred due to ``Post-F" fringe rotation}
\tablehead{
\colhead{Observation} & \colhead{Maximum} & \colhead{Frequency} & \colhead{Spectral points per} & \colhead{Maximum} \\
&\colhead{baseline (km)}&\colhead{(MHz)}&\colhead{16 MHz band} &\colhead{decorrelation (\%)}
}
\startdata
LBA low frequency continuum & 1400 & 1600 & 128 & 0.003 \\
LBA high frequency continuum & 1700 & 8400 & 128 & 0.13 \\
VLBA low frequency continuum & 8600 & 1600 & 128 & 0.12 \\
VLBA high frequency continuum & 8600 & 22200 & 128 & 21.1 \\
LBA water masers & 1700 & 22200 & 1024 & 47.6 \\
\enddata
\label{tab:decor}
\end{deluxetable}

\subsection[Cross--multiplication and accumulation]{Cross--multiplication and accumulation}
After each station has been individually delayed, fringe--rotated and corrected for
fractional sample error, the desired visibility term $S_{pq}(\nu)$\ can be obtained
through a complex multiplication of the corrected baseband station data, after conjugating the
second data stream:
\begin{equation}
S_{pq}(\nu) = s_{p}(\nu)s^{\ast}_{q}(\nu)
\end{equation}

This is repeated for all baselines, and the result is accumulated for a desired time interval, typically seconds for narrow--field VLBI.  If $p=q$, the result is an autocorrelation, which is necessarily 
real--valued.
At the end of each integration period, the visibility data is normalised and stored.
The normalisation can be performed in an a priori manner (in theory it is dependent only upon
the integration time, since the normalised autocorrelations should tend to unity), but quantisation 
threshold errors make corrections based on the measured autocorrelation necessary.  The corrections
can be calculated and applied at correlation time, or in later processing.

\section[Correlator implementations]{Correlator implementations}
\label{difx:impl}
\subsection[FX vs XF correlators]{FX vs XF correlators}
\label{difx:impl:fxxf}
As shown above, it is possible (and mathematically simplest) to form the visibility output of a
correlator by Fourier transforming the station data, and then cross--multiplying the corrected data
streams.  This approach of channelisation (F) followed by cross--multiplication (X) is known
as an FX--style correlator.  
Through the Fourier duality between convolution and multiplication, it is easy to 
see that another possible implementation of a correlator would be to form the convolution
of the time series for each baseline and Fourier transform the result.  This is the 
so--called XF style of correlator.  

Until relatively recently, all correlators have been implemented as XF 
correlators -- the FX concept was first
implemented by \citet{chikada87a}.  As the correlator developed for this thesis is of the FX type, details
of the implementation of an XF--style correlator will not be given here -- the interested reader is directed to \citet{romney99a} and references therein.  However, some of the functional differences between 
XF and FX correlators are:

\begin{enumerate}
\item In an XF correlator, it is possible (indeed, necessary, to make the design computationally viable)
to accumulate the lag values and perform a single Fourier transform on the accumulated lags
before storing the visibilities. 
\item XF correlators can operate at much lower precision than FX correlators and can generally
make use of optimised, low--precision integer arithmetic.  
The presence of a Fourier transformation before accumulation (and
the use of station--based fringe rotation) means that an FX--style correlator requires higher
precision operations, although the FX architecture requires less operations overall.  The use of 
low--precision operations (including fringe rotation) is generally necessary to make the 
architecture viable. 
\item Due to the accumulation performed before transformation to the frequency domain,
the fractional sample error cannot generally be approximated by a constant in an XF correlator,
which means that correction of fractional sample error in the frequency domain in a station--based
manner is not possible\footnote{Frequency domain fractional sample correction could be performed by
shortening the accumulation time, or the equivalent fractional sample correction could be 
performed in the time domain via a convolution, but neither approach is generally computationally 
viable}.  The usual alternative employed by XF correlators is baseline--based alignment of data,
fringe rotation and fractional sample correction.  This allows alignment of the data streams to within
half a sample, as opposed to one sample with station based alignment, which when 
coupled with coordinated fringe rotation allows minimal decorrelation due to fractional sample
errors (see \citealt{romney99a} for details).  Low precision fringe rotation (which cannot be employed
in station--based fringe rotation due to the possible of spurious correlation between harmonics
of the fringe function) can also be used in this approach for higher efficiency.
\end{enumerate}

\subsection[Hardware platforms]{Hardware platforms}
\label{difx:impl:hardplat}
Both XF-- and FX--style correlators have traditionally been highly application-specific 
devices, based on purpose-built integrated circuits (Application Specific Integrated Circuits -- 
ASICs).  In the last 20 years, Field 
Programmable Gate Arrays (FPGAs) have become popular components in correlator designs, 
with one prominent example being the Very Long Baseline Array (VLBA) correlator 
\citep{napier94a}.  FPGAs are reconfigurable or reprogrammable devices that offer more 
flexibility than application-specific integrated circuits (ASICs) while still being highly efficient.
However, FPGAs still require a greater familiarity with the underlying hardware than 
coding on a general purpose CPU, and configurations can be specific to the particular FPGA,
limiting portability.
Throughout the remainder of this thesis, correlators based on ASICs and/or FPGAs will be referred
to simply as ``hardware correlators".  Modern VLBI hardware correlators include the
NRAO Very Long Baseline Array 
correlator \citep{napier94a}; the Joint Institute for VLBI in Europe (JIVE) correlator 
\citep{casse99a}; the Canadian NRC S2 correlator \citep{carlson99a}; the Japanese VLBI 
Space Observatory Programme (VSOP) correlator \citep{horiuchi00a}; and the Australia 
Telescope National Facility (ATNF) S2 correlator \citep{wilson96a}

Generally speaking, hardware correlators allow the most efficient design possible, by tailoring
the hardware to the processing required.  However, this efficiency comes with two drawbacks --
the design of the correlator incurs large Non--Recurring Engineering (NRE) costs,
and upgrading or altering the completed correlator is difficult 
(although this second point is mitigated somewhat in FPGA--based correlators).

An alternative to a hardware correlator is to implement the correlation algorithm in a programming
language such as C or C++, and run the correlation program in a generic multi-processor 
computing environment.  Such a system will be hereafter referred to as a ``software correlator".
The first VLBI correlators of the 1960's were implemented in software 
\citep{bare67a,moran67a}, 
but data processing requirements soon outgrew the capabilities of existing mainframe
computers, and dedicated hardware systems were quickly developed.
In recent years, the pace of advances in commodity computing hardware have begun to outstrip 
the demand increases of VLBI data processing, and software correlators have again become
economically feasible.  The first modern, high data--rate software correlator was developed for
VLBI by the Communications Research Laboratory (CRL) in Japan in the late 1990s
\citep{kondo04a}.  Specialist software correlators have also been developed for purposes
such as processing VLBI observations of spacecraft, which were used to
track the Huygens probe as it entered the atmosphere of Titan \citep{avruch06a}.

Compared to a hardware correlator, a software correlator offers several advantages: 
chiefly the ease and speed of development (minimising NRE), the adaptability of the system, 
and the flexibility of making alterations after the
correlator has been deployed.  The correlation algorithm is ``embarrassingly parallel" and very 
well suited to commodity parallel computing architectures, such as Beowulf clusters with 
Gigabit ethernet interconnect.

Despite their obvious advantages in flexibility, by the mid--2000's no software correlators had 
yet been developed which
were compatible with all the major VLBI formats currently used world--wide. As discussed in 
Chapter~1, at this time the LBA 
was completing an upgrade which would provide the increased sensitivity necessary to undertake
the pulsar astrometry program planned for this thesis, but which required a new correlator.
Accordingly, a completely new, 
general--purpose software correlator was developed as part of this thesis.
The primary goal of this correlator was to enable
the LBA to undertake high--sensitivity observations and improve operational flexibility,
but a secondary aim was for the correlator to be usable for VLBI 
(and indeed connected element interferometry)
by many groups worldwide.  The resultant ``DiFX" software correlator is described in
Section~\ref{difx:difx} befow, and its testing and verification are described in Section~\ref{difx:deploy}.
Performance benchmarking of the DiFX code is described in Section~\ref{difx:perf}, and 
some of the additional applications it has enabled, beyond
the pulsar astrometry of this thesis project, are described in Section~\ref{difx:additional}.

\section[DiFX]{DiFX}
\label{difx:difx}
DiFX (Distributed FX)
is the name given to a code project encompassing a FX--style correlator (mpifxcorr) implemented
in C++ and utilising message passing conforming to the Message Passing Implementation 
(MPI\footnote{http://www-unix.mcs.anl.gov/mpi/})
standard, as well as related functionalities
such as geometric model generation (gencalc\_delays), a graphical front--end (DiFXGUI) and
various data and visibility inspection tools and plug--in packages.  The DiFX package
can be downloaded from 
\verb+http://astronomy.swin.edu.au/~adeller/software/difx/+.

DiFX requires an operating environment with a functional installation of MPI, and an
implementation of a vector library.  It was originally developed on an Intel cluster and by
default makes use of the Intel Performance Primitives 
(IPP\footnote{http://www.intel.com/cd/software/products/asmo-na/eng/perflib/ipp/index.htm})
library for vector calculations.  The IPP library is is optimised for modern 32 and 64 bit CPU
architectures and allows code acceleration of up to an order of magnitude compared to
unoptimised code. It can run on any Intel or AMD CPU, but will only dispatch optimised
code on Intel CPUs.  An equivalent AMD library, the AMD Performance Library
(APL\footnote{http://developer.amd.com/apl\_help/aa\_000\_frames.html}) is now
available, but has not been tested.

\subsection[The DiFX code]{The DiFX code}
\label{difx:difx:code}
\begin{figure}
\begin{center}
\includegraphics[angle=0, width=0.9\textwidth]{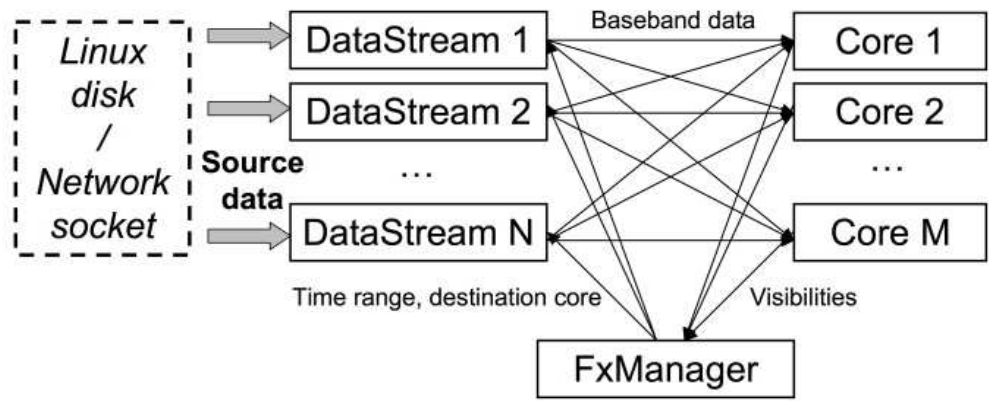}
\end{center}
\caption[Overview of the software correlator architecture]
{Overview of the software correlator architecture.  Data is loaded into memory 
from a disk or network connection by Datastream nodes.  These nodes are directed by 
a Master node to send data from given time ranges (typically several ms) to the 
processing elements (Core nodes).  The processed data
are sent to the master node for long-term accumulation and storage on disk.}
\label{fig:corrlayout}
\end{figure}

Figure \ref{fig:corrlayout} shows the high-level class structure of DiFX,
along with the data flow. The correlation is managed by a master node (FxManager),
which instructs data management nodes (Datastream) to send time ranges
of baseband data to processing nodes (Core).  The data are then processed
by the Core nodes, and the results sent back to the FxManager.
Double buffered, non-blocking communication is used to avoid latency delays
and maximise throughtput.  Both the Datastream and Core classes can be (and have
been) extended to allow maximum code re-use when handling different
data formats and processing algorithms.  The Core nodes make use of an allocatable number
of threads to maximise performance on a heterogenous cluster.

The Datastream nodes can read the baseband data into their memory buffers from a local disk,
a network disk or a network socket.  Once the data are loaded into the datastream buffer, the
remainder of the system is unaware of its origin.  This is one of the most powerful aspects of
this correlator architecture, meaning the same correlator can easily be used for
disk-based VLBI correlation and real-time eVLBI, where the data is transmitted in real
time from the telescopes to the correlator over optical fibre.  The use of DiFX for eVLBI is discussed
further in Section~\ref{difx:additional:evlbi}.

The implementation of the antenna-- and baseline--based operations described in
Section~\ref{difx:math} that are necessary to
form the visibility output of DiFX is described below.  The additional processing required
in the case of pulsar observations is also discussed.

\subsection[Antenna-based operations]{Antenna-based operations}
\label{difx:difx:antenna}
\subsubsection[Alignment of telescope data streams]{Alignment of telescope data streams}
\label{difx:difx:antenna:align}
To correlate data from a number of different telescopes, the changing delays between those telescopes must be calculated and used to align the recorded data streams at a predetermined point in space (in this case the geocentre) throughout the experiment. 

DiFX uses CALC 9\footnote{http://gemini.gsfc.nasa.gov/solve} to generate a
geometric delay model $\tau_{g}(t)$\ for each telescope in a given observation, at regular intervals (usually 1 second).  CALC models many geometric effects, including precession, nutation, ocean and atmospheric loading, and is used by many VLBI correlators including the VLBA and 
EVN correlators.  These delays are
then interpolated (using a quadratic approximation) to produce accurate delays ($\Delta\tau < 1 \times 10^{-15}$ seconds, compared to an exact CALC value) in double precision for any time during the course of the observation.   The estimated station clock offets and rates are added to the CALC--generated geometric delays.

The baseband data for each telescope are loaded into large buffers in memory, and the
interpolated delay model is used to calculate the accurate delay between each telescope and the centre of the Earth at any given time during the experiment.  This delay, rounded to the nearest sample, is the integer sample delay.  The difference between the delay and the integer sample delay is recorded as the antenna based fractional sample delay (up to $\pm$ 0.5 sample).  Note that the alignment of any two data streams (as opposed to a data stream alignment with the geocentre) is accurate to $\pm$ 1 sample.

The integer--sample delay is used to offset the data
pointer in memory and select the data to be correlated (some number of samples which is an
integer number of FFT windows, starting from the time of alignment).  The fractional sample error is retained to correct the phase as a function of frequency following alignment to within one sample, fringe rotation, and channelisation (Section~\ref{difx:difx:antenna:fft}).

Once the baseband data for each telescope have been selected, they are transferred to a processing node (``Core") using a non--blocking MPI send, and unpacked from the
coarsely quantised representation (usually a 2-bit representation) to a floating point (single precision) representation.  Presently supported VLBI formats include the Australian standard LBADR 
(Phillips et al., in preparation), Mark5A and Mark5B \citep{whitney03a}, 
and the Japanese format K5 \citep{koyama04a}.
If supported by the baseband format, missing or corrupted data is detected
during the unpacking, and a count of valid samples is maintained.  
Ancillary data such as the sampled geometric delays at the beginning of each FFT 
window are also transferred (in double precision).  From this
point on, all operations in the correlator are performed using floating point arithmetic,
in single precision unless otherwise specified.  Note that the data volume is expanded by a factor of 16 at this point.  The choice of single precision floats (roughly double the precision necessary) was dictated by the capabilities of modern CPUs, which process single--precision floating point numbers 
efficiently.  Using sufficient precision also avoids the small decorrelation losses incurred by optimised, low precision operations often used in hardware correlators.  This is a good example of the sacrifice of efficiency for simplicity and accuracy with a software correlator.

At this point all data streams from all telescopes are aligned to within $\pm$ 1 sample of each other and the fractional sample errors for each of the telescope data streams are recorded for later use.  A set number of samples from each telescope data stream have been selected and are awaiting 
unpacking and processing on a ``Core" node (e.g. a PC in a Beowulf cluster).

\subsubsection[Fringe rotation]{Fringe rotation}
\label{difx:difx:antenna:fringerot}
As shown in Section~\ref{difx:math:geocomp}, 
for ``pre--F" fringe rotation, the necessary complex fringe rotation function is assembled 
for each time sample by taking the sine and cosine of the
geocentric delay multiplied by the sky frequency $\nu_{0}$ in a vector operation; it is applied via
a vector complex multiplication for each telescope's data stream.  The geometric delay
at each time sample is obtained by interpolation between values supplied for the
start and end of each FFT segment.

Since the baseband data have already
been unpacked to a floating point representation by this stage, a floating point fringe
rotation is applied which yields no fringe rotation losses, compared, for example, to
a 6.25\% loss of signal to noise for three level digital fringe 
rotation in a two level complex correlator \citep{roberts97a}.

Implemented as such, fringe rotation represents a mixing operation and will result in a phase difference term which is quasi-stationary at zero phase (the desired term) and a phase sum term which has a phase rate of twice the fringe rotation function, $\sim4\pi\nu_{0}\tau(t)$.  The sum term vector averages to a (normally) negligible contribution to the correlator; for typical VLBI fringe rates (100s of kHz) and integration times (seconds) the relative magnitude of the unwanted contribution to each visibility point is $<10^{-5}$.  In a software correlator it would be simple to control the integration time so that the rapidly varying phase term is integrated over exactly an integral number of terms of phase, thus making no contribution to the correlator output.  This feature is not currently implemented in DiFX.


As noted in Section~\ref{difx:math:geocomp}, an alternative method of fringe rotation for
sufficiently low fringe rates is to apply a single correction per FFT window (``post--F" fringe
rotation).   In this case, no modulation is applied to the time domain data, but a single 
fringe--rotation value (appropriate for the midpoint of the FFT window) is calculated and retained
for application during fractional--sampled correction.
Post--F fringe rotation is desirable in situations where the fringe
rate is extremely low, when the double-frequency term introduced
by the mixing operation of pre--F fringe rotation is not effectively averaged to zero over the
course of an integration and makes a significant and undesirable contribution to the correlator output.  Switching from pre--F to post--F
fringe rotation would be beneficial for periods of time in
most experiments when the source traverses periods of low phase rate.  Sources near a celestial pole can have very low fringe rates for long periods of time.  Alternatively, if very short correlator integration times are used, the sum term may not integrate to zero when using pre--F fringe rotation.  Post-F fringe rotation would therefore be a natural choice in these circumstances.

It should be noted that it is possible to undertake the exact equivalent to pre--F fringe rotation in the frequency domain.  However, this would involve the Fourier transform of the fringe rotation function and a convolution in the frequency domain, which is at least as computationally intensive as the complex multiplication of the data and fringe rotation in the time domain.

DiFX implements pre--F or post--F fringe rotation as a user controlled option.

\subsubsection[Channelisation and fractional sample error correction]{Channelisation and fractional sample error correction}
\label{difx:difx:antenna:fft}
Once the data are aligned (and phase corrected, if pre--F fringe rotation is used), 
the time series data are converted into frequency series data (channelised), prior to cross multiplication.

Channelisation of the data can be accomplished using an FFT.
If pre--F fringe rotation has been applied, the data are already in
complex form, and so a complex-to-complex FFT is used.  The positive
or negative frequencies are selected in the case of upper or lower
sideband data respectively.  If post--F fringe rotation is to be
applied, the data are still real and so a more efficient
real--to--complex FFT may be used.  This is possible due to the
conjugate symmetry property of an FFT of a real data series.  In this
case, lower sideband data may be recovered by reversing and
conjugating the resultant channels.

The final station--based operation is fractional--sample correction.
The fractional sample error is assumed not to vary over the FFT window, which
is equivalent to the assumption made for post--F fringe rotation, but is considerably less 
stringent since the phase change is proportional to the bandwidth, rather than sky 
frequency as in the case of fringe rotation.  As shown in Section~\ref{difx:math:geocomp},
the frequency domain correction manifests itself as a slope in the phase as a function of 
frequency across the observed bandwidth.

Thus, after channelisation, a further vector complex multiplication is
applied to the channels, correcting the fractional sample error.  In
the case of post--F fringe rotation, the saved fringe rotation value is added
to the fractional--sample correction for each frequency channel and the two steps are performed
together.

\subsection[Baseline-based operations]{Baseline-based operations}
\subsubsection[Cross multiplication of telescope data streams]{Cross multiplication of telescope data streams}
For each selected baseline, the channelised, compensated data from the telescope pair are
cross-multiplied on a channel by channel basis (after forming the complex 
conjugate for the channelised data from one telescope), using a complex vector muliplication, 
to yield the frequency domain complex visibilities that are
the fundamental observables of an interferometer.  This is repeated
for each selected band/polarisation on each selected baseline.  If dual
polarisations have been recorded for any given band, the cross-polarisation
terms can also be multiplied, recovering polarisation information for the target 
source.  

\subsubsection[Integration of correlated output]{Integration of correlated output}
Once the above cycle of operations has been completed, it is repeated
and the resulting visibilities accumulated (complex added) until a set accumulation
time has been reached.  Generally, on each cycle the input time
increment is equal to the corresponding FFT length (twice the number of spectral points), 
but it is also possible to overlap FFTs.  This
allows more measurements of higher lags and greater sensitivity to
spectral line observations, at the cost of increased computation. At the Core node, this ``Short
Term Accumulation" (STA) is equal to the number of FFT windows sent by the Datastream nodes.
The STA results are then returned to the FxManager via a non--blocking MPI send, along
with ancillary information such as the number of ``good" samples for each baseline.  The
desired number of STA results are then accumulated at the FxManager node to reach the
desired integration time.

\subsubsection[Calibration for nominal telescope $T_{sys}$]{Calibration for nominal telescope $T_{sys}$}
Cross multiplication, accumulation and normalisation 
gives the complex cross power spectrum for each baseline, representing the correlated 
fraction of the geometric mean of the powers detected at each telescope.  
To obtain the correlated power in units of Jy, the cross power spectra (amplitude 
components) should be scaled by the geometric mean of the powers received at 
each telescope measured in Jy i.e. the $T_{sys}$ in Jy routinely measured at each antenna.  
Since the autocorrelation spectra are calculated concurrently with the cross--correlation
spectra, this correction can be made within the correlator, before  writing the visibilities to disk.
Calibration is applied with a vector multiplication to each array of visibility spectra.
Alternatively, the cross--correlation spectra can be simply be normalised by the number
of contributing samples, and the calibration of the data can be completed offline, once 
measured, rather than a priori, $T_{sys}$ information is available.  The application
of normalisation or a priori system calibration is a user selectable option in DiFX.

When the $T_{sys}$ for each telescope is applied to calibrate the visibilities, 
it is also necessary to apply a scaling factor to compensate for decorrelation due to 
the coarse quantisation.  This corrects the visibility amplitudes, but of course cannot 
recover the lost signal to noise. For the 2-bit data typically processed, this scaling factor 
is 1/0.88 in the low-correlation limit \citep{cooper70a}.  The relationship becomes 
non-linear at high correlation and the scaling factor approaches unity as the 
correlation coefficient approaches unity.  When applied on--line at the correlator, the
low--correlation regime is assumed, and a correction for high-correlation cases 
can be applied in post-processing if necessary.

\subsubsection[Export of visibility data]{Export of visibility data}
Once an accumulation interval has been reached, the visibilities must
be stored in a useful format.  DiFX can produce 
RPFITS\footnote{http://www.atnf.csiro.au/computing/software/rpfits.html} format data,
or an intermediate binary format which can be translated offline into
FITS-IDI\footnote{http://www.aoc.nrao.edu/aips/FITS-IDI.html}.  RPFITS and FITS-IDI files can
be loaded into analysis packages such as AIPS\footnote{http://www.aoc.nrao.edu/aips}, 
CASA\footnote{http://casa.nrao.edu/}, or 
MIRIAD\footnote{http://www.atnf.csiro.au/computing/software/miriad} for data
reduction.  Ancillary information is included in the FITS file along with the complex visibilities, time stamps, and ($u$,$v$,$w$) coordinates.  Presently, only FITS-IDI output supports the storage of
visibility weights, which are adjusted based on the count of  ``good" samples detected in the
unpacking process.

\subsection[Special processing operations: pulsar binning]{Special processing operations: pulsar binning}
\label{difx:difx:pulsar}
As discussed in Chapter~\ref{pulsars}, 
pulsed signals are dispersed as they travel through the interstellar
medium (ISM), resulting in a smearing of the pulse arrival time in
frequency.  In order to
correct for the dispersive effects of the ISM, DiFX
employs incoherent dedispersion \citep{voute02a}.  This allows the
visibilities generated by the correlator to be divided into pulse
phase bins.  Unlike hardware correlators which typically allow only a
single on/off bin, or else employ $2^{\mathrm{N}}$ bins of fixed width, DiFX
allows an arbitrary number of bins placed at
arbitary phase intervals.  The individual bins can be written out separately in
the FITS file format to enable investigation of pulse phase dependent effects,
or can be filtered within the correlator to maximise signal to noise,
based on a priori pulse profile information.

To calculate which phase bin a visibility at a given frequency and
time corresponds to, the software correlator requires information on
the pulsar's ephemeris, which is supplied in the form of one or more
``polyco" files containing a polynomial description of apparent pulse
phase as a function of time.  These are generated using the pulsar
analysis program TEMPO\footnote{http://pulsar.princeton.edu/tempo/reference\_manual.html}, and require prior timing of a pulsar.  Additional software has been produced to
verify the pulsar timing by using the generated polyco files and the baseband data 
from an experiment, allowing phase bins to be accurately set before correlation.

For VLBI observations of pulsars, it is usually desirable
to maximise the signal to noise of the observations by binning the
visibilities based on the pulse phase, and applying a filter to the
binned output based on the signal strength in that phase.  Typically
this filter is implemented as a binary on/off for each phase bin.  Using the
pulse profile generated from the baseband data of an observation,
however, DiFX allows a user-specified number of
bins to be generated and a filter applied based on pulse strength $\times$ bin
width, allowing the maximum theoretical retrieval of signal, as described below.  This
also reduces the output data volume, since only an ``integrated
on-pulse'' visibility is retained, rather than potentially many phase
bins.

Consider observing a single pulse, divided into $M$ equally spaced phase bins.
Let the pulsar signal strength as a function of phase bin
be $S(m)$, and the noise in single phase bin be $Z \times \sqrt{M}$, where $Z$ is the baseline sensitivity
for an integration time of a single pulse period.  When all bins are summed (effectively no binning), the S/N ratio will be:
\begin{equation}
\frac{\sum_{m=0}^{M}S(m)}{Z}
\end{equation}
\noindent
as the signal adds coherently while the noise adds in quadrature.
For a simple on/off gate accepting only bins $m_{1}$ to $m_{2}$, the S/N ratio will be:
\begin{equation}
\frac{\sum_{m=m_{1}}^{m_{2}}S(m)}{\sqrt{\sum_{m=m_{1}}^{m_{2}}\left( Z\times\sqrt{M}\right)^{2} }}
\end{equation}
Finally, for the case where each bin is weighted by the pulse signal strength in that bin,
the S/N ratio will be:
\begin{equation}
\frac{\sum_{m=0}^{M}\left(S(m)\right)^{2}}{\sqrt{\sum_{m=0}^{M}\left( S(m) \times Z \times \sqrt{M}\right)^{2} }}
\end{equation}
For a Gaussian shaped pulse, this allows a modest improvement in recovered signal to 
noise of 6\% compared to an optimally placed single on/off bin.
On a more complicated profile, such as a Gaussian main pulse with a Gaussian interpulse at half the amplitude, the improvement in recovered signal to noise increases to 21\%.
Throughout the remainder of this thesis, the term gate, when referring to a DiFX--correlated 
pulsar experiment, refers to a matched pulse filter in the sense described above.

\subsection[Operating DiFX]{Operating DiFX}
DiFX is controlled via an interactive Graphical User Interface (GUI), which
calls the various component programs and helper scripts.  The primary purpose of the GUI is to
facilitate easy editing of the text files which configure the
correlator, run external programs such as the delay model generator, 
and provide feedback while a job is running.  
Two files are necessary to run the actual correlator program.  The first
is an experiment configuration file, containing tables of stations,
frequency setups, etc, analogous to a typical hardware correlator job configuration script.
The second file contains the list of compute nodes on which the
correlator program will run.

While it is possible to run all tasks required to operate
the correlator manually, in practise they are organised via the GUI.
This consists of running a series of helper applications
from the GUI to generate the necessary input for the correlator.
These include a script to extract experiment information from the VLBI
exchange (VEX) file used to configure and schedule the telescopes at observe time, a delay and 
($u$,$v$,$w$) generator which makes use of CALC 9, and scripts to extract the
current load of available nodes.  Pulsar--specific information such
as pulse profiles and bin settings can also be loaded.  This information is presented via
the GUI and adjustments to the configuration, such as selecting 
computational resources to be used, can be made before launching a correlation job.

In the future it is planned to incorporate some real-time feedback of
amplitude, phase and lag information from the current correlation via the GUI.
This would be similar to the visibility spectra displays available continuously
at connected-element interferometers.

Alternative interfaces for configuring and launching DiFX have been developed by
other users of the software, but are generally site--specific and interface to existing 
correlator control software.  The foremost example of this approach is the VLBA, 
which is described in Section~\ref{difx:additional:vlba}.

\section[Deployment and verification]{Deployment and verification}
\label{difx:deploy}
In order to verify that the DiFX correlator was functioning correcting, three separate comparisons were made against existing, well--verified hardware correlators.  In each case, the comparison showed 
complete agreement once differences in correlator architecture were accounted for.
These comparisons are detailed below.

\subsection[Comparison to the LBA S2 correlator]{Comparison to the LBA S2 correlator}
\label{difx:deploy:s2comp}
Observations to provide data for a correlator comparison between DiFX and the ATNF 
S2 correlator were undertaken on March 12, 2006, with the following subset of the LBA: 
Parkes (64 m), ATCA (phased array of 5 $\times$ 22 m), Mopra (22 m), Hobart (26 m).

Data from these observations were recorded simultaneously to S2 tapes and the LBADR 
disks (Phillips et al. 2008, in preparation) during a 20 minute period, UT 02:30--02:50, 
corresponding to a scan on a bright quasar (PKS 0208$-$512).  Two 16 MHz  right circular 
polarisation (RCP) bands were recorded, in the frequency ranges 2252 $-$ 2268 MHz and 
2268 $-$ 2284 MHz.

The data recorded on S2 tapes were shipped to the ATNF LBA S2 correlator \citep{wilson96a} at 
ATNF headquarters and processed.  The S2 correlator is an XF--style hardware correlator.
The data recorded to LBADR disks were shipped to the 
Swinburne University of Technology supercomputer and processed using the software correlator.

At both correlators identical $T_{sys}$ values in Jy were specified for each antenna and applied 
in order to produce nominally calibrated visibility amplitudes,  and both correlators used 
identical clock models, in the form of a single clock offset and linear rate as a function of time 
per antenna.  The data were processed at each correlator using a two second 
integration time and 32 spectral channels across each 16 MHz band.

Different implementations of the CALC-based delay generation were used at each correlator, 
and thus small differences exist in the applied delay models, which leads to differences in the 
correlated visibility phase.  These delay model differences were calculated and the 
phase due to differential delay model has been subtracted in the following discussion.

\begin{figure}[!t]
\includegraphics[angle=270,width=\textwidth]{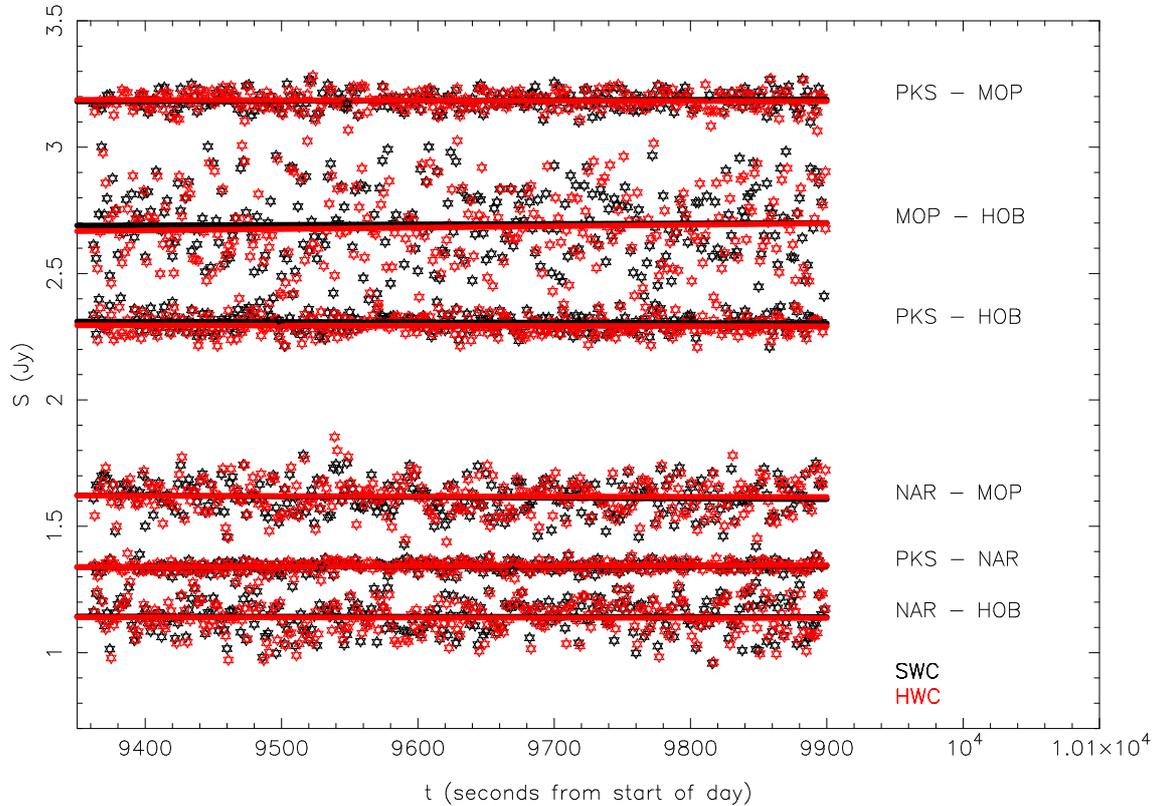}
\caption[Comparison of S2 and DiFX visibility amplitudes with time]
{S2 (red) and DiFX (black) visibility amplitude vs time for the 2252 -- 2268 MHz band on the source PKS 0208$-$512, as described in the text (PKS $=$ Parkes; MOP $=$ Mopra; HOB $=$ Hobart; NAR $=$ ATCA).  Symbols represent the actual visibilities produced by the correlators, while the lines represent linear least-squares fits to the visibilities (one line per dataset).}
\label{fig:s2amptime}
\end{figure}

\begin{figure}
\begin{center}
\includegraphics[angle=270,width=\textwidth]{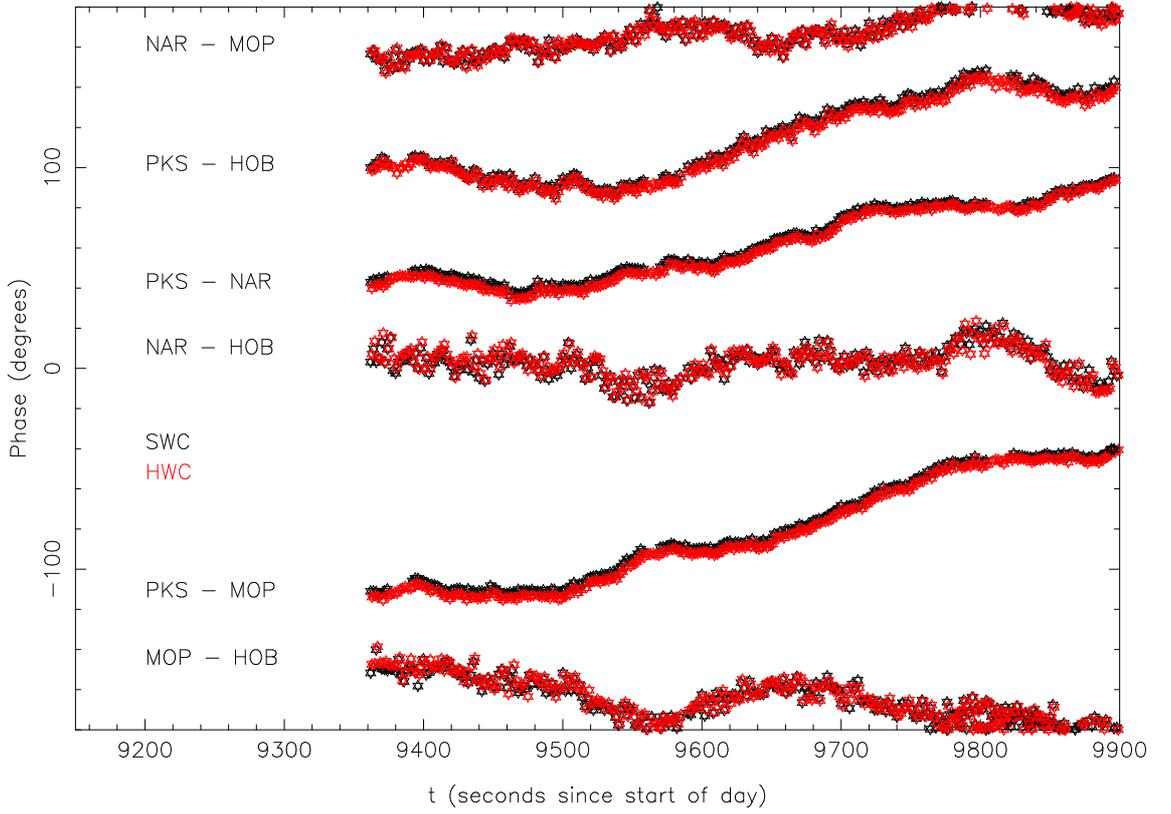}
\caption[Comparison of S2 and DiFX visibility phases with time]
{S2 (red) and DiFX (black) visibility phase vs time for the 2252 -- 2268 MHz band on the source 
PKS 0208$-$512, as described in the text.  Antenna labels are identical to 
Figure~\ref{fig:s2amptime} above.  The PKS-NAR baseline has been shifted by 
$-50\degrees$ for clarity.}
\label{fig:s2phasetime}
\end{center}
\end{figure}

\begin{figure}
\includegraphics[angle=270,width=\textwidth]{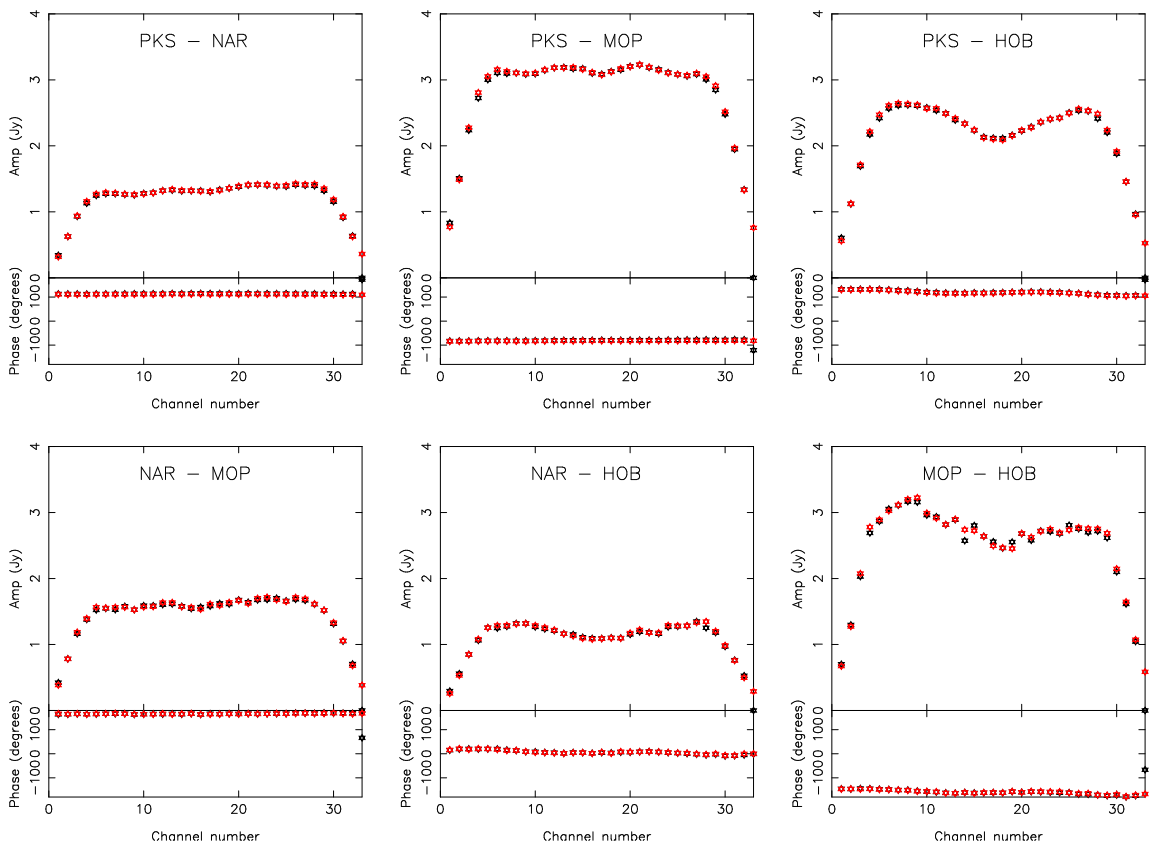}
\caption[Comparison of S2 and DiFX visibility amplitude and phase with frequency]
{S2 (red) and DiFX (black) visibility amplitude and phase vs frequency data for the 2252 -- 2268 MHz 
band on the source PKS 0208$-$512, as described in the text.  Antenna labels are identical to
Figure~\ref{fig:s2amptime} above.  The S2 data has been corrected for fractional-sample error 
decorrelation at the band edges as described in the text.}
\label{fig:s2ampfreq}
\end{figure}

From both correlators, RPFITS format data were output and loaded into the MIRIAD software package
\citep{sault95a} for inspection and analysis.  The data from the two correlators are compared 
in Figures \ref{fig:s2amptime} -- \ref{fig:s2ampfreq}.

Figure \ref{fig:s2amptime} shows the visibility amplitudes for all baselines from both correlators 
as a function of time, over the period 02:36:00 -- 02:45:00 UT, for one of the 16 MHz bands 
(2252 $-$ 2268 MHz).  These amplitudes represent the vector averaged data over the 
frequency channel range 10 $-$ 21 (to avoid the edges of the band).  The data for each 
baseline were fit to a first order polynomial model ($S(t)=\frac{dS}{dt}t + S_{0}$, where 
$S$ is the flux density in Jy, $t$ is the offset in seconds from UT 02:40:30, and $S_{0}$ 
is the extrapolated flux density at time UT 02:40:30, using a standard linear least squares 
routine.  The root mean square (RMS) variation around the best fit model was calculated 
for each baseline.  The fitted models are shown in Figure \ref{fig:s2amptime} and show 
no significant differences between the S2 correlator and the software correlator.  Further, 
the calculated RMS for each baseline agrees very well between DiFX and the S2 
correlator, as summarised in Table \ref{tab:s2comp}.

Figure \ref{fig:s2phasetime} shows the visibility phase as a function of time for each of the six 
baselines in the array.  Again the data represent the vector averaged correlator output over 
the frequency channel range 10 $-$ 21 within the 2252 $-$ 2268 MHz band.  As discussed 
above, small differences between the delay models used at each correlator have been taken 
into account as part of this comparison.  

\begin{deluxetable}{ccccc}
\tablecaption{Linear fit parameters for visibility amplitude vs time for DiFX and the LBA S2 correlator, with 95\% confidence limits}
\tabletypesize{\tiny}
\tablewidth{0pt}
\tablehead{
\colhead{Baseline} & \colhead{Offset$_{\rm DiFX}$ (Jy)} & \colhead{Offset$_{\rm LBA}$ (Jy)}  & \colhead{Slope$_{\rm DiFX}$ ($\mu \rm Jy$ s$^{-1}$)} &\colhead{Slope$_{\rm LBA}$ ($\mu \rm Jy$ s$^{-1}$)}
}
\startdata
PKS - NAR & $1.341 \pm 0.030$ & $1.343 \pm 0.028$ & $\phantom{-}10 \pm \phantom{0}13$ & $\phantom{-}14 \pm \phantom{0}12$\\
PKS - MOP & $3.185 \pm 0.058$ & $3.185 \pm 0.063$ & $\phantom{-}14 \pm \phantom{0}24$ & $-11 \pm \phantom{0}26$\\
PKS - HOB & $2.307 \pm 0.058$ & $2.293 \pm 0.061$ & $-12 \pm \phantom{0}24$  & $-\phantom{0}6 \pm \phantom{0}24$\\
NAR - MOP & $1.616 \pm 0.109$ & $1.619 \pm 0.114$ & $-27 \pm \phantom{0}43$ & $-10 \pm \phantom{0}45$\\
NAR - HOB & $1.142 \pm 0.111$ & $1.139 \pm 0.116$ & $-\phantom{0}3 \pm \phantom{0}44$ & $-\phantom{0}5 \pm \phantom{0}46$\\
MOP - HOB & $2.694 \pm 0.256$ & $2.681 \pm 0.257$ & $\phantom{-}18 \pm 101$ & $\phantom{-}56 \pm 101$\\
\enddata
\label{tab:s2comp}
\end{deluxetable}

Figure \ref{fig:s2ampfreq} shows a comparison of the visibility amplitudes and phases as a 
function of frequency in the 2252 $-$ 2268 MHz band.  The data represented here result from 
a vector average of the two datasets over a two minute time range, UT 02:40:00 $-$ 02:42:00.  
Since the S2 correlator is an XF--style correlator, it cannot exactly correct fractional sample 
error in the same manner as an FX correlator such as DiFX, as the channelisation is performed
after accumulation.  The coarse (post--accumulation) fractional sample correction leads to 
decorrelation at all points except the band center, up to a maximum of $\sim10\%$ at the band 
edges on long baselines where the geometric delay changes by a sample or more over an
integration period. Compensation has been performed for this band edge decorrelation 
in the S2 correlator amplitudes shown in Figure \ref{fig:s2ampfreq}.

\subsection[Comparison to the VLBA correlator]{Comparison to the VLBA correlator}
Data obtained as part of a regular series of VLBA test observations were used as a basis for 
a correlator comparison between DiFX and the VLBA correlator, which
is an FX--style hardware correlator
\citep{napier94a}.  The observations were made on August 5, 2006 using the Brewster, 
Los Alamos, Mauna Kea, Owens Valley, Pie Town, and Saint Croix VLBA stations.  
One bit digitised data sampled at the Nyquist rate for four dual polarisation bands, 
each of 8 MHz bandwidth, were recorded using the Mk5 system \citep{whitney03a}.  
The four bands were at centre frequencies of 2279.49, 2287.49, 2295.49, and 2303.49 MHz.  
The experiment code for the observations was MT628 and the source observed was
0923$+$392, a strong and compact active galactic nucleus.  Approximately two 
minutes of data recorded in this way was used for the comparison.

The Mk5 data were correlated on the VLBA correlator and exported to FITS format files. 
The data were also shipped to the Swinburne supercomputer and correlated using DiFX,
producing RPFITS format files.  In both cases, no 
scaling of the correlated visibility amplitudes by the system temperatures were made at the 
correlators.  The visibilities remained in the form of correlation coefficients for the purposes 
of the comparison -- i.e. a system temperature of unity was used to scale the amplitudes.  
Each 8 MHz band was correlated with 64 spectral points, and an integration time of 2.048 
seconds was used.

The VLBA correlator data were read into AIPS using FITLD with the parameter DIGICOR$=$1.  
The DIGICOR parameter is used to apply certain scalings to the visibility amplitudes for data 
from the VLBA correlator.  Further, to obtain the most accurate scaling of the visibility amplitudes, 
the task ACCOR was used to correct for imperfect sampler thresholds, deriving corrections to 
the antenna-based amplitudes of $\sim0.5\%$.  These ACCOR corrections were applied to 
the data, which were then written to disk in FITS format.

The software correlator data were read directly into AIPS and then written to disk in the 
same FITS format as the VLBA correlator data.  No corrections to amplitude or phase of 
the software correlated data were made in AIPS.

\begin{figure}[t!]
\includegraphics[angle=270,width=\textwidth]{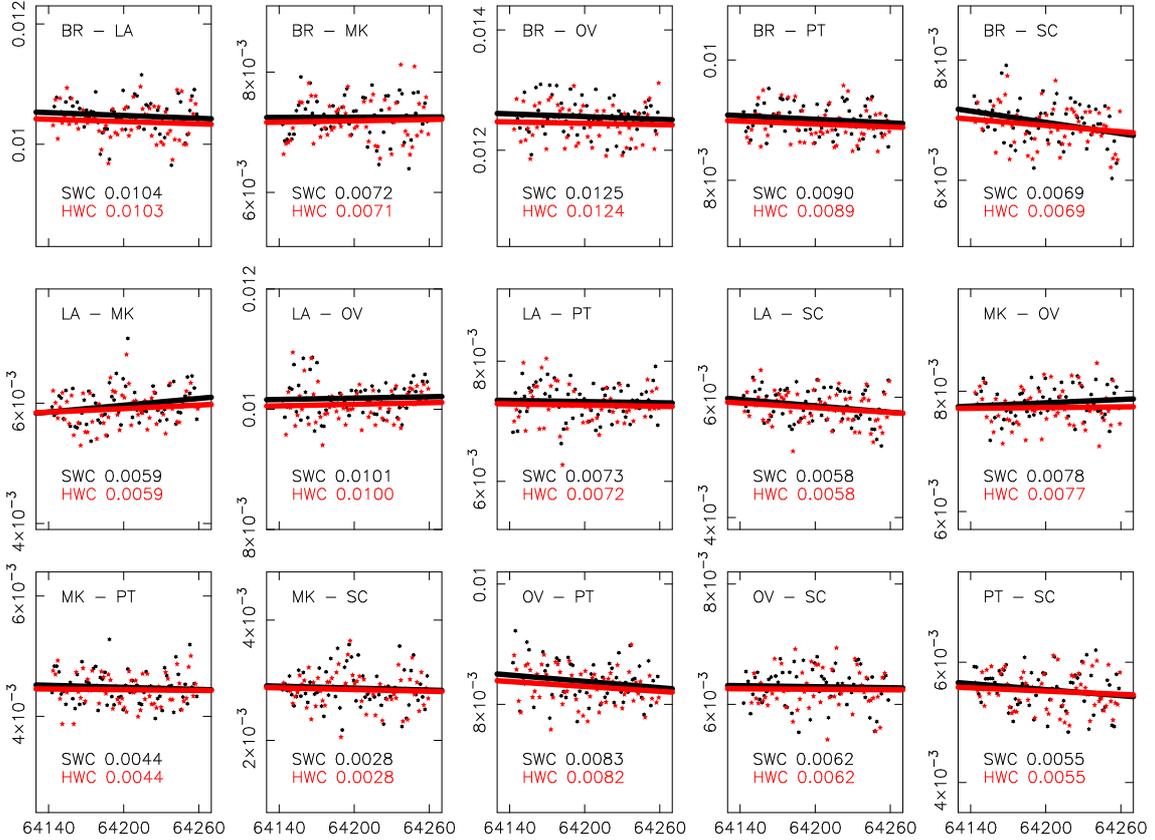}
\caption[Comparison of VLBA and DiFX visibility amplitudes with time]
{VLBA correlator (red) and DiFX (black) visibility amplitude vs time for the 2283.49 -- 2291.49 RCP band from the VLBA test observation MT628, as described in the text.  The units of time are seconds from UT 00:00:00, and the amplitude scale is correlation coefficient.  Symbols represent the actual visibilities produced by the correlators, while the lines represent linear least-squares fits to the visibilities.  The text annotation on each panel lists the average correlation coefficient amplitude for each correlator over the time period, as tabulated in Table \ref{tab:vlbacomp}.}
\label{fig:vlbaamptime}
\end{figure}

The VLBA correlator data and the software correlator data were both imported into 
MIRIAD for inspection and analysis, using the same software as used for the comparison 
with the LBA correlator described above.  RCP from the 2283.49 -- 2291.49 MHz band over 
the time range UT 17:49:00 $-$ 17:51:00 was used in all comparison plots below.

Since the delay models used by the VLBA and software correlators differ at the picosecond 
level, as is the case for the comparison with the LBA data in 
Section~\ref{difx:deploy:s2comp}, differences 
in the visibility phase exist between the correlated datasets.  As with the LBA comparison, 
compensation has been performed for the phase errors due to the delay model differences 
in the following comparison.

\begin{figure}
\includegraphics[angle=270,width=\textwidth]{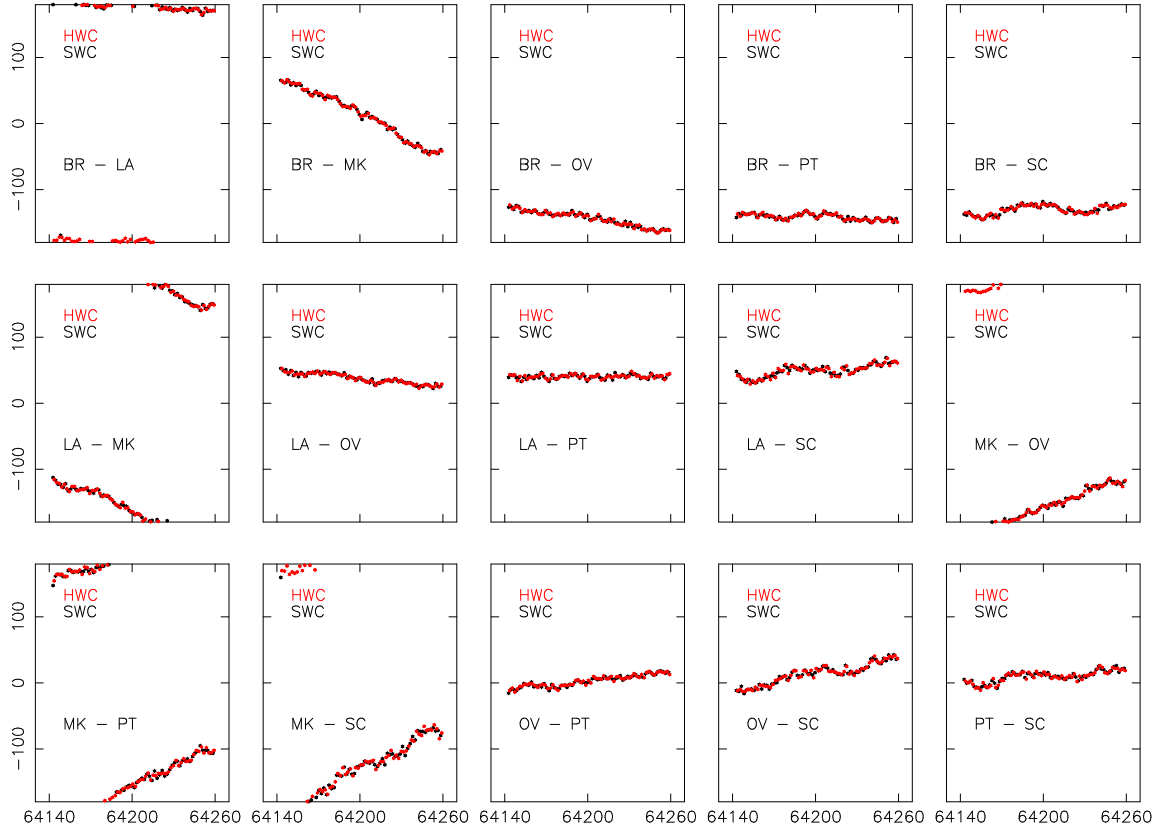}
\caption[Comparison of VLBA and DiFX visibility phases with time]
{VLBA correlator (red) and DiFX (black) visibility phase vs time for the 2283.49 -- 2291.49 RCP band from the VLBA test observation MT628, as described in the text.  The units of time are seconds from UT 00:00:00, and phase is displayed in degrees.}
\label{fig:vlbaphasetime}
\end{figure}

\begin{deluxetable}{ccccc}
\tablecaption{Linear fit parameters for visibility amplitude (in units of correlation coefficient) vs 
time for DiFX and the VLBA correlator, with 95\% confidence limits}
\tabletypesize{\tiny}
\tablewidth{0pt}
\tablehead{
\colhead{Baseline} & \colhead{Offset$_{\rm DiFX}$} & \colhead{Offset$_{\rm VLBA}$}  & \colhead{Slope$_{\rm DiFX}$ ($s^{-1} \times 10^{-6}$)} &\colhead{Slope$_{\rm VLBA}$ ($s^{-1} \times 10^{-6}$)}
}
\startdata
BR - LA & $0.0104 \pm 0.0004 $& $0.0103 \pm 0.0005$ & $-0.8 \pm 1.7$ & $-0.9 \pm 1.7$ \\
BR - MK & $0.0072  \pm 0.0005 $& $0.0071 \pm 0.0006$ & $\phantom{-}0.1 \pm 1.8$ & $\phantom{-}0.5 \pm 2.0$ \\
BR - OV & $0.0125  \pm 0.0005 $& $0.0124 \pm 0.0005$ & $-0.7 \pm 1.7$ & $-0.5 \pm 1.8$ \\
BR - PT & $0.0090  \pm 0.0004 $& $0.0089 \pm 0.0004$ & $-1.0 \pm 1.3$ & $-1.2 \pm 1.5$ \\
BR - SC & $0.0069  \pm 0.0005 $& $0.0069 \pm 0.0005$ & $-3.1 \pm 2.0$ &  $-2.5 \pm 1.8$ \\
LA - MK & $0.0059  \pm 0.0005 $& $0.0059 \pm 0.0005$ & $\phantom{-}1.9 \pm 1.7$ &   $\phantom{-}1.4 \pm 1.7$ \\
LA - OV & $0.0101  \pm 0.0005 $&  $0.0100 \pm 0.0005$ &  $\phantom{-}0.4 \pm 1.7$ & $\phantom{-}0.6 \pm 1.7$ \\
LA - PT & $0.0073  \pm 0.0005 $&  $0.0072 \pm 0.0005$ & $-0.3 \pm 1.7$ & $-0.5 \pm 1.8$ \\
LA - SC & $0.0058  \pm 0.0004 $& $0.0058 \pm 0.0004$ & $-1.8 \pm 1.5$ & $-1.9 \pm 1.5$ \\
MK - OV & $0.0078  \pm 0.0004 $&  $0.0077 \pm 0.0005$ & $\phantom{-}0.9 \pm 1.5$ & $\phantom{-}0.3 \pm 1.8$ \\
MK - PT & $0.0044  \pm 0.0004 $& $0.0044 \pm 0.0004$ & $-0.6 \pm 1.7$ & $ -0.3 \pm 1.5$ \\
MK - SC & $0.0028  \pm 0.0005 $&$ 0.0028 \pm 0.0005$ & $-0.6 \pm 1.8$ & $ -0.7 \pm 1.7$ \\
OV - PT & $0.0083  \pm 0.0005 $& $0.0082 \pm 0.0005$ & $-1.8 \pm 1.8$ & $ -1.9 \pm 1.7$ \\
OV - SC & $0.0062  \pm 0.0005 $& $0.0062 \pm 0.0005$ &$-0.3 \pm 1.8$ & $ -0.2 \pm 1.8 $\\
PT - SC & $0.0055  \pm 0.0005 $& $0.0055 \pm 0.0005$ & $-1.7 \pm 2.0$ & $ -1.3 \pm 1.8$ \\
\enddata
\label{tab:vlbacomp}
\end{deluxetable}

\begin{figure}
\includegraphics[angle=270,width=\textwidth]{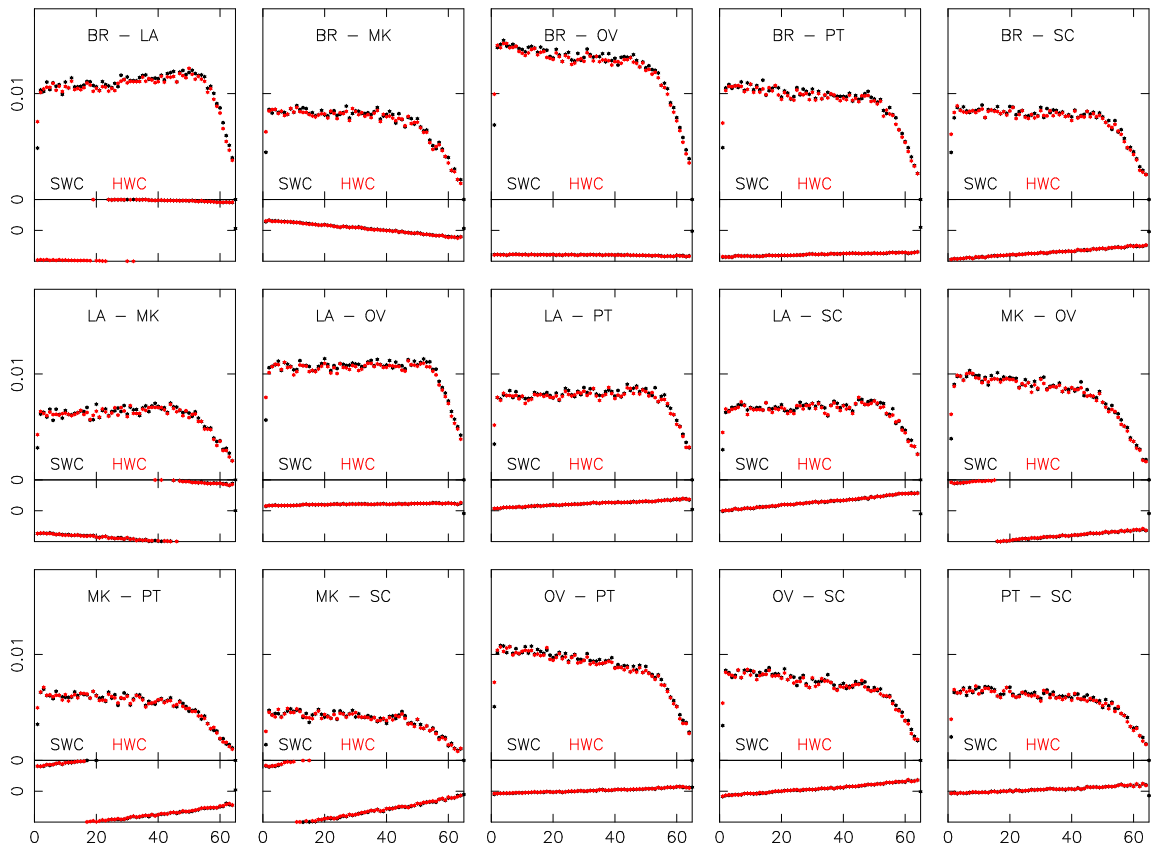}
\caption[Comparison of S2 and DiFX visibility amplitude and phase with frequency]
{VLBA correlator (red) and DiFX (black) visibility amplitude and phase as a function of frequency for the 2283.49 -- 2291.49 RCP band from the VLBA test observation MT628, as described in the text.  The vertical scale for correlation coefficient amplitude on each panel is 0 -- 0.018, while the phase scale spans $\pm 180\degrees$.  The horizontal scale for each panel displays channels 0--64.}
\label{fig:vlbaampfreq}
\end{figure}

Figure \ref{fig:vlbaamptime} shows the visibility amplitudes for all baselines from both correlators 
as a function of time.  These amplitudes represent the vector averaged data over the frequency 
channel range 10 $-$ 55 (to avoid the edges of the band).  The data for each baseline were fit 
to a first order polynomial model ($S(t)=\frac{dS}{dt}t + S_{0}$, where $S$ is the correlation 
coefficient, $t$ is the offset in seconds from UT 17:50:00, and $S_{0}$ is the extrapolated 
correlation coefficient at time UT 17:50:00) using a standard least squares regression.  
The RMS variation around the best fit model was calculated for each 
baseline.  The fitted models are shown in Figure \ref{fig:vlbaamptime} and show no significant 
differences between the VLBA correlator and DiFX.  Further, the calculated 
RMS for each baseline agrees very well between the VLBA correlator and DiFX.  
The results of the comparison are summarised in Table \ref{tab:vlbacomp}.

Figure \ref{fig:vlbaphasetime} shows the visibility phase as a function of time for each of the 
fifteen baselines in the array.  Again, the data represents the vector averaged correlator output 
over the frequency channel range 10 $-$ 55 within the band.  As discussed above, small 
differences between the delay models used at each correlator cause phase offsets between
the two correlators, and have been taken into account as part of this comparison. 

Figure \ref{fig:vlbaampfreq} shows a comparison of the visibility amplitudes and phases as a 
function of frequency in the band.  The data represented here results from a vector average of 
the two datasets over a two minute time range.  Figures \ref{fig:vlbaamptime}, \ref{fig:vlbaphasetime} 
and \ref{fig:vlbaampfreq} show that the results obtained by the VLBA correlator and DiFX agree 
to within the RMS errors of the visibilities in each case, as expected.  Since the VLBA correlator is
an FX--style correlator, like DiFX, there is no requirement to correct for band--edge decorrelation
as was required for the LBA correlator.

\subsection[Comparison to the MPIfR geodetic correlator]{Comparison to the MPIfR geodetic correlator}
\label{difx:deploy:bonncomp}
Unlike the S2 and VLBA correlators, the MarkIV correlator operated by the Max Planck 
Institut f\"{u}r Radioastronomie (MPIfR) in Bonn, Germany, is used primarily for geodetic
observations.  The MarkIV correlator, an XF--type hardware correlator \citep{whitney93a}, has been 
used heavily for geodetic VLBI processing for many years as part of the International 
VLBI Service (IVS; http://ivscc.gsfc.nasa.gov/) and is one of a very small number of 
trusted geodetic correlators around the world.

The comparison with this correlator is described in detail in \citet{tingay08a}, and the results
are summarised here.
The data used for the correlator comparison was a subset of the data collected for a geodetic
experiment (BM261) conducted with the NRAO VLBA on 2007 July 03.  One minute of data 
(11:23:00 UT $-$ 11:24:00 UT) from four antennas were selected for the purposes of the 
comparison:  Fort Davis (FD), Pie Town (PT), Owens Valley (OV), and Kitt Peak (KP).  
8 dual polarisation bands of bandwidth 8 MHz were recorded, with the following centre frequencies:
1350.55, 1358.55, 1366.55, 1374.55, 1382.55, 1390.55, 1398.55, 1406.55 MHz.  
For the comparison, the RCP data from the 1350.55 MHz band was used.  

The data at the VLBA antennas were recorded to Mark5 disk packs and transported to the MarkIV correlator at MPIfR, where correlation was performed with the MarkIV correlator.  
The comparison subset of data was then exported from the Mark5 units to
a standard filesystem and correlated using DiFX on a commodity cluster of linux machines
at MPIfR.  The data were correlated with 128 frequency channels across the band.
Prior to correlation, a clock model was derived for the experiment and applied identically
for each correlator.  As with the previous two comparisons, however, geometric models were derived
independently, and the model difference was calculated and subtracted from the following analysis.

\begin{figure}
\begin{center}
\includegraphics[angle=270,width=\textwidth]{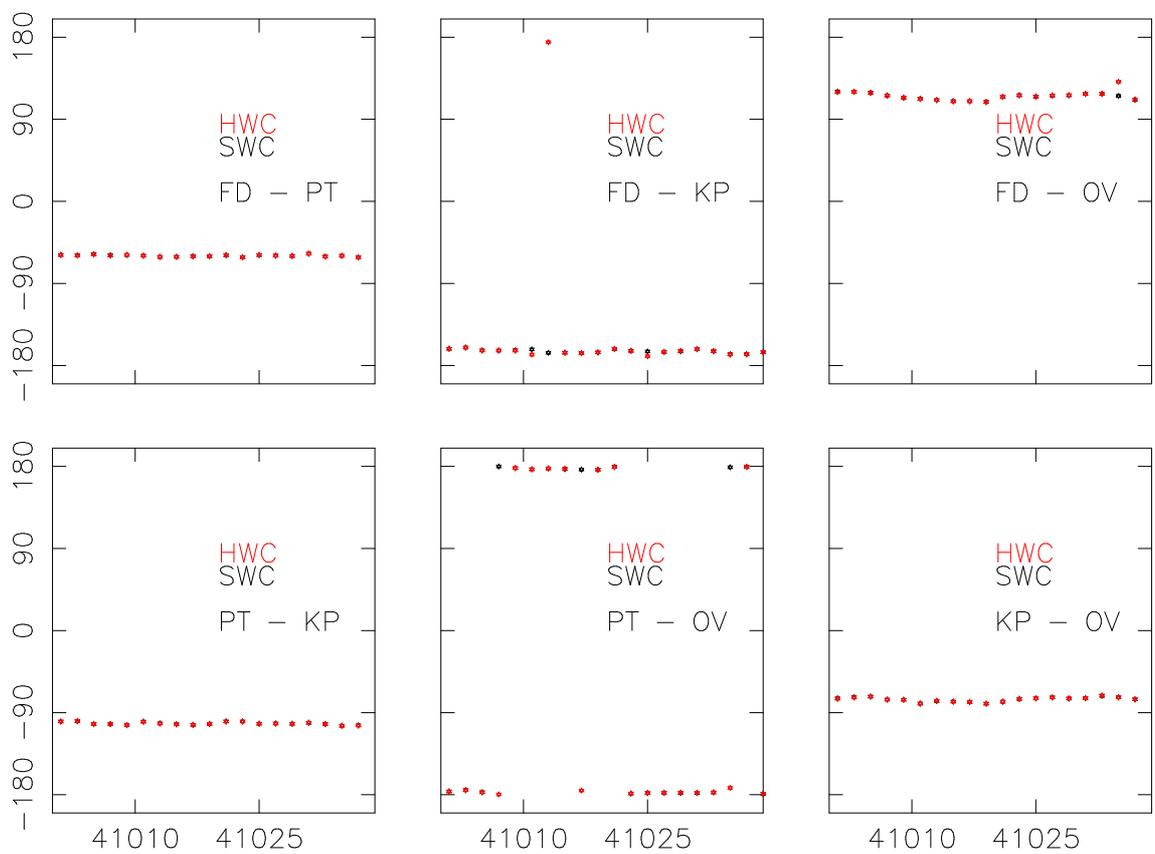}
\caption[Comparison of Bonn MkIV and DiFX visibility amplitudes with time]
{Geodetic correlator (red) and DiFX (black) visibility amplitude vs time for the 1346.55 -- 1354.55
RCP band from the geodetic observation BM261, as described in the text.  The units of time are seconds from UT 00:00:00, and the phase is displayed in degrees.}
\label{fig:bonnphasetime}
\end{center}
\end{figure}

\begin{figure}
\begin{center}
\includegraphics[angle=270,width=\textwidth]{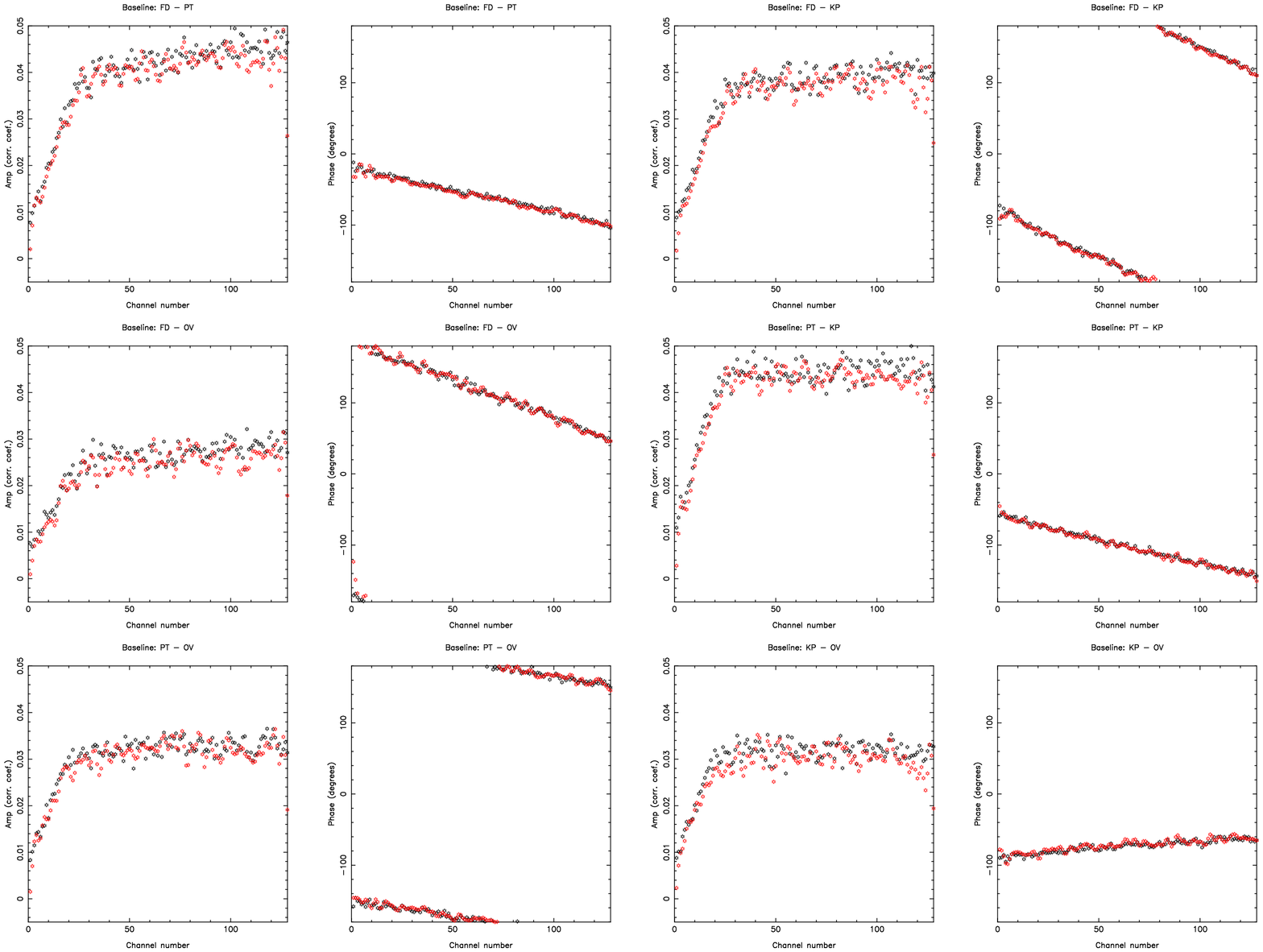}
\caption[Comparison of Bonn MkIV and DiFX visibilities with frequency]
{Geodetic correlator (red) and DiFX (black) visibility amplitude and phase as a function of frequency for the 1346.55 -- 1354.55
RCP band from the geodetic observation BM261, as described in the text.  The vertical scale for correlation coefficient amplitude on each panel is 0 -- 0.05, while the phase scale spans 
$\pm 180\degrees$.  The horizontal scale for each panel displays channels 0--128.}
\label{fig:bonnampfreq}
\end{center}
\end{figure}

Figure \ref{fig:bonnphasetime} shows the visibility phase as a function of time for each of the 
six baselines correlated.  As with the previous comparisons, 
the data represents the vector averaged correlator output from the centre of the band
(frequency channels 32 $-$ 96), and the phase offsets due to small 
differences between the delay models used at each correlator have been 
corrected in this analysis.
Figure \ref{fig:bonnampfreq} shows a comparison of the visibility amplitudes and phases as a 
function of frequency in the band.  The data represented here results from a vector average of 
the two datasets over the one minute comparison time range.

Figures \ref{fig:bonnphasetime} and \ref{fig:bonnampfreq} show that the results obtained by 
the MarkIV correlator and DiFX are in good agreement.  Since the MarkIV correlator is used 
primarily for geodesy and the amplitude correction signal path is not well documented, no attempt
has been made to improve the a priori amplitude corrections in the manner of the VLBA correlator 
comparison.  As the MarkIV correlator is an XF--style correlator
and the visibilities were not corrected for fractional--sample decorrelation at the band edge,
a noticeable difference can be seen at the band edge in amplitudes.  
Figure~\ref{fig:bonnphasetime} also shows several glitches in visibility phase in the MarkIV 
correlator output.  The origin of these glitches is unknown, but they are almost certainly 
an error in the MarkIV correlator output, as no glitches are seen to occur on correlators simultaneously
and the magnitude of the phase jump is many times the phase RMS. A possible cause of the phase
jumps is an error on the Mark5 Station Unit during playback at the correlator 
(C. Phillips, private communication).

\section[Performance benchmarking]{Performance benchmarking}
\label{difx:perf}
In order to keep every compute node used in the correlation fully loaded,
they must be kept supplied with raw data.  If this condition is satisfied, the
correlation is CPU--limited, and the addition of further nodes will result
in a linear performance gain.  In practise, however, at some
point obtaining data from the data source (network socket or disk) and 
transmitting it across the local network to the
processing nodes will no longer occur quickly enough,
and the correlation becomes data--limited rather
than CPU--limited.  Correct selection of correlation parameters, and
good cluster design, will minimise the networking overhead imposed on
a correlation job, and ensure that all compute nodes are fully utilised.  This is discussed in 
Section~\ref{difx:perf:network} below,
and performance profiles for the CPU-limited case are presented in Section~\ref{difx:perf:cpu}.

\subsection[Networking considerations]{Networking considerations}
\label{difx:perf:network}
As described in Section~\ref{difx:difx:code}, double-buffered communications to the processing 
nodes are used to ensure that nodes are never idle as long as sufficient aggregate 
networking capability is available.  The use of MPI communications adds a small 
but unavoidable overhead to data transfer, meaning the maximum throughput of the system 
is slightly less than the maximum network capacity on the most heavily loaded data path.

There are two significant data flows: out of each Datastream and into
the FxManager.  For any high speed correlation, there will be more
Core nodes than Datastream nodes, so the aggregate rate into a Core will be lower
than that out of a Datastream.  The flow out of a Core is a factor of $N_{\rm cores}$ times
lower than that into the FxManager node.

If processing in real time (when processing time equals observation time), the rate
out of each Datastream will be equal to the recording rate, which can be up to 1 Gbps with
modern VLBI arrays and is within the capabilities of modern commodity ethernet equipment.
The rate into the FxManager node will be equal to the product of the recording rate, the
compression ratio, and the
number of Cores, where the compression ratio is the ratio of data into
a Core to data out of a Core.  This is determined by the number of
antennas (since number of baselines scales with number of antennas
squared), the number of channels in the output cross-power spectrum, the number
of polarisation products correlated, and the integration time used before sending data back to
the FxManager node.

It is clearly desirable to maximise the size of data messages sent to a
core for processing, since this minimises the data rate into the
FxManager node for a given number of Cores.  However, if the messages are too large,
performance will suffer as RAM capacity is exceeded.  Network latency
may also become problematic, even with buffering.  Furthermore, it
should be apparent that in this architecture, the Cores act
as short-term accumulators (STAs), with the FxManager performing the
long term accumulation.  The length of the STA sets the minimum
integration time.  It is important to note, however, that the STA
interval is entirely configurable in the software correlator, to be as
short as a single FFT, although network bandwidth and latency are likely to be limiting
factors in this case.

For the majority of experiments it is possible to set a STA length
which satisfies all the network criteria and allows the Cores to be
maximally utilised.  For combinations of large numbers of
antennas and very high spectral and time resolution, however, it is
impossible to set an STA which allows a satisfactorily low return data
rate to the FxManager node.  In this case, real time processing
of the experiment is not possible without the installation of 
additional network and/or CPU capacity on the FxManager node.

It is important to emphasise that although it is possible to find
experimental configurations for which DiFX suffers
a reduction in performance, these configurations would be impossible
on existing hardware correlators.  If communication to the FxManager node
is limiting performance, it is also possible to parallelize a disk-based experiment by
dividing an experiment into several time ranges and processing these time ranges
simultaneously, allowing an aggregate processing rate which equals
real time.  This is actually one of the most powerful aspects of the
software correlator, and one which would allow scheduling of
correlation to always ensure the cluster was being fully utilised.

\subsection[CPU-limited performance]{CPU-limited performance}
\label{difx:perf:cpu}
\begin{figure}[t!]
\begin{center}
\includegraphics[angle=270,width=0.85\textwidth]{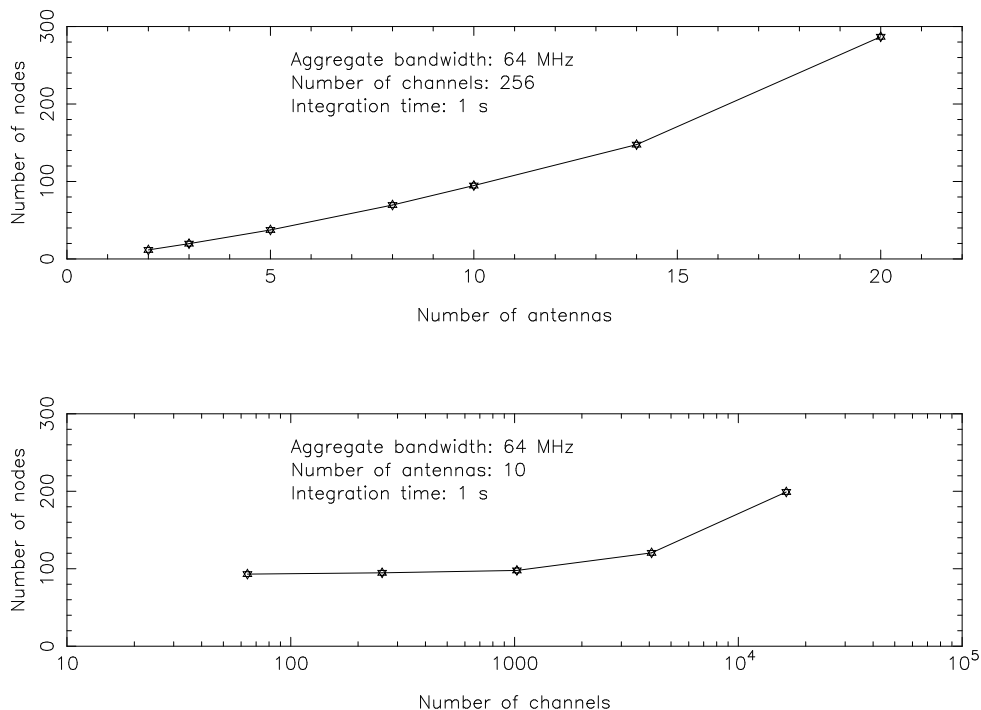}
\caption[Performance benchmarking of DiFX]
{Benchmark data showing the computational requirements of DiFX to correlate in real--time, as described in the text.  The nodes are single core 3.2 GHz Pentium processors with 1 GB RAM, and in both benchmarks 64 MHz of total bandwidth per station was correlated with a 1 second integration period.  Top panel shows the scaling of computational requirements with number of antenna, using 256 spectral points per 8 MHz subband.  Bottom panel shows the scaling of computional requirements with spectral points per subbband for a ten station array.}
\label{fig:benchmarks}
\end{center}
\end{figure}

Figure \ref{fig:benchmarks} shows the results of performance testing on the Swinburne
cluster available in 2006 (3.2 GHz Pentium 4 machines,Gigabit ethernet network) for different
array sizes and spectral resolutions.  The results shown in Figure \ref{fig:benchmarks} were obtained for data for which the aggregate bandwidth was 64 MHz, broken up into 8 bands each of 8 MHz bandwidth (4 $\times$ dual polarisation 8 MHz bands: data were 2-bit sampled: antenna data rate 256 Mbps).  Node requirements for real-time operation are extrapolated from the compute time on an 8 node cluster.  The correlation integration time is 1 second and all correlations provide all four polarisation products.
RAM requirements per node ranged from 10 -- 50 MB depending on spectral resolution, showing
that large amounts of RAM are unnecessary for typical correlations.
It can be seen that even a modestly sized commodity cluster can 
process a VLBI-sized array in real time at currently available data rates.

Recently, the Swinburne supercomputer cluster has been upgraded to multicore, 64 bit machines.
Testing on this architecture has shown that the correlation code performance 
scales near--linearly with the number of threads used per node (up to the number of CPU cores),
as expected for this embarrassingly parallel problem.  Data rates of up to 1 Gbps per antenna have
been correlated in real--time on these new machines, in the eVLBI experiments described
in Section~\ref{difx:additional:evlbi}.

\section[Additional applications]{Additional applications}
\label{difx:additional}
Whilst DiFX was developed with the primary aim of servicing the needs of the LBA, every effort was
made during development to ensure that DiFX would be adaptable to the needs of other 
interferometers that required an upgrade path from an existing correlator.  This approach has 
enabled the use of DiFX to correlate data from every major VLBI array in the world, as well as
facilitating the rapid development of several new instruments.  Those science highlights
which are external to this thesis are summarised briefly below.

\subsection[LBA eVLBI]{LBA eVLBI}
\label{difx:additional:evlbi}
Operational eVLBI modes have been tested using DiFX, using real-time data 
from the three ATNF telescopes (Parkes, ATCA, and Mopra), the University of Tasmania telescope
in Hobart, and international telescopes including the Kashima 34m telescope in Japan, and
the Shanghai 25m telescope in China. These observations have seen data rates up to 1 Gbps
per antenna using Australian--only antennas, and 512 Mbps per antenna 
in international experiments, equaling
the highest data rate experiments performed with the EVN \citep[highest published data
rate 256 Mbps\footnote{for unpublished higher data rate results, see 
http://www.expres-eu.org/512Mbps\_6tel.html};][]{paragi07a}.  Correlation has taken place
using computing resources at Swinburne University, the ATNF, and the University of 
Western Australia in Perth (a Cray XD-1 utilising Opteron processors and on-board Xilinx FPGAs).  
A more extensive description of the eVLBI activities of the LBA using DiFX can be found
in \citet{phillips07a}.

First science results from these eVLBI experiments have recently been published, and include
observations of the Galactic x--ray binary Circinus X--1 \citep{phillips07a} and the
supernova remnant of SN1987A \citep{tingay08b}.

\subsection[VLBA sensitivity upgrade]{VLBA sensitivity upgrade}
\label{difx:additional:vlba}
The VLBA is the only full--time VLBI instrument in the world, and has been responsible for much
of the rapid advance in VLBI science over the past decade.  When designed, it used the Mark4
tape system \citep{whitney93a}, and it has been upgraded to use the Mark5A disk--based
system \citep{whitney03a}.  Recently, a major upgrade to VLBA sensitivity has been planned, aiming 
to reach data rates of 4 Gbps, an 8--fold increase over the current maximum data rate.  
Information on the planned sensitivity upgrade is available at 
\verb+http://www.vlba.nrao.edu/memos/sensi/+.
In order to reach these increased data rates, the VLBA requires both a new digital backend and
a new correlator.  The digital backend, named Mark5C, is under development, and DiFX
has been chosen as the upgrade path for the correlator.  ``First light" with VLBA data on
a local installation of DiFX has been obtained in 2008, and production usage is expected to 
commence in late 2008.  Support and configuration tools have been developed to allow DiFX
to function as a ``drop--in" replacement for the existing hardware correlator, allowing
duplication of existing functionality at the same time as providing new capabilities.

\subsection[High Sensitivity Array observations of pulsar scintillation]{High Sensitivity Array observations of pulsar scintillation}
\label{difx:additional:scint}
As discussed in Chapter~\ref{pulsars}, the inhomogeneous distribution of the ISM leads to
diffractive and refractive scattering of signals from compact sources such as pulsars.  Whilst
single--dish studies of pulsar scintillation have already shed considerable light on the nature of the
structures in the ISM responsible for this scintillation 
\citep[see e.g.][and references therein]{cordes06a}, high frequency resolution VLBI observations of 
scintillating pulsars offer the possibility of directly imaging the ISM structures
at extremely high spatial resolution.

\begin{figure}
\includegraphics[width=1.0\textwidth]{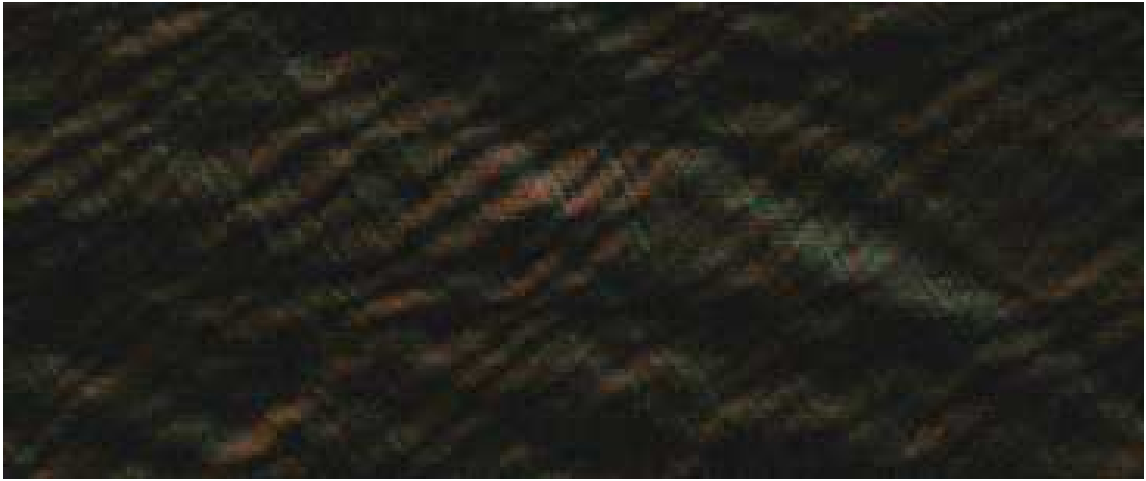}
\caption[Cross--power dynamic spectrum for the pulsar B0834$-$04 on the GBT--Arecibo baseline, 
correlated using DiFX]
{The cross--power dynamic spectrum showing scintillation variations for the pulsar B0834$-$04 on the Green Bank Telescope -- Arecibo baseline.  Brightness represents the visibility amplitude and colour represents the visibility phase.  Increasing frequency runs left to right and increasing time runs top to bottom.  This section of the dynamic spectrum represents just 5\% of the time span and 0.5\% of the bandwidth of the observation (330 seconds and 160 kHz).}
\label{fig:scintillation}
\end{figure}

The capabilities of DiFX for this type of analysis have recently been demonstrated with 
observations of PSR B0834$-$04 in 2005 (Brisken et al., in preparation).  
Elements of the High Sensitivity Array (HSA), comprising the NRAO Green Bank Telescope 
(GBT; 100 m), Westerbork (14 $\times$ 25 m), Jodrell Bank (76 m), and Arecibo (305 m) 
were used to provide an ultra-sensitive array at 327 MHz.  The data were recorded using the 
Mk5 system and correlated on a prototype implementation of DiFX.  The main requirement
 on the correlation was 250 Hz wide frequency channels,
over the broadest bandwidth available, to maximise signal to noise.  
For these observations a 32 MHz band was available.  The DiFX correlation used 
131,072 frequency channels, obtaining a frequency resolution of 
244 Hz.  Such extreme frequency resolution for a continuum experiment is beyond
the capabilities of any existing hardware correlator.  Figure \ref{fig:scintillation} shows a section 
of the dynamic spectrum from this observation which shows the scintillation structure as a
function of time and frequency on the GBT--Arecibo baseline.  A deconvolution process allows
the generation of a speckle image of the scattering disk using the interfeometric phase information
(Brisken et al., in preparation).  The effective resolution of the speckle image is better than 1 mas, 
revealing structure which is a factor of 40 smaller than the synthesised beam of the interferometer.

\subsection[EVN pulsar astrometry]{EVN pulsar astrometry}
DiFX has been used to correlate Mark5 data from the EVN with multiple pulsar gates and phase
centres to make an image of the pulsar distribution in the globular cluster M15 (Deller et al.,
in preparation).  Through phase referencing to a nearby quasar and taking an ensemble
average of pulsar motions over several epochs, it will be possible to obtain the proper motion
of the cluster with a higher signal/noise ratio than would be possible using a single pulsar.
It is also possible to further improve the calibration of the data by using the brightest pulsar
as an in--beam calibrator, improving the sensitivity to the other pulsars in the cluster and making
it possible to conduct extremely precise relative astrometry to probe the intra--cluster dynamics.

\subsection[Widefield VLBI]{Widefield VLBI}
DiFX is being used to correlate VLBA observations of the Chandra Deep Field South, using high
time and frequency resolution which is not available with the existing VLBA hardware correlator.
This project aims to obtain VLBI--resolution imaging over a $30\times30$\ arcminute field of view.
Post--correlation phase shifting to potential targets identified from lower--resolution observations
allows many targets to be imaged from a single correlator pass.  In the future, DiFX will support
multiple phase centres during correlation, eliminating the need for post--correlation shifting and 
averaging of visibilities before imaging, which will enable this and other 
larger--scale VLBI surveys to be carried out quickly and efficiently.

\subsection[Australian/New Zealand geodesy]{Australian/New Zealand geodesy}
A new 3--element, full time geodetic array is planned to commence operations in Australia
in 2009 as part of the AuScope project\footnote{http://www.auscope.org.au/}.  This array
will contribute to observations which address both local and global geodetic goals.  
Closely linked to the AuScope project is the development of a geodetic VLBI capacity in 
New Zealand, led by the Centre for Radiophysics and Space Science at Auckland University
of Technology\footnote{http://www.aut.ac.nz/about/faculties/design\_and\_creative\_technologies/faculty\_research\_office/\linebreak[4] centre\_for\_radiophysics\_and\_space\_research/mission.htm}.  The 
DiFX software correlator will be used for the correlation of these new geodetic arrays.

\subsection[Worldwide geodesy]{Worldwide geodesy}
As is the case with many areas of VLBI, the geodetic community is currently planning a
significant upgrade in capabilities.  This upgrade is spearheaded by a sensitivity increase 
driven by wider recording bandwidths, but also includes new hardware such as smaller, more rapidly
slewing antennas and new, improved digital backends.  The upgrade project is known as 
VLBI2010\footnote{http://ivs.nict.go.jp/mirror/about/wg/wg3/}, and aims to achieve an initial recording
rate of 2--4 Gbps with the potential for expansion to 8, 16 or even 32 Gbps.  As the Mark 4 hardware
correlators currently used for geodetic VLBI are limited to a 1 Gbps correlation rate, this project 
will necessitate new and improved correlator infrastructure \citep{niell05a}. 
At present, the use of DiFX in this role is being investigated by MPIfR, with activities including 
the correlator comparison shown in Section~\ref{difx:deploy:bonncomp}.
\chapter[OBSERVATIONS, DATA REDUCTION AND ANALYSIS]{ASTROMETRIC OBSERVATIONS, DATA REDUCTION AND ANALYSIS}
\label{techniques}

\section[Goals and target selection]{Goals and target selection}
\label{techniques:select}
The observational program of this thesis encompassed 8 pulsars, which are listed in 
Table~\ref{tab:targets}.  As already noted in Chapter~\ref{intro}, previous Southern Hemisphere 
VLBI pulsar astrometry programs \citep{dodson03a,legge02a,bailes90a} have yielded 
only two published pulsar parallaxes, so one 
primary motivation was to increase the number of southern VLBI parallaxes and hence
improve models of Galactic electron distributions at southern declinations.  Additionally,
as this is the first such large scale southern parallax study, target pulsars
of varying brightness and predicted distance were chosen in order to determine the types of targets 
that would be feasible for future southern VLBI studies.  
The brightness and predicted distance
of the selected targets are shown below in Table~\ref{tab:targets}.
Within the bounds of these criteria, however,
it was possible to choose target pulsars for which a VLBI parallax would enable a deeper
understanding of additional astrophysical phenomena.  

\begin{deluxetable}{lcccccc}
\tabletypesize{\tiny}
\tablecaption{Target pulsars}
\tablewidth{0pt}
\tablehead{
\colhead{Pulsar} & \colhead{$DM$ distance} & \colhead{1600 MHz} & \colhead{Pulsar} & Equivalent gated & \colhead{Reference} & \colhead{Calibrator/target} \\
\colhead{name} & \colhead{(pc)\tablenotemark{a}} & \colhead{flux (mJy)} & \colhead{gating gain} & 1600 MHz flux (mJy) & \colhead{source} & \colhead{separation (deg)} 
}
\startdata
J0108--1431 		& 130 	& 0.6 \tablenotemark{b} 	& 5.1 	& 3		& J0111--1317 	& 1.5 \\
J0437--4715 		& 140 	& 140 				& 6.25	& 875	& J0439--4522 	& 1.9 \\
J0630--2834 		& 2150 	& 23 					& 3.5 	& 81		& J0628--2805 	& 0.7 \\
J0737--3039A/B 	& 570 	& 1.6 				& 2.5 	& 4		& J0738--3025 	& 0.4 \\
J1559--4438 		& 1600 	& 40 					& 3.6 	& 144	& J1604--4441 	& 0.9 \\
J2048--1616 		& 640 	& 13 					& 3.6		& 47		& J2047--1639 	& 0.5 \\
J2144--3933 		& 180 	& 0.8\tablenotemark{b} 	& 10 		& 8		& J2141--3729 	& 2.1 \\
J2145--0750 		& 500 	& 8\tablenotemark{b} 	& 4.3 	& 34		& J2142--0437 	& 3.3 \\
\enddata
\tablenotetext{a}{Taken from Taylor \& Cordes (1993)}
\tablenotetext{b}{Pulsar suffers heavily from long timescale scintillation, so individual epochs vary considerably from the average value shown.}
\label{tab:targets}
\end{deluxetable}

The chosen targets can be divided into three broad categories: 
binary pulsars used for tests of GR and gravitational wave detection 
(\ptwo, \pfour\ and \peight), pulsars with unusual luminosity
in the radio (low luminosity pulsars \pone\ and \pseven) or x-ray 
(\pthree, whose x--ray luminosity is anomalously high), and ``technique check"
sources which are bright and predicted to be nearby (\pfive\ and \psix).
The results for each pulsar, and the implications of these results, are discussed individually
in Chapter~\ref{results}.  

The remainder of this chapter will be devoted to describing the observations and
data reduction applied to all pulsars, with a particular focus on the tools developed 
for pulsar astrometry reduction which address aspects of the data reduction process
not previously highlighted in the literature.  In order to illustrate the concepts, the results
for \pfive\ will be used, as this proved the superior of the two ``technique check" sources.

\section[Observations]{Observations}
\label{techniques:obs}
Eight observational epochs were spread over a two year period from May 2006 to February 2008.
Epochs were typically 24 hours in duration and a subset of the eight pulsars were observed,
depending on which were closest to parallax extrema.
The observing frequency was centred on 1400 MHz for the first observation, and 1650 MHz 
for the remaining seven observations.   PSR J0437--4715 was observed separately, with four epochs of
12 hours duration, centred on 8400 MHz.  Observations at this higher frequency were made possible
by the high flux and narrow pulse profile of PSR J0437--4715 .  Dual circular polarization was used at all
frequencies.

\begin{figure}
\begin{center}
\begin{tabular}{cc}
\includegraphics[width=0.4\textwidth]{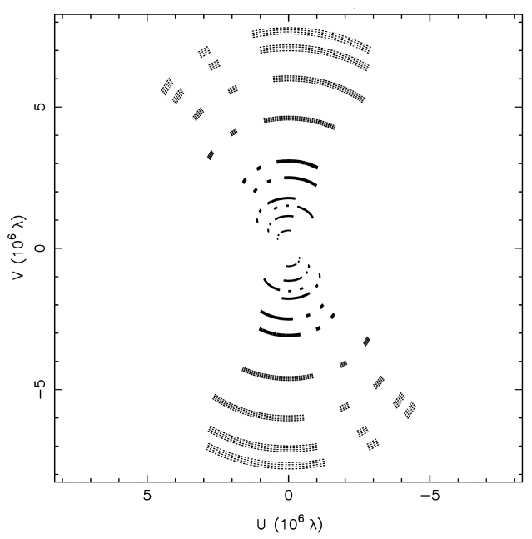} &
\includegraphics[width=0.4\textwidth]{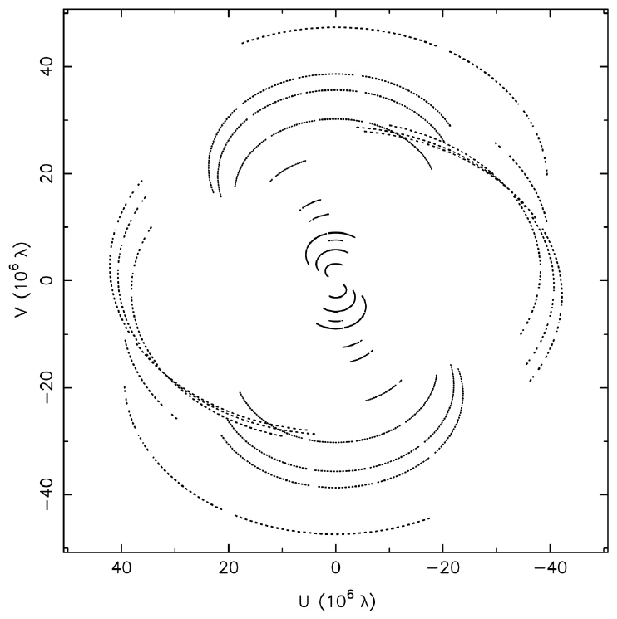} \\
\end{tabular}
\caption[Typical $uv$\ coverage for 1650 MHz and 8400 MHz observations]
{Typical $uv$\ coverage at 1650 MHz (no Ceduna; left panel) and at 8400 MHz (with Ceduna;
right).  The target sources are \pfive\ and \ptwo\ respectively. Without
Ceduna, the $uv$\ coverage is heavily biased north--south, and wide hour--angle coverage is 
necessary to gain acceptable $uv$ coverage.  The displayed plots are optimal $uv$\ coverage --
equipment failure or telescope commitments often meant only a subset of this coverage was obtained.}
\label{fig:uvcoverage}
\end{center}
\end{figure}

The Australian Long Baseline Array (LBA) consists of six antennas -- the 
Australia Telescope National Facility (ATNF) telescopes in New South Wales (Parkes, Australia Telescope Compact Array [ATCA], Mopra); the University of 
Tasmania telescopes at Hobart, Tasmania and Ceduna, South Australia; and the
NASA DSN facility at Tidbinbilla, Australian Capital Territory.
The Parkes, phased ATCA, Mopra, and Hobart telescopes participated in all experiments, 
but a Tidbinbilla antenna (70m or 34m) was used only when available, and Ceduna 
participated in observations of J0437-4715 only\footnote{Ceduna 
does not possess a 1600 MHz receiver, so could only participate in the higher frequency
experiments}.  The maximum baseline length with Ceduna is 1700 km, and without Ceduna 
is 1400 km.  Representative $uv$\ coverage at 1650 MHz and 8400 MHz is shown in 
Figure~\ref{fig:uvcoverage}.

All observations used the recently introduced LBADR disk-based recording system
(Phillips et al., in preparation).
At the three ATNF observatories, the presence of two Data Acquisition System (DAS) units allowed
a recording rate of 512 Mbps (8 $\times$ 16 MHz bands, Nyquist sampled at 2 bits), while the
non-ATNF stations recorded at 256 Mbps (4 $\times$ 16 MHz bands).  For epochs 
where dual--polarization
feeds were available at all antennas, two frequency bands were dropped at the non--ATNF
stations, but as some of the NASA DSN feeds are single polarization only, 1650 MHz 
epochs featuring the 70m NASA antenna and 8400 MHz epochs featuring the 34m NASA
antenna instead retained a single polarization of all frequency bands.  The observation dates, targets,
and participating antennas are summarised in Table~\ref{tab:obs}.

A phase reference 
cycle time of six minutes, apportioned equally to target and calibrator, was used for all observations.
As the LBA consists of disparate antennas ranging up to 70m in diameter (and, when phased, the 
ATCA has an equivalent diameter of hundreds of metres for the purposes of calculating 
field of view), it was not possible to utilise in--beam calibrators for any sources, unlike
recent pulsar astrometric programs using the VLBA \citep{chatterjee01a,chatterjee05a}.

The data were correlated using matched filtering on pulse profiles with the DiFX 
software correlator, producing RPFITS format
visibility data.  Table~\ref{tab:targets}
shows the predicted gain due to pulse profile filtering for each target source.  Pulsar ephemerides
were obtained using the ATNF Pulsar Catalogue \citep{manchester05a}, 
with the exception of the double pulsar \pfour, 
which was periodically updated with the latest published ephemeris. 
Two second integrations and 64 spectral points per 16 MHz band were 
used for all observations.

\begin{deluxetable}{lcccccc}
\tabletypesize{\tiny}
\tablecaption{Observation summary}
\tablewidth{0pt}
\tablehead{
\colhead{Observation} & \colhead{Duration} & 
\colhead{Observed pulsars} & \colhead{Participating telescopes} \\
\colhead{date (MJD)} & \colhead{(hours)} &&
}
\startdata
53868	& 12	& 0437--4715 				
		& Parkes, ATCA, Mopra, Hobart, Ceduna \\
53870	& 15	& 1559--4438, 2048--1616, 2144--3933, 2145--0750				
		& Parkes, ATCA, Mopra, Hobart \\
53970	& 24	& 0108--1431, 0630--2834, 0737--3039
		& Parkes, ATCA, Mopra, Hobart, DSS43 (70m) \\
	    	& & 1559--4438, 2048--1616, 2144--3933 & \\
54055	& 12	& 0437--4715 	
		& Parkes, ATCA, Mopra, Hobart, Ceduna\tablenotemark{A},\ \  DSS43 (70m) \\
54057	& 24	& 0108--1431, 0630--2834, 0737--3039
		& Parkes, ATCA, Mopra, Hobart, DSS43 (70m) \\
		& & 1559--4438, 2048--1616, 2144--3933, 2145--0750 & \\
54127	& 24 	& 0108--1431, 0630--2834
		& Parkes, ATCA, Mopra, Hobart, DSS43 (70m) \\
		& & 0737--3039, 1559--4438, 2144--3933 & \\
54181	& 12	& 0437--4715
		& Parkes, ATCA, Mopra, Hobart, Ceduna, DSS43 (70m)\tablenotemark{B} \\
54182	& 24	& 0630--2834, 0737--3039, 1559--4438
		& Parkes, ATCA, Mopra\tablenotemark{C}, Hobart, DSS43 (70m) \\
		& & 2048--1616, 2144--3933, 2145--0750 & \\
54307	& 24	& 0108--1431, 0630--2834
		& Parkes, ATCA, Mopra, Hobart, DSS43 (70m) \\
		& & 0737--3039, 1559--4438, 2144--3933 & \\
54413	& 24	& 0108--1431, 0630--2834, 0737--3039
		& Parkes, ATCA, Mopra, Hobart \\
		& & 1559--4438, 2048--1616, 2144--3933, 2145--0750 & \\
54417	& 12	& 0437--4715
		& Parkes, ATCA, Mopra, Hobart, Ceduna, DSS34 (34m)\tablenotemark{B} \\
54500	& 24	& 0108--1431, 0630--2834
		& Parkes, ATCA, Mopra, Hobart, DSS43 (70m)\tablenotemark{D} \\
		& & 0737--3039, 1559--4438, 2144--3933 & \\
\enddata
\tablenotetext{A}{Equipment failure meant no useful data was available from Ceduna for this session.}
\tablenotetext{B}{Limited time ($\sim\,4$\ hours) was available at the DSS telescopes in 
these sessions.}
\tablenotetext{C}{Equipment failure led to the loss of half of the Mopra data from this session.}
\tablenotetext{D}{Setup errors meant no useful DSS43 data was available from this session.}
\label{tab:obs}
\end{deluxetable}

\section[Data reduction]{Data reduction}
\label{techniques:datared}
Data reduction was performed principally in AIPS, 
using the python interface ParselTongue \citep{kettenis06a}.  The DIFMAP package 
\citep{shepherd97a}
was used for imaging and self calibration.  The data reduction was implemented
as a pipeline, with user interaction for imaging, editing of solution tables, and visibility flagging.  
All scripts used in the data reduction process are freely available
and can be downloaded from \linebreak[4] 
\verb+http://astronomy.swin.edu.au/~adeller/software/scripts/+. 
The individual stages of the pipeline are described below.

\subsection[Amplitude and weight calibration and flagging]{Amplitude and weight calibration and flagging}
Amplitude calibration using the measured telescope system temperatures was carried out using
the AIPS tasks APCAL and ANTAB.  Flagging based on predicted or (when available) logged
telescope off-source times due to slewing or failures was applied using UVFLG, while the first
and last 10 seconds of every scan was excised with the task QUACK.  For the ATCA,
the tied array infrastructure required flagging the 
first three correlator integrations (totaling 30 seconds) of each scan with QUACK. The weight of each 
visibility point, which is set to a constant value when the RPFITS format data is loaded into
AIPS, was initially scaled by the predicted baseline sensitivity using a ParselTongue script.  The effect of
data weighting is investigated further in Section~\ref{techniques:optweight}.

\subsection[Geometric model and ionospheric corrections]{Geometric model and ionospheric corrections}
\label{techniques:datared:geoiono}
At the low frequencies which are generally used for pulsar astrometry, ionospheric variations
usually make the dominant contribution to systematic error \citep[see e.g.][]{brisken02a}.  
Using ionospheric models based on Global Position System (GPS) data 
provided by the NASA Jet Propulsion Laboratory\footnote{available 
from the Crustal Dynamics Data Information Systems (CDDIS) archive: 
ftp://cddis.gsfc.nasa.gov/pub/gps/products/ionex/}, the AIPS task TECOR was used to 
correct phase variations due to the ionosphere.
This observing program roughly coincided with the solar minimum of 2006,
and consequently the ionospheric variations were generally at a minimum.
For \ptwo, which was observed at 8.4 GHz, the position shifts introduced by ionospheric correction
were very small ($\sim$100 $\mu$as) and did not improve the quality of the astrometric fit,
so this step was not applied for \ptwo.

Total Electron Content (TEC) models suffer at southern declinations due to the relatively
sparse distribution of GPS receivers at southern latitudes.  Consequently, the 
derived ionospheric corrections are much less reliable for the LBA than for similar 
Northern Hemisphere instruments.  The dispersive delay corrections generated 
by TECOR were inspected for each epoch, and the effectiveness
of different TEC maps is investigated in 
Section~\ref{techniques:check:tecor}.

While the observational program was underway, considerably more accurate station
positions were derived for several LBA antennas using archival geodetic observations and the 
OCCAM software \citep{titov04a}, a dedicated 22 GHz LBA geodetic experiment 
(Petrov et al., in preparation), 
and GPS survey measurements.  Additionally, more accurate positions for some calibrators were 
published in the 5th VLBA Calibrator Survey \citep[VCS5;][]{kovalev07a}.  The visibilities
were corrected to account for the revised positions using an AIPS SN table generated
with the Wizardry feature of ParselTongue, which stored the difference between the initial
and corrected geometric models.

For some pulsars, their proper motions ($> 100$ mas yr$^{-1}$)\ cause significant position shifts
over the course of a 24 hour observation, comparable in some cases to the epoch positional
accuracy.  As the geometric model generation used in DiFX
at the time of these observations could not account for proper motion, the visibility phases and
$uvw$\ values were corrected in AIPS using a SN table generated by ParselTongue, interpolating
between predicted postions for the pulsar at the start and end of the experiment.  This also
allowed the proper motion to be refined (or, in some cases, measured for the first time) over
the course of the observations before the final corrections were applied.

\subsection[Fringe-fitting and amplitude calibration refinement]{Fringe-fitting and amplitude calibration refinement}
\label{techniques:datared:fringefit}
\begin{figure}
\begin{center}
\begin{tabular}{cc}
\includegraphics[width=0.5\textwidth, angle=270, clip]{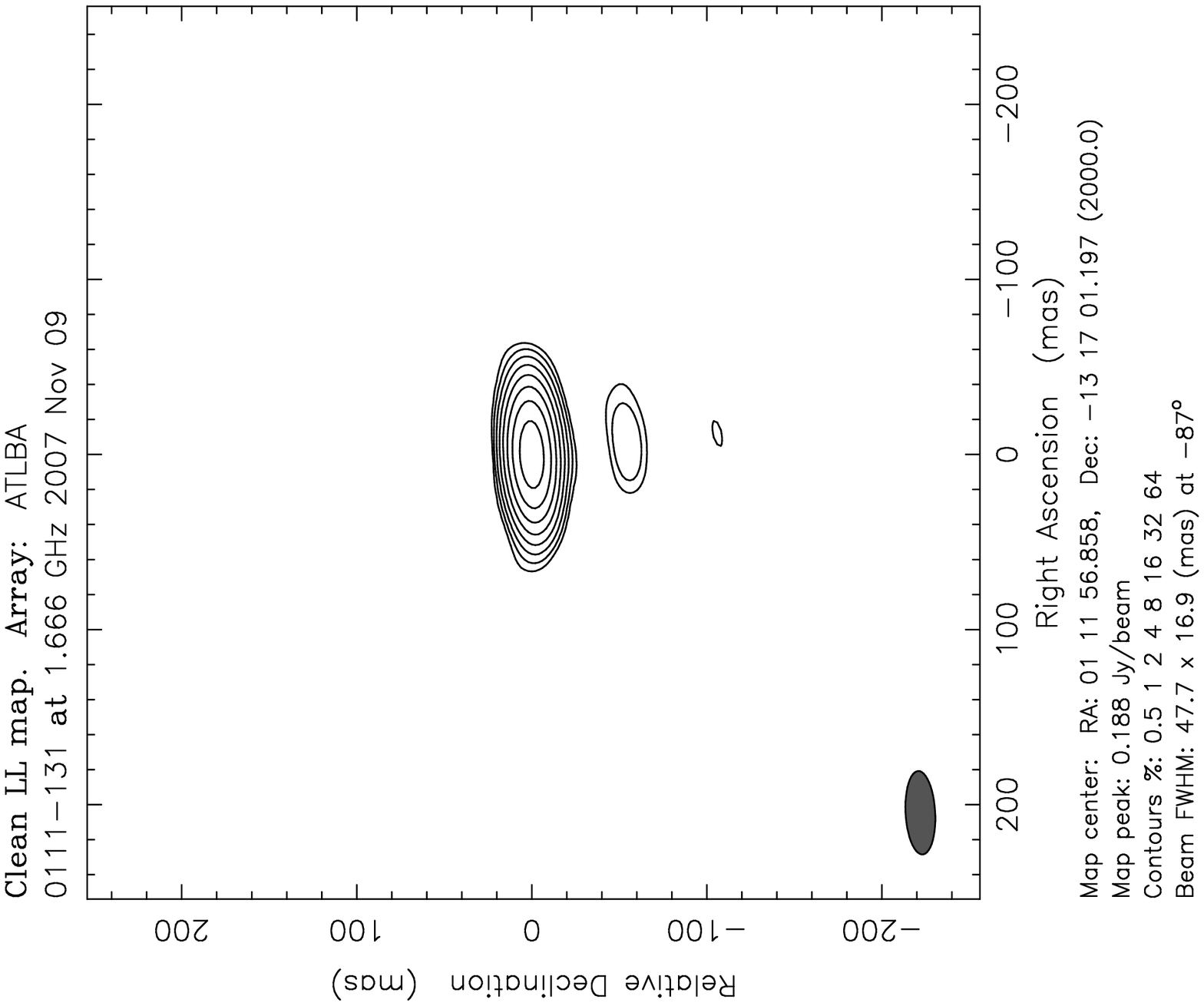} &
\includegraphics[width=0.5\textwidth, angle=270, clip]{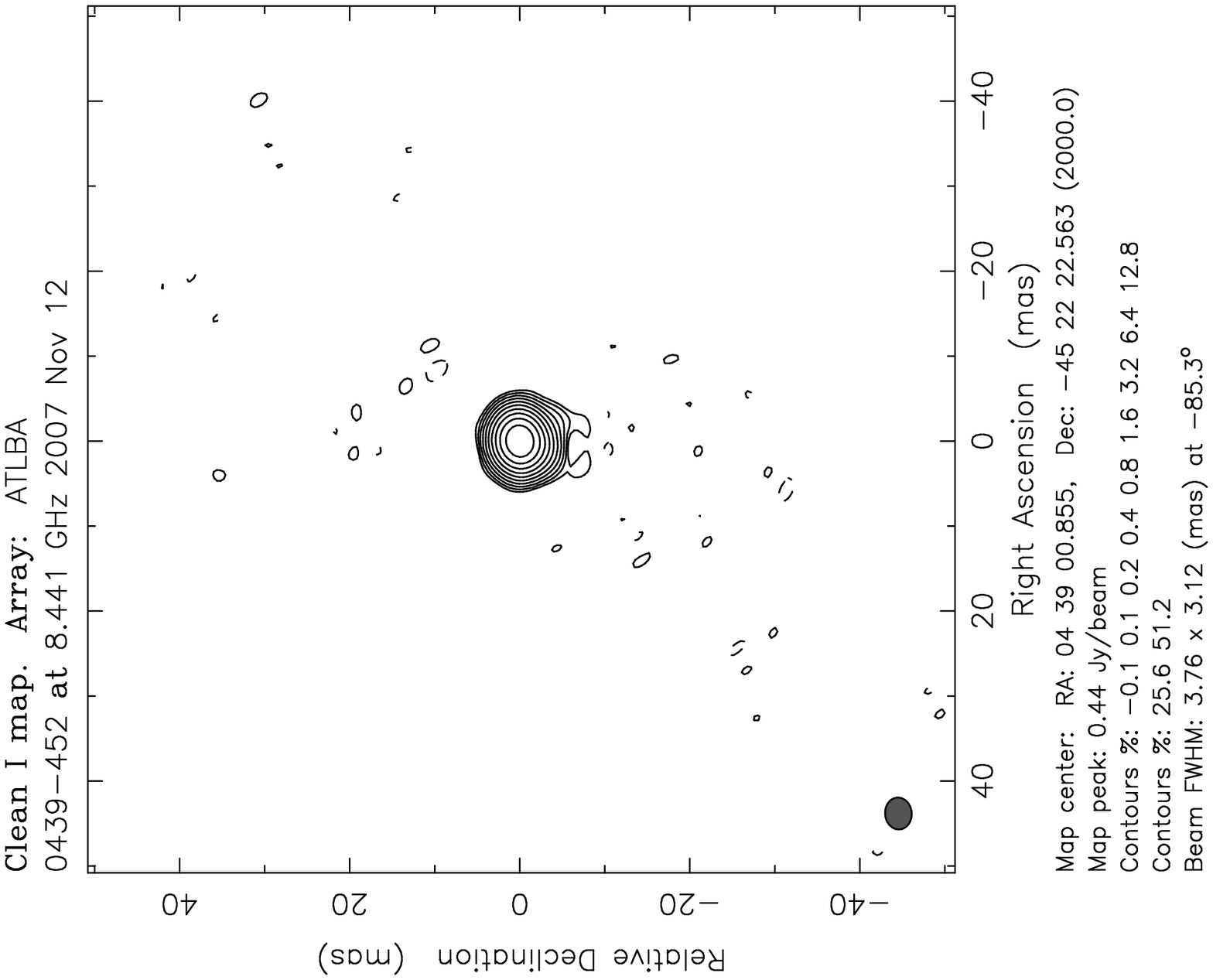} \\
\includegraphics[width=0.5\textwidth, angle=270, clip]{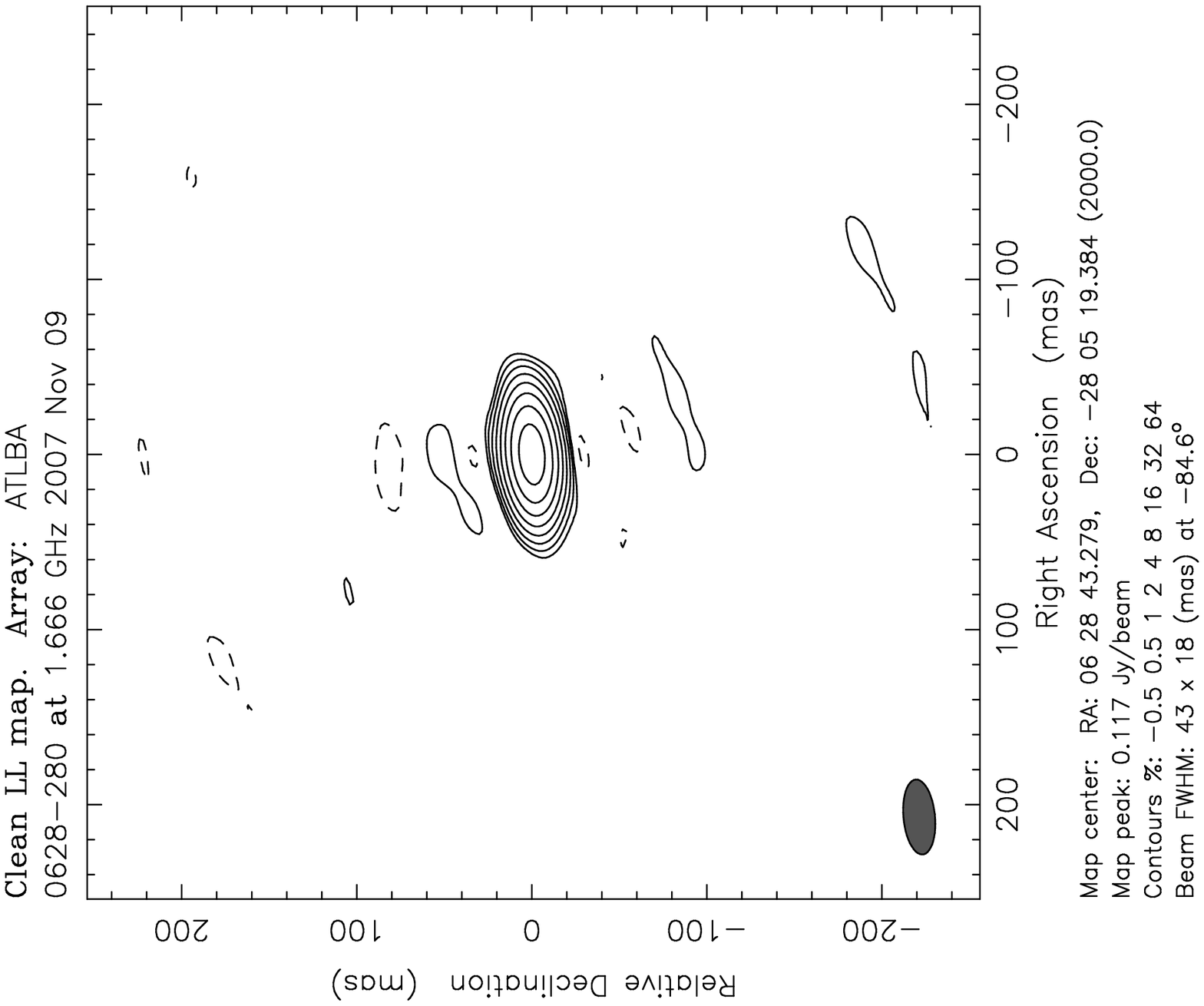} &
\includegraphics[width=0.5\textwidth, angle=270, clip]{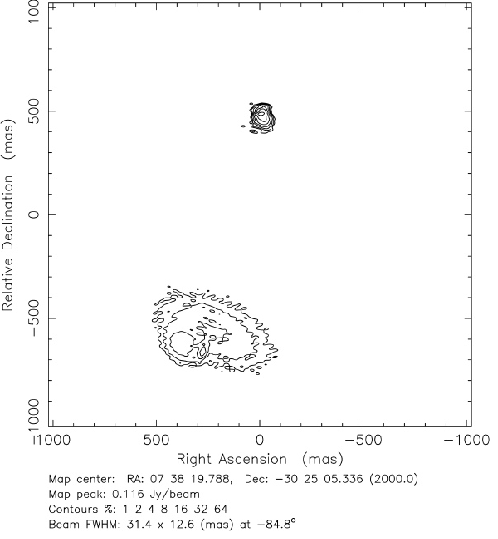} \\
\end{tabular}
\caption[LBA images of the VLBI phase reference sources]
{LBA images of the phase reference sources J0111--1317, J0439--4522, J0628--2805 and
B0736--303.  0.4 mas pixels were used for J0439--4522
(8 GHz observations) and 2 mas pixels used elsewhere.  Uniform weighting with visibility weights
raised to the power $-1$\ was used for all images.}
\label{fig:phaserefs1}
\end{center}
\end{figure}

\begin{figure}
\begin{center}
\begin{tabular}{cc}
\includegraphics[width=0.5\textwidth, angle=270, clip]{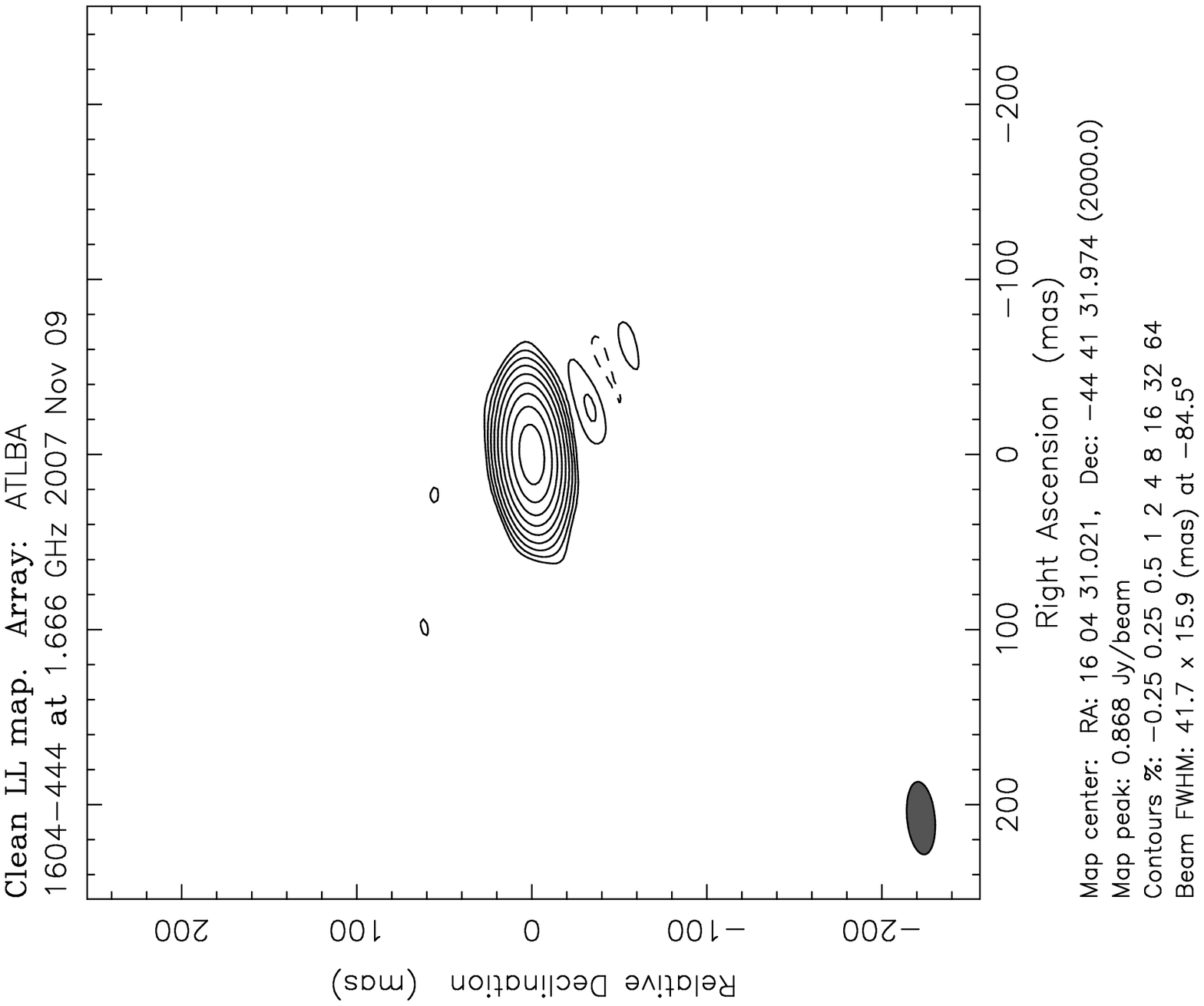} &
\includegraphics[width=0.5\textwidth, angle=270, clip]{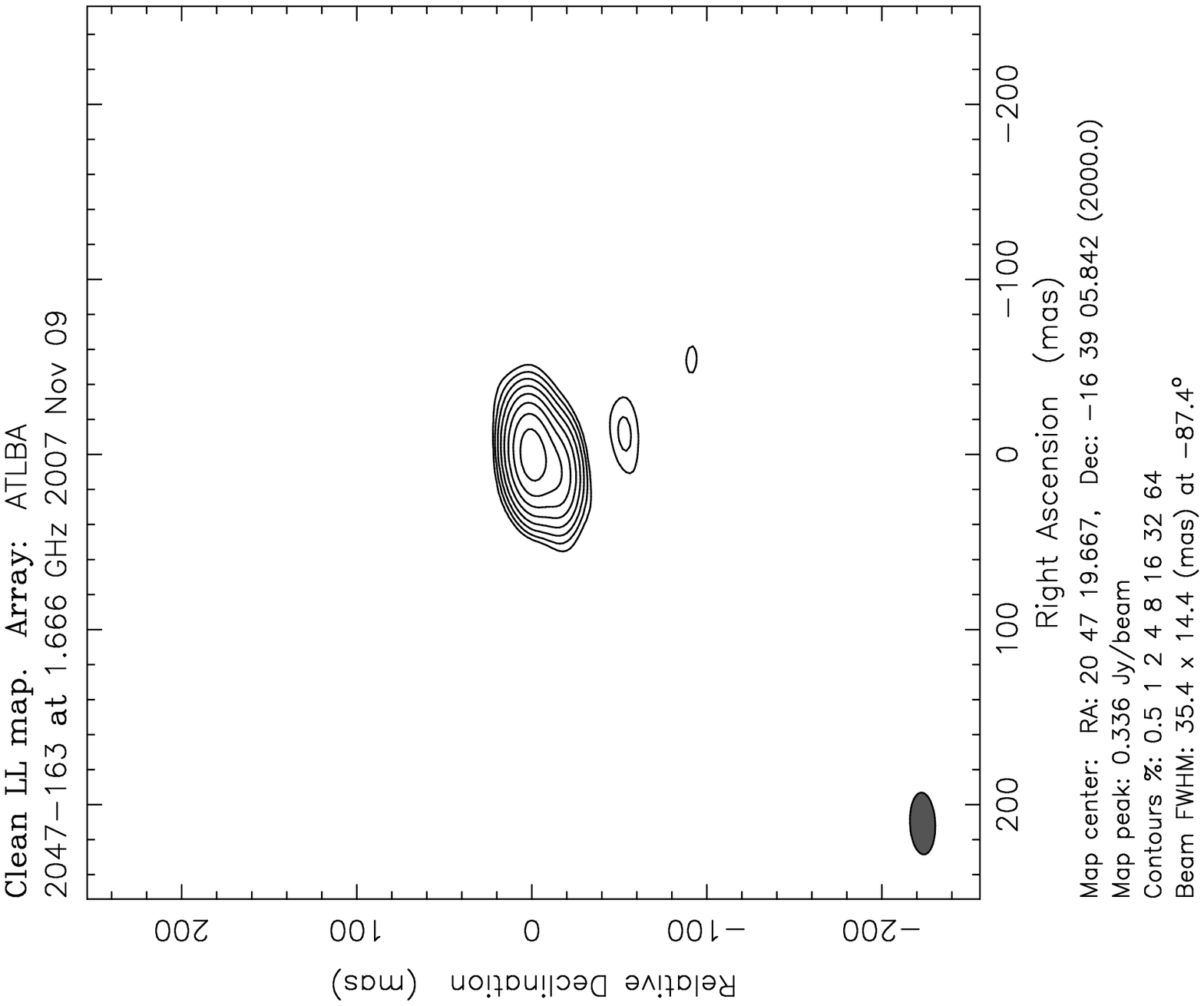} \\
\includegraphics[width=0.5\textwidth, angle=270, clip]{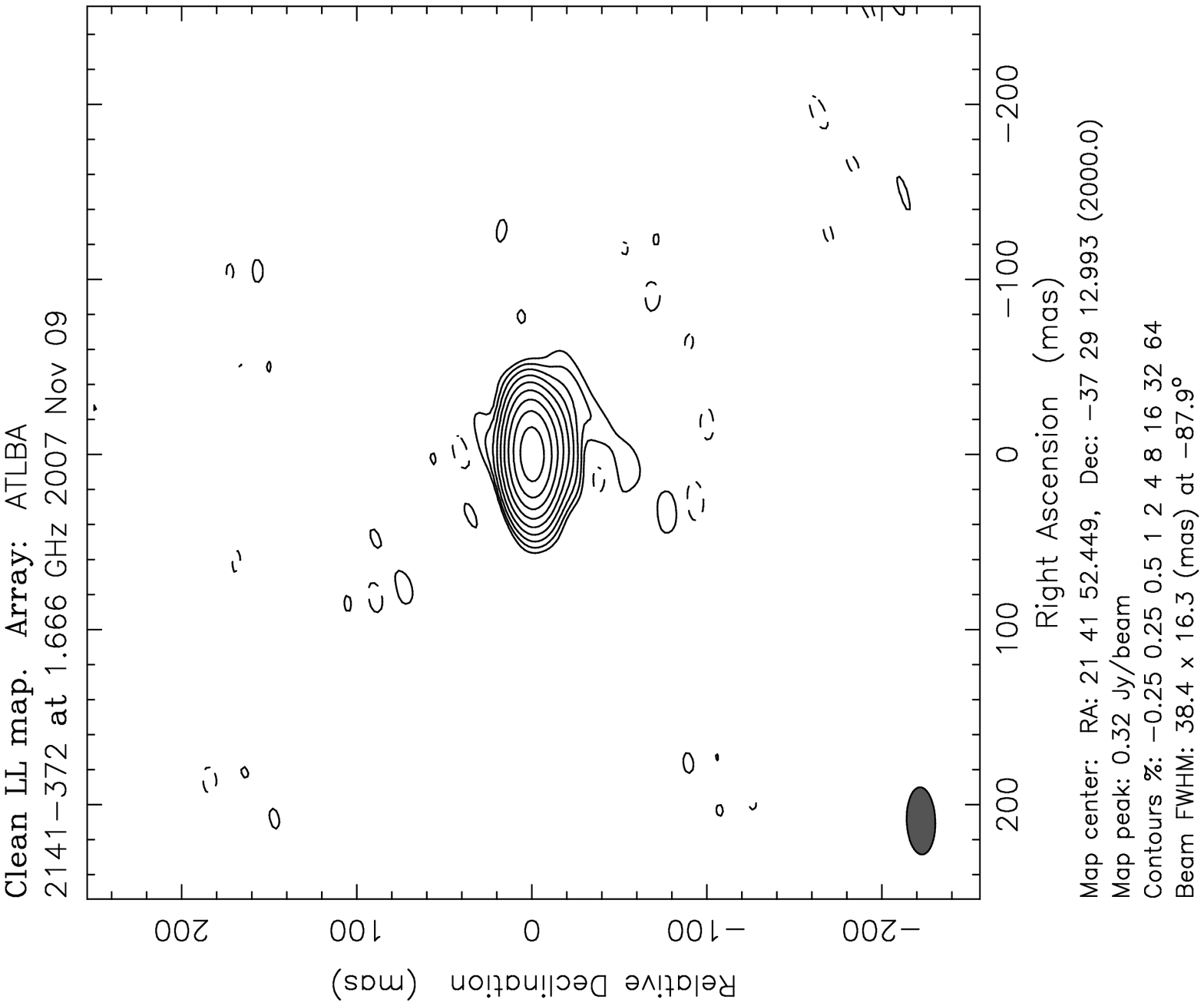} &
\includegraphics[width=0.5\textwidth, angle=270, clip]{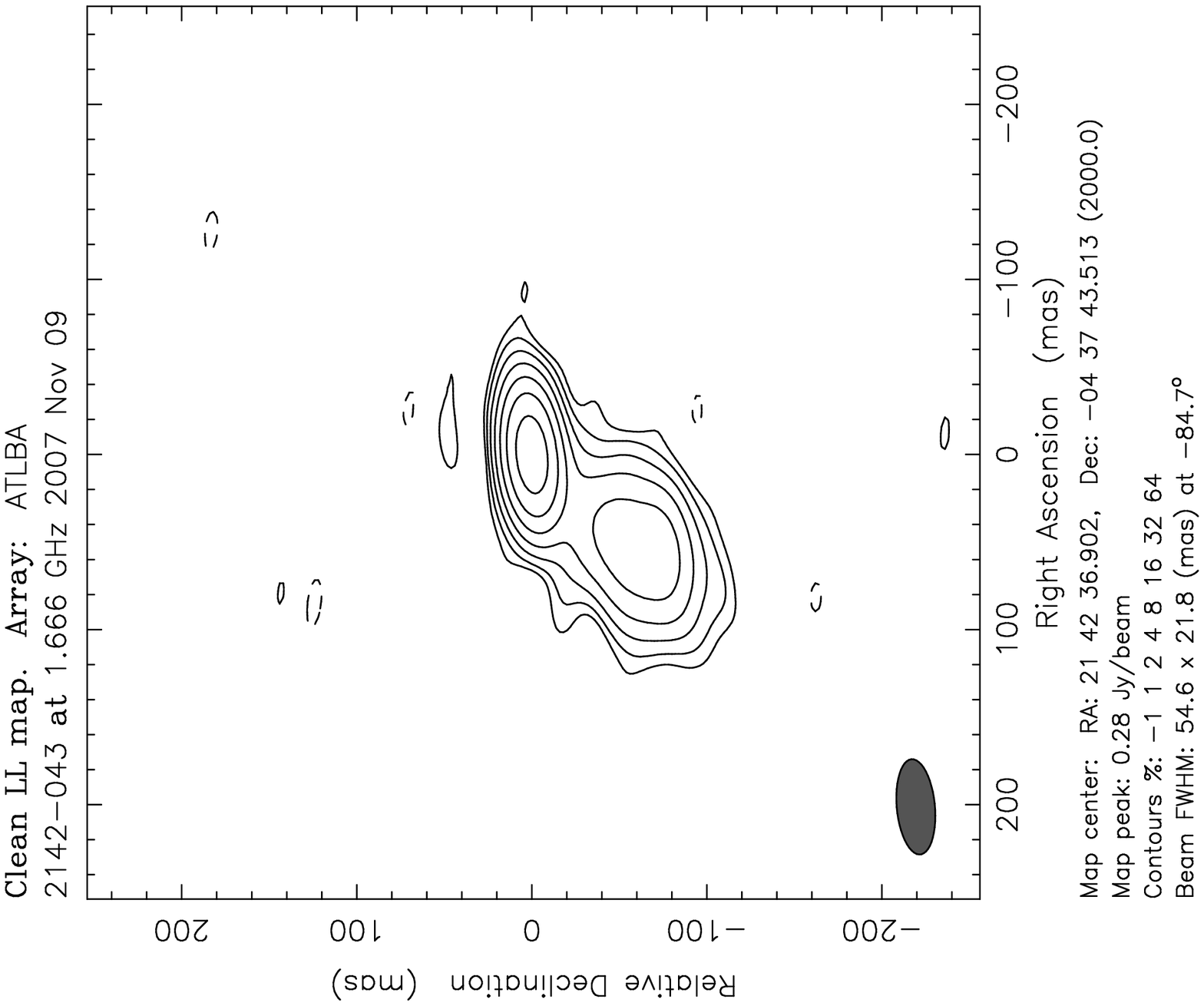} \\
\end{tabular}
\caption[LBA images of  the VLBI phase reference sources continued]
{LBA images of the phase reference sources J1604--4441,
J2047--1639, J2141--3729 and J2142--0437.  2 mas pixels and uniform weighting with 
visibility weights raised to the power $-1$\ were used for all images.}
\label{fig:phaserefs2}
\end{center}
\end{figure}

Fringe fitting was performed using the AIPS task FRING, using a point source model, on the phase reference calibrator data for each target pulsar.
Subsequently, a single structural model was produced for each calibrator using the combined 
datasets from all epochs.  With the exception of B0736--303, each source was modeled using 
a dominant component (delta or narrow
Gaussian) fixed at the phase center, and 0 -- 2 secondary components which were allowed to vary
in position\footnote{B0736--303 was modeled using a combination of CLEAN and modelfit 
components, and is discussed further in Section~\ref{results:binary:0737}}.  
The flux density of all components was allowed to vary.  Typically, the variation in flux 
density of secondary component(s) between epochs was $<1\%$ of the image peak flux density.  
Images were then generated using uniform weighting (bin size two pizels) and
baseline weights set to predicted noise RMS (difmap uvweight settings 2,-1).
The images of each phase reference source are shown in Figures~\ref{fig:phaserefs1} and 
\ref{fig:phaserefs2}.  With the exception of B0736--303 and
J2142--0437, the corrections due to reference source structure were very small.

The solutions were applied to each calibrator and the data averaged in frequency,
exported to disk and loaded into  DIFMAP.  The calibrator model was loaded and 
several iterations of self-calibration and modelfitting performed.  The difmap `modelfit'
command uses the Levenberg--Marquardt least--squares minimisation algorithm to fit the
free model parameters to the visibility points, incorporating the visibility weights.  The self calibration 
corrections were then written to disk as an AIPS SN table using the `cordump' patch to 
difmap\footnote{http://astronomy.swin.edu.au/{\char126}elenc/difmap-patches/}.  Additionally,
data points flagged in DIFMAP were collated in a Wizardry script and converted into a 
flag file suitable for the AIPS task UVFLG.


\clearpage

\clearpage

The amplitude corrections generated in this manner allowed compensation for 
imperfectly measured system temperature values, and were also applied to visibility weights.  
For bands which could not be self--calibrated, due to only being observed by the three 
ATNF antennas, a correction was estimated based on the nearest band with self
calibration solutions of the same polarization.  The self calibration solutions were 
loaded into AIPS using TBIN, and applied to the target pulsars using CLCAL.

The use of bandpass calibration was investigated but found to make insignificant difference to
the fitted target position.  The LBA Data Acquisition System (DAS) utilitizes digital filtering and
typical bandpass phase ripple was $<2$\ degrees.  When averaging in frequency, the lowest and
highest 10\% of the band was discarded and the central 80\% of the band averaged with uniform
weight assigned to each channel.

\subsection[Pulsar scintillation correction]{Pulsar scintillation correction}
\begin{figure}
\begin{center}
\begin{tabular}{cc}
\includegraphics[width=0.45\textwidth]{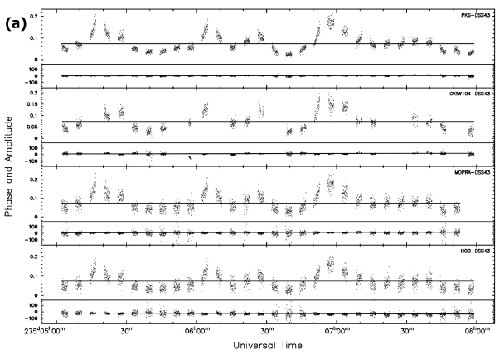} &
\includegraphics[width=0.45\textwidth]{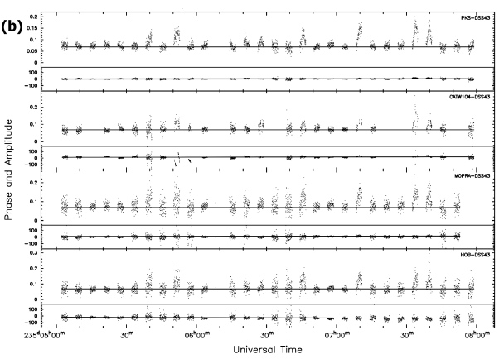} \\
\includegraphics[width=0.45\textwidth]{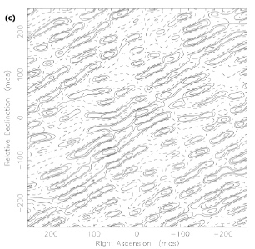} &
\includegraphics[width=0.45\textwidth]{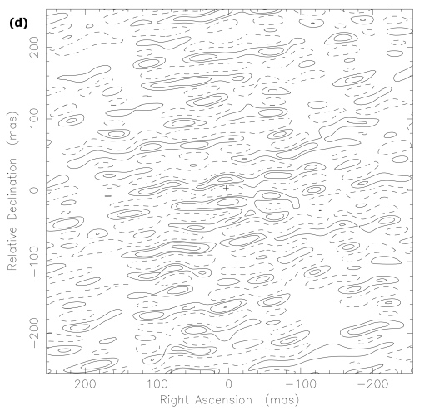} \\
\end{tabular}
\caption[Illustration of the effects of diffractive scintillation on pulsar observations]
{The effects of diffractive scintillation for \pfive.  Panel a) shows the uncorrected
visibility amplitude on baselines including the Tidbinbilla antenna from one experiment -- the 
scintillation timescale of several minutes is apparent.  Panel b) shows the same visibility amplitudes after
correcting for scintillation.  Panel c) shows the image residuals of the uncorrected dataset -- contours are 1,2,4 and 8 mJy/beam.  Panel d) shows image residuals after correcting for scintillation, with contours at 1,2 and 4 mJy/beam -- the improvement in image quality is obvious.}
\label{fig:scint}
\end{center}
\end{figure}

Nearby pulsars can suffer dramatic, and rapid, variations in visibility amplitude due to diffractive
scintillation.  The size of the scintillation pattern is typically much larger than the size of the Earth,
and so the amplitude variations are essentially independent of baseline length.  Maximal amplitude 
fluctuations (where the RMS is equal to the mean
flux density) are seen for pulsars in the strong scattering regime 
\citep[see e.g.][]{walker98a}, which can be predicted from Galactic
electron distribution models. The NE2001 model \citep{cordes02a}
predicts that strong scintillation should be observed for \pthree, 
\pfive, \psix, \pseven\ and \peight.  An example is shown for
\pfive\ in Figure~\ref{fig:scint}a, which shows the variation of visibility amplitude with time
for Tidbinbilla baselines over a 2 hour period.  Observed scintillation parameters for target
pulsars are shown in Table~\ref{tab:scint}. Assuming the material responsible for the
scintillation is turbulent, with a Kolmogorov distribution, the 
scintillation time $\tau_{\mathrm{scint}}$\ and scintillation bandwidth
$B_{\mathrm{scint}}$\ can be scaled to the frequencies used in these
observations with the relations $\tau_{\mathrm{scint}}\propto\nu^{1.2}$\ 
and $B_{\mathrm{scint}}\propto\nu^{4.4}$\ \citep{cordes86a}.

\begin{deluxetable}{lrrrrr}
\tabletypesize{\tiny}
\tablecaption{Observed pulsar scintillation parameters and estimated scattering disk sizes}
\tablewidth{0pt}
\tablehead{
\colhead{Pulsar} & \colhead{Scintillation} & \colhead{Scintillation} & \colhead{Observing} & 
\colhead{Reference} & \colhead{Scattering disk size\tablenotemark{a}} \\
\colhead{name} & \colhead{time (s)} & \colhead{bandwidth (MHz)} & \colhead{frequency (MHz)} & 
 & \colhead{(mas, 1650 MHz)} 
}
\startdata
J0108--1431 	& *\tablenotemark{b}		& *	& *		& \citet{johnston98a} & 0.003 \\
J0437--4715 	& 600		& 17.4		& 660 	& \citet{johnston98a} & 0.052\\
J0630--2834 	& 456		& 0.2 		& 327	& \citet{bhat99a} & 0.022\\
J0737--3039A 	& $>70$\tablenotemark{c}& 1.4\tablenotemark{d} & 1400 	& Coles et al. (2005) & 0.406 \\
J1559--4438 	& 77			& 0.16		& 660	& \citet{johnston98a} & 0.133 \\
J2048--1616 	& 138		& 0.54		& 327	& \citet{bhat99a} & 0.023 \\
J2144--3933 	& 1500		& 2.9			& 660	& \citet{johnston98a} & 0.126 \\
J2145--0750 	& 1510		& 1.48 		& 436	& \citet{johnston98a} & 0.046 \\
\enddata
\tablenotetext{a}{Calculated from decorrelation bandwidth where available, and taken from NE2001 model  where scintillation measurements are unavailable}
\tablenotetext{b}{Scintillation parameters have not been measured, but are believed to be extremely large, as expected for a very nearby pulsar}
\tablenotetext{c}{Varies considerable with orbital velocity -- values quoted for highest transverse velocity}
\tablenotetext{d}{Coles, private communication}
\label{tab:scint}
\end{deluxetable}

Left uncorrected, the variation in visibility amplitude with time scatters power throughout the image
plane, as shown in Figure~\ref{fig:scint}c, which shows the residuals for \pfive\ after fitting a single 
point source to the visibilities shown in Figure~\ref{fig:scint}a.  To overcome this, a ParselTongue
script was written to produce an AIPS SN table which would flatten visibility amplitudes
by averaging data over all sensitive baselines to 1/4 of the scintillation time, normalizing,
and taking the square root to obtain an antenna based correction.  The visibility weights were
then scaled by the inverse of the square of the correction, upweighting points when
the amplitude was high and downweighting points of low significance.  The effect of these 
corrections is shown in Figures~\ref{fig:scint}b and \ref{fig:scint}d, which show the visibility amplitudes 
and image residuals respectively.  In this example, the signal to noise ratio (SNR)
of the detection is improved by a factor of two when the visibility amplitudes are
corrected for scintillation, which reduces the position determination error 
by a corresponding factor.

\section[Positional determination and parallax fitting]{Positional determination and parallax fitting}
\label{techniques:posdet}
The calibrated visibilities for each pulsar were averaged in frequency, written to disk
and loaded into DIFMAP.  A single delta function model component was initialized
at the peak of the dirty image, and the modelfit command used to obtain the best fit
for the pulsar visibility data.  Each observing band, as well as the combined dataset, 
was then imaged separately using uniform weighting (as before with a bin size of 2 pixels 
and weights set to predicted baseline SNR) producing 8 images which were saved 
and read back into AIPS using the task FITLD.  
Variations to the weighting scheme are discussed in Section~\ref{techniques:check:weights}.
The AIPS task JMFIT was used to determine the pulsar position and formal 
(SNR--based) errors in the image plane.  
Systematic offsets of several (3--6) mas were observed at all epochs 
between the 4 bands that were only contributed to by the three ATNF 
antennas, and the 4 bands common to all antennas.  The magnitude and 
direction of these offsets varied between epochs, and so the ATNF--only 
bands (which additionally had formal errors of approximately five times 
the other bands, primarily due to the shorter baselines) were dropped.   
It is suspected that these offsets were caused by the lack of an accurate amplitude 
self--calibration solution for these bands.

\clearpage

For each pulsar, the best fit values of J2000 position (right ascension and declination), proper motion 
(right ascension and declination) and parallax were initially determined by iteratively 
minimising the error
function calculated by summing the value of predicted minus actual position
over all measurements, weighted by the individual measurement errors.  The iterative
minimisation code used is described in \citet{brisken02a}, and accounts for the covariances
between parameters when calculating the uncertainty of fitted parameters. A reference time
for the proper motion was chosen to be 31 Dec 2006 (MJD 54100) to minimize
proper motion uncertainty contibutions to be position uncertainty.

This approach yields error estimates for the 5 fitted parameters which are almost certainly
an underestimate, for two reasons:
\begin{enumerate}
\item There are systematic errors, varying from band to band within an epoch (intra--epoch
errors) such as the residual unmodeled differential ionosphere, bandpass effects etc, and 
varying from epoch to epoch (inter--epoch errors) caused by effects dependent on 
observing time, such as seasonal or diurnal ionospheric variations, and variations
in refractive scintillation image wander \citep{rickett90a}.  These should increase
the error on each individual measurement, but estimating the magnitude of the systematic
component for each individual measurement is poorly constrained.
\item Each measurement is assumed to be completely independent, whereas as noted above
correlated errors are expected between measurements from the same epoch.  In essence, this
approach overestimates the number of independent measurements, lowering the reduced
chi--squared and implying a better fit than the actual result.
\end{enumerate}
It is possible to make an estimate of the magnitude of the intra--epoch errors by comparing
the scatter in fitted positions from the individual bands.  Forming a weighted centroid
position for each epoch utilising all measurements from that epoch allows an estimation
of the likelihood that the measured points are consistent with that centroid, through the 
calculation of a reduced chi--squared value.  If the reduced chi--squared value exceeds unity,
the presence of unmodeled systematic errors can be inferred.

Since there is no a priori knowledge of the systematic error distribution,  
an equal systematic error was allocated to to each measurement, and added in
quadrature to the original measurement error. The errors in right ascension and declination 
are treated separately. This is necessarily an iterative procedure, since
the weighted centroid will be altered by the addition of these systematic error estimates.
In effect, this assumes a zero mean, gaussian distribution for the systematic errors.
Although this is unlikely to be the true distribution, it is the most reasonable assumption available, and
certainly more correct than assuming no systematic errors at all.  The intra--epoch systematic
error estimate for an epoch is obtained when the reduced chi--squared reaches unity.

Once a single position measurement and error has been calculated for each epoch,
the fit to position, parallax and proper motion was re--calculated, and 
the reduced chi--squared of the fit inspected again.   Since the degrees of freedom
were reduced, the addition of systematic errors did not always result in a lowering
of reduced chi--squared.  The presence of inter--epoch systematic errors was inferred if
the reduced chi--squared remained greater than unity.
Again, without knowledge of the underlying distribution of these errors, it was decided to
apportion an equal amount to each epoch, iterating until a reduced chi--squared of 
unity was obtained.  As with intra--epoch errors, right ascension and declination were 
treated separately.

Thus, the final astrometric dataset for each pulsar consisted of a single position measurement
for each epoch, with a total error equal to the weighted sum of the individual band
formal errors,  added in quadrature to the estimates of intra--epoch and inter--epoch
systematic errors.

For each pulsar, the robustness of error estimation after the inclusion of estimated systematic
contributions was checked by implementing a bootstrap technique.
Bootstrapping, which involves repeated trials on samples selected with replacement 
from the population of measured position points, is a statistical technique allowing the 
estimation of parameter errors without a complete knowledge of the underlying distribution
\citep{press02a}. It differs from Monte Carlo analysis in that bootstrapping uses the original
sample set as a population to draw from, whereas Monte Carlo techniques generate data
samples based upon assumed parameter distributions.
 In this instance, the original single band position measurements
were taken as the population, and N samples were drawn, where N was the original number
of measurements.  Each bootstrap consisted of 10,000 such trials, and the parameter
errors estimated from the variance of the resultant distributions.
A minimum of 3 different epochs needed to be included to ensure a fit was possible - on
the rare occasion that a trial did not satisfy this requirement, it was re--drawn.

Thus, three sets of fitted parameters and errors were obtained for each pulsar -- a ``naive" result
using the single--band positions (which generally possessed a reduced chi--squared 
greater than 1.0), a bootstrap result, and a more conservative estimate which
attempts to account for the impact of systematic errors (the ``inclusive" fit).  
In general, the estimated magnitude
of errors on fitted parameters increased through these three different schemes.  
Typically, the ratio in the errors on the inclusive fit to those on the naive fit ranged from 0.95 to 1.90.  
The final error values obtained from the inclusive fit are
the most accurate estimation possible, and are believed to be inherently conservative.  All quoted 
errors are 1$\sigma$ unless otherwise stated.

Pulsar positions are given in the ICRF frame, and position errors are the formal astrometric fit
errors only and do not include the component due to reference source position uncertainty.  
With the exception of B0736--303, all reference sources have been previously observed in ICRF
or VLBA Calibrator Series (VCS) observations, and possess position uncertainties ranging 
from 0.8 mas to 1.6 mas.  B0736--303 was observed using the VLBA in phase--referencing mode,
with ICRF source B0736--332 used as a calibrator. This relative astrometry obtained the position
of B0736--303 in the ICRF frame, but the precision was limited to tens of milliarcseconds, due to the
large differential ionospheric contribution at the low elevation of the sources for the VLBA.  

This position error is unacceptably large, and accordingly the 
position of B0736--303 was ``reverse--engineered" from the astrometric fit to the
position of \pfour.  Since the position of \pfour\ is known to better than one mas in the Solar System
reference frame, and the Solar System frame and ICRF are aligned at the several mas level, 
the VLBI position obtained from the astrometric fit could be compared to the
timing position and a calibrator offset deduced.  As with the VLBA phase referencing to B0736--332,
a small position offset due to phase referencing errors can be expected, but as \pfour\ is much closer
to B0736--303 ($\sim$20 arcminutes, as opposed to three degrees for B0736--332), and is observed
at higher elevation using the LBA, these
residual errors are much smaller.  As described in Section \ref{results:binary:0737:vlbi}, the
position of B0736--332 was adjusted to ensure a match between the timing and VLBI positions
for \pfour, resulting in a final positional accuracy for B0736--303 estimated at approximately five mas.
In hindsight, phase--referencing of B0736--303 to B0736--332 should also have been performed
using the LBA, but insufficient time was available for such an observation after this position error 
was discovered, and a single LBA observation would be unlikely to improve on the estimated
five mas accuracy eventually determined for B0736--303.

\section[Technique check: PSR J1559--4438]{Technique check: PSR J1559--4438}
\label{techniques:check}

\subsection[Initial results]{Initial results}
\label{techniques:check:init}
\begin{figure}
\begin{center}
\includegraphics[width=0.8\textwidth]{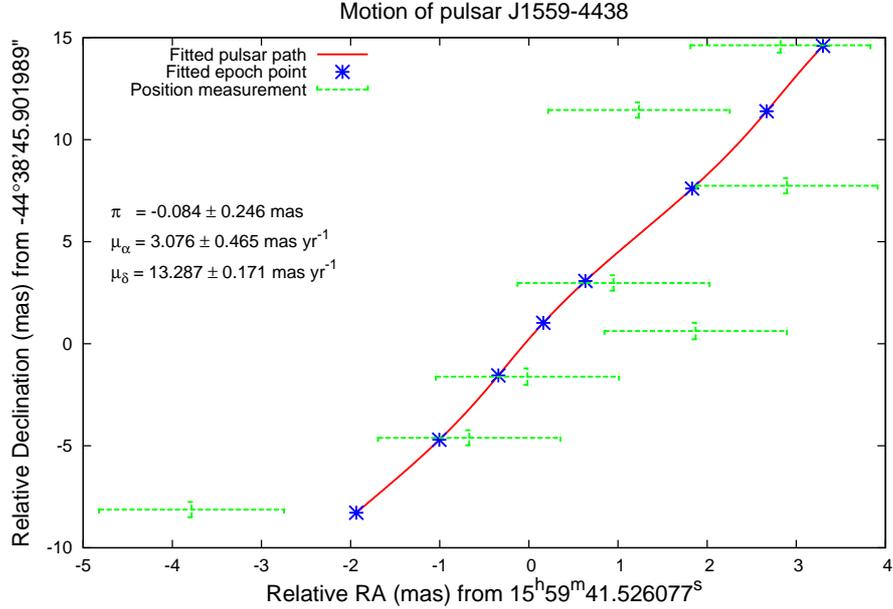}
\caption[Position fit for \pfive, using sensitivity--weighted visibilities]
{Motion of the pulsar in right ascension and declination, with measured positions
overlaid on the best fit.  Sensitivity--weighted visibilities were used. 
The motion of the pulsar is positive in right ascension and declination.
The first epoch (lower left) is clearly inconsistent.}
\label{fig:weightedfit}
\end{center}
\end{figure}

Using the techniques described above, initial results were obtained for \pfive\ (shown
in Table~\ref{tab:initial}).  The motion of \pfive\ in
right ascension and declination is shown in Figure~\ref{fig:weightedfit}, along with the fitted path
according to the inclusive fit.
The fit is clearly unsatisfactory, since it predicts a negative parallax (though consistent with zero).
The final column of Table~\ref{tab:initial} shows that systematic errors far exceed the nominal
single--epoch positional accuracies, and inspection of Figure~\ref{fig:weightedfit} shows that 
the first epoch (MJD 53870) is markedly discrepant with the remaining epochs.

\begin{deluxetable}{lrrr}
\tabletypesize{\tiny}
\tablecaption{Initial results (sensitivity--weighted visibilities) for \pfive}
\tablewidth{0pt}
\tablehead{
\colhead{Parameter} & \colhead{Naive fit} & \colhead{Bootstrap fit} & \colhead{Inclusive fit}
}
\startdata
Right ascension (J2000)	&  15:59:41.526092 $\pm$ 0.000005	
					& 15:59:35.526093 $\pm$ 0.000031	
					& 15:59:41.526077 $\pm$ 0.000026		\\
Declination (J2000) 	&  -44:38:45.902028 $\pm$ 0.000017	
				& -44:38:45.902034 $\pm$ 0.000062	
				& -44:38:45.901989 $\pm$ 0.000102	\\
$\mu_{\alpha}$	(mas yr$^{-1}$)	&  2.64 $\pm$ 0.07\phn\phn\phn\phn	
						& 2.65 $\pm$ 0.44\phn\phn\phn\phn		
						& 3.08 $\pm$ 0.46\phn\phn\phn\phn	   \\
$\mu_{\delta}$	(mas yr$^{-1}$)	&  13.36 $\pm$ 0.02\phn\phn\phn\phn	
						& 13.36 $\pm$ 0.06\phn\phn\phn\phn		
						& 13.29 $\pm$ 0.17\phn\phn\phn\phn	   \\
$\pi$ (mas)	 			&  -0.066 $\pm$ 0.038\phn\phn\phn	
						& -0.054 $\pm$ 0.193\phn\phn\phn	
						& -0.083 $\pm$ 0.245\phn\phn\phn  \\
Nominal distance (pc)		&  -15100 	\phs \phn\phd\phn\phn\phn\phn\phn\phn$\  $			
						& -18500 \phs \phn\phd\phn\phn\phn\phn\phn\phn$\  $		
						& -11900 \phs \phn\phd\phn\phn\phn\phn\phn\phn$\  $	\\
Nominal $v_{t}$ (km s$^{-1}$)	&  -974 \phs \phn\phd\phn\phn\phn\phn\phn\phn$\  $
						& -1190 \phs \phn\phd\phn\phn\phn\phn\phn\phn$\  $
						& -770  \phs \phn\phd\phn\phn\phn\phn\phn\phn$\  $ \\
Reduced chi--squared 		&  8.8 \phs \phn\phd\phn\phn\phn\phn\phn\phn$\  $
						&  \phs \phn\phd\phn\phn\phn\phn\phn\phn$\  $
						& 1.0 \phs \phn\phd\phn\phn\phn\phn\phn\phn$\  $	\\
Mean epoch fit error (mas)			&  \phs \phn\phd\phn\phn\phn\phn\phn\phn$\  $ 
									&  \phs \phn\phd\phn\phn\phn\phn\phn\phn$\  $
									& 0.198 \phs \phn\phd\phn\phn\phn\phn\phn\phn$\  $ \\
Intra--epoch sys. error (mas) 	&  \phs \phn\phd\phn\phn\phn\phn\phn\phn$\  $ 
									&  \phs \phn\phd\phn\phn\phn\phn\phn\phn$\  $
									& 0.097 \phs \phn\phd\phn\phn\phn\phn\phn\phn$\  $ \\
Inter--epoch sys. error (mas) 	&  \phs \phn\phd\phn\phn\phn\phn\phn\phn$\  $ 
									&  \phs \phn\phd\phn\phn\phn\phn\phn\phn$\  $
									& 1.070 \phs \phn\phd\phn\phn\phn\phn\phn\phn$\  $ \\
\enddata
\label{tab:initial}
\end{deluxetable}

As shown below, fine-tuning of the data reduction is required in order to obtain optimal results 
from each pulsar, in particular with regard to the details of the ionospheric corrections and the 
visibility weighting schemes using in imaging.  The steps taken for \pfive\ in are described 
below in Sections~\ref{techniques:check:tecor} and \ref{techniques:check:weights}.

\subsection[Ionospheric correction]{Ionospheric correction}
\label{techniques:check:tecor}

The obvious source of the large systematic errors present in the initial fit shown in 
Section~\ref{techniques:check:init} is the varying ionospheric correction 
between epochs.  Accordingly, as a first check, the position fits for each epoch were 
recalculated after subtracting the differential ionospheric correction 
between \pfive\ and its phase reference -- in effect, removing 
the applied ionospheric correction and leaving the data uncorrected
for ionospheric effects.  This was implemented
using a ParselTongue script which made a two--point interpolation between adjacent calibrator scans
to calculate the differential correction to the target (which was not absorbed into the fringe--fit),
which was stored in a CL table and subtracted using the AIPS task SPLAT.  The applied values
were saved for later analysis.

\begin{figure}[h!]
\begin{center}
\includegraphics[width=0.8\textwidth]{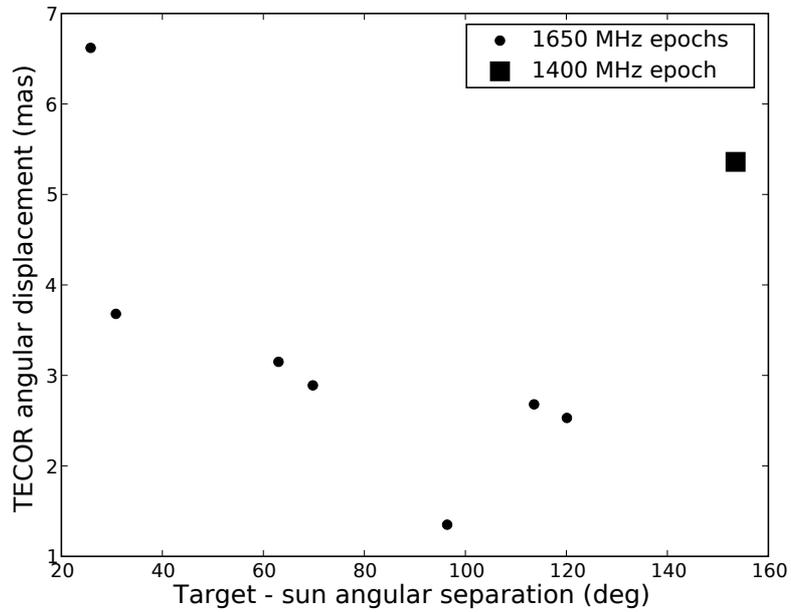}
\caption[Position shifts caused by ionospheric correction vs. pulsar--Sun angular separation]
{Shift in fitted position due to ionospheric correction for \pfive\ vs 
angular separation between the pulsar and the Sun.  The single 1400 MHz epoch has an
unusually large correction given the large angular separation between the pulsar and the Sun.}
\label{fig:tecor_sun}
\end{center}
\end{figure}

\begin{figure}
\begin{center}
\includegraphics[width=0.8\textwidth]{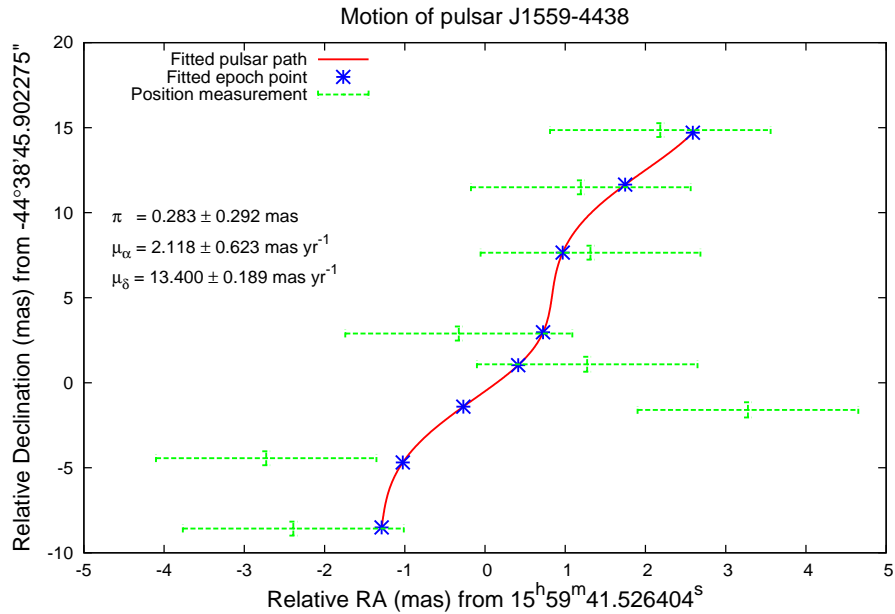}
\caption[Position fit for \pfive\ without ionospheric corrections]
{Motion of the pulsar in right ascension and declination, with measured positions
overlaid on the best fit, when ionospheric corrections have been removed.  
Sensitivity--weighted visibilities were used. 
The first epoch (lower left) is now more consistent, but the third epoch (during which the pulsar--Sun
angular separation was only 26 degrees) is now inconsistent.}
\label{fig:notecorfit}
\end{center}
\end{figure}

Intuitively, the largest ionospheric corrections would be expected when the angular displacement of
the pulsar from the sun is small, since this angular separation (along with the level of solar activity) 
is the largest influence on the TEC. 
The angular displacement at each epoch between the original fitted 
position and the position obtained when ionospheric correction was 
removed is presented in Table~\ref{tab:tecorshifts}, and
plotted against angular separation of the pulsar from the sun at the time of observation in
Figure~\ref{fig:tecor_sun}.  The revised astrometric fit obtained without ionospheric correction 
is plotted in Figure~\ref{fig:notecorfit}.

It is immediately apparent from Figure~\ref{fig:tecor_sun} that the first epoch 
(MJD 53870) is discrepant in that
the position shift due to ionospheric correction is unusually large, given the large angular
separation from the Sun.  As shown in Figure~\ref{fig:notecorfit}, 
this epoch becomes more consistent with the fit when the ionospheric 
correction is removed, this epoch
becomes more consistent with the fit, but the third epoch (MJD 54057) becomes much more
inconsistent.  This is unsurprising, however, since this epoch had the smallest pulsar--Sun separation
and the largest ionospheric corrections.  

To investigate whether the chosen ionospheric map was at fault, the first epoch was re--reduced with
all available maps from the NASA CDDIS 
archive\footnote{ftp://cddis.gsfc.nasa.gov/pub/gps/products/ionex/}, but no significant change 
was found in fitted position.  Given that the different TEC maps make use of many of the same
GPS stations, this is unsurprising.  This problem is exacerbated at southern declinations due
to the low density of GPS receivers at southern latitudes.  
Additionally, any errors in the TEC maps would have an
impact $\sim\,$40\% greater for this first epoch, due to its lower observing frequency of 1400 MHz,
compared with the 1650 MHz center frequency used for all subsequent observations.

\clearpage

\begin{deluxetable}{lccc}
\tabletypesize{\tiny}
\tablecaption{Average position shift due to ionospheric corrections for \pfive}
\tablewidth{0pt}
\tablehead{
\colhead{Epoch MJD} & \colhead{RA shift (mas)} & \colhead{Dec shift (mas)} & \colhead{Sun separation (deg)}
}
\startdata
53870	&  $-$5.36	& \phs0.27	& 153.5	\\
53970	&  $-$1.34	& \phs0.15	& \phn96.4	\\
54057	&  $-$6.61	& \phs0.39	& \phn25.8	\\
54127	&  $-$3.15	& $-$0.07		& \phn63.0	\\
54182	&  $-$2.66	& \phs0.35	& 113.6	\\
54307	&  $-$2.51	& \phs0.34	& 120.1	\\
54413	&  $-$3.68	& \phs0.20	& \phn30.8	\\
54500	&  $-$2.89	& \phs0.05	& \phn69.8	\\
\enddata
\label{tab:tecorshifts}
\end{deluxetable}

\clearpage

\begin{figure}
\begin{center}
\includegraphics[width=0.8\textwidth]{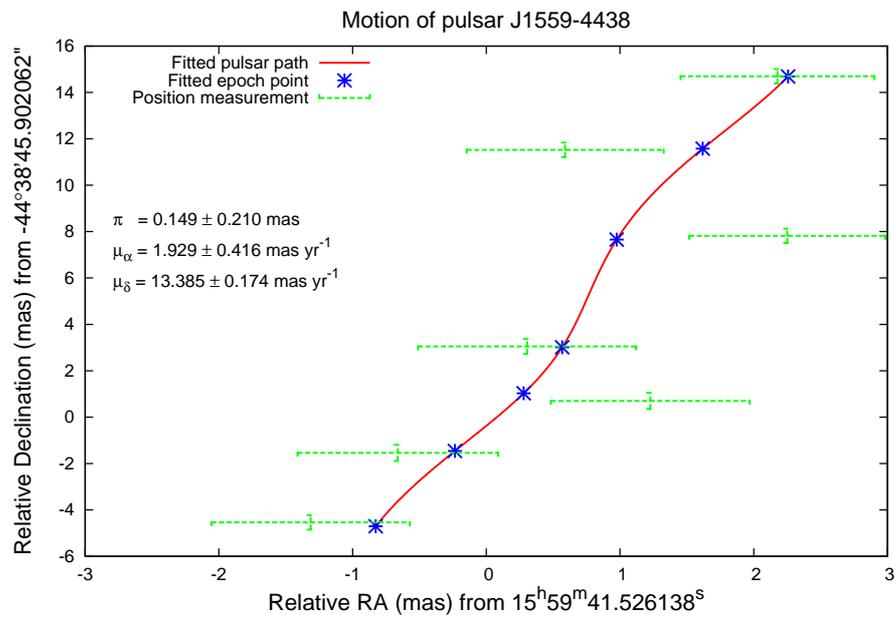}
\caption[Position fit for \pfive\ after dropping the first epoch]
{Motion of the pulsar in right ascension and declination, with measured positions
overlaid on the best fit, with ionospheric corrections reinstated but the first epoch dropped.  
Sensitivity--weighted visibilities were used. 
The fit is improved considerably.}
\label{fig:weighted_no_v190b}
\end{center}
\end{figure}

Thus, due to the probability of residual ionospheric errors for this epoch, and also
the potential for frequency--dependent calibrator source structure, the first epoch 
(MJD 53870, the only 1400 MHz epoch) was dropped from all further analysis.  
The fit to the remaining seven epochs, with
ionospheric corrections re--enabled, is shown in Figure~\ref{fig:weighted_no_v190b}.  While
a realistic fit is now obtained, the measurement of parallax is still not significant.

\clearpage

\subsection[Data weights]{Data weights}
\label{techniques:check:weights}
Initial pulsar imaging and position fitting used visibilities weighted according
to the best estimate of instantaneous baseline sensitivity.  Whilst this is theoretically optimal 
for data which consists only of signal $S$ and additive random thermal errors \etherm, 
it fails to account for the presence of multiplicative 
systematic errors $\esys=e^{i \phisys}$ caused by 
residual calibration errors.  Typically, these systematic
errors are dominated by atmospheric and ionospheric gradients, although other contributions
include antenna and calibrator source position errors, time--variation of antenna bandpasses, 
and instrumental phase jitter. \citet{fomalont05a}
presents a theoretical review of phase referencing errors, while \citet{pradel06a} presents a 
simulation--based approach.

If \phisys\ had zero--mean and was ergodic, its effect would be
indistinguishable from thermal noise and could be easily estimated, allowing the visibility weights
to be corrected.  For antenna/source position errors
and large--scale atmospheric/ionospheric structure, however, the residual errors are correlated
over long times, causing systematic shifts in the fitted position for the target.

When normal sensitivity--based 
weighting is employed in the presence of substantial and persistent systematic
phase errors, the systematic noise on the most sensitive baseline will be absorbed into the fit, at the cost 
of a poorer fit to the less sensitive baselines.  The magnitude of the induced error will be
dependent on the ratio of systematic to thermal errors, and the discrepancy in sensitivity
between the most and least sensitive baselines in the array.   For the LBA,
the most sensitive baseline (Parkes--DSS43: system equivalent flux density 30 Jy) 
is roughly 13 times more sensitive than 
the least sensitive (Hobart--Mopra: 380 Jy).  Thus, the LBA is particularly susceptible to 
the influence of systematic errors, due to the pronounced variation in baseline sensitivities.

\begin{figure}
\begin{center}
\includegraphics[width=0.8\textwidth]{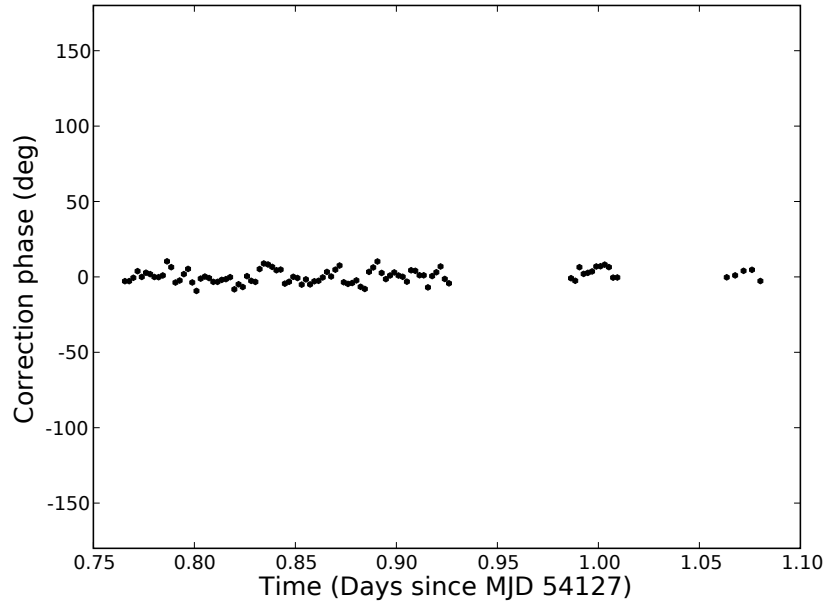}
\caption[Self--calibration corrections for the ATCA station on  \pfive]
{Self--calibration corrections, using a three--minute timescale, 
for the ATCA station on \pfive.  Clear systematic 
deviations are seen from the zero--mean distribution expected from purely thermal noise.
The RMS of the corrections exceeds those expected due to thermal noise by an order of magnitude.}
\label{fig:selfcal}
\end{center}
\end{figure}

The systematic errors can be crudely estimated (in a model--dependent fashion) 
by performing phase--only self--calibration on the target pulsar over a sufficiently 
long timescale to obtain sufficient SNR,
and comparing the magnitude of the corrections to those expected from thermal noise alone.
While this approach probes systematic
errors over a somewhat shorter time period than those which would dominate for inter--epoch errors
(tens of minutes, rather than hours), it is illustrative of the presence of systematic 
errors overall.
Such corrections are shown in Figure~\ref{fig:selfcal} for the ATCA station using a three--minute
solution interval during the observation on MJD 54127.  The corrections are clearly correlated 
over timescales of tens of minutes, and
the RMS deviation of $4.4\,^{\circ}$\ is an order of magnitude greater than the estimated thermal 
phase RMS of $0.4\,^{\circ}$\ (calculated in this high--SNR limit as station sensitivity divided
divided by target flux density, scaled by pulse filtering gain and converted from radians to degrees).  
Thus, for this observation, systematic errors $\gg$ thermal errors and weighting visibilities 
by sensitivity actually degrades the quality of the position fit.

If the average \esys\ could be accurately estimated for each baseline over the duration of an 
experiment, the baseline visibility weights could be adjusted by assuming that \esys\ is time--invariant.
Even more desirable would be the estimation of \esys\ as a function of time, allowing 
a time--variable adjustment of the visibility weights.
Given present instrumentation, there is no way to reliably estimate systematic 
error (time dependent or independent) in a model independent fashion.
In the limit where $\esys\gg\etherm$, however,  
the visibility weight for each baseline (regardless of sensitivity) 
will be dominated by the systematic error contribution, resulting in
approximately equal weights for all visibilities.  Accordingly, visibility weights for all
baselines were reset to an equal, constant value and data reduction repeated.  
The results are discussed below in Section~\ref{techniques:check:final}.

\subsection[Final results]{Final results}
\label{techniques:check:final}
\begin{figure}[h!]
\begin{center}
\includegraphics[width=0.8\textwidth]{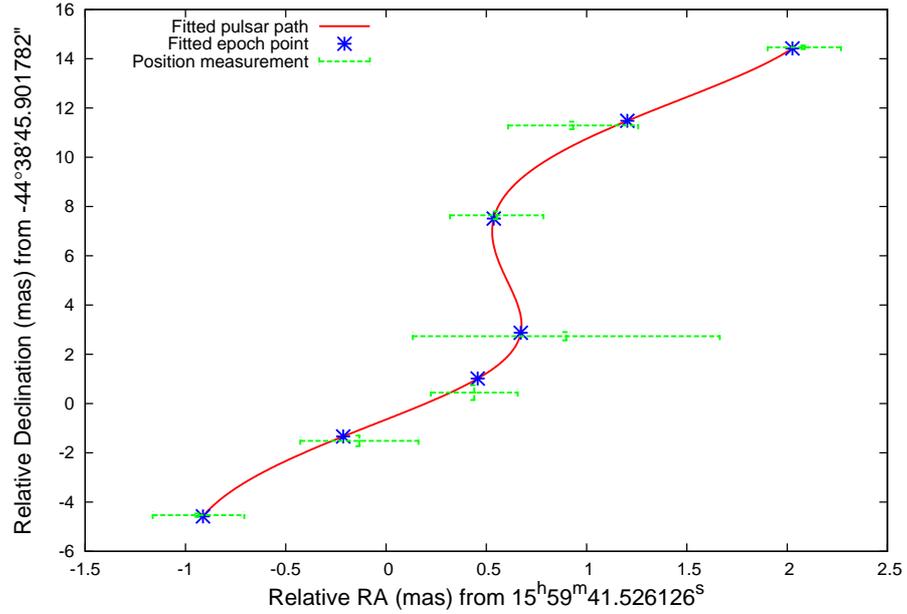}
\caption[Position fit for \pfive\ using equally weighted visibilities]
{Motion of the pulsar in right ascension and declination, with measured positions
overlaid on the best fit, with ionospheric corrections reinstated but the first epoch dropped.  
Equally weighted visibilities were used, mitigating systematic errors and allowing for the first
time a significant measurement of the parallax for this pulsar.}
\label{fig:final}
\end{center}
\end{figure}

The revised fit obtained using equally weighted visibility data is described in Table~\ref{tab:final} and
plotted in Figure~\ref{fig:final}.  Through comparison of Table~\ref{tab:final} with Table~\ref{tab:initial},
it can be seen that the average fit error for a single epoch has increased by 20\%, but the 
inter--epoch systematic error has decreased by 95\%.  Thus, while using equally weighted data
incurs a small sensitivity penalty, it benefits significantly through the reduced susceptibility
to systematic errors.  The fitted distance of 2600 pc is consistent with the NE2001 distance
of 2350 pc, as well as HI absorption measurements which imply a lower distance limit of
$2000 \pm 500$\ pc \citep{koribalski95a}.
The implications of this distance measurement are discussed in Section \ref{results:isolated:1559}.

The use of natural weighting, as opposed to uniform weighting, was investigated but found
to produce inferior results.  Fitted parameters remained relatively constant but errors on
the fitted parameters increased by $\sim\,$50\%.  This is unsurprising, since the use of
natural weighting promotes a larger beamsize due to the concentration of more visibility points 
at small {\it uv} distances.   Natural weighting is obviously more likely
to prove useful when $\etherm\gg\esys$, and its use for the weakest target pulsars is shown in
Chapter~\ref{results}.

\section[Optimal data weighting]{Optimal data weighting}
\label{techniques:optweight}
From the results shown in Sections~\ref{techniques:check:weights} and 
\ref{techniques:check:final}, it is clear that 
for \pfive\ the astrometric error budget is dominated by systematic errors, and that the
use of equally weighted visibility points is optimal.  However, Table~\ref{tab:targets} shows that 
this may not be the case for other pulsars in the target sample, as some are orders of magnitude
fainter than \pfive.  Accordingly,
the conditions under which sensitivity--weighted visibilities give superior results to
equally--weighted visibilities were investigated.  This was carried out by adding simulated
thermal noise of varying RMS to the existing dataset.

Three ``noisy" datasets $D_{A}$, $D_{B}$, and $D_{C}$\ were constructed by adding zero--mean,
gaussian--distributed noise to the real and imaginary visibility components of the original 
observations.  Since the theoretical single epoch SNR for sensitivity--weighted data should be 
$\sim800$ (a factor of 10 greater than the typically obtained SNR), 
the RMS of the added noise in the three datasets was set to 
predicted baseline sensitivity scaled by a factor of 20, 40, and 80, which should allow
a maximum single--epoch SNR of 40, 20 and 10 respectively.
The results of fitting the modified datasets (using the inclusive fit approach only) with
and without the use of sensitivity weighting are presented in Tables \ref{tab:optweight_with} and
\ref{tab:optweight_without} respectively.

\begin{deluxetable}{lrrr}
\tabletypesize{\tiny}
\tablecaption{Final results (equally weighted visibilities) for PSR J1559--4438}
\tablewidth{0pt}
\tablehead{
\colhead{Parameter} & \colhead{Naive fit} & \colhead{Bootstrap fit} & \colhead{Inclusive fit}
}
\startdata
Right ascension (J2000)	& 15:59:41.526121 $\pm$ 0.000006	
					& 15:59:35.526121 $\pm$ 0.000008	
					& 15:59:41.526126 $\pm$ 0.000008		\\
Declination (J2000) 	& -44:38:45.901849 $\pm$ 0.000020	
				& -44:38:45.901859 $\pm$ 0.000072	
				& -44:38:45.901778 $\pm$ 0.000035	\\
$\mu_{\alpha}$	(mas yr$^{-1}$)	& 1.62 $\pm$ 0.08\phn\phn\phn\phn	
					& 1.60 $\pm$ 0.16\phn\phn\phn\phn		
					& 1.52 $\pm$ 0.14\phn\phn\phn\phn	   \\
$\mu_{\delta}$	(mas yr$^{-1}$)	&  13.20 $\pm$ 0.02\phn\phn\phn\phn	
					& 13.21 $\pm$ 0.08\phn\phn\phn\phn		
					& 13.15 $\pm$ 0.05\phn\phn\phn\phn	   \\
$\pi$ (mas)	 		&  0.280 $\pm$ 0.048\phn\phn\phn	
					& 0.291 $\pm$ 0.111\phn\phn\phn	
					& 0.384 $\pm$ 0.081\phn\phn\phn  \\
Nominal distance (pc)	& 3570 	\phs \phn\phd\phn\phn\phn\phn\phn\phn$\  $			
					& 3440 \phs \phn\phd\phn\phn\phn\phn\phn\phn$\  $		
					& 2600 \phs \phn\phd\phn\phn\phn\phn\phn\phn$\  $	\\
Nominal $v_{t}$ (km s$^{-1}$)	&  225 \phs \phn\phd\phn\phn\phn\phn\phn\phn$\  $
					& 217 \phs \phn\phd\phn\phn\phn\phn\phn\phn$\  $
					& 164  \phs \phn\phd\phn\phn\phn\phn\phn\phn$\  $ \\
Reduced chi--squared 	&  3.6 \phs \phn\phd\phn\phn\phn\phn\phn\phn$\  $
					&  \phs \phn\phd\phn\phn\phn\phn\phn\phn$\  $
					& 1.0 \phs \phn\phd\phn\phn\phn\phn\phn\phn$\  $	\\
Mean epoch fit error (mas)			&  \phs \phn\phd\phn\phn\phn\phn\phn\phn$\  $ 
									&  \phs \phn\phd\phn\phn\phn\phn\phn\phn$\  $
									& 0.242 \phs \phn\phd\phn\phn\phn\phn\phn\phn$\  $ \\
Intra--epoch sys. error (mas) 	&  \phs \phn\phd\phn\phn\phn\phn\phn\phn$\  $ 
									&  \phs \phn\phd\phn\phn\phn\phn\phn\phn$\  $
									& 0.259 \phs \phn\phd\phn\phn\phn\phn\phn\phn$\  $ \\
Inter--epoch sys. error (mas) 	&  \phs \phn\phd\phn\phn\phn\phn\phn\phn$\  $ 
									&  \phs \phn\phd\phn\phn\phn\phn\phn\phn$\  $
									& 0.055 \phs \phn\phd\phn\phn\phn\phn\phn\phn$\  $ \\
\enddata
\label{tab:final}
\end{deluxetable}
\clearpage

\begin{deluxetable}{lrrr}
\tabletypesize{\tiny}
\tablecaption{Noise--added fits for PSR J1559--4438, sensitivity--weighted visibilities}
\tablewidth{0pt}
\tablehead{
\colhead{Parameter} & \colhead{$D_{A}$} & \colhead{$D_{B}$} & \colhead{$D_{C}$}
}
\startdata
Right ascension (J2000)	& 15:59:41.526163 $\pm$ 0.000027	
					& 15:59:35.526086 $\pm$ 0.000025	
					& 15:	59:41.526025 $\pm$ 0.000041		\\
Declination (J2000) 	& -44:38:45.902054 $\pm$ 0.000154	
				& -44:38:45.902131 $\pm$ 0.000163	
				& -44:38:45.901760 $\pm$ 0.000258	\\
$\mu_{\alpha}$	(mas yr$^{-1}$)	& 2.03 $\pm$ 0.46\phn\phn\phn\phn	
					& 1.44 $\pm$ 0.59\phn\phn\phn\phn		
					& 2.83 $\pm$ 0.79\phn\phn\phn\phn	   \\
$\mu_{\delta}$	(mas yr$^{-1}$)	& 13.39 $\pm$ 0.24\phn\phn\phn\phn	
					& 13.40 $\pm$ 0.21\phn\phn\phn\phn		
					& 13.00 $\pm$ 0.44\phn\phn\phn\phn	   \\
$\pi$ (mas)	 		& 0.109 $\pm$ 0.278\phn\phn\phn	
					& 0.270 $\pm$ 0.373\phn\phn\phn	
					& 0.621 $\pm$ 0.579\phn\phn\phn  \\
Nominal distance (pc)	& 9200 \phs \phn\phd\phn\phn\phn\phn\phn\phn$\  $			
					& 3700 \phs \phn\phd\phn\phn\phn\phn\phn\phn$\  $		
					& 1610 \phs \phn\phd\phn\phn\phn\phn\phn\phn$\  $	\\
Nominal $v_{t}$ (km s$^{-1}$)	& 590 \phs \phn\phd\phn\phn\phn\phn\phn\phn$\  $
					& 237 \phs \phn\phd\phn\phn\phn\phn\phn\phn$\  $
					& 102  \phs \phn\phd\phn\phn\phn\phn\phn\phn$\  $ \\
Reduced chi--squared 	& 1.0 \phs \phn\phd\phn\phn\phn\phn\phn\phn$\  $
					& 1.0 \phs \phn\phd\phn\phn\phn\phn\phn\phn$\  $
					& 0.5 \phs \phn\phd\phn\phn\phn\phn\phn\phn$\  $	\\
Mean epoch fit error(mas)			& 0.441 \phs \phn\phd\phn\phn\phn\phn\phn\phn$\  $ 
									& 0.904 \phs \phn\phd\phn\phn\phn\phn\phn\phn$\  $
									& 1.308 \phs \phn\phd\phn\phn\phn\phn\phn\phn$\  $ \\
Intra--epoch sys. error (mas) 	& 0.504 \phs \phn\phd\phn\phn\phn\phn\phn\phn$\  $ 
									& 1.412 \phs \phn\phd\phn\phn\phn\phn\phn\phn$\  $
									& 0.863 \phs \phn\phd\phn\phn\phn\phn\phn\phn$\  $ \\
Inter--epoch sys. error (mas) 	& 0.769 \phs \phn\phd\phn\phn\phn\phn\phn\phn$\  $ 
									& 0.315 \phs \phn\phd\phn\phn\phn\phn\phn\phn$\  $
									& 0.0 \phs \phn\phd\phn\phn\phn\phn\phn\phn$\  $ \\
Average single--epoch SNR 	& 33  \phs \phn\phd\phn\phn\phn\phn\phn\phn$\  $
						& 17  \phs \phn\phd\phn\phn\phn\phn\phn\phn$\  $
						& 9  \phs \phn\phd\phn\phn\phn\phn\phn\phn$\  $\\
\enddata
\label{tab:optweight_with}
\end{deluxetable}

\begin{deluxetable}{lrrr}
\tabletypesize{\tiny}
\tablecaption{Noise--added fits for PSR J1559--4438, equally weighted visibilities}
\tablewidth{0pt}
\tablehead{
\colhead{Parameter} & \colhead{$D_{A}$} & \colhead{$D_{B}$} & \colhead{$D_{C}$}
}
\startdata
Right ascension (J2000)	& 15:59:41.526142 $\pm$ 0.000026
					& 15:59:35.526115 $\pm$ 0.000042	
					& 15:59:41.526268 $\pm$ 0.000084		\\
Declination (J2000) 	& -44:38:45.901745 $\pm$ 0.000076	
				& -44:38:45.902220 $\pm$ 0.000234	
				& -44:38:45.901798 $\pm$ 0.000420	\\
$\mu_{\alpha}$	(mas yr$^{-1}$)	& 0.91 $\pm$ 0.38\phn\phn\phn\phn	
					& 0.22 $\pm$ 0.99\phn\phn\phn\phn		
					& -0.04 $\pm$ 1.20\phn\phn\phn\phn	   \\
$\mu_{\delta}$	(mas yr$^{-1}$)	&  13.18 $\pm$ 0.10\phn\phn\phn\phn	
					& 13.56 $\pm$ 0.32\phn\phn\phn\phn		
					& 12.65 $\pm$ 0.65\phn\phn\phn\phn	   \\
$\pi$ (mas)	 		& 0.544 $\pm$ 0.235\phn\phn\phn	
					& 0.760 $\pm$ 0.597\phn\phn\phn	
					& 2.092 $\pm$ 0.787\phn\phn\phn  \\
Nominal distance (pc)	& 1840 \phs \phn\phd\phn\phn\phn\phn\phn\phn$\  $			
					& 1320 \phs \phn\phd\phn\phn\phn\phn\phn\phn$\  $		
					& 480 \phs \phn\phd\phn\phn\phn\phn\phn\phn$\  $	\\
Nominal $v_{t}$ (km s$^{-1}$)	& 115 \phs \phn\phd\phn\phn\phn\phn\phn\phn$\  $
					& 85 \phs \phn\phd\phn\phn\phn\phn\phn\phn$\  $
					& 29  \phs \phn\phd\phn\phn\phn\phn\phn\phn$\  $ \\
Reduced chi--squared 	& 0.8 \phs \phn\phd\phn\phn\phn\phn\phn\phn$\  $
					& 1.0 \phs \phn\phd\phn\phn\phn\phn\phn\phn$\  $
					& 0.9 \phs \phn\phd\phn\phn\phn\phn\phn\phn$\  $	\\
Mean epoch fit error(mas)			& 0.618 \phs \phn\phd\phn\phn\phn\phn\phn\phn$\  $ 
									& 0.937 \phs \phn\phd\phn\phn\phn\phn\phn\phn$\  $
									& 1.853 \phs \phn\phd\phn\phn\phn\phn\phn\phn$\  $ \\
Intra--epoch sys. error (mas) 	& 0.791 \phs \phn\phd\phn\phn\phn\phn\phn\phn$\  $ 
									& 1.433 \phs \phn\phd\phn\phn\phn\phn\phn\phn$\  $
									& 1.531 \phs \phn\phd\phn\phn\phn\phn\phn\phn$\  $ \\
Inter--epoch sys. error (mas) 	& 0.000 \phs \phn\phd\phn\phn\phn\phn\phn\phn$\  $ 
									& 0.380 \phs \phn\phd\phn\phn\phn\phn\phn\phn$\  $
									& 0.000 \phs \phn\phd\phn\phn\phn\phn\phn\phn$\  $ \\
Average single--epoch SNR 	& 20  \phs \phn\phd\phn\phn\phn\phn\phn\phn$\  $
						& 10  \phs \phn\phd\phn\phn\phn\phn\phn\phn$\  $
						& 5  \phs \phn\phd\phn\phn\phn\phn\phn\phn$\  $ \\
\enddata
\label{tab:optweight_without}
\end{deluxetable}

Tables~\ref{tab:optweight_with} and \ref{tab:optweight_without} show that while the equally weighted
dataset performs better for $D_{A}$, when the average epoch SNR is still high, its performance
rapidly deteriorates as the epoch SNR decreases. 
In $D_{C}$, the pulsar was not detected in several
epochs using equally--weighted data.  The reduction in performance is
less marked for the weighted datasets, although they are clearly still affected by
systematic errors.  However, if the pulsar was closer and the parallax larger, these systematics
would be less dominant, and weighted datasets would allow measurement of a parallax
when equally weighted datasets may be overwhelmed by thermal errors, to the point of not detecting
the pulsar in a single epoch.

As noted in Section~\ref{techniques:check:weights}, 
the use of weighted visibility points would always be
optimal if the weights could include an estimate of the systematic error contribution to that
visibility.  In the absence of such an estimate, the results obtained in this investigation lead to the 
guideline that for the LBA (with typical observing
conditions and calibrator throws), the transition region from systematic to thermal error dominated
astrometry occurs when the single--epoch detection SNR falls to approximately 10 for
equally--weighted visibilities.  This is
shown by the similarity of result quality for $D_{B}$, where the average epoch SNR was approximately 
10 for the equally--weighted visibilities.
Alternatively, both weighting regimes can be used and average total single--epoch error 
(formal + systematic) can be compared to estimate the optimal weighting scheme.
Again, this is borne out in the simulated datasets, where a transition
from systematic errors dominating to epoch fit errors dominating is seen as more noise is added.
These guidelines have been applied to the results obtained in Chapter~\ref{results}.

\section[Contributions to systematic error]{Contributions to systematic error}
\label{techniques:errorbudget}
The major contributions to systematic error in VLBI astrometry 
include geometric model errors (station/source
position, Earth Orientation Parameters), residual ionospheric and tropospheric errors, 
variable phase reference source structure, and image
wander due to refractive scintillation.  Of these, at 1650 MHz residual ionospheric errors would be
expected to dominate, despite the a priori ionospheric calibration employed 
\citep[e.g.][]{brisken02a}.

It is difficult to estimate the magnitude of the potential
correlation between epochs of the residual ionospheric
errors, due to the model--dependence of the ionospheric correction.  Whilst some correlation is
likely, it is unlikely to dominate, due to the significant variability inherent in the ionosphere.
Residual tropospheric errors
also have the potential for some correlation from epoch to epoch, for example due to imperfect
modeling of seasonal atmospheric variations.  Residual tropospheric errors, however, should be
considerably smaller than the ionospheric errors.

Geometric model errors cause relative astrometric errors which increase with calibrator throw.  
Earth Orientation Parameters (EOPs) are well determined
by geodetic observations and make minimal contributions to astrometric errors \citep{pradel06a}.
Similarly, the mean position of well-determined calibrators makes a minimal contribution.  
Some LBA stations, however, have position uncertainties of several cm, 
which could make a several hundred $\mu$as 
contribution to systematic error.  The magnitude of the error
depends on the source declination and the calibrator--target separation.  Given similar $uv$\ coverage
at different epochs, however, the offset will be largely constant with time and is absorbed into
the reference position of the target.
Future planned geodetic observations
will continue to improve LBA station positions, and reduce this systematic contribution.  Additionally,
as noted in Section~\ref{techniques:datared:geoiono}, small station position errors can be 
corrected post--correlation, which offers the potential to further improve previous position fits.

Refractive image wander is caused by large--scale fluctuations in the ISM, and can
be estimated based on the strength of the pulsar scattering and the scattering disk size
\citep[e.g.][]{rickett90a}.  For strong scattering, which includes PSR J1559--4438, the image
wander is less than the scattering disk size, if Kolmogorov turbulence is assumed for the scattering
material \citep{rickett90a}.  Thus, since Table~\ref{tab:scint} shows that scattering disk of 
\pfive\ is estimated to be only 133 $\mu$as at 1650 MHz, 
the maximum refractive image wander is $\ll 100\ \mu$as, and can be discounted 
as a source of systematic error.  Table~\ref{tab:scint} shows that refractive scintillation is unlikely
to be significant for any of the pulsars targeted in this thesis.

The variability of calibrator structure with time depends on the source chosen, but all compact
extragalactic radio sources are expected to show some variability, with typical
RMS values of 100~$\mu$as \citep{fomalont05a}.  
This image wander may be correlated from epoch to epoch over short time periods 
and absorbed into proper motion fits, but over long times (which could be longer than
as astrometric observing program), the mean apparent position will be constant.
Some sources, such as B2201+315 and B1739+522 (an ``other" and ``candidate" member of the
ICRF, respectively), have shown apparent motion of many hundreds of mas per year, which
can persist for up to several years \citep{titov07a}. As detailed long--term information
is not available for any of the reference sources used in this work, the impact on proper motion
due to source variability is difficult to quantify, but the stability of the fitted models used
for amplitude refinement (as discussed in Section~\ref{techniques:datared:fringefit}) suggests
that any reference source variability is likely to be small.  As shown in the following chapter, 
formal errors for proper motion measurements are only accurate to several hundred $\mu$as per year
for most pulsars, and so the contribution of reference source variability is likely to be minimal in
most cases.

The magnitude of observed systematic errors, and the correlation of systematic errors with
calibrator structure and angular separation from the target, are investigated further with the
full pulsar set in Section~\ref{results:sys_astro} and used to estimate the astrometric
accuracy which can be obtained with present and future VLBI arrays.
\chapter{ASTROMETRIC RESULTS AND INTERPRETATION}
\label{results}

The data obtained for each pulsar were processed using the pipeline described in 
Chapter~\ref{techniques}, applying the guidelines developed to minimise the total
astrometric error budget.  Table~\ref{tab:allresults} shows the fitted parameters
for each target pulsar, as well as estimates of mean thermal and systematic errors.
Unless otherwise stated, all errors are 1$\sigma$.


The implications of the measured values are discussed for each pulsar in turn 
in Sections~\ref{results:binary} and \ref{results:isolated} below,
with the analysis for \ptwo\ closely following that presented in \citet{deller08b}, the
analysis for \pfour\ closely following that presented in \citet{deller08c}, and the analysis
for \pfive\ closely following that presented in \citet{deller08a}.
In Sections~\ref{results:distvel} and \ref{results:sys_astro}, the revised distance and
velocity measurements for all pulsars are used to estimate the accuracy of pulsar distance
and velocity models, and predict the astrometric accuracy that will be achievable with future
VLBI observations.

\section[Binary millisecond pulsars]{Binary millisecond pulsars}
\label{results:binary}
As discussed in Section~\ref{pulsars:binary}, a variety of fascinating physics can be probed using 
binary millisecond pulsars, 
such as tests of GR and neutron star formation events.  The three binary pulsars 
targeted in this program were chosen because uncertainty regarding their true distance was
limiting the science which could otherwise be achieved with the system.  While significant parallaxes
were observed for \ptwo\ ($120\sigma$) and \pfour\ ($6\sigma$), no parallax could be
detected for \peight, which had previously been unsuccessfully targeted with the VLBA 
\citep{brisken02a}.  The implications of the distance and velocity measurements for
each pulsar are discussed below.

\begin{deluxetable}{lrrrr}
\tabletypesize{\tiny}
\tablecaption{Astrometric fits for all target pulsars}
\tablewidth{0pt}
\rotate
\tablehead{
\colhead{Parameter} & \colhead{PSR J0108-1431} & \colhead{PSR J0437--4715} & 
\colhead{PSR J0630--2834} & \colhead{PSR J0737--3039}
}
\startdata
Right ascension (J2000)		& 01:08:08.347016 $\pm$ 0.000088
                        				& 04:37:15.883250 $\pm$ 0.000003
                        				& 06:30:49.404393 $\pm$ 0.000043
						& 07:37:51.248419 $\pm$ 0.000026               \\[0.5ex]
Declination (J2000)		& -14:31:50.187139 $\pm$ 0.001069
                     	   		& -47:15:09.031863 $\pm$ 0.000037
                        			& -28:34:42.778813 $\pm$ 0.000372
					& -30:39:40.714310 $\pm$ 0.000099               \\[0.5ex]
$\mu_{\alpha}$  (mas yr$^{-1}$) 	& 75.05 $\pm$ 2.26\phn\phn\phn\phn
							& 121.679 $\pm$ 0.05\phn\phn\phn\phn
							& -46.30 $\pm$ 0.99\phn\phn\phn\phn
							& -3.82 $\pm$ 0.62\phn\phn\phn\phn	\\[0.5ex]
$\mu_{\delta}$  (mas yr$^{-1}$) 	& -152.54 $\pm$ 1.65\phn\phn\phn\phn
							& -71.820 $\pm$ 0.09\phn\phn\phn\phn
							& 21.26 $\pm$ 0.52\phn\phn\phn\phn
							& 2.13 $\pm$ 0.23\phn\phn\phn\phn	\\[0.5ex]
$\pi$ (mas)	& 4.170 $\pm$ 1.421\phn\phn\phn
			& 6.396 $\pm$ 0.054\phn\phn\phn
			& 3.009 $\pm$ 0.409\phn\phn\phn
			& 0.872 $\pm$ 0.143\phn\phn\phn	\\[0.5ex]
Distance (pc) 	& $240^{+124}_{-61}$\phs \phn\phd\phn\phn\phn\phn\phn\phn$\  $
			& $156.3^{+1.3}_{-1.3}$\phs \phn\phd\phn\phn\phn\phn\phn\phn$\  $
			& $332^{+52}_{-40}$\phs \phn\phd\phn\phn\phn\phn\phn\phn$\  $
			& $1150^{+220}_{-160}$\phs \phn\phd\phn\phn\phn\phn\phn\phn$\  $	\\[0.5ex]
$v_{t}$ (km s$^{-1}$)	& $194^{+104}_{-51}$\phs \phn\phd\phn\phn\phn\phn\phn\phn$\  $
					& $104.7^{+1.0}_{-1.0}$\phs \phn\phd\phn\phn\phn\phn\phn\phn$\  $
					& $80^{+15}_{-11}$\phs \phn\phd\phn\phn\phn\phn\phn\phn$\  $
					& $24^{+9}_{-6}$\phs \phn\phd\phn\phn\phn\phn\phn\phn$\  $	\\[0.25ex]
Visibility weighting		& Sensitivity\phs \phn\phd\phn\phn\phn\phn\phn\phn$\  $
					& Equal\phs \phn\phd\phn\phn\phn\phn\phn\phn$\  $
					& Equal\phs \phn\phd\phn\phn\phn\phn\phn\phn$\  $
					& Sensitivity\phs \phn\phd\phn\phn\phn\phn\phn\phn$\  $	\\[0.5ex]
Image weighting		& Natural\phs \phn\phd\phn\phn\phn\phn\phn\phn$\  $
					& Uniform\phs \phn\phd\phn\phn\phn\phn\phn\phn$\  $
					& Uniform\phs \phn\phd\phn\phn\phn\phn\phn\phn$\  $
					& Natural\phs \phn\phd\phn\phn\phn\phn\phn\phn$\  $		\\[0.5ex]
Average epoch mean fit error (mas)	& 1.232\phs \phn\phd\phn\phn\phn\phn\phn\phn$\  $
							& 0.059\phs \phn\phd\phn\phn\phn\phn\phn\phn$\  $
							& 0.765\phs \phn\phd\phn\phn\phn\phn\phn\phn$\  $
							& 0.747\phs \phn\phd\phn\phn\phn\phn\phn\phn$\  $	\\[0.5ex]
Average intra--epoch systematic error (mas)	& 2.477\phs \phn\phd\phn\phn\phn\phn\phn\phn$\  $
									& 0.068\phs \phn\phd\phn\phn\phn\phn\phn\phn$\  $
									& 0.839\phs \phn\phd\phn\phn\phn\phn\phn\phn$\  $
									& 0.939\phs \phn\phd\phn\phn\phn\phn\phn\phn$\  $ 
									\\[0.5ex]
Average inter--epoch systematic error (mas)	& 4.310\phs \phn\phd\phn\phn\phn\phn\phn\phn$\  $ 
									& 0.103\phs \phn\phd\phn\phn\phn\phn\phn\phn$\  $ 
									& 1.205\phs \phn\phd\phn\phn\phn\phn\phn\phn$\  $
									& 0.0\phs \phn\phd\phn\phn\phn\phn\phn\phn$\  $	
									\\[0.5ex]
Average single--epoch SNR	& 8\phs \phn\phd\phn\phn\phn\phn\phn\phn$\  $
						& 21\phs \phn\phd\phn\phn\phn\phn\phn\phn$\  $
						& 15\phs \phn\phd\phn\phn\phn\phn\phn\phn$\  $
						& 17\phs \phn\phd\phn\phn\phn\phn\phn\phn$\  $	\\[1.0ex]
\hline\hline
&&&&\\[-4pt]
\multicolumn{1}{c}{\tiny Parameter} & 
\multicolumn{1}{c}{\tiny PSR J1559--4438} & 
\multicolumn{1}{c}{\tiny PSR J2048--1616} & 
\multicolumn{1}{c}{\tiny PSR J2144--3933} & 
\multicolumn{1}{c}{\tiny PSR J2145--0750\tablenotemark{A}} \\[1.0ex]
\hline
Right ascension (J2000)		& 15:59:41.526126 $\pm$ 0.000008
						& 20:48:35.640637 $\pm$ 0.000040
						& 12:44:12.060404 $\pm$ 0.000045
						& 21:45:50.461901 $\pm$ 0.000098                \\[0.5ex]
Declination (J2000)		& -44:38:45.901778 $\pm$ 0.000035
					& -16:16:44.553501 $\pm$ 0.000147
					& -39:33:56.885041 $\pm$ 0.000316
					& -07:50:18.462388 $\pm$ 0.000558                \\[0.5ex]
$\mu_{\alpha}$  (mas yr$^{-1}$) 	& 1.52 $\pm$ 0.14\phn\phn\phn\phn
							& 114.24 $\pm$ 0.52\phn\phn\phn\phn
							& -57.89 $\pm$ 0.88\phn\phn\phn\phn
							& -15.43 $\pm$ 2.07\phn\phn\phn\phn	\\[0.5ex]
$\mu_{\delta}$  (mas yr$^{-1}$) 	& 13.15 $\pm$ 0.05\phn\phn\phn\phn
							& -4.03 $\pm$ 0.24\phn\phn\phn\phn
							& -155.90 $\pm$ 0.54\phn\phn\phn\phn
							& -7.67 $\pm$ 0.81\phn\phn\phn\phn	\\[0.5ex]
$\pi$ (mas)	& 0.384 $\pm$ 0.081\phn\phn\phn
			& 1.712 $\pm$ 0.909\phn\phn\phn
			& 6.051 $\pm$ 0.560\phn\phn\phn
			& --\phs \phn\phd\phn\phn\phn\phn\phn\phn$\  $		\\[0.5ex]
Distance (pc) 	& $2600^{+690}_{-450}$\phs \phn\phd\phn\phn\phn\phn\phn\phn$\  $
			& $580^{+660}_{-200}$\phs \phn\phd\phn\phn\phn\phn\phn\phn$\  $
			& $165^{+17}_{-14}$\phs \phn\phd\phn\phn\phn\phn\phn\phn$\  $
			& --\phs \phn\phd\phn\phn\phn\phn\phn\phn$\  $		\\[0.5ex]
$v_{t}$ (km s$^{-1}$)	& $163^{+44}_{-29}$\phs \phn\phd\phn\phn\phn\phn\phn\phn$\  $
					& $317^{+362}_{-111}$\phs \phn\phd\phn\phn\phn\phn\phn\phn$\  $
					& $130^{+14}_{-12}$\phs \phn\phd\phn\phn\phn\phn\phn\phn$\  $
					& --\phs \phn\phd\phn\phn\phn\phn\phn\phn$\  $		\\[0.5ex]
Visibility weighting		& Equal\phs \phn\phd\phn\phn\phn\phn\phn\phn$\  $
					& Equal\phs \phn\phd\phn\phn\phn\phn\phn\phn$\  $
					& Equal\phs \phn\phd\phn\phn\phn\phn\phn\phn$\  $
					& Sensitivity\phs \phn\phd\phn\phn\phn\phn\phn\phn$\  $		\\[0.5ex]
Image weighting		& Uniform\phs \phn\phd\phn\phn\phn\phn\phn\phn$\  $
					& Uniform\phs \phn\phd\phn\phn\phn\phn\phn\phn$\  $
					& Natural\phs \phn\phd\phn\phn\phn\phn\phn\phn$\  $
					& Natural\phs \phn\phd\phn\phn\phn\phn\phn\phn$\  $		\\[0.5ex]
Average epoch mean fit error (mas)	& 0.242\phs \phn\phd\phn\phn\phn\phn\phn\phn$\  $
							& 0.517\phs \phn\phd\phn\phn\phn\phn\phn\phn$\  $
							& 1.025\phs \phn\phd\phn\phn\phn\phn\phn\phn$\  $
							& 2.136\phs \phn\phd\phn\phn\phn\phn\phn\phn$\  $	\\[0.5ex]
Average intra--epoch systematic error (mas)	& 0.259\phs \phn\phd\phn\phn\phn\phn\phn\phn$\  $
									& 1.282\phs \phn\phd\phn\phn\phn\phn\phn\phn$\  $
									& 1.450\phs \phn\phd\phn\phn\phn\phn\phn\phn$\  $
									& 0.0\phs \phn\phd\phn\phn\phn\phn\phn\phn$\  $	
									\\[0.5ex]
Average inter--epoch systematic error (mas)	& 0.055\phs \phn\phd\phn\phn\phn\phn\phn\phn$\  $
									& 0.105\phs \phn\phd\phn\phn\phn\phn\phn\phn$\  $
									& 0.875\phs \phn\phd\phn\phn\phn\phn\phn\phn$\  $
									& 0.0\phs \phn\phd\phn\phn\phn\phn\phn\phn$\  $	
									\\[0.5ex]
Average single--epoch SNR	& 50\phs \phn\phd\phn\phn\phn\phn\phn\phn$\  $
						& 23\phs \phn\phd\phn\phn\phn\phn\phn\phn$\  $
						& 10\phs \phn\phd\phn\phn\phn\phn\phn\phn$\  $
						& 8\phs \phn\phd\phn\phn\phn\phn\phn\phn$\  $	\\[0.5ex]
\enddata
\tablenotetext{A}{Based on two detections, with parallax fixed at 0.  Variation of the parallax value 
between 0 and 2 mas results in less than 100 $\mu$as yr$^{-1}$\ difference in derived proper motion.}
\label{tab:allresults}
\end{deluxetable}
\clearpage

%

\subsection[\ptwo]{\ptwo}
\label{results:binary:0437}
\ptwo\ is the brightest and nearest observed millisecond pulsar, and has also been 
studied in the optical \citep{bell93a}, ultraviolet \citep{kargaltsev04a}, and X--ray 
\citep{zavlin02a} bands since its discovery by \citet{johnston93a}.  The high rotational 
stability and close proximity of this pulsar--white dwarf binary system make it an excellent
probe of General Relativity (GR) and alternate forms of gravitational theories.  The measurement
of its Shapiro delay by \citet{van-straten01a} is one such test which has
shown consistency with GR predictions.  The search for the low frequency stochastic gravitational
wave background (GWB) using pulsar timing arrays 
\citep[e.g.][]{jenet05a} is another test of GR which is 
facilitated in part by timing of \ptwo. 

\subsubsection[VLBI results]{VLBI results}

\begin{figure}
\begin{center}
\begin{tabular}{cc}
\includegraphics[width=0.45\textwidth]{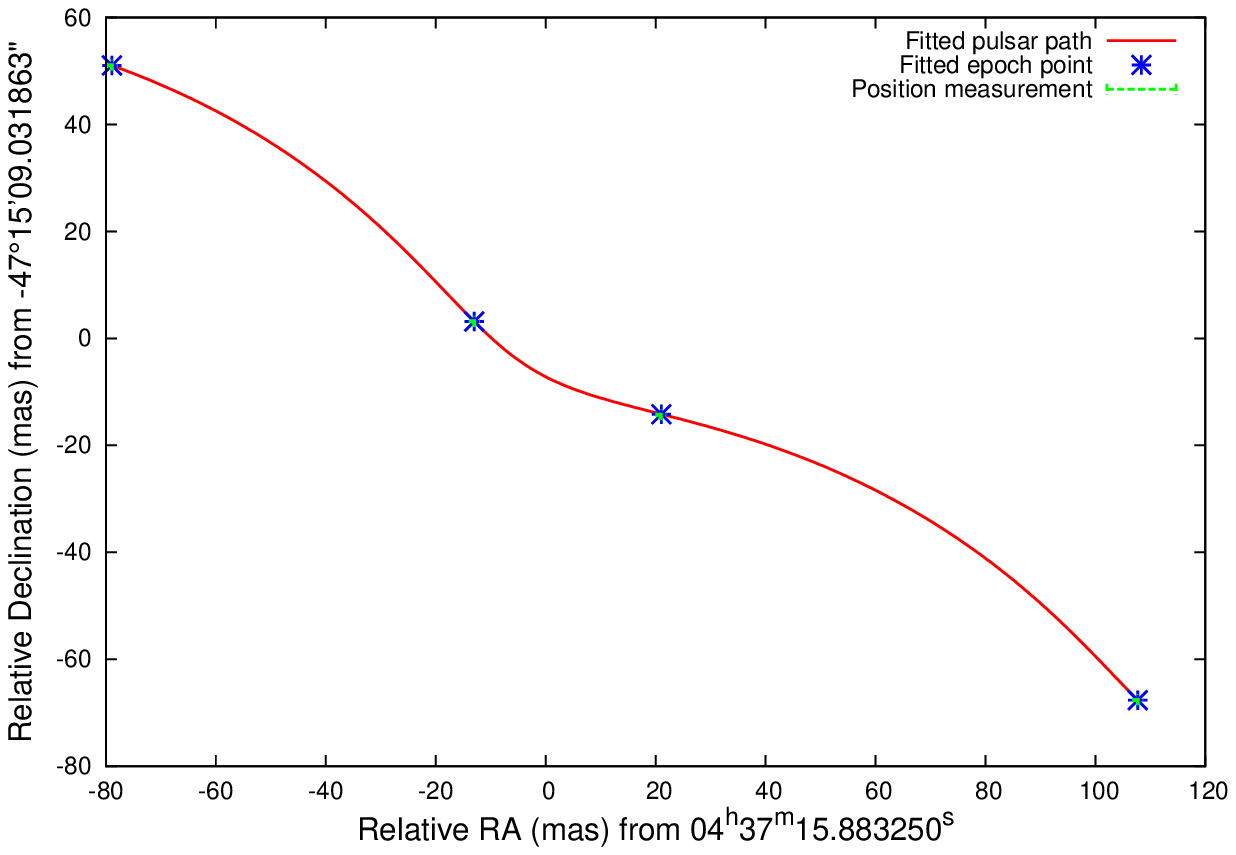} &
\includegraphics[width=0.45\textwidth]{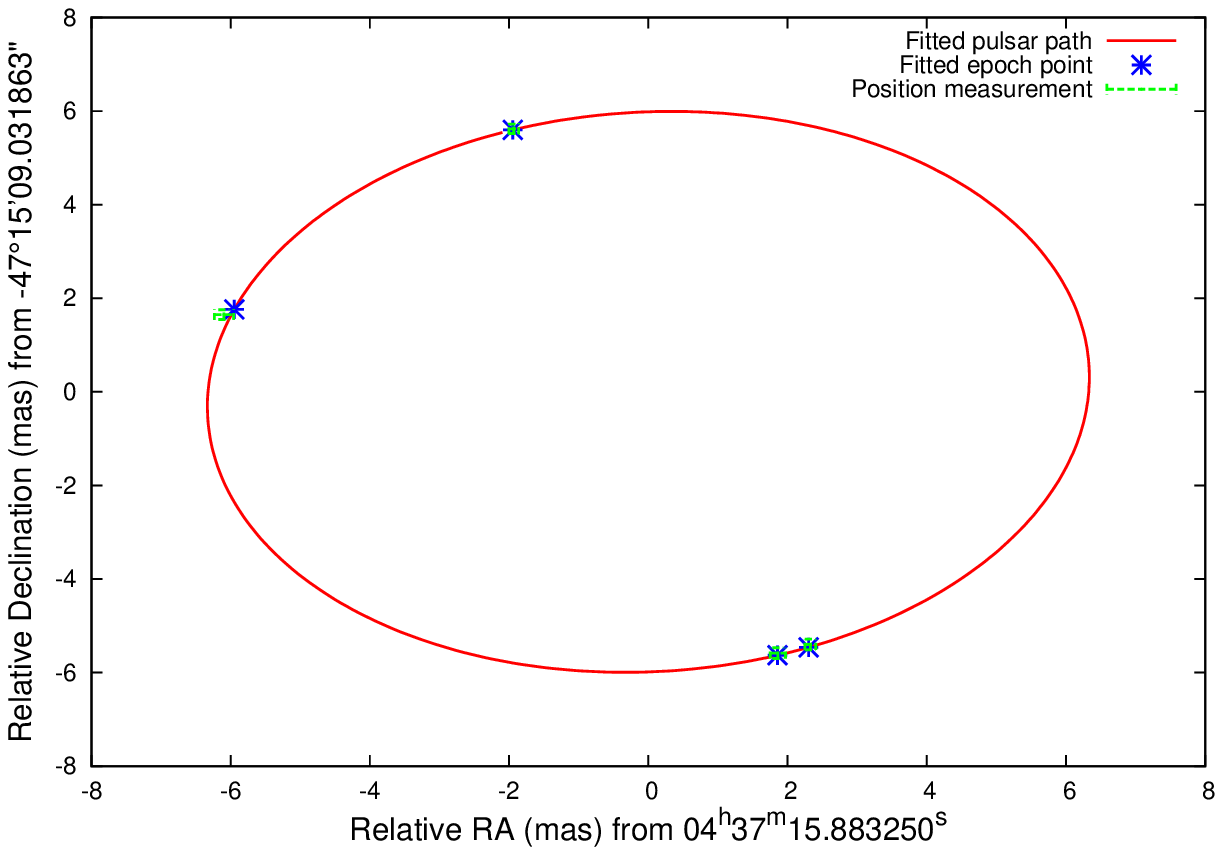} \\
\includegraphics[width=0.45\textwidth]{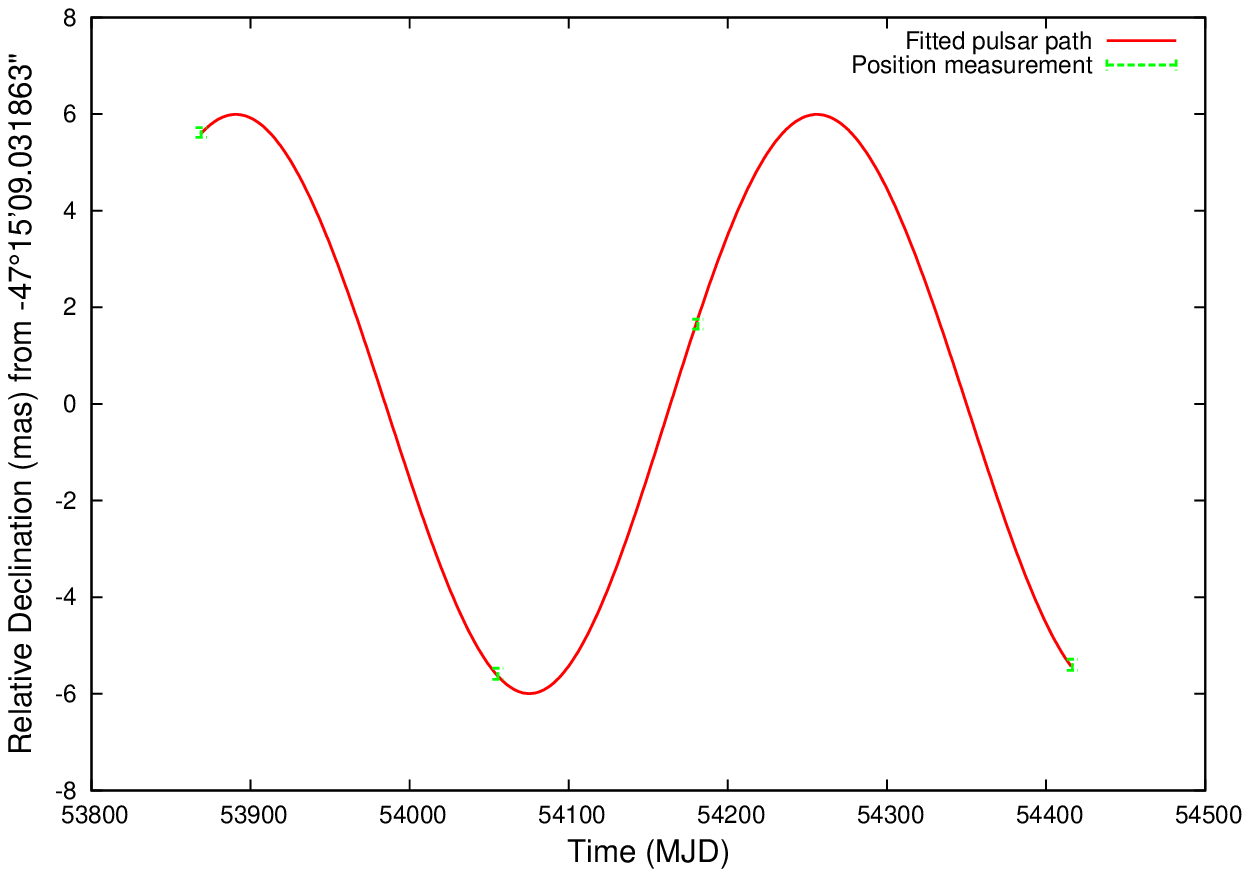} &
\includegraphics[width=0.45\textwidth]{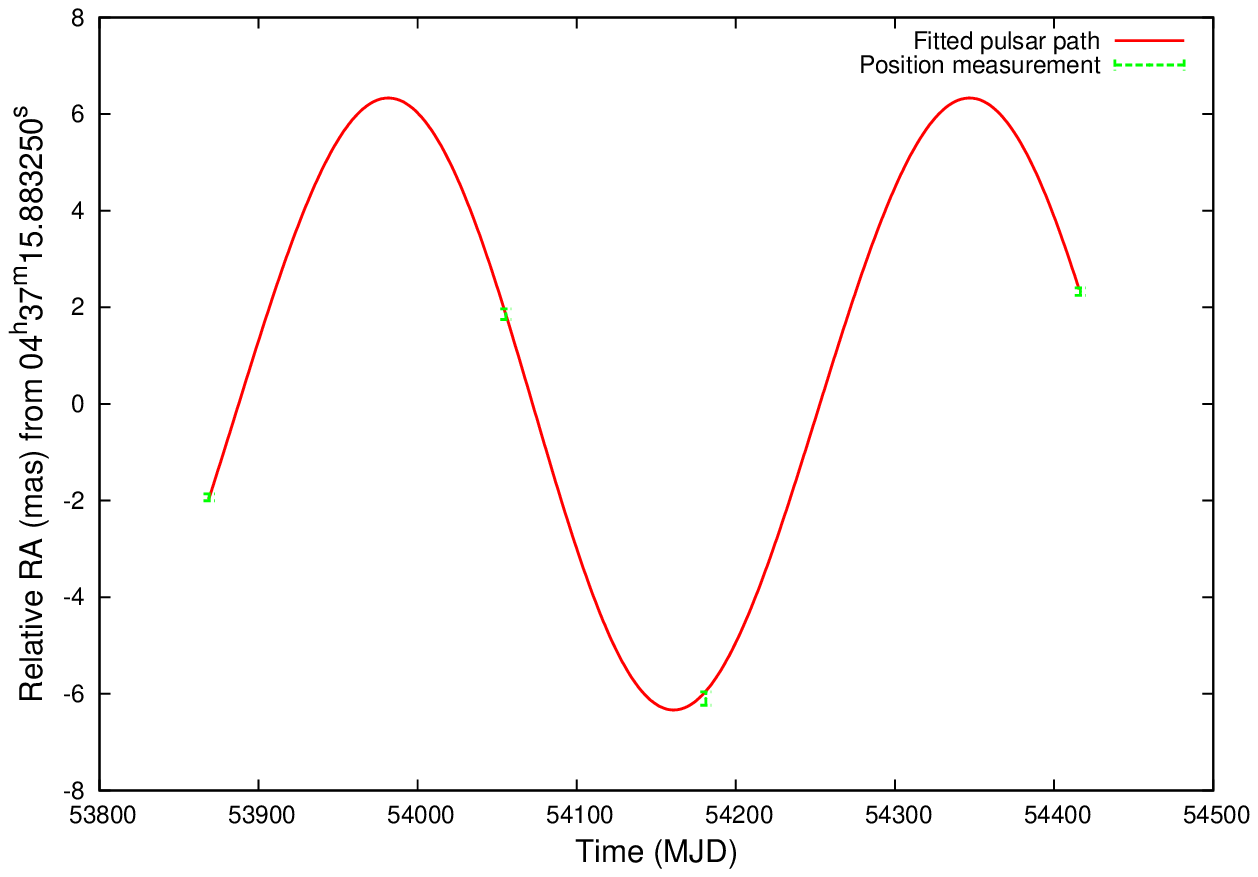} \\
\end{tabular}
\caption[Motion of \ptwo, with measured positions overlaid on the best fit]
{Motion of \ptwo, with measured positions
overlaid on the best fit.  Clockwise from top left: Motion in declination vs right ascension; as before 
but with proper motion subtracted; right ascension vs time with proper motion subtracted; and
declination vs time with proper motion subtracted.}
\label{fig:0437fit}
\end{center}
\end{figure}

Figure~\ref{fig:0437fit} shows the fitted and measured positions for \ptwo, and the full set of
fitted parameters is shown in Table~\ref{tab:allresults}.
The distance of $156.3 \pm 1.3$\,pc derived for \ptwo\ from these observations is the most 
accurate distance measurement (in both absolute and fractional distance) 
for a pulsar to date and approaches the most accurate distance measurements 
made of objects outside the Solar System \citep[T Tauri, $147.6 \pm 0.6\,{\mathrm{pc}}$:][]{loinard07a}.  
Previously, the highest--precision VLBI pulsar distance determinations were 
those made by \citet{brisken02a}, who measured the distance of PSR J0953+0755 to 
an accuracy of 5 pc, along with 8 other Northern Hemisphere pulsars.
The two previous parallax measurements made using a southern array, of PSR J0835--4510 
\citep{dodson03a} and PSR J1456--6843 \citep{bailes90a}, had 1$\sigma$\,distance
errors of 19 and 70 pc respectively.

The distance to \ptwo\ estimated from the TC93 model (140 pc) and the NE2001 model (189 pc) 
is within $\sim$20\% of the measured distance of $156.3 \pm 1.3$\,pc in both cases.  Previous timing
parallaxes of \ptwo\ \citep{hotan06a,van-straten01a,sandhu97a} 
have yielded considerable variation in measured parallax, the cause of which is ascribed to
the inaccuracy of earlier Solar System ephemerides \citep{verbiest08a}.  These observations,
which are independent of the Solar System ephemeris, provide confirmation that the earlier
measurements of timing parallax were inaccurate.

Previous measurements of the transverse speed of \ptwo\ using scintillation observations 
have also provided widely varying results.  \citet{johnston98a} measured a scintillation speed
of 170\,\kms\ assuming a distance of 180 pc, while \citet{gothoskar00a} measured 
231\,\kms, also assuming a distance of 180 pc.  Neither of these values are
consistent with the well--determined VLBI value of $104.7\pm1.0$\,\kms.  As discussed in 
Section~\ref{results:distvel:new}, errors in scintillation speed estimates of this magnitude
are not uncommon, and may often be due to the simplifying assumptions made about the
scattering material.

\subsubsection[Comparison to timing astrometry]{Comparison to timing astrometry}
\label{results:binary:0437:timingcomp}
To compare the VLBI and timing positions, the timing data
of \citet{verbiest08a} have been re--fitted to obtain the position of 
PSR J0437--4715 at the VLBI reference epoch (MJD 54100).
Table~\ref{tab:0437comparison} shows that the timing and VLBI positions at MJD 54100 differ by
over two mas, many times the formal errors shown. However, the formal errors are negligible
compared to the uncertainty in the VLBI phase reference calibrator position
\citep[0.8 mas;][]{ma98a}, and the potential constant offsets due to phase--referencing
errors such as station position errors \citep[known to exist at the cm level for the LBA;][]{deller08a}.  
Discrepancies between interferometric and timing positions of even larger magnitudes have
been found using the DE200 frame \citep{bartel96a}, 
and differences at the mas level still exist for the position
PSR J0437--4715 calculated using the newer DE414 solar system ephemeris, as compared to the
DE405 ephemeris used by \citet{verbiest08a}. Thus, it is concluded that the VLBI and timing 
position difference is consistent with the uncertainty in the calibrator position 
and the offset between the solar system frame and the ICRF.

\begin{deluxetable}{lrr}
\tabletypesize{\tiny}
\tablecaption
{Fitted VLBI results for PSR J0437--4715 and comparative timing values (positions
re--referenced to the VLBI proper motion epoch)}
\tablewidth{0pt}
\tablehead{
\colhead{Parameter} & \colhead{Fitted value and error} & \colhead{\citet{verbiest08a} timing values\phm{\tablenotemark{A}}}}
\startdata
Right Ascension (J2000)	&  	04$^{\mathrm h}$37$^{\mathrm m}$15.883250$^{\mathrm s}$ $\pm$ 0.000003	& 
				04$^{\mathrm h}$37$^{\mathrm m}$15.883185$^{\mathrm s}$ $\pm$ 0.000006\phm{\tablenotemark{A}} 	\\
Declination (J2000) 	&  	$-$47\degrees15'09.031863" $\pm$ 0.000037	& 
				$-$47\degrees15'09.034033" $\pm$ 0.000070\phm{\tablenotemark{A}}	\\
$\mu_{\alpha}$	(mas/yr)	&  	121.679 $\pm$ 0.052\phn\phn\phn	&
						121.453 $\pm$ 0.010\phn\phn\phn\phm{\tablenotemark{A}}	\\
$\mu_{\delta}$	(mas/yr)	&  	$-$71.820 $\pm$ 0.086\phn\phn\phn	&
						$-$71.457 $\pm$ 0.012\phn\phn\phn\phm{\tablenotemark{A}}	\\
Parallax $\pi$ (mas)	 		&  	6.396 $\pm$ 0.054\phn\phn\phn	&
						6.65 $\pm$ 0.51\phn\phn\phn\phn\phm{\tablenotemark{A}}	\\
Distance (pc)	&  	156.3 $\pm$ 1.3\phn\phn\phn\phn\phn		&
				157.0 $\pm$ 2.4\tablenotemark{A}\phn\phn\phn\phn\phn\\		
Transverse velocity $v_{\mathrm T}$ (km/s)	&  	104.71 $\pm$ 0.95\phn\phn\phn\phn	&
				104.9 $\pm$ 1.6\tablenotemark{A}\phn\phn\phn\phn\phn	\\
Reduced chi--squared 	&  	1.0 \phs \phn\phd\phn\phn\phn\phn\phn\phn$\  $ &
					 \phs \phn\phd\phn\phn\phn\phn\phn\phn$\  $\phm{\tablenotemark{A}}\\
Average epoch mean fit error (mas)			& 0.059 \phs \phn\phd\phn\phn\phn\phn\phn\phn$\  $ &
					 \phs \phn\phd\phn\phn\phn\phn\phn\phn$\  $\phm{\tablenotemark{A}}\\
Average intra--epoch systematic error (mas) 	& 0.068 \phs \phn\phd\phn\phn\phn\phn\phn\phn$\  $ &
					 \phs \phn\phd\phn\phn\phn\phn\phn\phn$\  $\phm{\tablenotemark{A}}\\
Average inter--epoch systematic error (mas) 	& 0.103 \phs \phn\phd\phn\phn\phn\phn\phn\phn$\  $  &
					 \phs \phn\phd\phn\phn\phn\phn\phn\phn$\  $\phm{\tablenotemark{A}}\\
Reference epoch for proper motion	(MJD)	& 54100.0 \phs \phn\phd\phn\phn\phn\phn\phn\phn$\  $&
			54100.0 \phs \phn\phd\phn\phn\phn\phn\phn\phn$\  $\phm{\tablenotemark{A}}\\
\enddata
\tablenotetext{A}{Derived from the kinematic distance obtained from \Pbdot, 
not the less precise parallax values}
\label{tab:0437comparison}
\end{deluxetable}

The parallax value obtained from VLBI is consistent with that derived from timing, and yields a
distance which is consistent with the kinematic distance of $157.0 \pm 2.4$\ pc 
derived from the orbital period derivative \Pbdot.  However, the
VLBI parallax measurement is an order of magnitude more precise than the timing measurement,
and yields a distance which is a factor of two more precise than the kinematic distance.

Finally, the values obtained for proper motion from VLBI differ by $\sim4\sigma$\ in both right
ascension and declination from the values estimated from pulsar timing.  
A likely cause for this discrepancy is small changes in the
centroid position of the phase reference source due to intrinsic source variability, which would be
absorbed into the astrometric fit.  If the centroid position change is roughly linear over the timescale
probed, the effects would largely be absorbed into the measured proper motion. 
\citet{titov07a} show that some ICRF sources exhibit apparent 
proper motions of hundreds of $\mu$as yr$^{-1}$ for periods of several years.
Whilst the calibrator used here (J0439--4522) has not previously been the target of detailed
variability studies such as those undertaken by \citet{titov07a}, it is possible to use archival 
observations to show that centroid position changes are a plausible explanation for the
proper motion discrepancy, as shown below.

First, the VLBI data obtained during this thesis on J0439--4522 can be inspected directly for evidence of
variability.  The model used for J0439--4522 in this thesis consisted of 
a $\sim$\,$1\,$mas FWHM Gaussian, with two additional 
delta components within 5 mas of strength 2\% and 0.2\% of peak flux.  There is no gross 
evidence of variability over the four observational epochs, as the width and positions of
the primary and secondary components remained constant
to within $\sim200\,\mu$as, and secondary fluxes remained constant 
to $\sim$\,$1\,$mJy. However, as the source is only barely resolved (the beamsize is 
$\sim$\,$3\,$mas) positional variability at the $\sim100\,\mu$as level would be difficult
to detect.  An indirect indicator of the past variability of the source can instead be obtained
from calibrator measurements using the 
ATCA\footnote{http://www.narrabri.atnf.csiro.au/cgi-bin/Calibrators/calfhis.cgi?source=0437-454\&band=3cm}.  These show that the total flux density on ATCA baselines has declined
by a factor of three in a near--linear manner over a four year period from 2004 to 2007.  
On the shortest timescales probed by this archival data (several months) departures from the
linearly decreasing trend on the order of 10\% of the total flux are seen.

The commensurate ATCA observations taken simultaneously with VLBI observations during this 
thesis are unsuited to accurate flux monitoring, due to the compact array configurations used and
the lack of ATCA flux calibrator scans, which are unnecessary for VLBI.
Thus, the archival ATCA observations
provide the best available means for estimating source variability.  The steady decrease in flux level is 
consistent with a near--linear change in the source centroid, which would be absorbed into the 
proper motion estimate.  Smaller non--linear centroid changes are also possible, which would affect
not only the proper motion, but also the parallax estimated from the VLBI measurements.  However,
the limited amount of short--term variability seen is consistent with a mostly linear change in
position centroid.  Thus, the archival ATCA data supports the theory that centroid changes in the
phase reference calibrator are responsible for the difference between the VLBI and timing 
measurements of proper motion, and indicates that the effect of the reference source variability
on parallax is likely to be considerably smaller.

The higher proper motion precision 
obtained with the timing data (a factor of 5--7 times better than the VLBI results) reflects 
the fact that the timing data spans a time baseline 7 times longer than the VLBI dataset.

\subsubsection[Limits on anomalous accelerations]{Limits on anomalous accelerations}
\label{results:binary:0437:limits}
The newly measured parallax of $\pi = 6.396 \pm 0.054$\,mas allows an improved 
measurement of any anomalous acceleration of either the Solar System or 
PSR J0437--4715. Specifically, the apparent acceleration due to time variability of 
Newton's gravitational constant $G$ and the mass of an undetected trans-Neptunian 
planet near the line of sight to the pulsar can be limited.

As first described by \citet{damour91a}, a precise measurement of a binary pulsar's orbital 
period derivative, \Pbdot, can be used to constrain a variation of the 
gravitational constant, $G$. However, as \citet{bell96a} pointed out, for 
PSR J0437$-$4715 a precise distance needs to be known in order to 
correct \Pbdot\ for the Shklovskii acceleration \citep{shklovskii70a} caused by its proper motion. 
This analysis has been performed, based exclusively on timing data, by
\citet{verbiest08a}, whose limit was dominated by the uncertainty in their 
parallax measurement. This VLBI parallax value improves their limit by 
a factor of nearly 10 down to $\dot{G}/G = (-5 \pm 26)\times 10^{-13}$\ yr$^{-1}$\ at 
95\% certainty. This value compares well to the most stringent limit currently 
published: $(4\pm 9) \times 10^{-13}$\ yr$^{-1}$, derived through Lunar Laser 
Ranging \citep[LLR;][]{williams04a}. Since $\pi$ and \Pbdot\ are now both measured to similar
precision, both measurements will have to be improved for a further 
significant increase in \Gdot\ sensitivity.  Additional VLBI observations and continued timing
could see this limit improve upon the existing LLR limit early in the next decade.

An alternative source of anomalous acceleration is heavy planets in a 
wide orbit around the Sun or the pulsar. Building upon the initial analysis of 
\citet{zakamska05a}, this parallax measurement can be combined with the timing results 
from \citet{verbiest08a} to derive the following result: 
$a_{\odot}/c = (3 \pm 16 ) \times 10^{-20}\,{\mathrm s}^{-1}$\ at the $2 \sigma$ level. 
This improves the limit published in \citet{verbiest08a} by an order of magnitude and 
makes PSR J0437--4715 a more sensitive Solar System accelerometer than PSR J1713+0747, 
the most precise pulsar listed by \citet{zakamska05a}. From this, the limit for 50\% of the 
sky\footnote{within 60\degrees\,of the line of sight towards and away from PSR J0437--4715} 
can be calculated as: 
$|a_{\odot, 50\% {\mathrm{sky}}}/c| \leq 3.9\times 10^{-19}\,{\mathrm s}^{-1}$ (95\% certainty).
This acceleration limit can be used to exclude massive bodies within
a given radius of the Sun; for example, at Kuiper--belt radii (50 AU) it
excludes a planet more massive than Uranus over 50\% of the sky, while  
Jupiter--mass planets are excluded within 226\,AU over 50\% of the sky.

\subsubsection[Impact of the stochastic GWB]{Impact of the stochastic GWB}
\label{results:binary:0437:gwb}
Using tools recently developed by \citet{hobbs08a}, the effect of a stochastic GWB 
on the observed value of \Pbdot\ from pulsar timing has been simulated. 
The characteristic strain spectrum of the GWB was set to that of the best published GWB 
limit \citep{jenet06a}, with dimensionless amplitude $A=1.1\times10^{-14}$\ at a period 
of one year and power--law dependence on frequency with exponent 
$\alpha=-2/3$ (as predicted for a GWB dominated by
black hole--black hole mergers). The simulated GWB causes the kinematic distance to be 
inconsistent with the VLBI parallax distance at the 2$\sigma$\, level in 
$\sim50\%$ of trials.  Thus, although these observations cannot improve 
upon the present GWB limit, they are consistent with it.
Simulations with a GWB with amplitude of $1.1\times 10^{-13}$ show inconsistencies 
between the kinematic and VLBI distances at the $2\sigma$ level in $95\%$\ of trials, 
providing an independent exclusion of a GWB with an amplitude at or above this value.

It is also interesting to note that the precise limit on $\dot{G}$\ presented above
would be impossible in a Universe with a strong GWB. In the simulations with GWB amplitude 
$1.1\times 10^{-13}$, the observed \Gdot\ value is inconsistent with 0 in $99\%$\ of cases, 
merely due to the GWB-induced corruption of the timing measurements.  Thus, the stochastic 
GWB must eventually limit the accuracy of measurements of \Gdot\ in the fashion outlined here.

\subsection[\pfour]{\pfour}
\label{results:binary:0737}
The double pulsar system \pfour\ \citep{burgay03a,lyne04a}
is one of only eight known double neutron star (DNS) systems, and the only system in which both 
neutron stars are visible as pulsars.
The discovery of the mildly recycled ``A" pulsar was reported by \citet{burgay03a},
but emission from the companion ``B" pulsar was not observed at the time, and it was not
discovered until follow--up observations some time later \citep{lyne04a}.
\pfour\ is the most relativistic known binary system,
with an orbital period of only 2.5 hours and a coalescence time (due to orbital energy loss to 
gravitational radiation) of 85 Myr.
\pfour\ provides the best current tests of General Relativity (GR),
with \citet{kramer06a} recently showing that measurements of the post--Keplerian parameter 
``s" agreed with the GR prediction to within 0.05\%.
The discovery of PSR J0737--3039A also led to a marked upward revision in the estimated 
Galactic merger rate of DNS systems \citep{kalogera04a}, although merger rate estimates
are still very uncertain due to the poor constraints available on the characteristics of the 
DNS population.

One of the major outstanding uncertainties surrounding \pfour\ is its distance.
\citet{kramer06a} detect a marginally significant timing parallax of $3\pm2$\ mas, but this is no 
more accurate than what is typically assumed for distances estimated from dispersion
measures in the NE2001 distance model.  The NE2001 $DM$ distance 
estimate for \pfour\ is 480 pc, while the earlier TC93 model estimates the
distance to be 570 pc.  Many previous studies \citep[e.g.][]{kramer06a,burgay03a} have assumed
the rounded distance of 500 pc when calculating luminosity and kinematic values.  As discussed
below, the accurate distance and velocity information for this system presented here 
enables tests of GR to proceed to greater precision in the future, as well as providing insight 
into its formation and x--ray emission.


\subsubsection[VLBI results]{VLBI results}
As shown earlier in Table~\ref{tab:targets}, \pfour\ was 
clearly the most challenging target undertaken in this program -- with an equivalent 
gated 1600 MHz flux density\footnote{Since pulsar A dominates the radio emission, gating was 
performed using the pulsar A ephemeris only} 
of 4 mJy, it is almost as faint as \pone, but with a $DM$\ distance four times 
greater, a factor of four greater precision was required to attain an equivalent parallax accuracy.
Whilst a very nearby (angular separation $\sim\,20$\ arcminutes) calibrator was available
(B0736--303), Figure~\ref{fig:phaserefs1}
shows that it presented a particular challenge, as it was weak (correlated flux density 
$\sim$30$\,$mJy on 
1000 km baselines), possessed complicated structure, and its position was not known precisely
in the ICRF.  As noted in Section~\ref{techniques:posdet}, the position of B0736--303 was 
refined using a phase--referenced VLBI observation, and finally determined to approximately
five mas accuracy by comparison of the VLBI position for \pfour\ with its published timing 
position \citep{kramer06a}. Whilst this is a relatively large positional uncertainty, the short 
calibrator throw means that the resultant differential errors are small.

\begin{figure}
\begin{center}
\begin{tabular}{cc}
\includegraphics[width=0.5\textwidth, angle=270, clip]{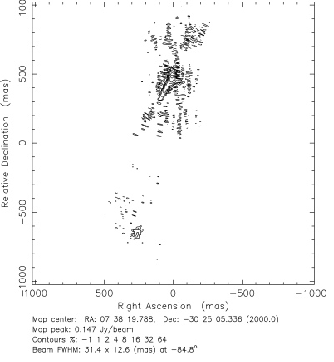} &
\includegraphics[width=0.5\textwidth, angle=270, clip]{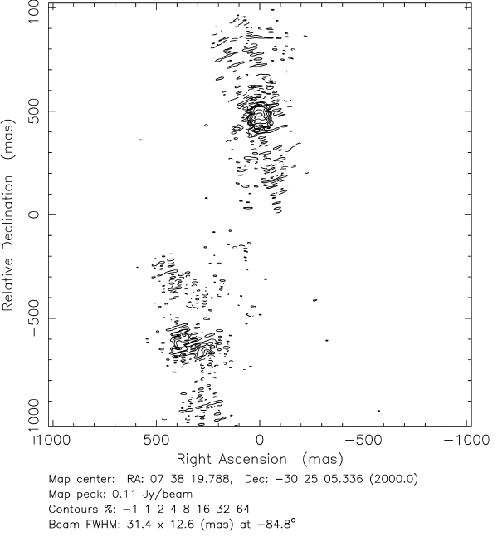} \\
\includegraphics[width=0.5\textwidth, angle=270, clip]{0736.newmod.clean.bbcrop.eps} & \\
\end{tabular}
\caption[Images of B0736--303, using different source models]
{Images of B0736--303, using Gaussian model components only (model A, top left), CLEAN
components only (model B, top right) and a combination of Gaussian and CLEAN components
(model C, bottom left).  Images contours are at 1, 2, 4, 8, 16, 32 and 64\% of peak flux density
in each case.  Model C clearly gives the best fit to the data.}
\label{fig:0736alternatives}
\end{center}
\end{figure}

The complex structure and positional uncertainty in the phase reference source introduces 
an extra element of model--dependence to the results obtained for \pfour.  The final result
obtained for parallax differed by 2$\sigma$\ from the initial result, which was obtained when 
the calibrator position was
in error by approximately 50 mas.  After the calibrator position was updated, 
several different calibrator models were produced (model A, which used
Gaussian model components; model B, which used standard CLEAN components;  and model C,
which used a combination of the two) and the results compared.  In each case, the model was held
constant over all epochs.  The clean images of the combined dataset for B0736--303 
(visibilities from all epochs concatenated into a single file)
are shown in Figure~\ref{fig:0736alternatives} for models A, B and C (which is the model shown in 
Figure~\ref{fig:phaserefs1}, but is shown again here for clarity).

As shown in Figure~\ref{fig:0736alternatives}, model C gives a significantly better fit to the data,
resulting in lower image residuals than model A or model B.  The data reduction for \pfour\ 
was repeated three times using models A, B and C, and the resultant position datasets
were fit for parallax and proper motion.  The results of this analysis showed that model C gave
a significantly better fit ($\pi = 0.87\pm0.14\,$mas) than model A ($\pi = 1.06\pm0.18\,$mas)
or B ($\pi = 0.76\pm0.20\,$mas).

Model C was thus clearly the best source model, and was chosen for the final results
presented in Table~\ref{tab:allresults}. Since any residual model--dependent errors are 
likely to be small (and difficult to quantify in any case, but on the order of the formal errors or smaller)
the error given for \pfour\ in Table~\ref{tab:allresults} and the discussion below
is simply the ``inclusive" astrometric fit error described in Section~\ref{techniques:posdet} and 
does not include the uncertainty due to calibrator structure discussed here.  However, it is worth
noting that the some uncertainty due to the calibrator source structure remains, and if the
model could be improved by additional observations with better $uv$\ coverage in the future
then the astrometric data could be re--reduced, with the potential for a slightly improved fit.

\label{results:binary:0737:vlbi}
\begin{figure}
\begin{center}
\begin{tabular}{c}
\includegraphics[width=0.55\textwidth]{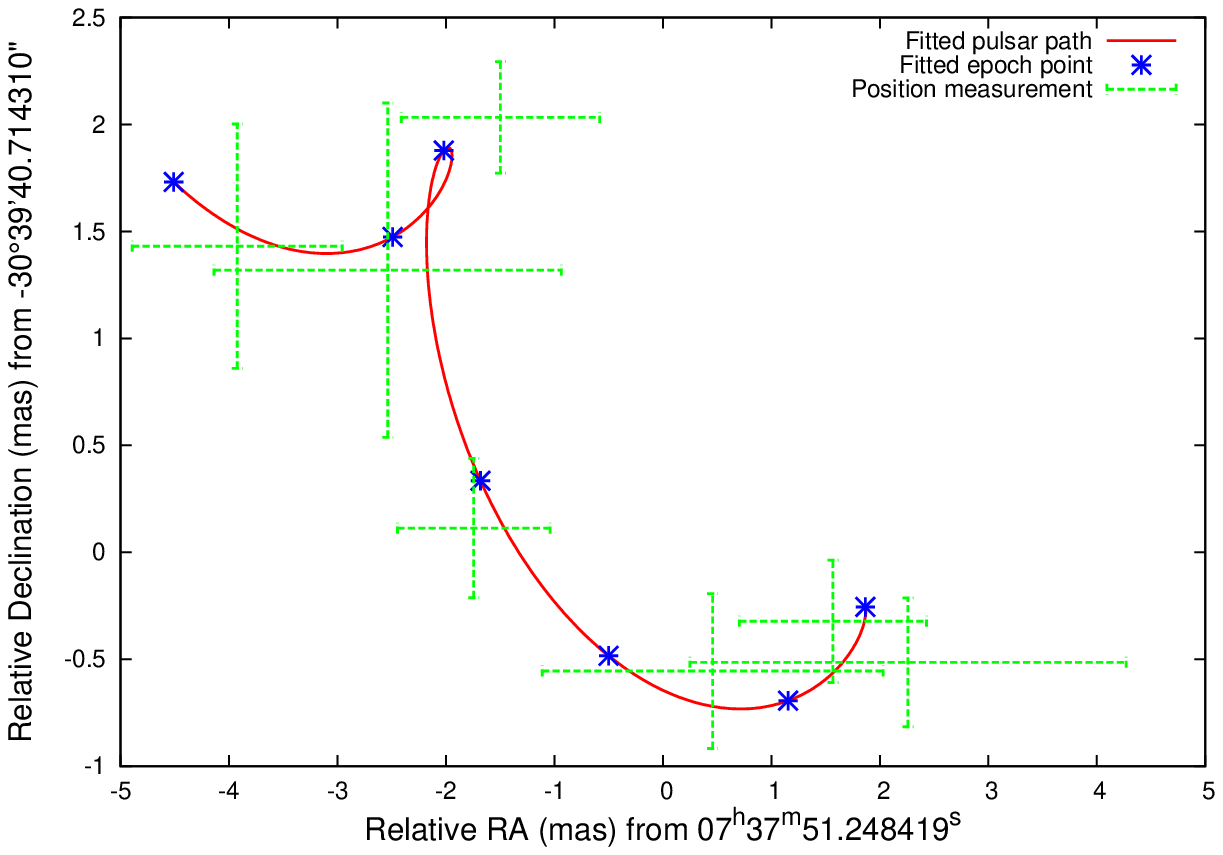} \\
\includegraphics[width=0.55\textwidth]{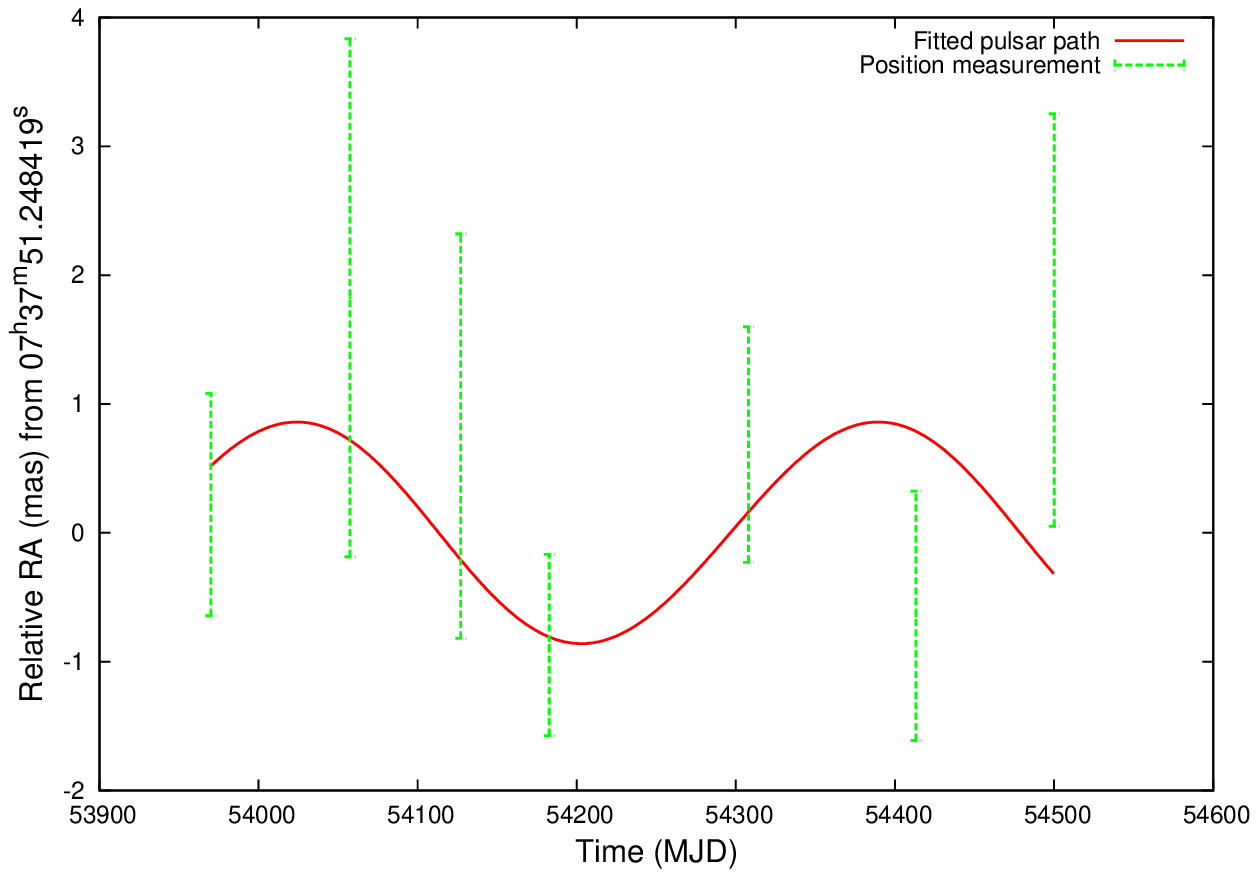} \\
\includegraphics[width=0.55\textwidth]{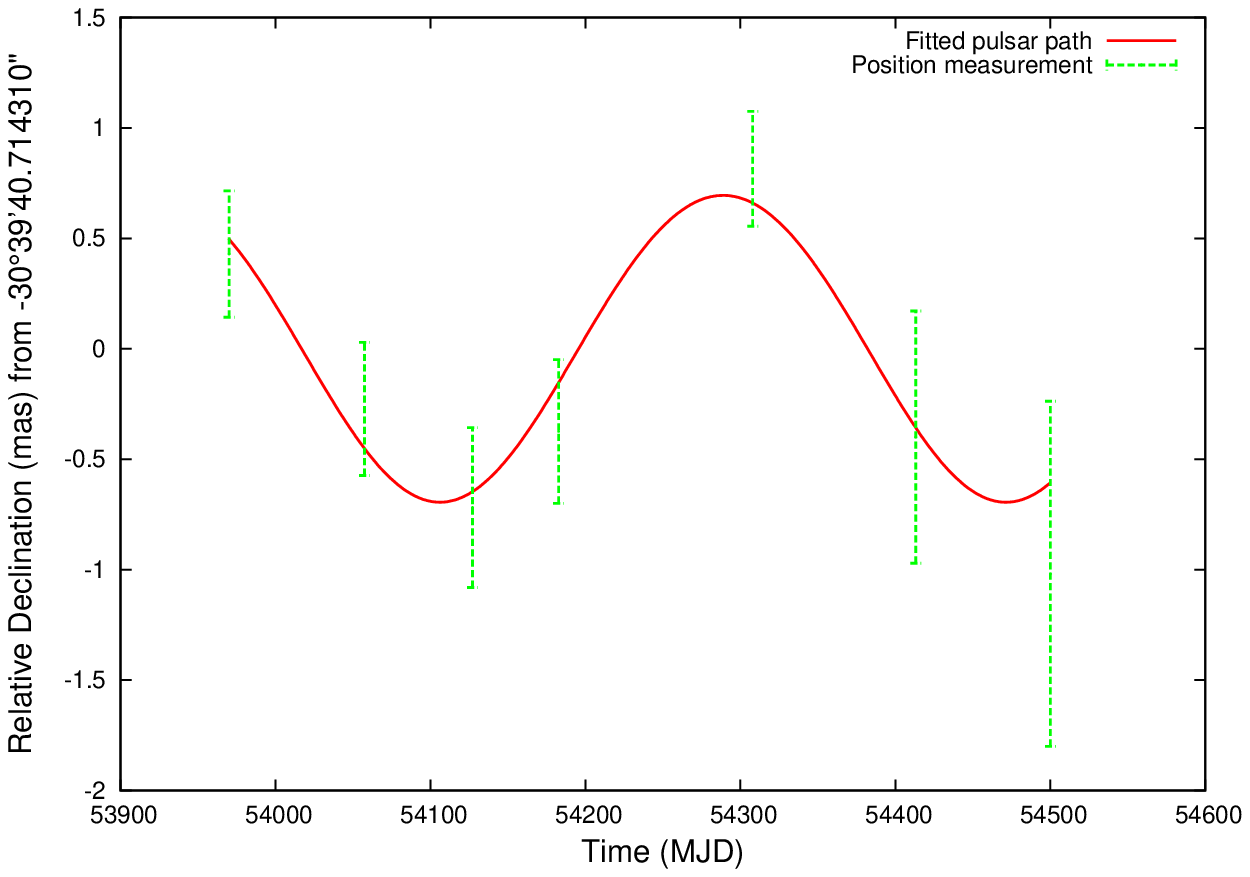} \\
\end{tabular}
\caption[Motion of \pfour, with measured positions overlaid on the best fit]
{Motion of \pfour, with measured positions
overlaid on the best fit.  From top: Motion in declination vs right ascension; 
right ascension vs time with proper motion subtracted; and
declination vs time with proper motion subtracted.}
\label{fig:0737fit}
\end{center}
\end{figure}

Figure~\ref{fig:0737fit} shows the final fitted and measured positions for \pfour.  
The measured parallax for \pfour\ ($0.872\pm0.143\,$mas)
corresponds to a distance of $1150^{+220}_{-160}$\,pc.  The successful
measurement of parallax and proper motion by this program represents the first significant
parallax and proper motion determination for a DNS system.
At 95\% confidence, the lower limit for the distance is 860\,pc -- still considerably in excess of 
the predicted $DM$\ distance for the system.  The revised value of electron density along the 
line of sight to \pfour\  is 0.043 cm$^{-3}$, which suggests the influence of the Gum nebula
along this sightline is less than originally thought.

The measured transverse velocity for \pfour\ is $24^{+9}_{-6}$\,\kms.  
However, when the best estimates of peculiar motion of the Solar System
and Galactic rotation \citep[e.g.][]{mignard00a} are subtracted in order to obtain a velocity in the
local standard of rest, the resultant transverse velocity is $9^{+6}_{-3}$\,\kms.  
This is well within the range of transverse velocities expected for the massive stars which
are progenitors for NS--NS binaries \citep[$\sim20\,\kms$; see e.g.][]{feast65a}, and such a low 
value is extremely unlikely unless both kicks imparted on the system by the supernova events 
which formed the pulsars were extremely small.  If the \pfour\ system does have a large
velocity, it must be predominantly in the radial direction -- a possibility discussed in detail 
below where possible formation models for the system are considered.

\subsubsection[Comparison to timing astrometry]{Comparison to timing astrometry}
The parallax measurement made through these VLBI observations is consistent with
the timing value of $3\pm2$\ mas obtained by \citet{kramer06a}, but it is considerably more precise.
The values obtained for proper motion ($\mu_{\alpha} = -3.82\pm0.62\,$mas yr$^{-1}$, 
$\mu_{\delta} = 2.13\pm0.23\,$mas yr$^{-1}$) are also consistent with those obtained
from timing ($\mu_{\alpha} = -3.3\pm0.4\,$mas yr$^{-1}$, $\mu_{\delta} = 2.6\pm0.5\,$mas yr$^{-1}$).

The measured transverse velocity of $24^{+9}_{-6}$\,\kms\ is over twice that obtained by 
\citet{kramer06a}, almost entirely due to the distance revision.  However, this value is still considerably
lower than those obtained through measurements of scintillation velocity by 
\citet{ransom04a}, who obtained $141\pm8\,\kms$, and the revised value of \citet{coles05a},
who obtained $66\pm15\,\kms$. Both of these studies assumed a distance to  \pfour\ close to the
NE2001 value (570 pc). Thus, as noted by \citet{kramer06a}, it is likely that the usual
scintillation velocity assumption of a thin scattering screen with an isotropic turbulence distribution
is not appropriate in this case.  Accounting for anisotropy in the ISM turbulence gives a 
measurement of scintillation velocity that is
lower still, and more consistent with the VLBI transverse velocity (W. Coles et al., in preparation).

Since the timing position of \pfour\ was used to refine the VLBI calibrator position, no meaningful
comparison can be made between the timing and VLBI positions.  When corrected for proper motion
to the same epoch, the timing and VLBI positions have been aligned to approximately one mas, which
is the the level of errors due to proper motion uncertainty.

Finally, it is appropriate to note that the continuation of timing observations of \pfour\ should see a 
determination of parallax and proper motion to equal or greater precision than these VLBI results within 
the next decade.

\subsubsection[Implications for tests of GR]{Implications for tests of GR}
One of the tests of GR possible with the \pfour\ system is the rate of change of orbital period (\Pbdot) due
to the loss of energy to gravitational radiation.  In order to perform this test, the observed \Pbdot\ must
be measured as accurately as possible (the current significance is 70$\sigma$; \citealt{kramer06a}), 
and contributing factors to \Pbdot\ other than GR must be estimated as accurately as possible.
For \pfour, the major contributing factors are differential Galactic rotation ($\Pbdot^{\mathrm{rot}}$), 
acceleration towards the plane of the Galaxy ($\Pbdot^{\mathrm{z}}$), and the apparent
acceleration caused by the transverse motion of the system 
(the Shklovskii effect $\Pbdot^{\mathrm{Shk}}$; 
\citealt{shklovskii70a}). These will collectively be referred to as galactic and kinematic contributions
($\Pbdot^{\mathrm{gk}}$).  At the newly calculated distance of 1150 pc, the magnitude of 
these effects can be calculated using Equations~2.12 and 2.28 from \citet{damour91a} as:

\begin{eqnarray}
\frac{\Pbdot^{\mathrm{rot}}}{\Pb} & = & -\frac{v_{0}^{2}}{c\,R_{0}}\times\left( \cos{l} + \frac{\beta}{\sin^{2}{l}+ \beta^{2}} \right) \label{eq:Pbdotrot}\\
\frac{\Pbdot^{\mathrm{z}}}{\Pb} & = & -\frac{K_{z}}{c}\ \sin{b}\\
\frac{\Pbdot^{\mathrm{Shk}}}{\Pb} & = &  \frac{\mu^{2}d}{c} 
\label{eq:Pbdotshk}
\end{eqnarray}

\noindent where $l$, $b$\ and $z$\ are the pulsar's Galactic longitude, latitude and height respectively,
$d$\ and $\mu$ are pulsar distance and proper motion, 
$v_{0}$\ and $R_{0}$\ are the Galactic speed and radius of the 
Solar System (taken to be 7.5 kpc and 195 \kms\ respectively; e.g. \citealt{dias05a}), 
$K_{z}$\ is the vertical gravitational potential of the Galaxy (taken from 
\citealt{holmberg04a} as 0.45 km$^{2}$ s$^{-2}$\ pc$^{-1}$ at the height of \pfour) 
and $c$\ is the speed of light.  The dominant uncertainties are  
in $d$\ (16\%) and $\mu$\ (15\%), and $R_{0}$, $v_{0}$\ and $K_{z}$, whose errors are typically 
estimated at $\sim$10\%.

Equations~\ref{eq:Pbdotrot} -- \ref{eq:Pbdotshk} give 
$\Pbdot^{\mathrm{rot}}/\Pb = (-4.3\pm 0.7) \times10^{-20}$\,s$^{-1}$, 
$\Pbdot^{\mathrm z}/\Pb = (3.8 \pm 0.8)\times10^{-21}$\,s$^{-1}$, and 
$\Pbdot^{\mathrm{Shk}}/\Pb = (5.3 \pm 1.8)\times10^{-20}$\,s$^{-1}$.  Combining these terms gives
$\Pbdot^{\mathrm{gk}}/\Pb = (1.5 \pm 2.1) \times 10^{-20}$\,s$^{-1}$, and multiplying by the observed
orbital period $\Pb = 8834.5\,$s \citep{kramer06a} yields the net effect of these terms on
the observed orbital period derivative: $\Pbdot^{\mathrm{gk}} = (1.3\pm1.8)\times10^{-16}$.


These contributions to orbital period derivative are four orders of magnitude below the 
GR contribution, and two orders of magnitude below the current measurement error
($\Pbdot^{\mathrm{obs}} = (-1.252\pm0.017)\times10^{-12}$; \citealt{kramer06a}).
Thus, with the current measurement accuracy of distance and transverse velocity,
GR tests with \pfour\ using \Pbdot\ can ultimately be made to the 0.01\% level, 
although of order ten years of continued precision timing will be 
required to reach this accuracy\footnote{since the \Pbdot\ error scales with $t^{-5/2}$, and the current
significance of 70$\sigma$\ was reached in 2.5 years \citep{kramer06a}}.  
As noted by \citet{kramer06a}, measuring \Pbdot\ at
this level will place stringent requirements on the class of gravitational theories which predict
significant amounts of dipolar, as opposed to quadrupolar, gravitational radiation, exceeding the
best solar system tests. Measuring the moment of inertia of pulsar A via a measurement of 
the spin--orbit coupling of the system, however, would require 
a further order of magnitude improvement in the measurement precision of \Pbdot\ \citep{kramer06a}.
In the near future, additional VLBI and/or timing measurements can be expected to reduce both the 
distance and velocity errors to \pfour\ below 10\%; however, even with negligible
error in these parameters, the existing accuracy of measurements of $R_{0}$, $v_{0}$\ and $K_{z}$
would limit the accuracy of \Pbdot\ measurements in this system to 0.004\%.  To attain the $10^{-5}$
precision necessary to measure the neutron star moment of inertia, 
the constants $R_{0}$ and $v_{0}$ must be measured to a precision approaching 1\%.
Such measurements may be possible within a decade via continued optical observations of stars near
the Galactic center \citep{ghez08a}, although attaining this precision is likely to be challenging.

\subsubsection[Implications for formation models]{Implications for formation models}
\label{results:binary:0737:formation}
As noted in Chapter~\ref{pulsars}, most neutron stars
are believed to receive large kicks at birth.  Indeed, the
discovery of the first binary pulsar PSR B1913+16 with its large
orbital eccentricity ($e=0.617$) supported this view and led to a
``standard model'' in which the newly-formed neutron stars 
in compact binaries received
large kicks and were often disrupted by this and the
associated mass loss \citep{bhattacharya91a}

\pfour, however, does not support the standard model. Its low
orbital eccentricity of just 0.08 \citep{kramer06a} and low transverse velocity suggest 
that very close binaries might influence the evolution of the progenitor stars and prevent large
mass loss and kicks during the supernova events \citep{podsiadlowski05a, stairs06a}. 
\citet{willems06a} have argued that the observed properties of \pfour\ can
still be explained by large kicks and significant
mass loss, however \citet{piran05a} suggest almost no
mass loss took place beyond the binding energy.  The refinement of the local transverse
velocity of the \pfour\ system presented here allows a
comment on the competing formation models for the system.

Since, as shown above, the transverse velocity of \pfour\ is no
greater than the likely peculiar velocity of its progenitor system, if the
\pfour\ system received a large velocity kick, it must possess a large radial
velocity. However, in a DNS system there are no
observational methods available to determine the radial velocity. Due to the accurate 
measurement of its Shapiro delay, \pfour\ is known to lie edge--on 
\citep{kramer06a}, and if the only kick provided to the system was
provided by the loss of binding energy during the supernova explosion,
the resultant 3D space velocity should be on the order of $\sim$50 km s$^{-1}$,
estimated from the system's observed eccentricity and orbital velocity  
\citep[e.g.][]{blaauw61a}.
This velocity would be constrained to the plane of the orbit and simple
geometry allows the estimation of the probability of observing 
a transverse velocity less than 10 km s$^{-1}$ to be about one in
eight, which is small, but not unreasonable.

Conversely, if the double pulsar had received a large kick \citep[e.g.][]{willems06a}, 
then the odds of observing such a low transverse velocity become
increasingly remote. Not only would the radial velocity
have to be increasingly large, but the inclination angle of the system must
not be altered by the kick.
Hence, the transverse velocity results shown here reinforce those of
\citet{kramer06a} and strongly favour the interpretation of \citet{piran05a},
who argue for almost no mass loss and kick in the case of \pfour.

An intriguing aside to the implication of low kick velocities in \pfour--like
systems is the possible, albeit speculative, explanation this offers to the formation of the
newly-discovered binary pulsar PSR J1903+0327, a fast, heavy, and highly recycled
millisecond pulsar (mass 1.8 solar masses, period 2.15 ms) 
with a one solar mass main sequence companion
\citep{champion08a}.  The formation of this system has confounded existing models, since 
it possesses an intermediate orbital eccentricity of $e=0.44$, and the orbit of a highly
recycled pulsar such as PSR J1903+0327 should have been circularised during the mass--transfer
phase to a very high degree \citep{alpar82a}.
 
If the progenitor system to \pfour\ had originally possessed a solar mass
companion in a wide orbit as part of a triple system, then the kicks due to the
\pfour\ supernovae would not have been sufficient to
disrupt the triple system. In such a situation, PSR J0737--3039A and B would coalesce in
85 Myr due to the emission of gravitational radiation, creating
either a heavy millisecond pulsar or a black hole that continues to orbit the companion star.
\citet{van-den-heuvel84a} proposed such a merger mechanism for
the creation of the first millisecond pulsar shortly after its
discovery.  In this scenario, the majority of the binding energy of
the relativistic binary will be lost in just a few seconds, and the resultant
heavy millisecond pulsar or a black hole is left in an eccentric orbit 
with the remaining main sequence star. The eccentricity would be
determined by the amount of matter lost. 

Such a system would strongly resemble PSR J1903+0327, but only
if the coalescing neutron stars formed a millisecond pulsar with a low
magnetic field, and approximately one solar mass was lost from
the coalescing system in much less than an orbital period of the outer star.
The \pfour\ system possesses ample angular momentum to form a
fast millisecond pulsar, but the enormous energy lost as the neutron stars
are disrupted may help regenerate the magnetic field of
the resultant heavy neutron star. Nonetheless, the low space velocity of the
\pfour\ system implies that such a formation pathway is realistic, and
that it could potentially explain the PSR J1903+0327 system.

\subsubsection[Implications for DNS merger rates]{Implications for DNS merger rates}
The size of the Galactic population of DNS systems, and hence DNS merger rate estimates, 
depends upon a host of assumptions including the
luminosity functions and beaming fractions of recycled pulsars.
As already noted, the poor constraints on the characteristics of recycled pulsars 
mean that the merger rate estimates remain uncertain.
In many recent merger rate studies \citep[e.g.][]{kalogera04a} the luminosity 
function of recycled pulsars (and in particular, its faint tail) 
has been inferred from the entire pulsar population, avoiding the small--number 
problem which would be inherent in attempting to model the faint end of the 
recycled pulsar luminosity function directly.  However, this approach would not 
be valid if the luminosity function of recycled pulsars differs
significantly from that of the general pulsar population. 
Typically, multiple population models have been 
considered, leading to the widely varying merger rate estimates. 

The distributions of beaming fraction and pulse shape for recycled pulsars are known to differ 
considerably from those for slower pulsars \citep[e.g][]{kramer98a}, and hence it is plausible that
their luminosity distribution also differs markedly.  The observational constraints on the luminosity
functions for normal and recycled pulsars are presented in \citet{lorimer08a}, which highlights
the considerable extrapolation from the small observed sample of faint recycled pulsars.
The revised distance presented here shows 
that the radio luminosity of \pfour\ is considerably greater (by a factor of five) 
than previously assumed.  If this revision were to markedly influence the assumed luminosity
distribution for recycled pulsars in DNS systems, then the assumed space density of DNS systems 
would be reduced, since there must be fewer systems in the Galaxy to match the 
number of observed systems. This would demand that estimates of the Galactic DNS merger rate
and corresponding detection rates for LIGO and Advanced LIGO such as those presented in
\citet{kalogera04a} also be reduced.  However, it is infeasible to make conclusive statements on the
basis of the single measurement presented here, and the true recycled pulsar luminosity function 
is likely to remain the object of intense study for some time to come.

\subsubsection[Implications for x--ray production]{Implications for x--ray production}
A final area in which this distance determination can contribute to understanding of the \pfour\ system
is its x--ray emission.
\pfour\ has been observed intensively in x--rays 
\citep{mclaughlin04a,chatterjee07a,possenti08a,pellizzoni08a}, with the x--ray 
luminosity measured at  $(2.3-2.4)\times10^{30}$\,erg s$^{-1}$, assuming 
the $DM$\ distance of 500 pc.  The newly measured VLBI distance of 1150 pc 
means this luminosity is revised upwards to $1.2\times10^{31}\,$erg s$^{-1}$.
The majority of the x--ray emission is seen to be modulated at the spin period of the A pulsar
\citep{chatterjee07a}.
However, there has been considerable debate 
over where and how the observed x--rays are generated, given the small binary separation
and interaction of the pulsar wind of the A pulsar with the magnetosphere of the B pulsar
\citep[see e.g.][]{lyutikov04a} which provide alternate x--ray generation mechanisms to the commonly 
assumed magnetospheric and thermal origins \citep[e.g.][]{becker00a}.  
Generation of x--rays due to such wind---magnetosphere interactions would
be expected to show orbital modulations, but no such orbital modulation had been seen until
a tentative detection by \citet{pellizzoni08a} of x--ray emission modulated at the period of the B
pulsar over part of the orbit.

Magnetospheric x--ray emission is believed to occur in many pulsars that display non--thermal 
x--ray spectra \citep{becker99a}, and there is no reason to believe that magnetospheric x--ray 
emission should not be generated in the \pfour\ system.  Since pulsar A has a spin--down luminosity
orders of magnitude higher than pulsar B ($5.9\times10^{33}$\,erg s$^{-1}$, compared to 
$1.7\times10^{30}$\,erg s$^{-1}$) 
it would be expected to dominate any magnetospheric emission.
\citet{possenti08a} state that the if the x--ray luminosity of the \pfour\ system is dominated by
magnetospheric emission from pulsar A, the predicted x--ray luminosity (based on relation
$L_{X}\propto \dot{E}^{1.1}$\ from \citealt{grindlay02a}) is 
$5\times10^{30}\,$erg s$^{-1}$.  Given the typical scatter of an order of magnitude about 
the $L_{X}-E$ relation, they concluded that this is consistent with the then--presumed
x--ray luminosity of $2.3\times10^{30}\,$erg s$^{-1}$.
Counting against the magnetospheric origin, on the other hand, was the fact that at
the $DM$\ distance of 500 pc, the neutral hydrogen column density
calculated from the absorbed fit ($N_{\mathrm H} \sim 1.5\times10^{21}$ cm$^{-2}$) 
was higher than expected from the pulsar's presumed location in the Gum nebula
(a nearby, large HII region which is modeled at a distance of $\sim$\,500 pc 
with a depth of several hundred pc in the NE2001 model).

However, as noted by \citet{possenti08a}, the value of 
$N_{\mathrm H}$\ is consistent with the usual average of 10 neutral hydrogen atoms for every
free electron along the line of sight.  The revised distance estimate presented here 
places \pfour\ beyond the Gum nebula,
making the $N_{\mathrm H}$\ measurements consistent with expectations.
The revised value of x--ray luminosity ($1.2\times10^{31}\,$erg s$^{-1}$) is approximately a factor
of two from the predicted value, which is no more discrepant than the original estimate.
Hence, this revised distance measure strongly supports a power--law model of
magnetospheric origin (from pulsar A) for the bulk of the x--ray emission from the \pfour\ system.

Whilst the original spin--down luminosity to x--ray luminosity conversion efficiency of 0.04\% was 
consistent with expectations, the revised conversion 
efficiency of 0.2\% implied by this new VLBI distance is the highest
amongst observed millisecond pulsars \citep[see e.g.][]{zavlin06a}.  There is considerable
scatter in the observed values of x--ray conversion efficiency amongst millisecond pulsars, however, 
and so this is not taken as evidence of an unusual x--ray production mechanism.  It is also
worth noting that at the 95\% confidence lower limit of distance (860 pc) the
x--ray conversion efficiency would be 0.12\%, less than PSR B1937+21 and PSR J0218+4232
\citep{zavlin06a}.

\subsection[\peight]{\peight}
\label{results:binary:2145}
\peight\ is a binary pulsar with spin period 16.05 ms and orbital period 6.84 days, which was 
discovered by \citet{bailes94a}.   Optical observations by \citet{bell95a} tentatively identified the 
expected white dwarf companion to \peight, which was confirmed by \citet{lundgren96a}.  
The scintillation velocity of \peight\ was measured to be $31 \pm 25$\,\kms\ by \citet{nicastro95a}, 
and a revised value of 51\,\kms\ (no error given) was published by \citet{johnston98a}.
Scintillation observations by \citet{gothoskar00a} suggested a considerably higher transverse
velocity of 113 \kms, but this was based on a single observation and may have been biased
by refractive scintillation effects.  All scintillation studies used the TC93 distance of 500 pc
for this pulsar.

Timing astrometry using a 10 year dataset by \citet{lohmer04a} measured a 
parallax of $2.0 \pm 0.6$\,mas,
and a proper motion of $14.1\pm0.4\,$mas yr$^{-1}$, corresponding to a transverse velocity of 
$33\pm9$\,\kms.  The implied distance of $500^{+215}_{-115}$\,pc is consistent with the $DM$--based
estimates of 500 pc (TC93) and 566 pc (NE2001).
However, \citet{hotan06a} reported no detection of parallax and an upper
limit of 0.9 mas, at 95\% confidence, albeit with a shorter (2.5 year) dataset, in addition to 
a proper motion measurement of $13.5\pm6.0\,$mas yr$^{-1}$.  
A previous attempt has been made to measure the VLBI parallax of this system 
by \citet{brisken02a} using the VLBA, but the pulsar was not detected and dropped
from the observing program.  Accordingly, the distance to this system is still controversial and 
confirmation of a previous result is keenly sought.

\begin{figure}
\begin{center}
\includegraphics[width=0.8\textwidth]{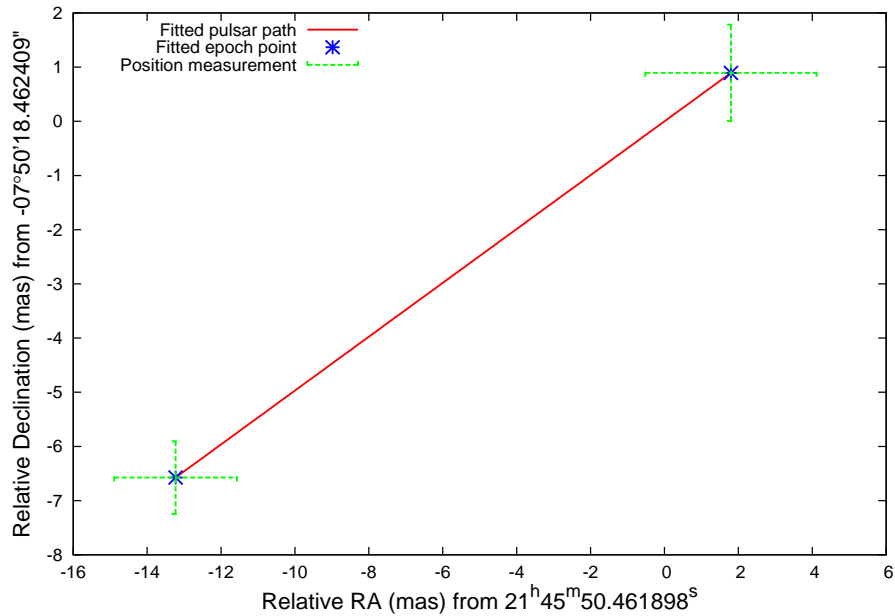}
\caption[Motion of \peight, with measured positions overlaid on the best fit]
{Motion of \peight, with measured positions overlaid on the best fit.  Since only two
significant measurements were made, the parallax of the pulsar was held fixed at zero, and
the fit to position and proper motion has no free parameters.  Variation
of the parallax between zero and two mas (corresponding to a distance $\geq500$\,pc) made 
an insignificant ($\lesssim100\,\mu$as yr$^{-1}$) difference to fitted proper motion.}
\label{fig:2145fit}
\end{center}
\end{figure}

From the four observations made of \peight\ during this observing program, significant detections
were made on only two occasions.  As was the case for the VLBA program of \citet{brisken02a}, 
it is believed that refractive scintillation is the major cause of the non--detections, as the pulsar's flux 
varies significantly over long timescales.  Although measurement of the parallax of \peight\ was
thus not possible, by holding parallax fixed at zero mas it was possible to measure a proper motion
of $\mu_{\alpha} = -15.4 \pm 2.1\,$mas yr$^{-1}$, $\mu_{\delta} = -7.7 \pm 0.80\,$mas yr$^{-1}$.  
Assuming a distance of 500 pc (both the TC93 estimate and the \citet{lohmer04a} timing parallax
measurement), this corresponds to a transverse velocity of 40 \kms, 20\% higher than the 
result given by the \citet{lohmer04a} timing.
Varying the parallax between zero and two mas results in a change in 
proper motion of less than 0.1 mas yr$^{-1}$, much smaller than the formal errors. As the proper
motion and position have been derived from only two measurements, there are no degrees
of freedom to the fit and no estimate of systematic errors, 
and so the result should be treated with some caution.  The 
fitted and measured positions of \peight\ are shown in Figure~\ref{fig:2145fit}.

The observed total proper motion of $17.2\pm2.2$\,mas yr$^{-1}$\ is consistent with the value of
$13.5\pm6.0\,$mas yr$^{-1}$ obtained by \citet{hotan06a}, but only consistent at the 2$\sigma$ level
with the value of $14.1\pm0.4\,$mas yr$^{-1}$ obtained by \citet{lohmer04a}\footnote{since 
\peight\ lies close to the ecliptic, timing observations obtain large covariances between proper 
motion in right ascension and declination, and so the comparison of total proper motion is more useful}.
As these VLBI results are derived from only two measurements, and there is no way of quantifying 
the systematic error, it is likely that the timing results are more reliable for this pulsar.

\section[Isolated pulsars]{Isolated pulsars}
\label{results:isolated}
Whilst isolated pulsars do not generally allow the tests of GR which can be made using binary pulsars,
many are interesting for other reasons.  Apart from providing more (and generally easier) targets for
studying the accuracy of $DM$\ distance models, as discussed in Section~\ref{pulsars:obs:timing},
isolated pulsars can also be used to investigate the accuracy of scintillation velocity measurements
and the alignment of velocity and rotation vectors in the general pulsar population,
as described by \citet{johnston05a}.  Another 
comparison that can be made is that of ``kinematic age" (the time it would have taken the pulsar
to reach its present Galactic latitude, given its present proper motion) to characteristic age, which
can yield some insight into the birth location and/or spin period of pulsars.  However,
apart from the ``technique check" pulsars \pfive\ and \psix, the pulsars included in this sample were
chosen primarily for their unusual luminosity characteristics.  In each case, an error in the assumed
$DM$ distance measurement could have potentially explained these unusual luminosities, and 
these VLBI observations offered a mechanism to remove this uncertainty.

At the outset of this analysis, it is appropriate to note the importance of geometric considerations in
determining pulsar luminosities, particularly in the radio.  The extremely anisotropic nature of pulsar
radio emission, due to the (non--uniform) beamed emission over a finite solid angle, makes direct comparison between pulsars problematic.   For any given pulsar, the emission beam may sweep 
directly over the observer, or the edge of the beam may only graze the line of sight, with an obvious 
impact on the apparent luminosity.  Pulsar apparent luminosities are generally quoted in mJy 
kpc$^{2}$, where the pulsar flux is simply scaled by distance squared, and no attempt is made to 
account for beaming effects.  Whilst beaming effects make radio luminosity analyses problematic for
individual pulsars, studies of pulsar radio luminosity functions are nevertheless extremely important
as the pulsar luminosity function is an essential component of pulsar simulations.

The results obtained for each isolated pulsar observed in this program are discussed in turn below.

\subsection[\pone]{\pone}
\label{results:isolated:0108}
\pone\ was first reported by  \citet{tauris94a}, and was postulated to be the nearest observed
radio pulsar -- its $DM$ of 2.38 remains the lowest of any known pulsar.  In the TC93 
Galactic electron model, \pone\ was predicted to lie 130 pc from Earth -- in the newer
NE2001 model \citep{cordes02a}, the distance was revised to 184 pc, which if correct would 
double the apparent radio luminosity, to a value of 300 $\mu$Jy kpc$^{2}$\ at 400 MHz (the 
lowest known value is that of PSR J0006+1834, which at a $DM$--derived 
distance of 700 pc has an apparent 400 MHz luminosity of 100 $\mu$Jy kpc$^{2}$).

\begin{figure}
\begin{center}
\begin{tabular}{cc}
\includegraphics[width=0.55\textwidth]{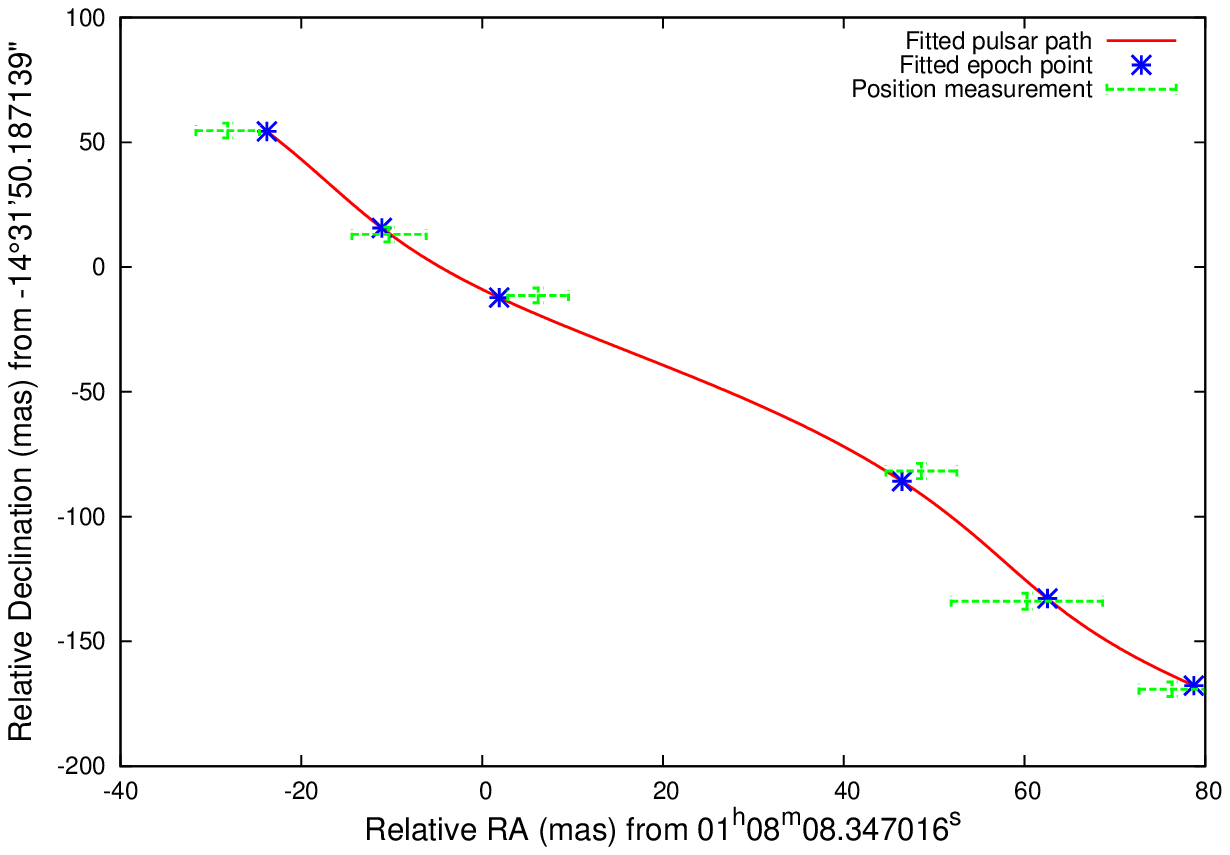}  \\
\includegraphics[width=0.55\textwidth]{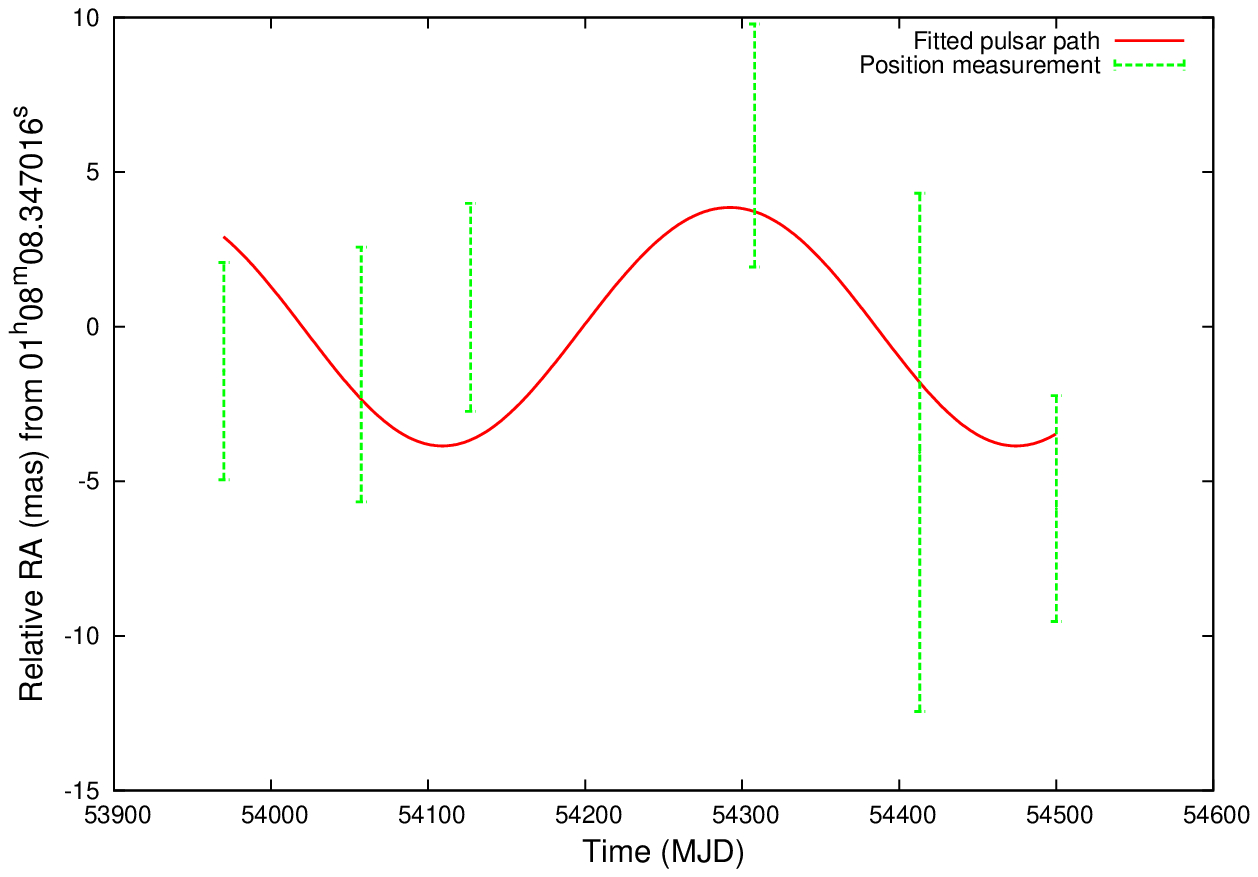} \\
\includegraphics[width=0.55\textwidth]{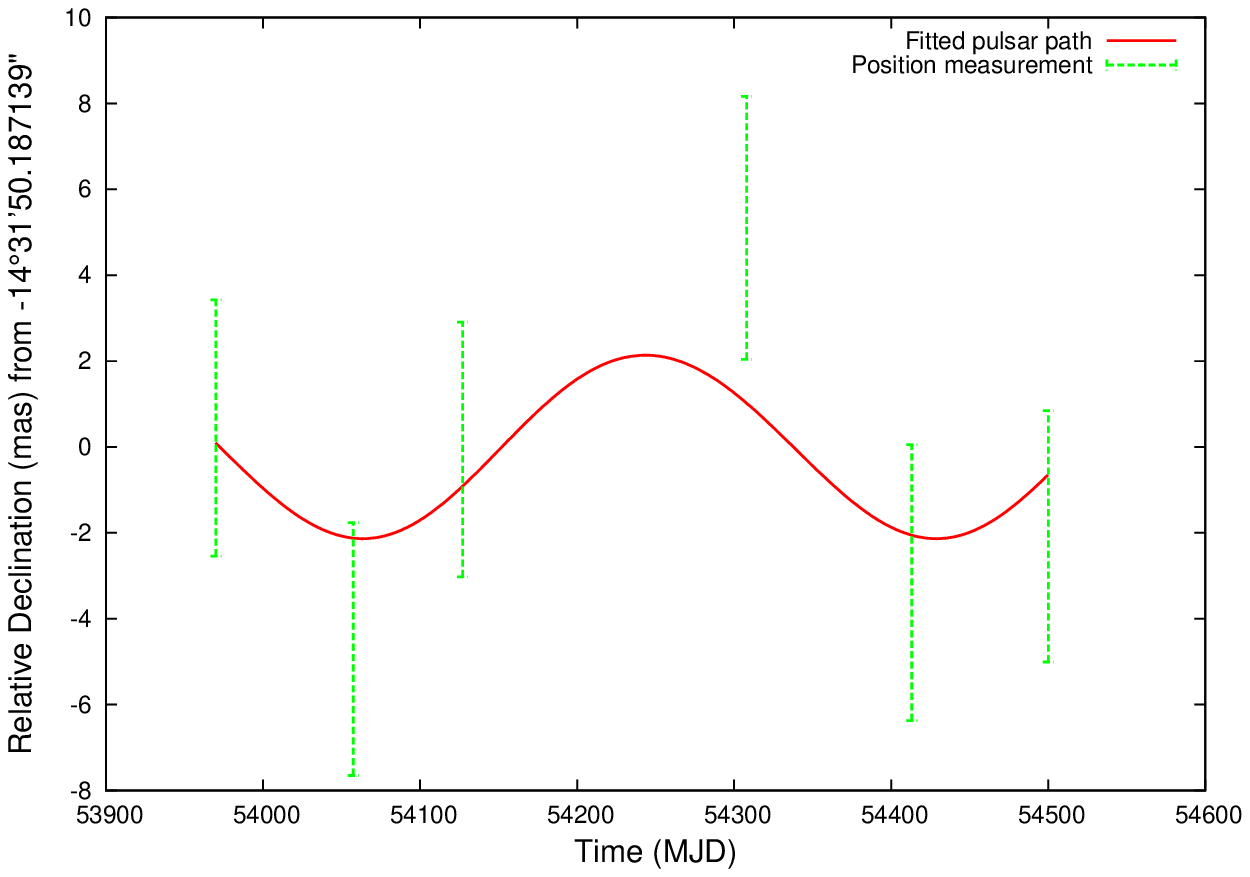} \\
\end{tabular}
\caption[Motion of \pone, with measured positions overlaid on the best fit]
{Motion of PSR J0108--1431, with measured positions
overlaid on the best fit.  From top: Motion in declination vs right ascension;
right ascension vs time with proper motion subtracted; and
declination vs time with proper motion subtracted.}
\label{fig:0108fit}
\end{center}
\end{figure}

Figure~\ref{fig:0108fit} shows the fitted and measured positions for \pone.  As shown in 
Table~\ref{tab:allresults}, the measured distance of  $240_{-61}^{+124}$\,pc is 
consistent with the NE2001 distance, but inconsistent at 95\% confidence with the earlier
TC93 distance.  It confirms that \pone\ is more distant than originally suspected, and while
its apparent radio luminosity is still low, it is no longer remarkably so.  In fact, 5 other pulsars 
(PSR J0006+1834, 
PSR J1829+2456, PSR J1918+1541, PSR J2015+2524, and PSR J2307+2225) have lower
400 MHz luminosities than the revised value of 510 $\mu$Jy kpc$^{2}$\ for \pone.  Of course, none
of these pulsars have an independent distance measure, so it is possible that one or more of
them may also be more distant than the model predictions, and hence also have greater 
luminosities.  In addition, as noted above, beaming effects mean that these measured apparent 
luminosities may differ considerably from the true luminosities for these pulsars.

This higher distance, however, does imply an even lower value of electron density than that predicted
by the NE2001 model: 0.0099 cm$^{-3}$.  This is amongst the lowest values amongst pulsars within
one kpc of the Solar System, and is the lowest value for pulsars nearer than 500 pc.

The detection of a parallax and proper motion enables for the first time a calculation of the
Shklovskii correction for \pone,  as shown:

\begin{eqnarray}
\frac{\dot{P}_{\mathrm{Shk}}}{P} &=& \frac{\mu^{2}D}{c} \\
&=& \frac{170^{2}\times240}{2.998\times10^{8}}\,\mathrm{mas}^{2}\,\mathrm{yr}^{-2}\,\mathrm{pc}\,\mathrm{s}\,\mathrm{m}^{-1} \nonumber \\ 
&=& 2.31\times10^{-2} \times 7.28\times10^{-16}\,\mathrm{s}^{-1} \nonumber \\ 
&=& 1.68\times10^{-17}\,\mathrm{s}^{-1} \nonumber \\ 
\dot{P}_{\mathrm{Shk}} &=& 0.808\times1.68\times10^{-17} = 1.37\times10^{-17}
\label{eq:shklovskii}
\end{eqnarray}
\noindent where $D$ is the pulsar distance, $\mu$\ is the total pulsar proper motion and $c$\ is 
the speed of light.

Since the measured \Pdot\ for \pone\ is $7.704\times10^{-17}$\ \citep{hobbs04a}, the 
Shklovskii term contributes 18\% of the observed rate of change of spin period for this 
pulsar.  The consequent downward revision to the intrinsic spin--down  luminosity of \pone\ is
important in the interpretation of the optical and x--ray conversion efficiency for \pone, 
as shown below.  Additionally, the characteristic age of \pone\ is revised
upwards from $166\times10^{6}\,$yr to $200\times10^{6}\,$yr.

\pone\ was recently detected in x--rays by \citet{pavlov08a}, who found that the observed emission
was compatible with a magnetospheric origin, but that a thermal component from heated
polar caps could not be excluded.  The spin--down luminosity
to x--ray luminosity conversion efficiency was 0.4\% when assuming a distance of 130 pc.  This value
is much larger than for young pulsars (typically 10$^{-5}$ to 10$^{-4}$, \citealt{pavlov08a} 
and references therein), but comparable to most other old, nonrecycled pulsars.  
At the best--fit distance for \pone, the increased x--ray luminosity coupled with the downward revision
to spin--down luminosity increases the conversion efficiency
to 1.7\%.  This would make \pone\ the most efficient known pulsar in terms of converting spin--down 
luminosity into x--ray luminosity, as
the distance to the previously most efficient pulsar, \pthree, has been revised dramatically 
downwards by the measurements of this thesis.  \pthree\ is discussed further 
in Section~\ref{results:isolated:0630} below.  However, at the 95\% confidence lower distance
estimate (143 pc) the x--ray conversion efficiency estimate is little changed at 0.5\%.  
A improvement on the parallax measurement would allow more useful constraints to be placed on 
the x--ray conversion efficiency.

Deep optical searches for \pone\ have been made using the VLT \citep{mignani03a} and while 
the pulsar was not detected in the original observation, a reanalysis based on the x--ray detection
of \citet{pavlov08a} produced a viable candidate \citep{mignani08a}.  Extrapolating the VLBI
position and proper motion back to the date of the optical data (July/August 2000) provides
further confirmation of the optical detection (VLBI position: RA 01:08:08.311(1), 
dec $-$14:31:49.12(1); optical position: RA 01:08:08.301(26), dec $-$14:31:49.15(27)).
\citet{mignani08a} show that the measured optical emission is compatible with
purely thermal emission from the bulk of the neutron star, with a surface temperature of
$(7-10)\times10^{4}$\,K, assuming a distance of 130 pc and a neutron star radius of 13 km.  
They note, however, that a determination
of the actual neutron star surface temperature is dependent on an accurate distance measurement.
Taking the lower limit of $7\times10^{4}$\,K from above, and combining with the 95\% confidence
lower limit for distance to \pone\ (143 pc), a simple re--scaling shows that these VLBI 
measurements imply that the temperature of \pone\ must be greater than 
$8.5\times10^{4}$\,K. At the best--fit distance of 240 pc, the implied temperature is in the range
$(2.4-3.4)\times10^{5}$\,K.  

Since the temperature of a neutron star with the age of 
\pone\ (Shklovskii--corrected characteristic age $200\times10^{6}$\,yr) should be less than
$3\times10^{4}$\,K if it had cooled according to standard models without any internal reheating
\citep[see Figure 3 of][]{schaab99a}, these results imply that
some reheating of isolated, slowly rotating neutron stars must occur.  While 
\pone\ could be younger than its characteristic age, assuming standard cooling to a
temperature at or higher than the 95\% confidence lower limit ($8.5\times10^{4}$\,K) implies 
an age less than $10\times10^{6}\,$yr, which is extremely unlikely.  
More accurate temperature measurements for a number of pulsars would be required to differentiate 
the possible reheating models 
\citep[see][for a review of reheating models]{schaab99a} but for \pone\ most reheating
processes do not have sufficient efficiency.  Alternate models based on neutron star 
interiors composed of quark matter suggest a slower cooling rate and higher temperatures
for old neutron stars \citep{alford05a}, but the 
surface temperature implied by these \pone\ observations is challenging to explain even in
that framework.  Alternatively, as noted in \citet{mignani08a}, a component 
of the optical emission may be non--thermal, although in this case the efficiency of the non--thermal 
process would be higher than in other optically--detected neutron stars 
such as PSR B0950+08 \citep{zharikov04a}.

The predicted velocity of 194 \kms\ is unremarkable, and the pulsar's high Galactic latitude of
$-76.8$\degrees\ makes is difficult to place good constraints on the true vertical component of velocity.
This renders a comparison of characteristic to kinematic age impossible for \pone, but as the pulsar
would likely have already completed more than one oscillation through the Galactic disk, such a 
comparison is not useful in any event.  The proper
motion is compatible with, but much more significant, than the value of 
$\mu_{\alpha} = 92 \pm 44$, $\mu_{\delta} = 176 \pm 70$\ mas yr$^{-1}$\ derived 
by \citet{pavlov08a}.

\subsection[\pthree]{\pthree}
\label{results:isolated:0630}
\pthree\ is a middle--aged pulsar which has been observed extensively as part of scintillation
studies \citep[see e.g.][]{cordes86b, bhat99a}, 
and would be undistinguished if not for its remarkable apparent luminosity
in x--rays.  \citet{becker05a} observed \pthree\ with XMM--Newton and found an x--ray luminosity
of $8.4\times10^{30}$\,erg s$^{-1}$, based on the NE2001 distance of 1.45 kpc. 
This implied that the pulsar was converting
$\sim\,$16\% of its spin--down luminosity into x--rays -- an order of magnitude more than any other
old or middle--aged pulsar.  They suggest that the most likely explanation
is that the distance to the pulsar is over--estimated, a theory which has been proven correct by 
these VLBI observations.  

\begin{figure}
\begin{center}
\begin{tabular}{c}
\includegraphics[width=0.55\textwidth]{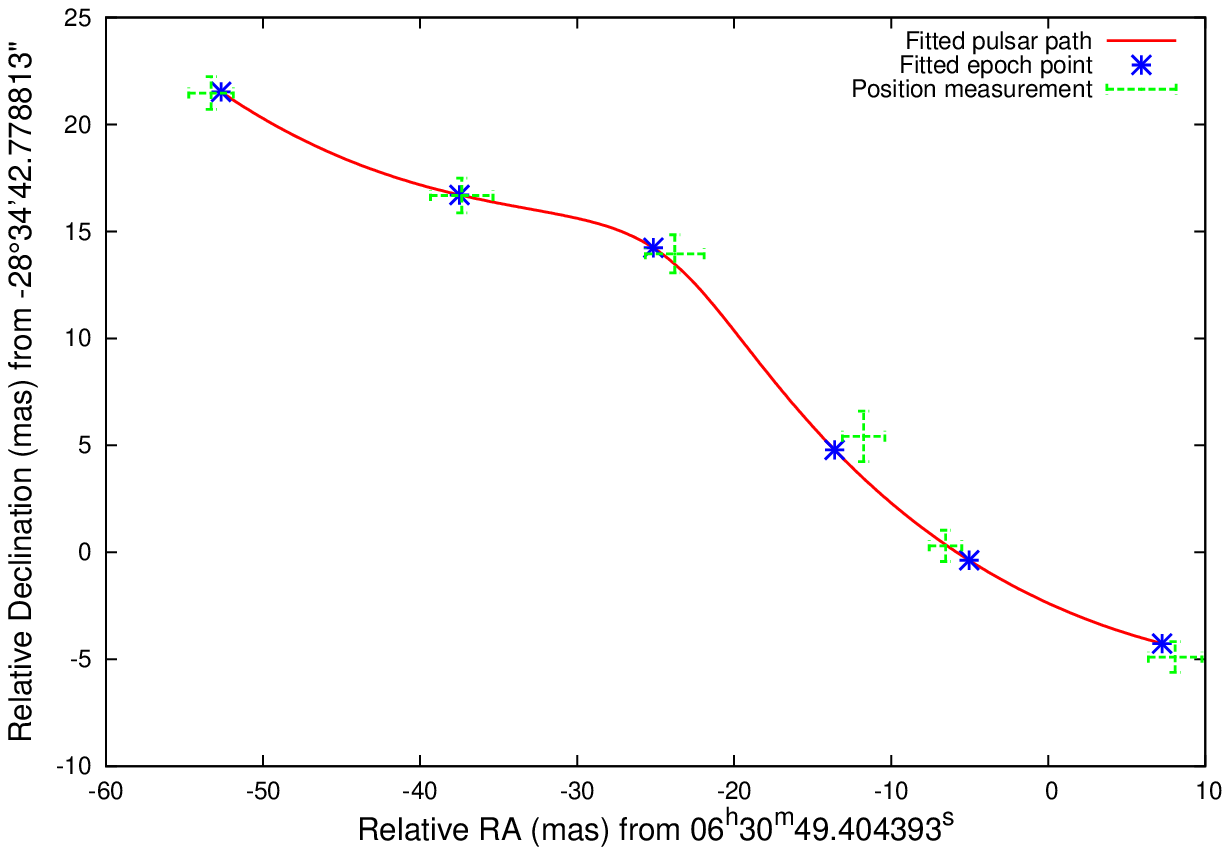} \\
\includegraphics[width=0.55\textwidth]{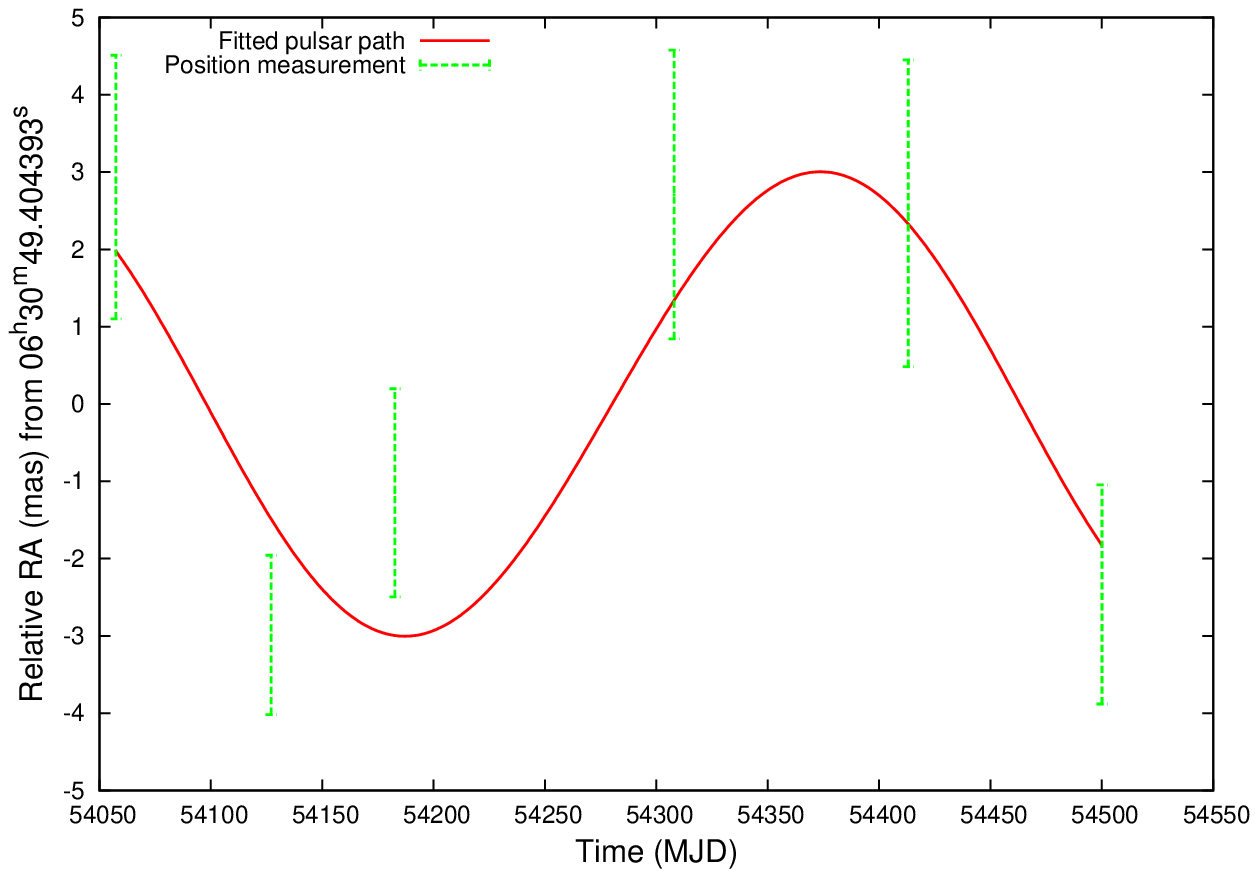} \\
\includegraphics[width=0.55\textwidth]{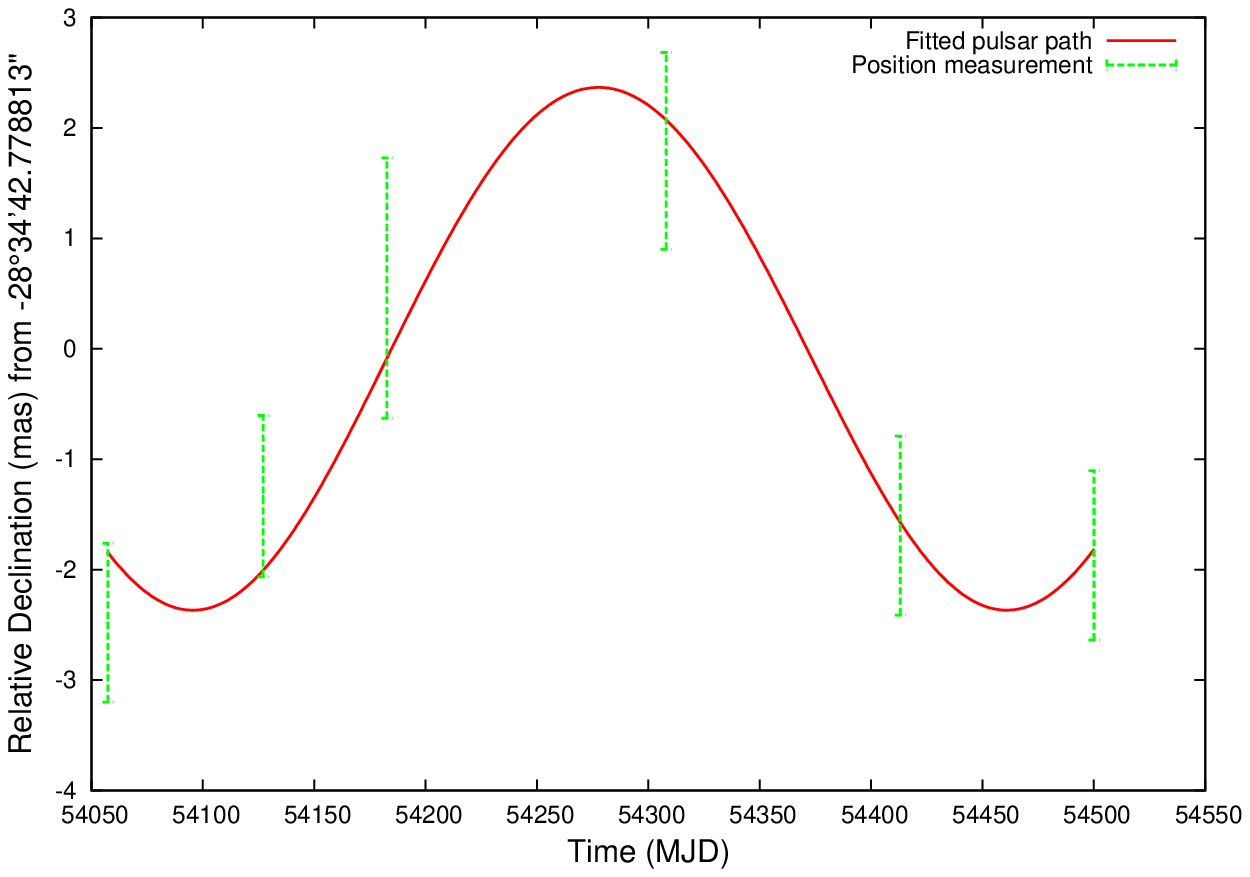} \\
\end{tabular}
\caption[Motion of \pthree, with measured positions overlaid on the best fit]
{Motion of \pthree, with measured positions
overlaid on the best fit.  From top: Motion in declination vs right ascension; 
right ascension vs time with proper motion subtracted; and
declination vs time with proper motion subtracted.}
\label{fig:0630fit}
\end{center}
\end{figure}

The measured and fitted positions of \pthree\ are shown in Figure~\ref{fig:0630fit}.  
The measured distance of $332^{+52}_{-40}$\ pc, as shown in Table~\ref{tab:allresults},
means that the actual x--ray conversion efficiency is a much less surprising $0.8\%$.  This is
within the $1\sigma$\ error bars of the efficiency derived for \pone\ in 
Section~\ref{results:isolated:0108} above.  The best--fit distance of 332 pc
is less than a quarter of the $DM$--derived distance in the NE2001
model, and a factor of seven smaller than the distance predicted by the TC93 model 
(2150 pc).  This large discrepancy is discussed further in Section~\ref{results:distvel}.

The proper motion of \pthree\ has previously been measured using the VLA by \citet{brisken03b},
and the measured values from this work 
($\mu_{\alpha} = -46.3\pm 0.99$\,mas yr$^{-1}$, $\mu_{\delta} = 21.26 \pm 0.52$\,mas yr$^{-1}$) 
are consistent with, but marginally
more precise, than the VLA results ($\mu_{\alpha} = -44.6\pm 0.9$\,mas yr$^{-1}$, 
$\mu_{\delta} = 19.5 \pm 2.2$\,mas yr$^{-1}$).  The resultant Shklovskii correction is less than 0.04\%
of the observed period derivative ($7.12\times10^{-15}$). The improved proper 
motion accuracy makes a marginal impact on the measurement of the alignment of the rotation 
and velocity vectors by \citet{johnston05a}, reducing the uncertainty in velocity position angle from 
3\degrees\ to 1\degrees.  This halves the error in the position angle difference measurement and
confirms that the velocity and rotation vectors of \pthree\ appear to be very highly aligned.

\citet{bhat99a} measured the scintillation velocity of \pthree\ to be $170\pm15$\,\kms, considerably 
higher than the earlier measurement of $60\pm13$\,\kms\ by \citet{cordes86b}.  However, the estimated
distance used by \citet{bhat99a} was the TC93 distance (2150 pc), which considerably biases the
estimated velocity.  \citet{cordes86b} uses a pre--TC93 distance estimate of 1240 pc.  When 
recalculated using the VLBI distance of 332 pc, the \citet{bhat99a} and \citet{cordes86b} speeds
are revised to $67\pm6$\,\kms\ and $31\pm7$\,\kms\ respectively. The transverse
velocity of $80^{+15}_{-11}$\,\kms\ obtained from these VLBI measurements is thus consistent with 
the \citet{bhat99a} measurements, but inconsistent with the earlier \citet{cordes86b} measurement.

The characteristic age of \pthree\ is $2.77\times10^{6}$\ years, while its kinematic age can
be calculated from its Galactic latitude $b=-16.758$\degrees\ and proper motion in Galactic 
latitude $\mu_{b}=-34.8$\,mas yr$^{-1}$\ as $1.73\times10^{6}$\ years.  At the best--fit distance of 
332 pc, the Galactic height of \pthree\ is $-$96 pc, and thus the characteristic and kinematic
ages are consistent, and imply that the progenitor for \pthree\ resided within 60 pc of the plane.

\subsection[\pfive]{\pfive}
\label{results:isolated:1559}
\begin{figure}
\begin{center}
\begin{tabular}{c}
\includegraphics[width=0.55\textwidth]{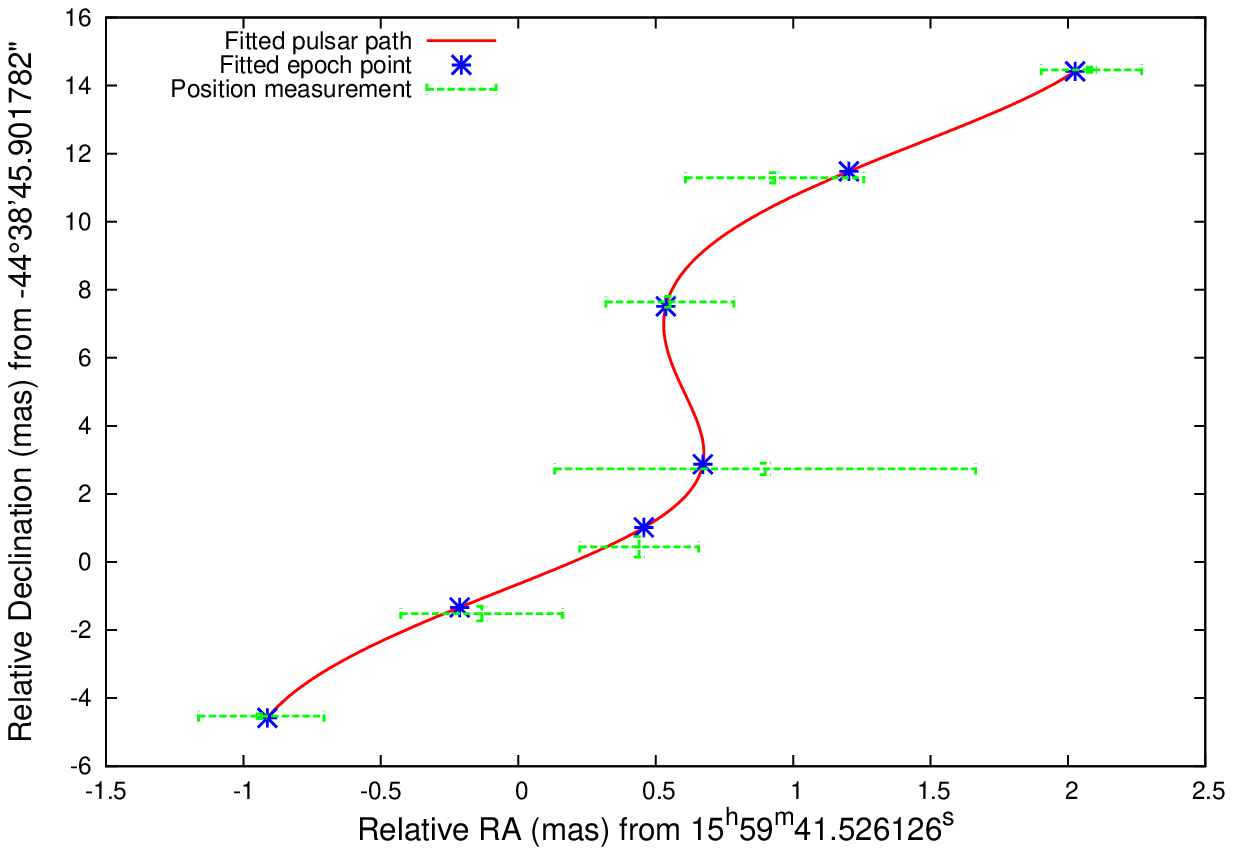} \\
\includegraphics[width=0.55\textwidth]{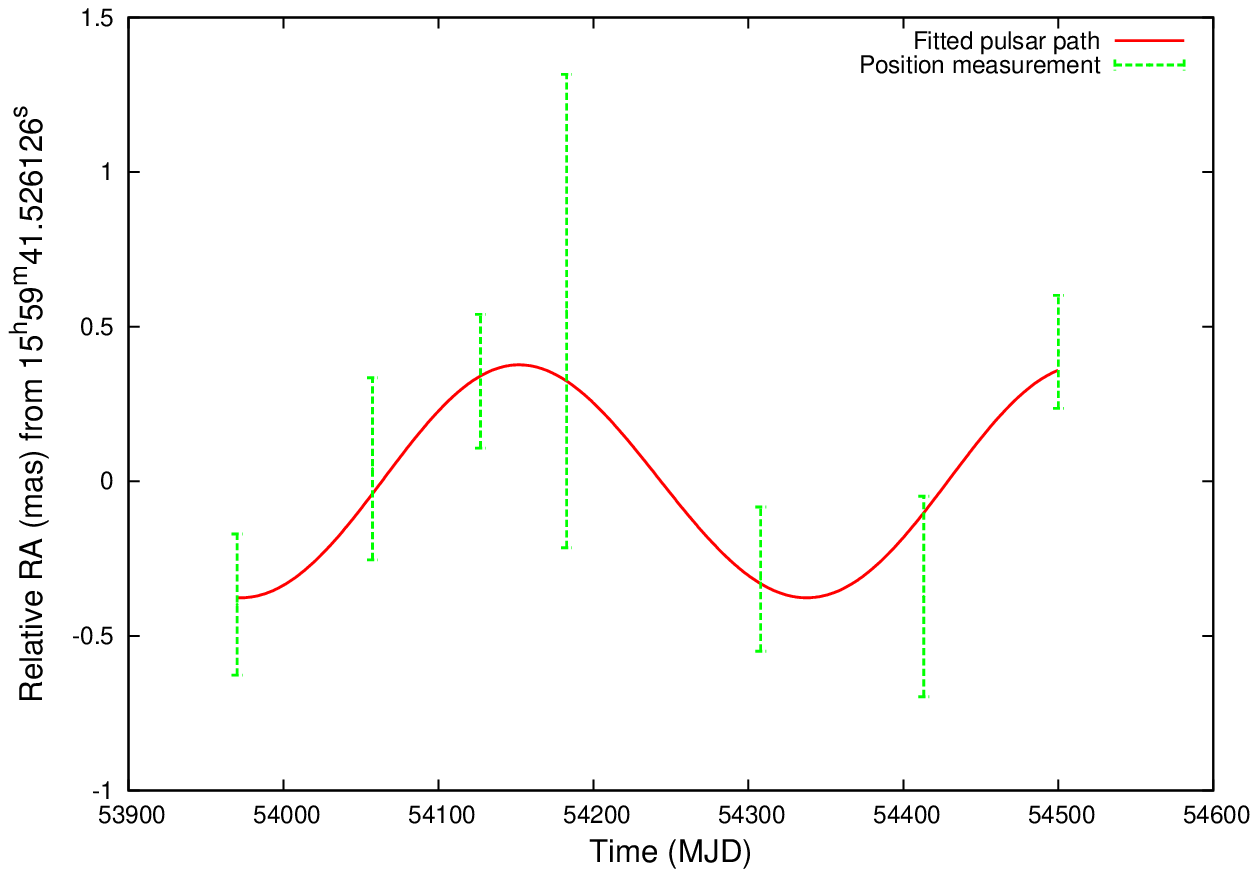} \\
\includegraphics[width=0.55\textwidth]{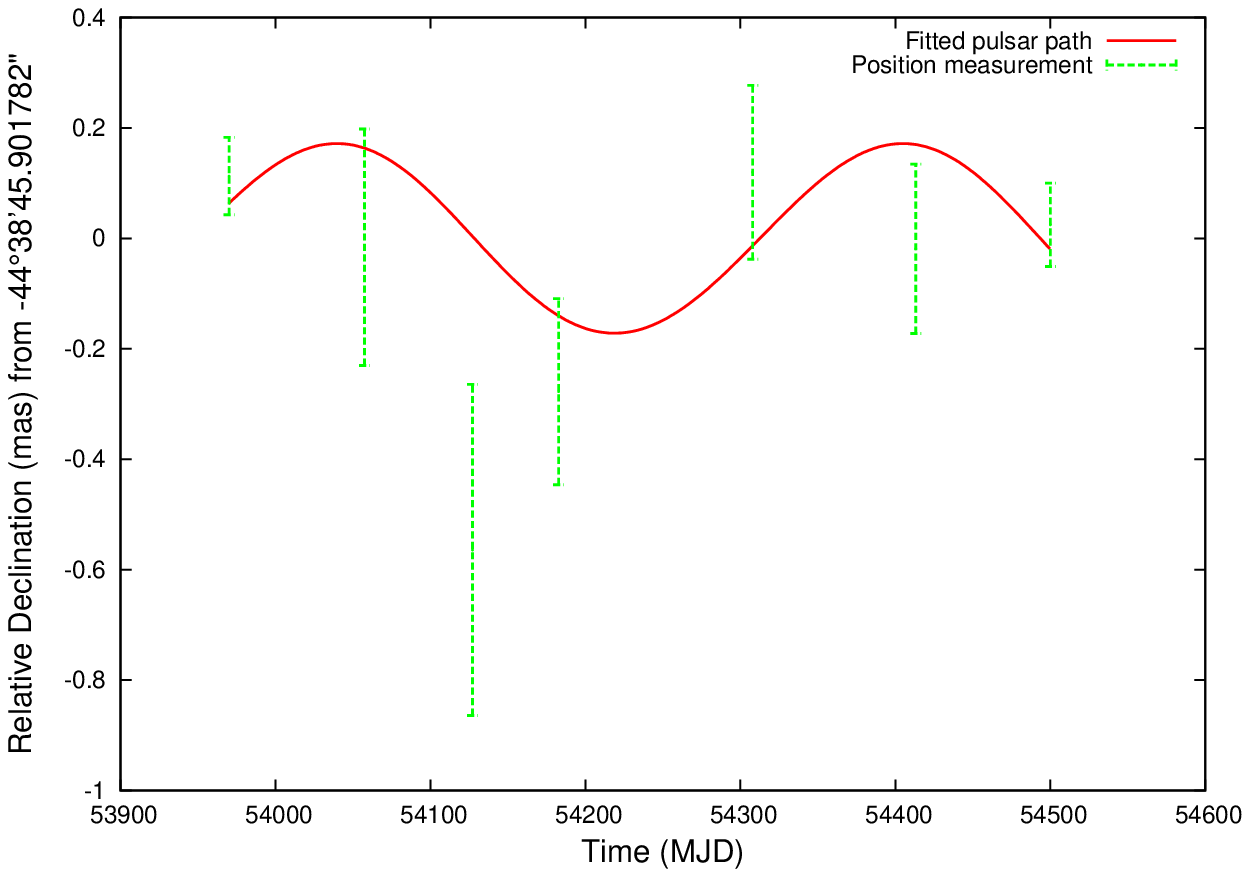} \\
\end{tabular}
\caption[Motion of \pfive, with measured positions overlaid on the best fit]
{Motion of \pfive, with measured positions
overlaid on the best fit.  From top: Motion in declination vs right ascension; 
right ascension vs time with proper motion subtracted; and
declination vs time with proper motion subtracted.}
\label{fig:1559fit}
\end{center}
\end{figure}

\pfive\ is a relatively unremarkable, middle--aged, isolated pulsar, which was included in this
observing program as the primary ``technique check" source.  Being relatively bright and
a strong scintillator, it provided an excellent opportunity to check the data reduction tools
developed in this thesis in a relatively high SNR regime.  However, as it has been well
studied in the past due to its relative proximity to the Solar System and brightness, a number
of interesting comparisons to previous studies can be made.  Although plots of the fitted
and measured positions of \pfive\ have been shown previously in Chapter~\ref{techniques}, they
are shown here again in Figure~\ref{fig:1559fit} for completeness.

The measured distance of $2600^{+690}_{-450}$\,pc is consistent with the NE2001 prediction 
\citep[2350 pc][]{cordes02a}, which differed considerably from the earlier 
TC93 distance estimate of 1580 pc.  
It is also consistent with the lower distance estimate of $2.0 \pm 0.5$ kpc made using HI
line absorptions by \citet{koribalski95a}. The measured values of
proper motion ($\mu_{\alpha} = 1.52 \pm 0.14$\,mas yr$^{-1}$, 
$\mu_{\delta} = 13.15 \pm 0.05$\,mas yr$^{-1}$) 
are consistent with the VLA observations of \citet{fomalont97a}, who measured
$\mu_{\alpha} = 1 \pm 6$\,mas yr$^{-1}$, $\mu_{\delta} = 14 \pm 11$\,mas yr$^{-1}$.
Neglecting acceleration from the Galactic potential, the kinematic age of the pulsar can be estimated 
from its Galactic latitude $b = 6.367\,^{\circ}$\ and
proper motion perpendicular to the Galactic plane $\mu_{b} = 8.93$\ mas yr$^{-1}$ as 
$2.57 \pm 1.02$ Myr, assuming a birth location within 100 pc of the plane \citep{faucher-giguere06a}.  
Under the standard assumption of a braking index
of 3, the observed period $P=257$\,ms and period derivative 
$\dot{P} = 1.01916\times10^{-15}$\ \citep{siegman93a} imply a birth period 
$P_{0}$\ between 35 and 151 ms -- longer than is often assumed for normal pulsars
\citep[see e.g.][]{migliazzo02a}, but similar to the calculated value of $P_{0} =139.6$\,ms 
for PSR J0538+2817 \citep{kramer03a}.  The correction to the observed \Pdot\ due 
to the Shklovskii effect is only 0.03\% of the observed value.

The transverse velocity measured for \pfive\ (164 km s$^{-1}$, 95\% confidence upper limit of 
287 km s$^{-1}$) is inconsistent with the 400 km s$^{-1}$ estimated from scintillation by 
\citet{johnston98a}, who assumed a 2 kpc distance to \pfive\ and a scattering screen midway
to the pulsar. If this discrepancy is interpreted as an error in the scattering screen location, 
then the scattering screen must reside considerably closer to the pulsar than to the solar system.
Alternatively, the distribution of turbulence in the scattering disk may be anisotropic, as appears to 
be the case for \pfour.

With an accurate proper motion now calculated, the position angle of the proper motion
for \pfive\ can be compared to the position angle of the emission polarisation, which tests
the alignment of the rotation and velocity vector, as described by \citet{johnston07a}.  If
the pulsar emits predominantly parallel or perpendicular to the magnetic field lines,
then the angle between the velocity and polarisation position angles is expected to be
0\degrees\ or 90\degrees\ respectively.  \citet{johnston07a} found plausible alignment
in 7/14 pulsars of similar ages to \pfive.  From
Table~\ref{tab:allresults}, it is easy to calculate the velocity position angle as 
PA$_{v} = 6.6 \pm 0.6$\degrees.  The polarisation position angle for \pfive\ given in 
\citet{johnston07a} is $71\pm3$, and so there does not appear to be a strong case for alignment
of the velocity and rotation axes for \pfive.  However, the polarisation profile of \pfive\ (as shown in
Figure~5 of \citealt{johnston07a}) is complicated, and it is possible that the original determination of
the magnetic field orientation from the polarisation position angle was incorrect.  
Thus, the alignment of the rotation and velocity vectors for \pfive\ cannot be excluded, and more
detailed polarisation studies are required.

\subsection[\psix]{\psix}
\begin{figure}
\begin{center}
\begin{tabular}{cc}
\includegraphics[width=0.55\textwidth]{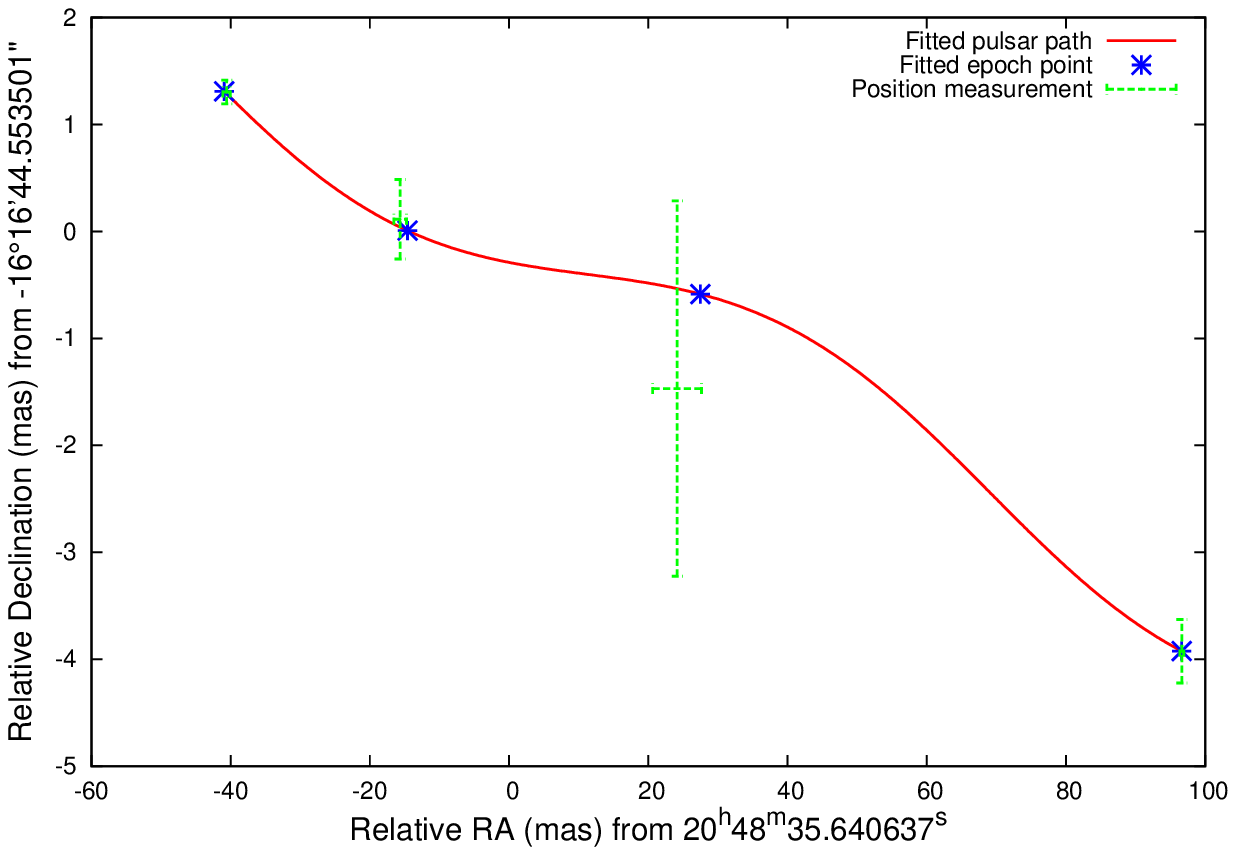} \\
\includegraphics[width=0.55\textwidth]{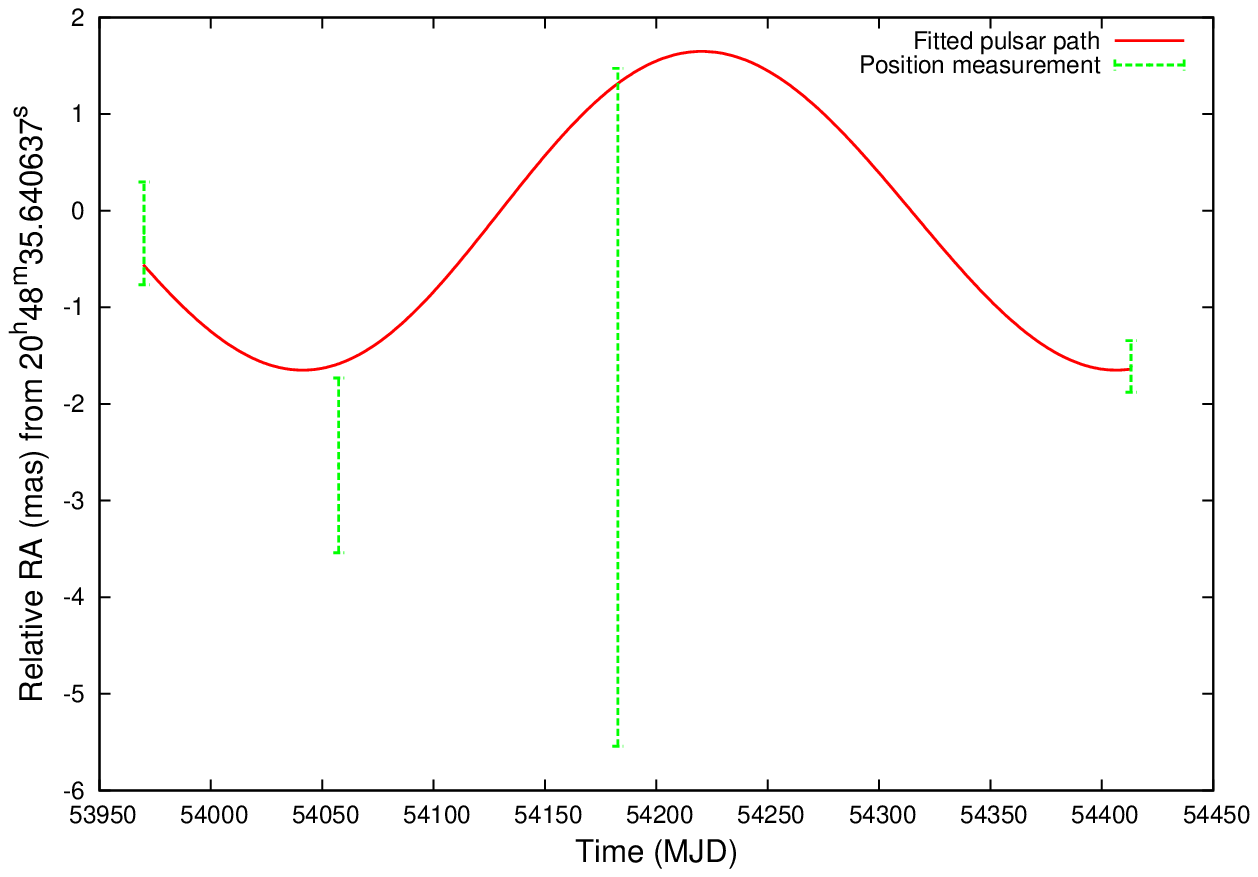} \\
\includegraphics[width=0.55\textwidth]{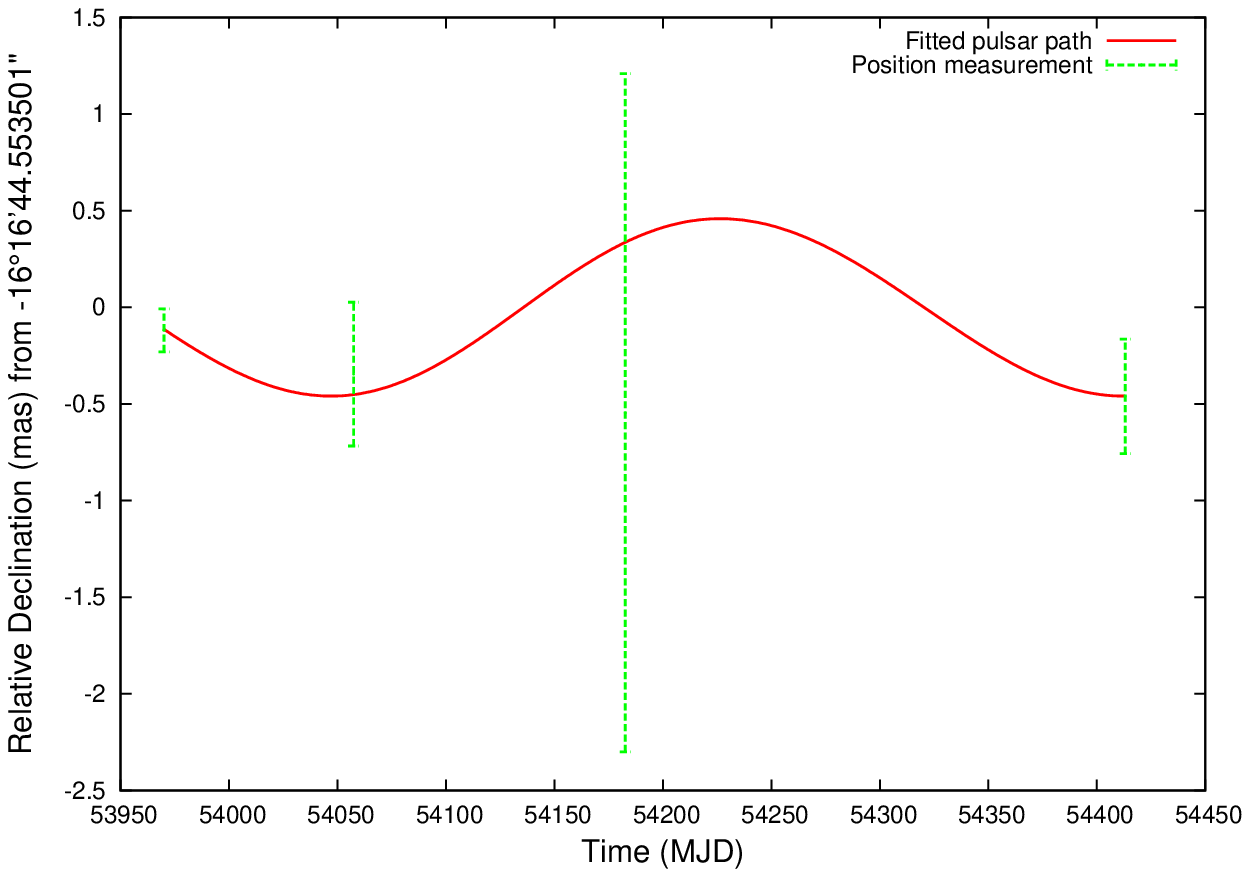} \\
\end{tabular}
\caption[Motion of \psix, with measured positions overlaid on the best fit]
{Motion of \psix, with measured positions
overlaid on the best fit.  From top: Motion in declination vs right ascension; 
right ascension vs time with proper motion subtracted; and
declination vs time with proper motion subtracted.}
\label{fig:2048fit}
\end{center}
\end{figure}

\psix\ was the second ``technique check" source included in the observing program, after \pfive.  It
is less bright than \pfive, but was predicted to be somewhat closer (640 pc in the TC93 model,
560 pc in the NE2001 model).  As shown in Table~\ref{tab:allresults}, whilst a measurement
of parallax was made for \psix, it was not significant (1.9$\sigma$).  Nevertheless, the best--fit distance of
580 pc is consistent with the $DM$--based distance estimates, and an accurate measurement
of proper motion was made.  Figure~\ref{fig:2048fit} shows the fitted and measured positions
of \psix.

It is apparent from Figure~\ref{fig:2048fit} that the large position error of the third epoch (MJD 54182) is
the primary reason for the poorly constrained fit.  Equipment failure at the Mopra telescope and limited
time on the Tidbinbilla telescope during this epoch reduced the number of telescopes on--source during
observations of \psix\ to three, and thus the large errors are unsurprising.  The covariance between 
proper motion and parallax was the largest for \psix, and hence adding even a single 
future position measurement at the appropriate parallax extrema would almost certainly allow
a significantly improved measurement of parallax.

The measured proper motion of $\mu_{\alpha} = 114.24 \pm 0.52$\,mas yr$^{-1}$, 
$\mu_{\delta} = -4.03 \pm 0.24$\,mas yr$^{-1}$ is 
consistent with but considerably more precise than the VLA measurement of \citet{fomalont97a},
which was $\mu_{\alpha} = 117 \pm 5$\,mas yr$^{-1}$, $\mu_{\delta} = -5 \pm 5$\,mas yr$^{-1}$.  
This allows an improved 
measurement of the transverse velocity vector position angle of $92.0 \pm 0.15$\degrees.  Since
the error in the alignment of polarisation and velocity position angle calculated by \citet{johnston07a}
was dominated by the error in polarisation position angle (5\degrees), this improvement
is not currently useful, although future improvements in polarisation measurements will allow
a more precise measurement of the position angle difference and hence the alignment
of the rotation and velocity vectors.  The Shklovskii component of \Pdot\ for \psix\ can be calculated
from the proper motion and best--fit distance to be $3.5\times10^{-17}$, or less than
0.4\% of the observed value of $\dot{P} = 1.0958\times10^{-14}$.

\citet{bhat99a} measured the scintillation velocity of \psix\ to be $501 \pm 29$\,\kms\ assuming
the TC93 distance of 640 pc, which cannot
currently be ruled out due to the distance uncertainty.  
However, if \psix\ does indeed reside at the best--fit
distance of 580 pc (which is consistent with the $DM$  estimates) then, as with \pfive, the scintillation
velocity has been overestimated by a factor of $\sim2$, and as with \pfive, this would support the
presence of a scattering screen closer to the pulsar than to the solar system.

The characteristic age of \psix\ is $2.84\times10^{6}$\ yr$^{-1}$, while the kinematic age can be
calculated from the Galactic latitude of $-33.077$\degrees and proper motion in Galactic latitude 
($-107\,$mas yr$^{-1}$) as $1.113\times10^{6}$\ yr.
As with \pfive, assuming a braking index of 3 and taking the observed period $P=1961.6$\,ms 
and period derivative $\dot{P} = 1.0958\times10^{-14}$ allows the calculation of birth period
$P_{0} = 1192\pm243$\,ms, assuming the pulsar progenitor resided within 100\,pc of the plane.  

Since the distance to \psix\ is still quite uncertain from these measurements, the most conservative
estimates for $P_{0}$\ and/or birth height 
can be obtained by assuming the pulsar distance is the minimum allowed value.
At the 95\% confidence lower limit of distance (283 pc), the Galactic height of \psix\ is 155 pc,
and the lower limit upon $P_{0}$, assuming a birth event 100\,pc above the plane, becomes
695\,ms.  Alternately, if the pulsar was born with a negligible spin period, the Galactic height of 
its progenitor must have been greater than 240\,pc (at 95\% confidence).

While, as noted in Chapter~\ref{pulsars}, considerable debate still exists as to the true distribution 
of initial pulsar spin periods, values greater than half a second are difficult to obtain with standard
pulsar birth models.  The second alternative, a progenitor residing considerably further above the
Galactic plane than usual, is possible but unlikely.  The remaining alternative to reconcile the 
characteristic and kinematic ages of \psix\ is that its braking index is greater than the canonical
value of 3.0.  Substituting the 95\% confidence 
upper bound on kinematic age ($1.83\times10^{6}$\,yr, assuming the minimum distance to \psix\ and
a progenitor 100 pc above the plane) into Equation~\ref{eq:charage} and assuming zero
initial spin period yields a minimum braking index of 4.1. 
As noted by \citet{yue07a}, however, none of the six reliable measurements of braking index 
made to date have been greater than 3.0, which would make such a high value surprising. 
Clearly, \psix\ is somewhat unusual in terms of generally assumed pulsar birth and evolution
parameters, and and further improvements to the parallax distance to \psix\ would be 
extremely valuable in constraining the birth parameters and spin evolution of this pulsar.

\clearpage
\subsection[\pseven]{\pseven}
\label{results:isolated:2144}
\pseven\ was discovered in the Parkes Southern Pulsar Survey \citep{manchester96a}, and 
was initially misidentified with a period of 2.83 seconds.  \citet{young99a} reported that
the true period is in fact 8.5 seconds, making \pseven\ the longest period radio pulsar known.
\pseven\ lies below the traditionally assumed pulsar ``death line" \citep[see e.g.][]{chen93a}, 
and hence its true luminosity is extremely important for models of pulsar emission and evolution.  
The discovery of \pseven\ prompted alternative models of pulsar emission in which 
long period pulsars such as \pseven\ could remain luminous in the radio \citep[e.g.][]{zhang00a}.
\pseven\ possesses a steeper spectral index than average ($-2.4$), 
and while several other pulsars appear less
luminous at 400 MHz, \pseven\ is the least luminous pulsar observed at 1400 MHz 
(apparent luminosity 24 $\mu$Jy kpc$^{2}$\ at assumed distance of 180 pc; average 
1400 MHz flux density of 0.75 mJy calculated from archival Parkes observations).

\begin{figure}
\begin{center}
\begin{tabular}{c}
\includegraphics[width=0.55\textwidth]{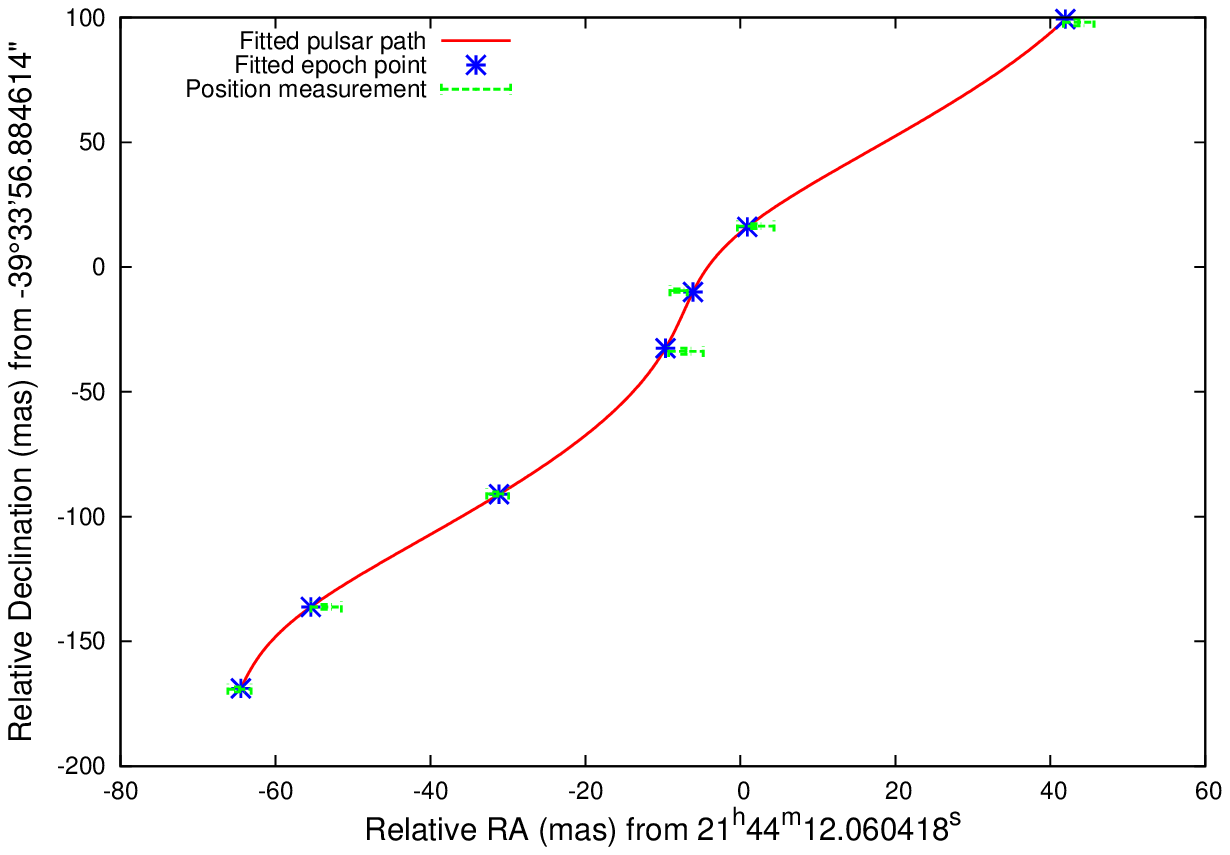}  \\
\includegraphics[width=0.55\textwidth]{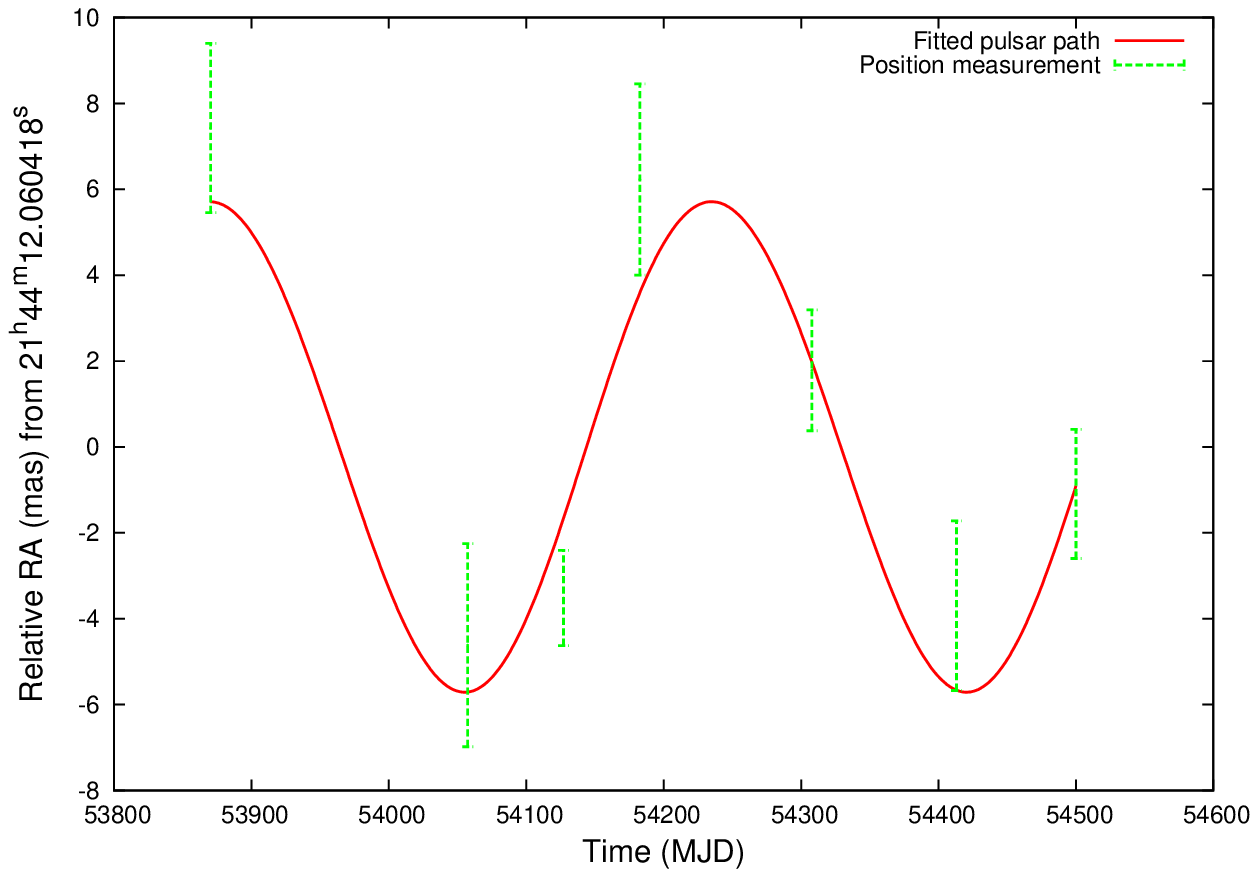} \\
\includegraphics[width=0.55\textwidth]{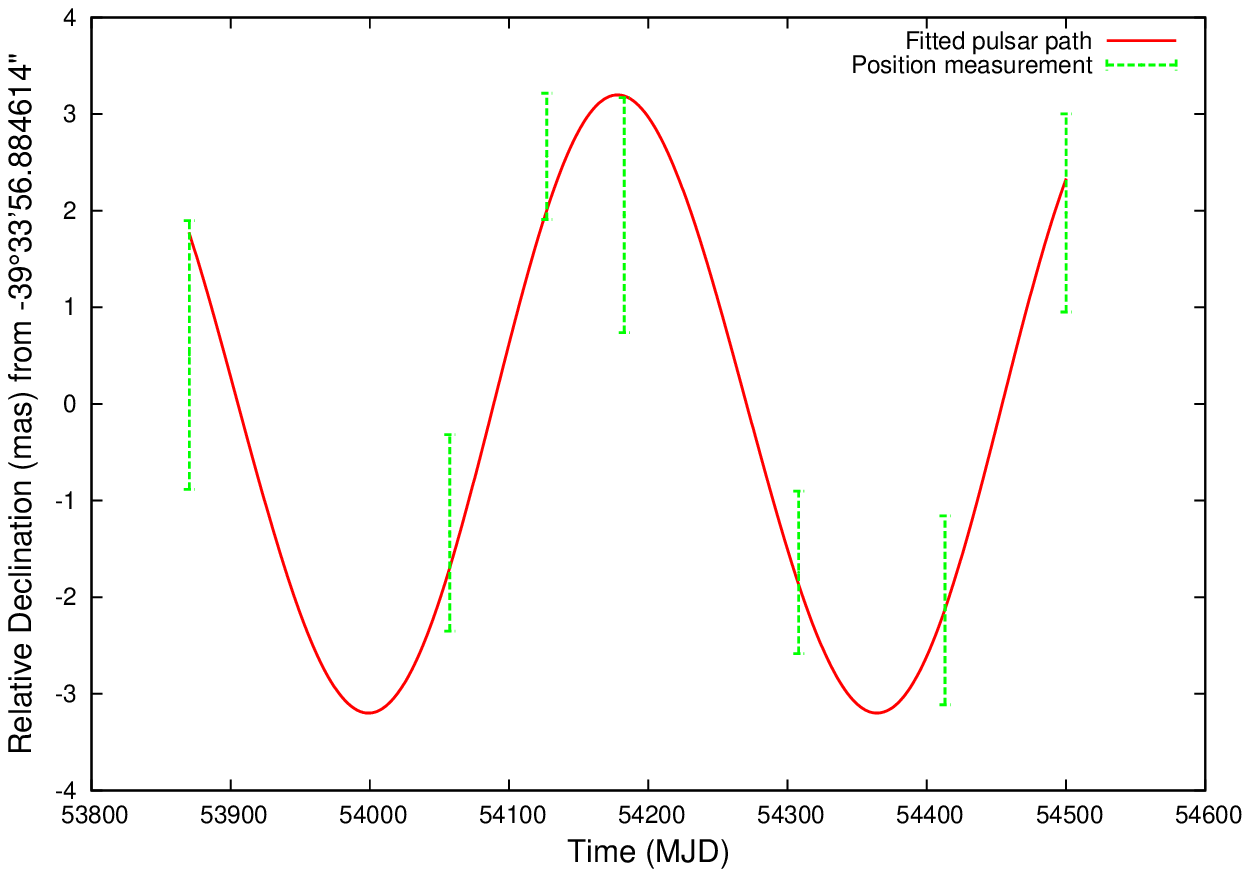} \\
\end{tabular}
\caption[Motion of \pseven, with measured positions overlaid on the best fit]
{Motion of \pseven, with measured positions
overlaid on the best fit.  From top: Motion in declination vs right ascension; 
right ascension vs time with proper motion subtracted; and
declination vs time with proper motion subtracted.}
\label{fig:2144fit}
\end{center}
\end{figure}

The fitted and measured positions of \pseven\ are shown in Figure~\ref{fig:2144fit}.  As shown in
Table~\ref{tab:allresults}, a highly significant parallax was detected, corresponding to a 
distance of $165^{+17}_{-14}$\,pc.  Given the generally assumed errors on $DM$ distances,
this is consistent with the TC93 value (180 pc), but not the NE2001 value (264 pc).  This
confirms that the apparent radio luminosity of \pseven\ is extremely low -- 15\% lower than the
previously assumed value of 24 $\mu$Jy kpc$^{2}$.

The proper motion measurement allows for the first time a calculation of the Shklovskii correction
for the period derivative of \pseven\ -- the Shklovskii contribution to \Pdot\ is $9.4\times10^{-17}$,
or approximately 19\% of the observed \Pdot\ of $4.96\times10^{-16}$.  The true spin--down luminosity
of \pseven\ is thus further reduced from the assumed value of $3.2\times10^{28}\,$erg s$^{-1}$\ to
$2.6\times10^{28}\,$erg s$^{-1}$, the smallest known spin--down luminosity of any pulsar.  
PSR J0343--3000, discovered in
the Parkes High--Latitude survey by \citet{burgay06a}, has the next lowest spin--down luminosity,
which is a factor of 5 higher than the revised value for \pseven.  In addition, this revision to the 
\Pdot\ value for \pseven\ places it even further past the assumed pulsar ``death line".

\begin{figure}
\begin{center}
\begin{tabular}{cc}
\includegraphics[width=0.49\textwidth]{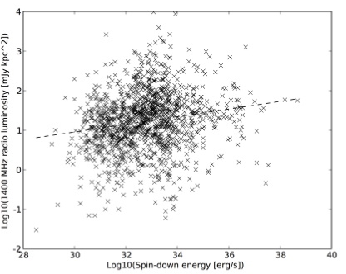} &
\includegraphics[width=0.49\textwidth]{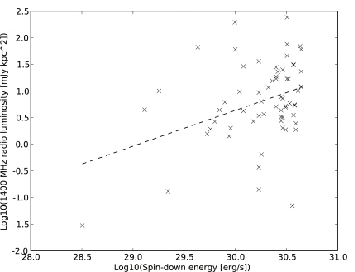} \\
\end{tabular}
\caption[Correlation between 1400 MHz radio luminosity and spin--down luminosity]
{Correlation between apparent 
radio luminosity and spin--down luminosity for all pulsars (left panel), and 
pulsars with periods near or below the traditional death line 
($B/P^{2} < 3.4 \times 10^{11} \mathrm{G}/\mathrm{s}$; right panel).  Whilst neither correlation
is particularly strong, the $r^{2}$\ value is higher for the ``death valley" pulsars in the right panel 
(0.136) compared to the whole population shown in the right panel (0.033).}
\label{fig:spinradiolum}
\end{center}
\end{figure}

The correlation between observed pulsar radio luminosities ($L$) 
and their spin--down luminosity ($\dot{E}$) has been the subject of considerable debate.
\citet{lyne98a} argued for a model in which the pulsar luminosity was independent of the 
other known physical pulsar parameters, but \citet{faucher-giguere06a} and \citet{malov06a}, 
amongst others, have shown evidence for a correlation  with the pulsar's spin--down luminosity,
with a dependence ranging from $L\propto\dot{E}^{1/3}$\ to $L\propto\dot{E}^{1/2}$.  More general
models with a dependence upon $P$\ and \Pdot\ are considered by \citet{faucher-giguere06a} but
are not investigated here.
Since the radio luminosity of ordinary pulsars is believed to be powered by their rotational 
energy, it is logical to assume that the radio and spin--down luminosities are related, 
but as already noted, complications such as differing beaming geometries make it plausible that 
this relationship could be difficult to identify observationally.  Additionally, since the radio emission
of pulsars makes up only a small component of their total emission, there is a potential for a large 
scatter in any relationship between radio luminosity and spin--down luminosity, and pulsar
spectral indices can vary considerably, meaning measurements based on a single frequency
introduce a further source of uncertainty.  Certainly, spin--down luminosity and 
observed 1400 MHz radio luminosity appears to be uncorrelated
when the entire population of pulsars is considered, as shown in the left panel of
Figure~\ref{fig:spinradiolum}. 

However, it is interesting to note that \pseven\ has both the lowest spin--down luminosity 
and the lowest apparent 1400 MHz radio luminosity of all known
pulsars.   \citet{malov06a} show that the radio efficiency of old pulsars ($P>0.1$\,s) appears 
to increase with period, with the conversion efficiency for 2 second pulsars on average
being an order of magnitude greater than for pulsars with period 0.2 seconds.
 \pseven\ is the longest period pulsar known and challenges standard pulsar emission
models \citep[e.g][]{zhang00a}, so it is reasonable to investigate the
possibility that pulsars with similar characteristics might show a stronger correlation between
spin--down luminosity and radio luminosity.  The vacuum gap
model of pulsar emission \citep[e.g.][]{bhattacharya92a} predicts a radio death line given by:
\begin{equation}
B/P^{2} < 1.7 \times 10^{11} \mathrm{G}\ \mathrm{s}^{-2}
\end{equation}

The right panel of 
Figure~\ref{fig:spinradiolum} plots apparent radio luminosity against spin--down luminosity
for pulsars that lie near or below this radio death line.
Since only 21 pulsars lie below this death line, and 1400 MHz luminosities are available for
only 14 of these, the $B/P^{2}$\ cut--off was chosen to be twice the vacuum gap death line value 
($3.4\times10^{11}\,$G\ s$^{-2}$).  This resulted in a sample of 65 pulsars.

As shown in Figure~\ref{fig:spinradiolum}, the correlation between spin--down
luminosity and apparent 1400 MHz radio luminosity is stronger 
for pulsars that lie near the traditional pulsar
death valley, towards the lower right corner of the $P$--\Pdot\ diagram.  The line of best fit
shown in the right panel of Figure~\ref{fig:spinradiolum} was obtained using standard linear
regression, and is given by the equation 
$\log_{10}{L} = 0.67\log_{10}{\dot{E}} - 19.57$.  The 95\% confidence interval for the power--law
exponent is [0.24, 1.1], and the $r^{2}$\ value of 0.136 is inconsistent with 
no correlation at greater than 99\% confidence for this sample size of 65 pulsars.
These results support the 
$L\propto\dot{E}^{0.5}$\ model presented by \citet{malov06a}, but models with a considerably
steeper or shallower exponent cannot be excluded.

As already noted, the unknown effect of the beaming geometry for each pulsar is bound to impose
a considerable scatter upon the apparent radio luminosities.  Furthermore, all spin--down 
luminosities are uncorrected for the Shklovskii effect and, as discussed further in 
Section~\ref{results:distvel:errors}, errors in pulsar distances could be corrupting the
radio luminosities and further obscuring a more obvious correlation.
The results presented here should not be 
interpreted as definitive evidence of a correlation between integrated pulsar radio luminosities and
spin--down luminosities -- rather, they suggest that such a correlation is more likely for pulsars
near the death valley, and that more detailed modeling could help confirm or deny
the existence of a correlation such as that proposed by \citet{malov06a}.  Such modeling would
necessarily include obtaining accurate distances and spectral indices and 
modeling beam geometries to estimate the integrated radio luminosity, and obtaining proper 
motions to calculate Shklovskii corrections to spin--down luminosity, 

As noted in the discovery paper of \pseven\  by \citet{young99a}, the low luminosity and narrow
beaming fraction ($\sim\,0.01$) of \pseven\ imply that many such objects exist in the Galaxy.  The 
distance measurement provided by this VLBI astrometric program has confirmed and 
slightly strengthened this hypothesis, since the pulsar is even closer and fainter than previously
thought.  Assuming the steep spectral index of \pseven\ is typical for similar long-period, 
low--luminosity objects, and that the Galactic population of such objects can be estimated
from \pseven\ alone (which yields a population of 100,000 objects; \citealt{young99a}), many 
such objects would be expected to be detected by 
LOFAR\footnote{http://www.lofar.org/}.  LOFAR, which is a low--frequency
aperture array telescope presently being constructed in the Netherlands, will search for pulsars
at frequencies of approximately 120 MHz.  

While the spectrum of \pseven--like objects
may exhibit a turn--over before this point, assuming the measured spectral index of $-2.4$ continues
to 120 MHz yields a 120 MHz
radio luminosity of 8 mJy kpc$^{2}$.  While the scope of LOFAR has been adjusted several times
recently, in pulsar survey mode it is expected to have a point source sensitivity in the 120 MHz
band of $\sim\,0.2$\,mJy\footnote{assuming a five minute integration, 24 MHz bandwidth, and a 
10\% pulsar duty cycle -- see http://www.lofar.org/p/astronomy\_spec.htm}.  Thus, LOFAR would
be expected to detect objects like \pseven\ up to distances of several kpc, beyond which distance
interstellar scattering is likely to limit detections more than sensitivity.  
Thus, potentially thousands of \pseven--like objects could
be expected to be detected with LOFAR.  Low--frequency observations of \pseven\ would
resolve the question of a potential spectral turnover, and help refine estimates of LOFAR
detection rates.

While this is the first measurement of proper motion for \pseven, \citet{johnston98a} have measured 
a scintillation velocity of 48 \kms\ assuming a distance of 330 pc, considerably lower than the
VLBI--derived value of $130^{+14}_{-12}$\,\kms.  Since the scintillation speed was underestimated, 
despite the fact that the distance to the pulsar was overestimated, a possible explanation is that the 
scattering screen resides closer to the solar system than the pulsar. 
Alternatively, the scintillation measurement could have been 
affected by refractive scintillation, or the thin--screen, isotropic turbulence model may not be applicable
in this case.

No detections of \pseven\ in optical or x--ray wavebands have been published.  With a 
Shklovskii--corrected characteristic
age of $336\times10^{6}$ years, it is even older than \pone, and might be expected to possess
similarly high conversion efficiency from spin--down luminosity to optical and x--ray luminosity.
However, even though it is considerably closer than \pone, the fact that its spin--down luminosity
is a factor of almost 50 smaller means that it would be challenging to detect in wavebands
other than radio.   A simple scaling of the optical observations of \pone\ to account
for the different distance and spin--down luminosity of \pseven\ reveals that
the U band optical magnitude would be 27.7.  which would be detectable with large ground--based
telescopes, although challenging.  The x--ray luminosity, assuming the 1.7\% conversion efficiency
seen in \pone, would be $5.5\times10^{26}$ erg s$^{-1}$.  Scaling the count rate seen by
\citet{pavlov08a} on \pone\ by the relative distance and luminosity of \pseven\ compared 
to \pone\ yields the extremely low
photon count rate of 0.08 counts ks$^{-1}$, and thus x-ray detection of \pseven\ is unlikely without
an extremely long exposure.

\section[Analysis of distance and velocity models]{Analysis of distance and velocity models}
\label{results:distvel}
Galactic electron distribution models are extremely important for all fields of pulsar research,
since their prediction of pulsar distances based on $DM$\ has the potential to bias estimates
of luminosity, velocity, and various timing terms for the vast majority of pulsars without an
independent distance constraint.  Hence, continual improvement of these models, and characterisation
of errors in the existing models (and the impact of these errors) is a crucial task.

Early attempts to construct simple Galactic electron models for the purpose of estimating pulsar
distances were made by \citet{manchester81a} and \citet{lyne85a}, but these were superseded by
the more comprehensive TC93 model of \citet{taylor93a}.  While the much more complex NE2001
model of \citet{cordes02a} is now available, distance estimates based on the TC93 model are still
in common usage and both models are considered below.

The six new (and one refined) pulsar parallaxes measured in this thesis make a substantial 
addition to the 29 published (18 VLBI and 11 timing) pulsar parallaxes\footnote{
Only the most accurate parallax measurement has been considered 
where multiple measurements exist, and
measurements less significant than 1.5$\sigma$ have been excluded} 
and hence a review of the accuracies of the TC93 and NE2001 models is timely.  
It is appropriate to note at the outset of this analysis  
that large distance errors may be over--represented in existing astrometric results. 
One reason is the selection effect of anomalous pulsars being chosen for study, which
is certainly the case for the \pthree\ system observed during this thesis.
Another potential factor is that most astrometric distance determinations to date have been for 
relatively nearby pulsars, where there is less opportunity for underdensities and overdensities 
in the ISM along the line of sight to cancel, and hence it is plausible that distance models are 
on average more reliable for more distant pulsars.

\begin{figure}
\begin{center}
\includegraphics[width=0.9\textwidth]{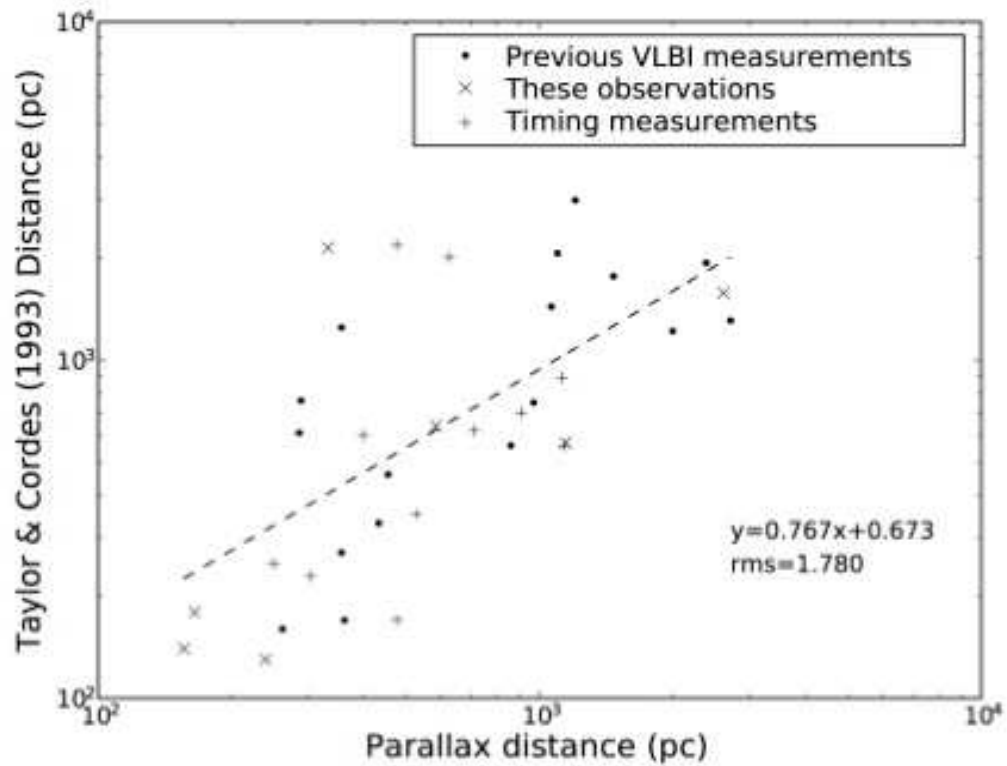}
\caption[Pulsar parallax distance versus TC93 distance]
{Parallax distance versus TC93 distance for all pulsars with a published
parallax.  The line of best fit was estimated from a Monte Carlo simulation of the true pulsar
distances based on parallax measurements and errors, averaging the results obtained from
standard linear regression. The TC 93 model is particularly uncertain for nearby pulsars,
with underestimates and overestimates by factors of up to six.}
\label{fig:pxdist_vs_tc93dist}
\end{center}
\end{figure}

\begin{figure}
\begin{center}
\includegraphics[width=0.9\textwidth]{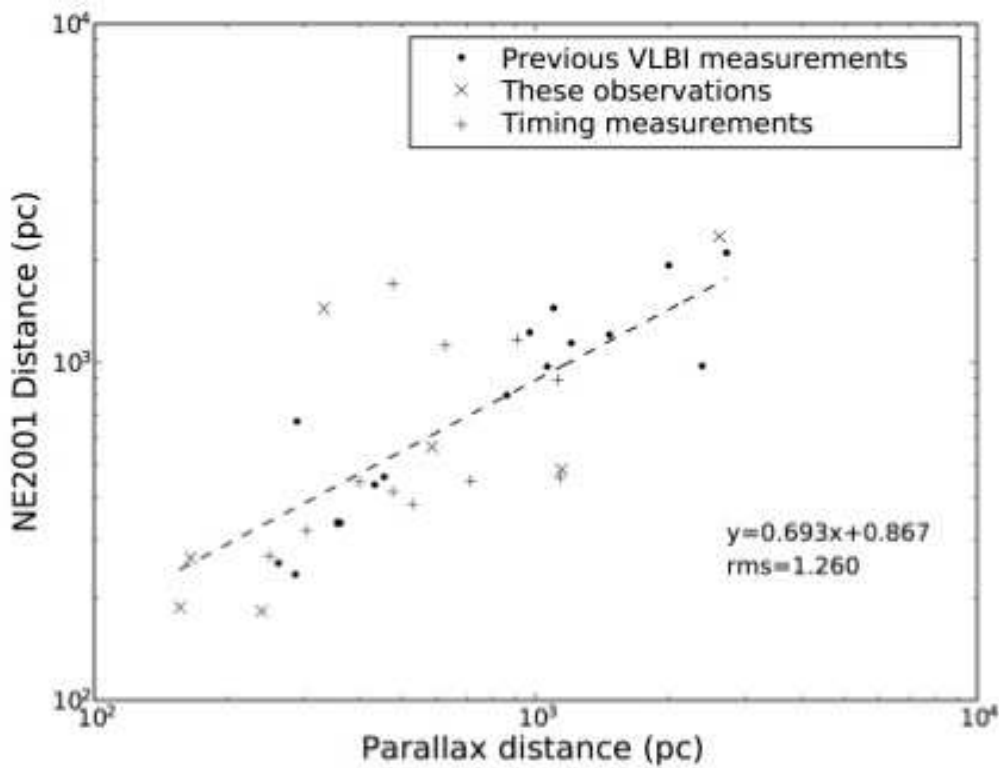}
\caption[Pulsar parallax distance versus NE2001 distance]
{Parallax distance versus NE2001 distance for all pulsars with a published
parallax.  As with Figure~\ref{fig:pxdist_vs_tc93dist}, the line of best fit was estimated using 
linear regression in a Monte Carlo simulation to account for parallax measurement errors.
The RMS error is substantially reduced compared to the TC93 predictions, but as some pulsar
distances had been measured prior to 2001 and were used to constrain the NE2001 model, 
part of the observed improvement must be attributed to this knowledge.}
\label{fig:pxdist_vs_ne2001dist}
\end{center}
\end{figure}

Figure~\ref{fig:pxdist_vs_tc93dist} plots the TC93 
distance for all 35 pulsars against the parallax distance, along with a line of best fit.
Figure~\ref{fig:pxdist_vs_ne2001dist} repeats this plot for NE2001 distances. The best fit line was
generated by averaging the linear regression results obtained from a 10,000 trial Monte Carlo 
simulation, where ``actual" pulsar distances were calculated for each trial based on the observed
parallax and parallax error.  Throughout this section, errors will be referred to in decibels (dB), which
is the most convenient representation for measurements where large underestimates or 
overestimates are
common.  The error in dB is defined as 
$10\log_{10}{\frac{\mathrm{model\ distance}}{\mathrm{parallax\ distance}}}$,
and hence positive values represent overestimates, and negative values represent underestimates.

\begin{figure}
\begin{center}
\includegraphics[width=0.9\textwidth]{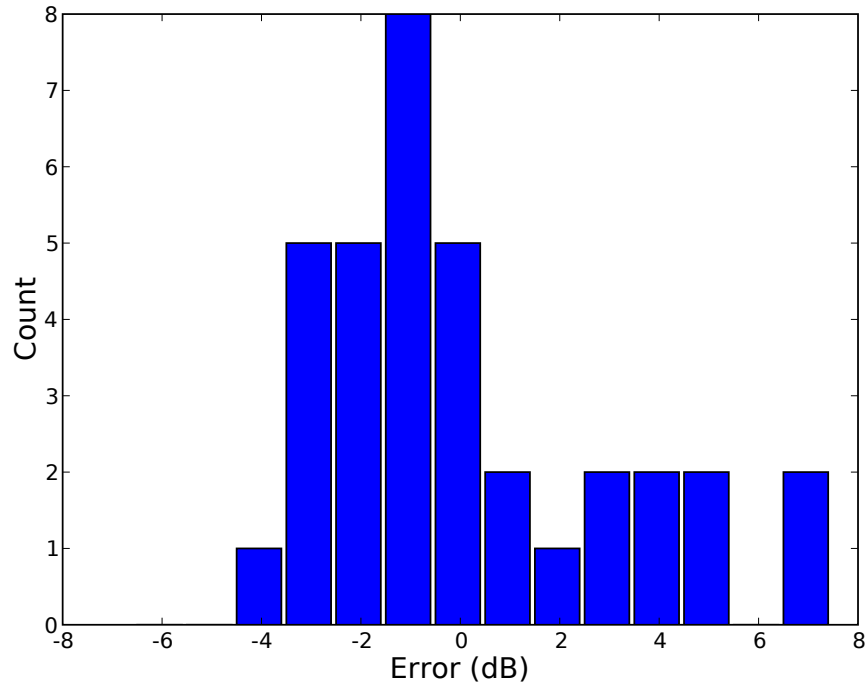}
\caption[Histogram of TC93 errors for pulsars with measured parallaxes]
{Histogram of TC93 errors for pulsars with measured parallaxes, with errors binned in 1 dB increments.
Underestimates of distance are more common in this model, but the largest errors are made when the
distance is overestimated.  The standard deviation of the errors is approximately 3 dB.}
\label{fig:tc93hist}
\end{center}
\end{figure}

\begin{figure}
\begin{center}
\includegraphics[width=0.9\textwidth]{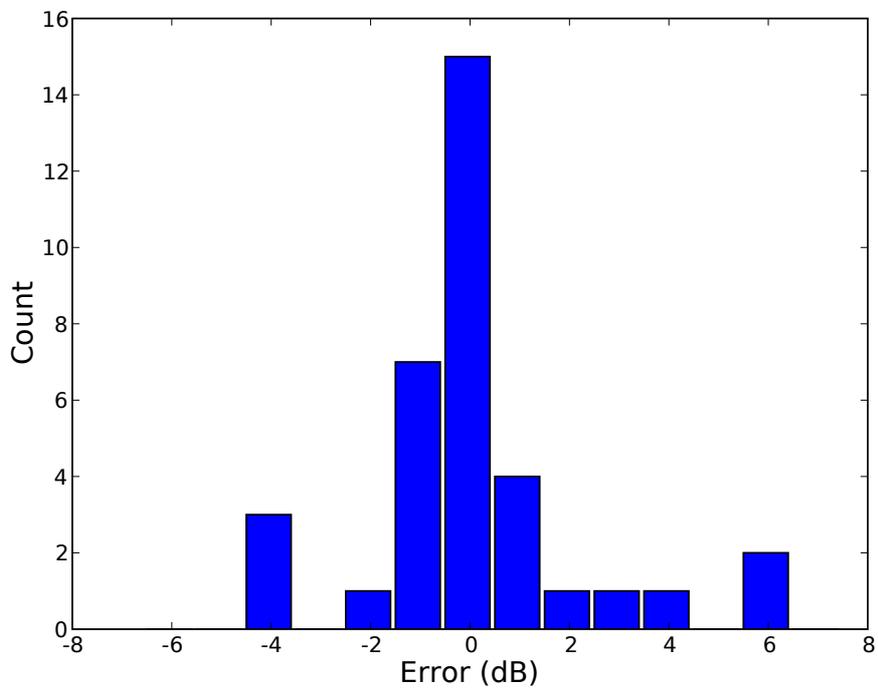}
\caption[Histogram of NE2001 errors for pulsars with measured parallaxes]
{Histogram of NE2001 errors for pulsars with measured parallaxes, with errors binned in 1 dB 
increments.  Unlike the TC93 model, there are no discernable systematic errors, but the distribution
is still clearly non--Gaussian, with a long tail of errors up to 6 dB.  The occurrence of errors with a 
magnitude greater than 3 dB has been somewhat reduced from the TC93 model 
(five, compared to eight for TC93).  
The standard deviation of the errors is approximately 2 dB, compared to the 3 dB seen for TC93.}
\label{fig:ne2001hist}
\end{center}
\end{figure}

Figures~\ref{fig:pxdist_vs_tc93dist} and \ref{fig:pxdist_vs_ne2001dist} show that the NE2001 model
has considerably improved upon the TC93 model, but significant errors are still made for
individual pulsars.  This can be highlighted more clearly by binning the errors, as shown in
Figure~\ref{fig:tc93hist} for the TC93 model and Figure~\ref{fig:ne2001hist} for the NE2001 model.
Figure~\ref{fig:tc93hist} shows a clear systematic offset in the median error of the TC93 model 
towards underestimated
distances, but the largest errors are seen when distances are overestimated.
Figure~\ref{fig:ne2001hist} shows that systematic bias has been removed by the NE2001 model,
but the distribution of errors still cannot be well approximated by a single Gaussian, with long wings
towards large errors. 
For both the TC93 and NE2001 models, the largest errors are seen when 
the distance is overestimated.  The effect of these high--error ``outliers" on population studies
of neutron stars is considered further in Section~\ref{results:distvel:errors}.

\subsection[Analysis of newly--measured pulsars]{Analysis of newly--measured pulsars}
\label{results:distvel:new}
Table~\ref{tab:distvel} shows the measured distances and velocities for the pulsars in this
survey, along with previous estimates of their distance from the TC93 and NE2001 electron
models, and (where available) estimates of their speed from scintillation studies.  \ptwo\ has
been excluded from the analysis of $DM$\ distance reliability below, since it contributed to 
the NE2001 model.  Since no parallax was detected for \peight, the VLBI velocity is shown 
assuming the NE2001 distance of 566 pc, but given the consequent uncertainty 
it is not included in the error analysis of scintillation velocity estimates.  Naturally, \peight\ cannot 
contribute to the estimates of $DM$\ distance reliability.

\begin{deluxetable}{lrrrrr}
\tabletypesize{\tiny}
\tablecaption{Comparison of distance and velocity to non VLBI--derived estimates}
\tablewidth{0pt}
\tablehead{
\colhead{Pulsar} & \colhead{VLBI distance} & \colhead{VLBI velocity} & 
\colhead{TC93 distance} & \colhead{NE2001 distance} & 
\colhead{Scintillation speed} \\
& \colhead{(pc)} & \colhead{(\kms)} & \colhead{(pc)} & \colhead{(pc)} & \colhead{(\kms)}
}
\startdata
\pone	& $240^{+124}_{-61}$	& $194^{+104}_{-51}$	& 130	& 184	& -- \\
\ptwo	& $156.3^{+1.3}_{-1.3}$	& $104.7^{+1.0}_{-1.0}$	& 140	& 189	
		& 170\tablenotemark{A}, \ 231\tablenotemark{B}\\
\pthree	& $332^{+52}_{-40}$	& $80^{+15}_{-11}$		& 2150	& 1444	
		& $60\pm13$\tablenotemark{C}, \ $170\pm15$\tablenotemark{D} \\
\pfour	& $1150^{+220}_{-160}$	& $24^{+9}_{-6}$		& 570	& 483	
		& $140.9\pm6.2$\tablenotemark{E}, \ $66\pm15$\tablenotemark{F} \\
\pfive	& $2600^{+690}_{-450}$	& $163^{+44}_{-29}$	& 1580	& 2384	
		& 400\tablenotemark{A} \\
\psix		& $580^{+660}_{-200}$	& $317^{+362}_{-111}$	& 640	& 562	
		&  $501 \pm 29$\tablenotemark{D} \\
\pseven	& $165^{+17}_{-14}$	& $130^{+14}_{-12}$	& 180	& 264	
		& 48\tablenotemark{A} \\
\peight	& --					& $46^{+6}_{-6}$		& 500	& 566	
		& $31 \pm 25$\tablenotemark{G}, \ 51\tablenotemark{A}, \ 113\tablenotemark{B} \\
\enddata
\tablenotetext{A}{\citet{johnston98a}, using TC93 distances, except for \pfive\ where 2 kpc 
(HI lower limit) was used and \pseven\ where 330 pc (unknown source) was used}
\tablenotetext{B}{\citet{gothoskar00a}, using TC93 distances}
\tablenotetext{C}{\citet{cordes86b}, using a distance of 1240 pc}
\tablenotetext{D}{\citet{bhat99a}, using TC93 distances}
\tablenotetext{E}{\citet{ransom04a}, using a distance of 600 pc}
\tablenotetext{F}{\citet{coles05a}, using a distance of 500 pc}
\tablenotetext{G}{\citet{nicastro95a}, using the TC93 distance}
\label{tab:distvel}
\end{deluxetable}

Table~\ref{tab:distvel} shows that for this admittedly limited sample
size, the NE2001 model is only slightly improved from the TC93 model, with a 
mean distance error of 2.1 dB compared
to 2.5 dB.  This is smaller than is seen for the entire population, where the NE2001 model shows a 1 dB
improvement over the TC93 model.  This may be partly due to the presence of \pthree, which 
dominates the error budget for both the TC93 and NE2001 models, as
the models overestimated the distance to this pulsar by a factor
of 6.5 and 4.3 respectively.  Clearly, for individual pulsars, $DM$ distances alone 
cannot be relied upon to provide luminosity, velocity or space density estimates.

Table~\ref{tab:distvel} shows that scintillation speed estimates are also generally unreliable for any
given system -- of the nine estimates available, only two are consistent with the VLBI--derived
velocities, and one of those (\psix) is likely to also be inconsistent, but cannot presently be ruled
out due to the low significance of the parallax detection.  In some cases (such as \pthree) the incorrect 
distance estimate to the pulsar was the main reason for the scintillation speed estimate errors, but
for most cases the $DM$\ distance error did not significantly contribute to the scintillation speed error.
The mean absolute error is 3.8 dB, with the largest error
(on \pfour) being a factor of 7.  Amongst this limited sample, scintillation speeds have consistently
overestimated the true transverse speed, on average by a factor of 1.75.  This is in contrast
to previous authors such as \citet{gupta95a}, who found that scintillation speeds tended to
underestimate the proper motion speed, at least for pulsars with a large Galactic height.
These findings are little changed when the scintillation speed estimates are corrected by 
the VLBI distance measurements -- the mean overestimate of speed becomes a factor of 1.5,
and the mean absolute error is 4.1 dB.

Most previous studies of scintillation speed have assumed a single, thin, turbulent scattering screen
midway between the pulsar and the Solar System.  The turbulence within the screen is generally 
assumed to be isotropic.  Recent results from scintillation studies
of \pfour\ (W. Coles et al., in preparation) and PSR B0834+06 (W. Brisken et al., in preparation)
suggest that anisotropy of ISM turbulence is the norm, rather than the exception.  When 
coupled with the uncertainty of the scattering screen location, this suggests that 
errors in scintillation speeds have typically been underestimated in the past.  Whilst the scintillation
speed method remains a valid approach for obtaining {\em population} velocity measurements
(subject to the same caveats highlighted in Section~\ref{results:distvel:errors}),
the natural conclusion is that too many uncertainties exist for the method to be useful for {\em 
individual} objects, at least when no information on the location and structure of the 
scattering material is available.

\subsection[Impact of $DM$\ distance errors]{Impact of $DM$\ distance errors}
\label{results:distvel:errors}
It has already been shown that although there is no evidence for systematic bias 
in the NE2001 distance model, errors exceeding 4 dB still exist for individual pulsars. Such 
errors can impact upon population analysis of pulsars, artificially creating tails of extremely high
or low values of distance--dependent parameters such as luminosity and velocity.  This can
affect estimates of the mean values of these quantities, as well
as confusing attempts to explain the underlying physics by generating false outliers.  As an example,
the case of neutron star transverse velocities is considered below.

%

\begin{figure}
\begin{center}
\includegraphics[width=0.9\textwidth]{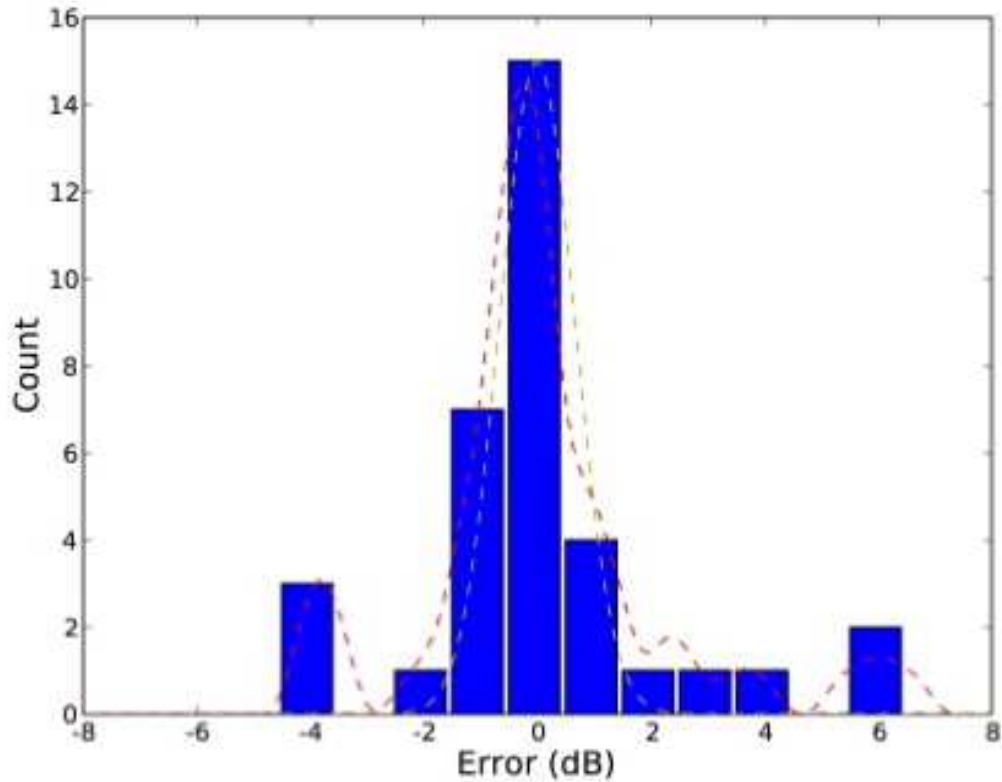}
\caption[Histogram of NE2001 errors, with models of error distribution]
{Histogram of NE2001 errors for pulsars with measured parallaxes, with error binned in 1 dB 
increments.  The ``binned" error model described in the text is shown as a dashed red line, while
the single Gaussian ``traditional" model which approximates the 20\% errors commonly assumed 
for $DM$ distances is shown as a yellow dashed line.}
\end{center}
\label{fig:ne2001histmodel}
\end{figure}

In order to model the effect of distance errors on transverse velocity estimates, it is necessary
to estimate the form of the error distribution shown in Figure~\ref{fig:ne2001hist}.
Since the small number of samples makes an estimation of the true distribution difficult, the error
distribution has instead been modeled by binning the errors in increments of 0.1 dB and smoothing
with a 2 dB Hanning window.  This is shown overlaid on the original binned error distribution in 
Figure~\ref{fig:ne2001histmodel}, and is hereafter referred to as the binned error model. 
As already noted in Section~\ref{results:distvel}, large errors are likely to be somewhat 
over--represented due to selection effects, and thus this model can safely be considered to be
a ``worst--case" representation of $DM$\ distance reliability.
Figure~\ref{fig:ne2001histmodel} also shows a single Gaussian
component model for the error distribution, with a standard deviation of 0.66 dB.  In this Gaussian
distribution, 75\% of errors are smaller than a factor of 1.2, and hence
it approximates the 20\% errors typically assumed for $DM$\ models.  This
distribution is hereafter referred to as the traditional error model.

\citet{hobbs05a} examine a large sample of pulsar proper motions and conclude that the distribution 
of 3D space velocities of young pulsars (characteristic age $< 3$\,Myr) is well fit by a Maxwellian
distribution with mean 431 \kms\ and one--dimensional RMS 265 \kms.  This velocity distribution
is used as the starting point for the simulations that follow.  

\begin{figure}
\begin{center}
\includegraphics[width=0.9\textwidth]{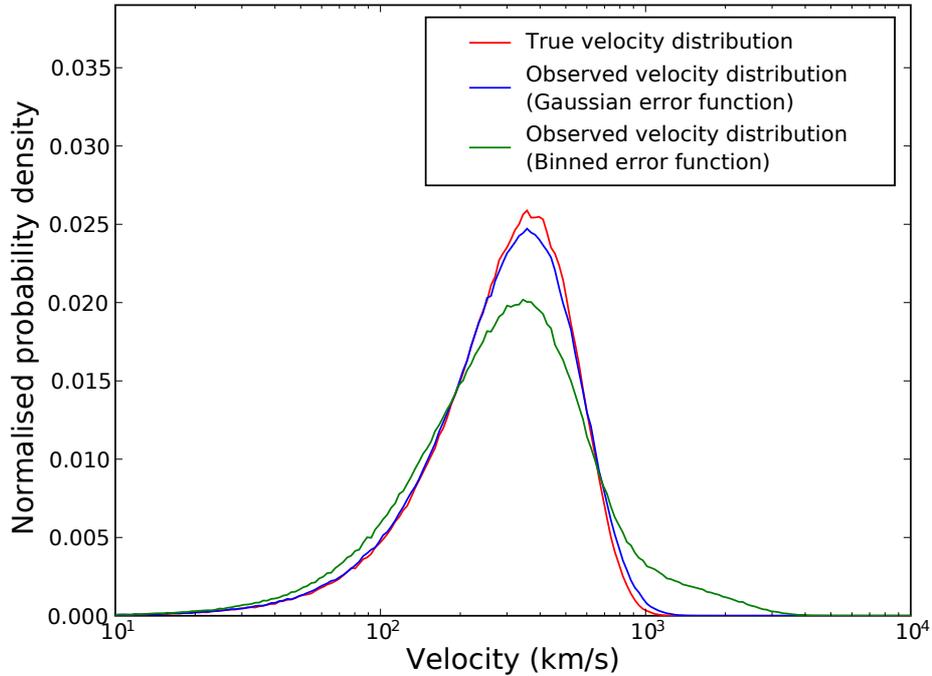}
\caption[Synthetic  2D velocity distribution]
{2D velocities for a synthetic pulsar population, as observed with no distance error (red), the
traditional distance error model (blue) and the
binned distance error model described in the text (green).  The effect
of the binned distance error model is to broaden the observed velocity distribution, 
particularly at large velocities.}
\label{fig:disterror_velocity}
\end{center}
\end{figure}

A Monte Carlo simulation was performed, creating a synthetic pulsar catalogue of 1 million
pulsars, where each pulsar's actual velocity was drawn from the 
distribution described above.  These are referred to as the unperturbed 
velocities.   The velocities of each
pulsar were then perturbed according to the binned error function and the traditional
error function described above, and the 
observed 2D velocity was recorded for each case.  
The observed 2D velocity distribution for the binned error function and the traditional error function,
along with the unperturbed 2D velocity distribution, is plotted in Figure~\ref{fig:disterror_velocity}.

It is immediately apparent that the broadening of the observed velocity distribution is
considerably greater for the binned error model, compared to the traditional error model, which  
is particularly pronounced at the high end of the velocity distribution.
The mean unperturbed 2D velocity is 332 \kms, which is increased by only 1\% to 336\,\kms\ in
the traditional error model, but by 18\% to 392 \kms\ in the binned error model.
This increase exceeds the $\pm40$\,\kms\ errors quoted for the mean 2D velocities of young
pulsars in \citet{hobbs05a}.

Even more important than the mean effect on pulsar velocities, however, is the effect on the high end 
of the pulsar velocity distribution.  \citet{hobbs05a} note that distance model inaccuracies could
be responsible for the sparse tail of very high velocity pulsars, but conclude that with the small
sample size available, the observed high--velocity pulsars are consistent with the tail of 
continuous velocity distribution.  
In considering the impact of distance model errors, it is useful to examine the frequency
of occurrence of pulsars with observed transverse speeds greater than 1000\,\kms\ in the synthetic
catalogue.  These pulsars will be described as very high velocity (VHV) pulsars.  The 
synthetic catalogue contains only one VHV pulsar in every 1000 in the unperturbed velocity 
distribution, but the occurrence rises to three in 1000 for the traditional error model, and 50 in 1000 
for the binned error model.


Thus, distance model errors can add a significant high--velocity tail to the observed pulsar 
velocity distribution, an effect which is significantly enhanced when the distance errors ``outliers"
are accounted for.  These results suggest that the mean pulsar 3D space velocity may have been 
previously overestimated, and that the majority of VHV pulsars may in fact simply be 
pulsars with incorrect distance estimates. As already stated, the binned error model used here is likely
to overestimate the true frequency of large errors, and hence the true impact of $DM$\ distance
errors is likely to lie somewhere in between the binned error model and the traditional error model.  
Additionally, there is clear evidence that at least some VHV pulsars are present in the Galaxy, with
\citet{chatterjee05a} using VLBA astrometry to show that the transverse velocity of PSR B1508+55
exceeds 1000\,\kms, and observations of PSR B2224+65 showing scintillation, proper motion and
bow shock results consistent with a velocity close to 
1000 \kms\ \citep{cordes86b,harrison93a,cordes93a}. Nonetheless,
the binned error model is useful in illustrating that the largest impact of occasional, rare distance errors
is at the high end of the pulsar velocity distribution.
A larger, unbiased sample of pulsar
distance measurements would enable a more accurate estimation of the true distance error
function for NE2001, which in turn would enable the effects of distance errors to be ``deconvolved"
from the measured pulsar velocity distribution with some confidence.

\section[Systematic errors and astrometric accuracy]{Systematic errors and astrometric accuracy}
\label{results:sys_astro}
Previous astrometric VLBI programs such as those carried out by \citet{chatterjee04a} have 
attempted to estimate the effect of various observing factors on the resultant astrometric accuracy.   
Two of the most important factors are the brightness of the pulsar, which determines the accuracy
to which a centroid position can be estimated within the synthesised beam, and the
angular separation of the pulsar from the calibrator (the ``calibrator throw"), which
determines the extent to which atmospheric and ionospheric gradients contribute
to systematic errors.  Broadly, these could be considered as dominating the thermal (random)
error component, and the systematic (non--random) error component to a position measurement.
It is appropriate to note at the outset of this analysis that the small number of measured parallaxes
and the potential for selection biases (for example, some ``easier" pulsars have been targeted by
earlier surveys with less sensitivity, leading to larger errors compared to what might be expected) 
mean the trends presented here should be interpreted with caution.

\begin{figure}
\begin{center}
\includegraphics[width=0.9\textwidth]{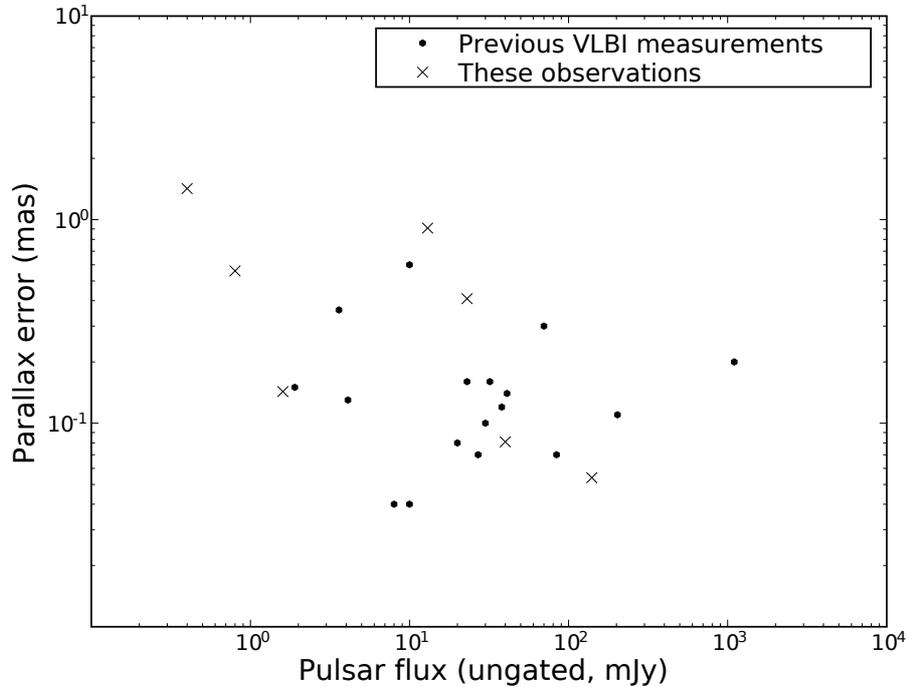}
\caption[VLBI parallax error plotted against pulsar flux density]
{Parallax error plotted against ungated pulsar flux density (mJy, 1400 MHz) for all pulsars with a 
published VLBI parallax.  Whilst stronger pulsars typically have smaller errors,  there is
a large scatter.  The three faintest pulsars with measured parallaxes to date were part of
this observing program.}
\label{fig:pxe_vs_flux}
\end{center}
\end{figure}

Figure~\ref{fig:pxe_vs_flux} shows the ungated pulsar flux density (at 1400 MHz) plotted against
parallax errors for all pulsars with a published VLBI parallax.  No attempt was made to account for 
the effect of varying calibrator throw.  Where several parallax
measurements have been made, the most accurate is used. As expected, considerable 
variation is seen, but there is a general trend towards lower parallax errors for brighter pulsars.

\begin{figure}[t!]
\begin{center}
\includegraphics[width=0.9\textwidth]{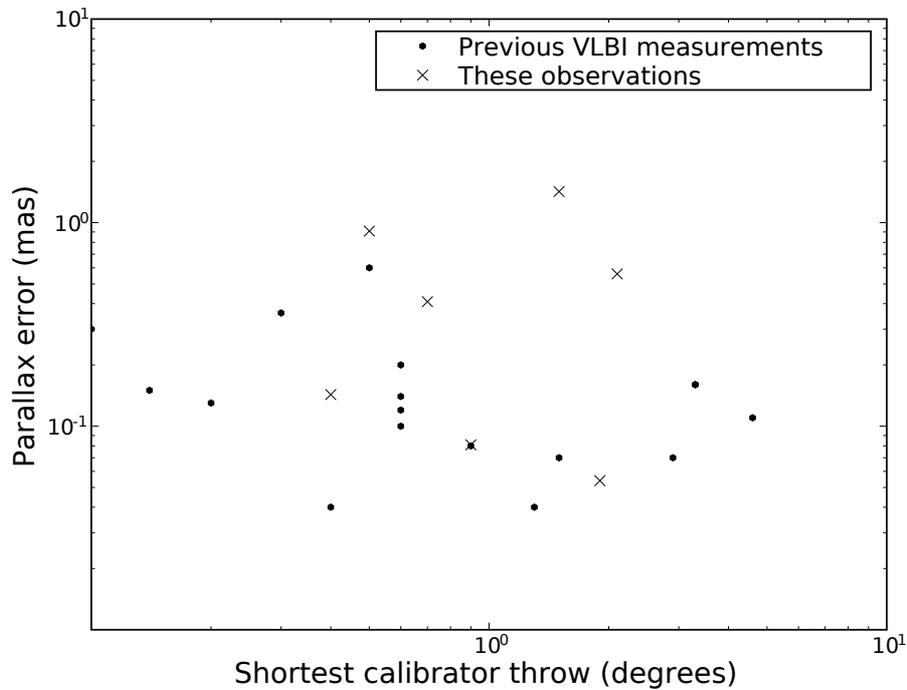}
\caption[VLBI parallax error plotted against calibrator throw in degrees]
{Parallax error plotted against calibrator throw in degrees, for all pulsars with a published
VLBI parallax.  Little trend is apparent with decreasing calibrator throw for both this work and
previous surveys, which is surprising
given the assumed dependence of systematic errors on calibrator throw.}
\label{fig:pxe_vs_calthrow}
\end{center}
\end{figure}

Figure~\ref{fig:pxe_vs_calthrow} shows the parallax errors plotted against calibrator throw
for all pulsars with a published VLBI parallax.   No attempt was made to correct for pulsar flux.  
As with Figure~\ref{fig:pxe_vs_flux}, the most 
accurate measurement was taken when  multiple measurements where available.
Where multiple calibrators were used, the
shortest calibrator throw was taken (including in--beam calibrators when used).  Surprisingly
little trend is seen with decreasing calibrator throw, which differs from predominantly linear 
decrease seen in the results of \citet{chatterjee04a}, and the simulation predictions of \citet{pradel06a}.

As previously stated, the sample used here consists of measurements made 
with many different VLBI arrays, at different recording rates and hence sensitivities.  This approach
differs from previous studies such as \citet{chatterjee04a}, who considered single objects with 
different calibrators, or multiple objects with different calibrator throws observed with the same array.  
The in--beam calibrators used also vary considerably in brightness, 
which may limit the attainable precision to less than the expected value
from the spatial interpolation alone.   The variation of pulsar flux amongst the different target objects, 
however, is likely to be the most important factor obscuring the expected correlation of
parallax error with calibrator throw.  This would imply that the majority of pulsar astrometry programs
to date have been sensitivity--limited, rather than being limited by systematic errors.

Ideally, a joint analysis would be made of pulsar flux and calibrator throw on
the parallax error, but the current sample size is too small for such an analysis to be meaningful.
More tightly controlled and larger samples of parallax measurements will become available in future
years, enabling a better analysis of the relative contribution of calibrator throw to parallax error.  

The pulsars observed in this thesis varied widely in flux and calibrator throw, and so different 
limiting factors are seen for the different targets. As shown in Figure~\ref{fig:pxe_vs_flux}, the three
faintest pulsars with measured VLBI parallaxes were observed in this thesis, and the results for
these three are partially (\pseven) or predominantly (\pone, \pfour) sensitivity--limited.  
\peight, which was only detected in two epochs, was also severely sensitivity limited.  
Thus, half of the target sample was dominated by thermal,
rather than systematic errors.  For the remaining four pulsars 
(\ptwo, \pthree, \pfive\ and \psix) systematic errors dominated thermal errors
and equally weighted
visibilities were used to minimise the impact of systematic errors on the fitted positions, as discussed
in Section~\ref{techniques:check:weights}\footnote{\pseven, which was moderately 
sensitivity--limited, used equally weighted visibilities, but used natural weighting for imaging to
maximise sensitivity, rather than the uniform weighting used for the systematic--dominated pulsars}.

The systematic error budget is likely to be dominated by residual ionospheric errors, since as noted in Section~\ref{techniques:datared:geoiono}, the relatively low density of GPS 
receivers in the Southern Hemisphere makes the GPS--based ionospheric correction less 
accurate for the LBA than for the VLBA or EVN.  As shown in Section~\ref{techniques:check:tecor}, 
the median position shift due to the application of ionospheric correction is approximately 3 mas
for \pfive.  Since the accuracy of GPS--derived TEC maps ranges from 3--10\% in the Northern
Hemisphere \citep{sekido04a}, the accuracy in the Southern Hemisphere is presumably 
$\sim\,$10\% or worse, and hence systematic errors due to residual ionospheric effects
would be expected to be hundreds of $\mu$as for pulsars observed at 1600 MHz, comparable
to the observed systematic error estimates.  As discussed
in Section~\ref{results:binary:0437}, the application of ionospheric corrections made no significant
difference for the 8 GHz observations, and residual tropospheric errors are expected to dominate
in this case.

\clearpage
\subsection[The future of pulsar astrometry]{The future of pulsar astrometry}
\label{results:sys_astro:future}
Over the next decade, the construction of new radiotelescopes and the upgrading of existing
interferometers to larger bandwidths will lead to a revolution in astrometry at radio wavelengths.
The most ambitious, and most distant, project is the Square Kilometre Array 
(SKA\footnote{http://www.skatelescope.org/}), which when completed will combine 
nanoJansky sensitivity with excellent spatial resolution\footnote{in its current specification, 
which includes baselines to 3000 km} of $\sim$1--10 mas (in the high frequency band).
LOFAR, which has already been discussed in 
Section~\ref{results:isolated:2144}, should be completed much earlier than the SKA and will
offer excellent sensitivity at low frequencies, providing astrometric capabilities for faint, nearby
objects, especially steep--spectrum pulsars. Given its low observing frequencies, LOFAR will
suffer from relatively low resolution and increased 
ionospheric effects.  For the faintest steep--spectrum objects, however, LOFAR will enable 
astrometry that would not otherwise have been possible.

In addition to these new facilities, large--scale bandwidth upgrades to 
the VLA\footnote{http://www.vla.nrao.edu/}, MERLIN\footnote{http://www.merlin.ac.uk/}, and 
VLBA\footnote{http://www.vlba.nrao.edu/} intruments are bringing
$\mu$Jy sensitivity to instruments with baselines of tens, hundreds and thousands of kilometres 
respectively.  Due to its shorter baselines, the VLA is unlikely to be able to compete
in terms of astrometric accuracy generally, but may be useful for high frequency observations
of flat--spectrum sources.

The increased sensitivity of these instruments will allow fainter pulsars to be targeted, but will 
not improve astrometric accuracy for brighter pulsars beyond what can presently be obtained
unless improved calibration methods can be applied.  As shown in 
Section~\ref{results:binary:0437} for \ptwo, systematic errors already dominate
the error budget with existing instrumentation for the brightest pulsars, even with the comparatively 
modest sensitivity of the existing LBA, and at frequencies where the influence of the ionosphere
are minimal.

One approach which has already been successfully 
demonstrated is the use of wide bandwidths to measure the quadratically
dependent (upon frequency) component of visibility phase induced by the ionosphere, using 
self--calibration on the target source \citep{brisken02a}.  Whilst this technique is likely to be
particularly valuable for pulsar astrometry with future instruments, it cannot account
for tropospheric errors and will still not be feasible for the faintest pulsars.  Thus, new methods
of calibration will be required. 

Ideally, calibration will minimise the error in the estimated path delay and complex gain in the direction
of the target source at all times, across the entire observing bandwidth.   This  
removal of the temporal and spatial interpolation which has traditionally been required in VLBI
phase--referencing experiments is possible using a technique which is commonly known as 
``field calibration", which is analogous to the technique of adaptive optics in optical astronomy.
Whilst the coherence time is much longer at radio wavelengths than optical, the density of 
strong radio sources is much lower than optical sources, meaning that this technique could not
be applied to the narrow--field, low sensitivity VLBI arrays which have been available to date.
The advent of wide--bandwidth systems will provide the sensitivity required to make this 
technique possible on the planned and upgraded facilities mentioned above.
\citet{bhatnagar08a} give an overview of the algorithms used for field calibration.

Thus, for astrometric purposes, the availability of more calibrator sources is potentially the largest
advantage of the improved sensitivity of new and upgraded instruments.  Of the instruments listed 
above, the upgraded MERLIN (eMERLIN) is unique in that it is a heterogeneous array -- all of
the other arrays are composed of identical elements. The presence of several large telescopes
in the eMERLIN array will make the field calibration described above more difficult, since the 
field of view of the array is limited by these large telescopes.  Thus, it will be difficult for eMERLIN
to reach the limiting accuracies that should be possible with the upgraded VLBA.

Since field--based calibration routines have not yet been tested in 
detail, it is difficult to predict the limiting accuracy that can be attained.  
Ultimately, the ability to model the (time--variable) structure of each source
used as an in--beam calibrator is likely to determine the precision achieved.
\citet{fomalont04a} estimate that the SKA can obtain single--epoch
accuracies of several $\mu$as for mJy targets, meaning sub--$\mu$as parallax
accuracies should be attainable.  However, as noted in \citet{fomalont04a}, the
final specification of the SKA instrument, in particular the amount of collecting area
devoted to long baseline stations, could significantly impact these estimates.  The upgraded
VLBA, unlike the SKA, will still be sensitivity limited for mJy sources -- nevertheless, 
$\mu$as parallax accuracies should be attainable, at least for some sources.

The accuracies for these radio interferometers can be contrasted with the next generation of
optical instrumentation.  The dedicated orbiting astrometric telescope 
GAIA\footnote{www.esa.int/science/gaia},
scheduled to launch in 2011, will derive accurate positions for over a billion stars.
The astrometric accuracy will depend on target brightness, but is expected to range from $\sim$7 
$\mu$as for bright stars, to $\sim25\,\mu$as for 15th magnitude stars, and sub--mas parallaxes for
20th magnitude stars \citep{lindegren08a}.  This will allow significant distance determination 
for these stars at distances as great as the Galactic centre.  
Clearly, GAIA will allow the measurement of a much greater number of
sources than will be possible with VLBI, even in the SKA era.   However, the most accurate 
determination of individual objects is almost certain to still be achieved using VLBI.

Since GAIA will be limited to objects much brighter than pulsars, it will not be able to contribute to 
pulsar astrometry specifically except for binary pulsars with an optically visible companion,
such as PSR B1259-63 \citep{johnston94a}.   While many pulsar companions are faint white 
dwarf stars that would be challenging targets for optical astrometry, the detection of even a few
of the brightest optical pulsar companions would enable an improved frame tie between the optical
reference frame and the ICRF.  Since the frame tie between the Solar System frame and the ICRF is
already good, thanks to comparisons of VLBI and timing astrometry, this would also serve to enhance the compatibility of the optical and Solar System frames.
\chapter{CONCLUSIONS}
\label{conclusions}
As described in Chapter~1, this thesis aimed to develop the tools necessary to enable
high--sensitivity pulsar observations with the Australian Long Baseline Array, allowing the
largest pulsar astrometry program in the Southern Hemisphere to date to be undertaken.
These tools included a general purpose software correlator (DiFX)
and a flexible astrometric data reduction pipeline which implements several novel
techniques for pulsar VLBI astrometry.  DiFX has been extensively
tested against three existing hardware correlators and is now used full--time
for the LBA National Facility; it is also presently being adopted by other VLBI arrays, including the
VLBA.  The astrometry project has been successfully concluded
with the measurement of seven pulsar parallaxes, more than tripling the number of Southern
Hemisphere pulsars with a VLBI--measured parallax.  Thus, the twin goals set at the outset of 
the thesis have been satisfactorily accomplished.  The contributions of this thesis to, and possible
future directions of, the main thesis areas of VLBI instrumentation and VLBI pulsar astrometry 
are considered in more detail below.

\section[VLBI instrumentation]{VLBI instrumentation}
The DiFX software correlator developed during this thesis continues to expand in capability
and scope, and is expected to become the primary correlator for a minimum of two additional 
arrays (the VLBA and the new Australian/NZ geodetic array funded under AuScope) during 
2009.  Unique experiments such as extremely high spectral resolution pulsar scintillation 
studies, using high sensitivity VLBI, have been correlated using DiFX.  These experiments
would not have been feasible using existing correlator infrastructure.  The first eVLBI observations
within Australia, and the first Australian 
eVLBI observations incorporating Asian antennas, have been
made using DiFX.  As more telescopes are connected to high--speed networks, eVLBI
offers the possibility of dramatically reducing operating costs for VLBI networks, by reducing costs
associated with maintaining and shipping large quantities of physical storage.

For the LBA, the bandwidth upgrade enabled by DiFX has resulted in a significant
increase in the number of proposals to use the instrument.  Other substantial improvements
have been made to LBA functionality during this thesis,
such as reductions in the uncertainty of antenna positions and refinement of the
correlator geometric model.  These improvements are reflected by the number of 
ongoing astrometric LBA programs, and the number of current astrometric LBA proposals.

Future developments in DiFX will see the addition of capabilities such as single pass correlation
of multiple phase centres, adaptation to new recording and output formats and increased 
ease of use.  These features, coupled with other instrumental improvements such as 
wide--bandwidth digital backends, will allow existing VLBI arrays such as the LBA and VLBA
to begin to conduct SKA--style wide--area, high angular resolution surveys at high sensitivity.
As well as dramatically improving knowledge of the radio sky at high resolution, this will 
help pave the way for SKA observations through development of wide--field imaging techniques
and calibration.

As the power and efficiency of digital electronics continue to grow in the future,
an ever--larger region of interferometer parameter space will become economic to correlate
with software--based, as opposed to hardware--based, solutions.  In the three and a half years of
this thesis, the computing capacity of the Swinburne supercomputer has increased four--fold,
and most correlations are now run using only the CPUs available on the machines which host 
the storage media.  For existing and planned VLBI arrays (excluding the SKA), a software--based
solution is feasible now, and will become more so in the future.  This offers the potential to continue
progress towards the standardisation of VLBI formats, which with attendant improvements to 
support software and usability will help make VLBI observations accessible to a wider 
cross--section of the astronomical community.

\section[VLBI pulsar astrometry]{VLBI pulsar astrometry}
Of eight pulsars observed in this thesis, parallax and proper motion 
measurements have been made for seven,
with a proper motion determination for the final object.  This is comparable to the largest 
published Northern Hemisphere parallax program conducted 
using the VLBA \citep{brisken02a}, where
nine parallaxes were observed from ten objects.  Ironically, the offending pulsar was the same in 
both instances -- \peight.  

\clearpage

The ensemble of independent distance and velocity measurements for these seven pulsars
has been used along with previously published results 
to analyse the capabilities and shortcomings of the NE2001
Galactic electron distribution model used to predict pulsar distances, the accuracy of estimates of
pulsar transverse velocities using scintillation, and the presence and relative contribution of
different sources of systematic error in relative astrometry.  This analysis has shown that while the
NE2001 distance model offers significant improvements over the previous TC93 distance model, 
appearing to eliminate the systematic bias seen in the latter, it still possesses a long tail of large 
distance errors.  A simple model of these infrequent, large distance errors shows they
systematically bias pulsar velocities derived using the NE2001 model, leading
to an 18\% increase in the mean observed 2D transverse pulsar 
velocity and a fifty--fold increase in the the number of very fast moving pulsars (transverse velocities
greater than 1000 \kms). However, this simple model is itself likely to be somewhat biased
by selection effects in the available independent pulsar distance measurements, and thus 
these results are almost certainly an upper limit to the errors induced in velocity 
estimates.  An unbiased parallax sample would aid in quantifying the true impact of distance model
errors.
Scintillation velocity estimates have been shown to be unreliable for individual pulsars under the
typical assumptions of a thin scattering screen with an isotropic turbulence distribution.

In addition, the distance and velocity 
measurements have allowed a number of important conclusions for the individual objects, with the
most significant being:

\begin{itemize}
\item The radio luminosity of \pone, while low, is a factor of 3.4 higher than previously thought.
Interestingly, though, its recently measured x--ray luminosity, coupled with its newly measured
distance and proper motion implies that it is the most efficient converter of spin--down luminosity
to x-ray luminosity amongst rotation--powered pulsars, and that reheating must
occur for isolated neutron stars.
\item The most precise pulsar distance determination made to date, for \ptwo, constrains the 
time rate of change of Newton's gravitational constant $G$\ to be less than 3 parts in 10$^{12}$,
and precludes the presence of an unseen Jupiter--mass planet within 226 AU of the Sun (within
50\% of the sky).
\item The parallax of \pthree\ shows that the pulsar is more than four times closer than predicted
by the NE2001 distance model, revising its x--ray luminosity down by a factor of 19 and proving
that the system does not require an unusually efficient method of producing x--rays.
\clearpage
\item The first significant measurement of the parallax of the double pulsar J0737--3039A/B shows
that the system is more than twice as distant as the $DM$--derived predictions, and enabled 
calculation of the kinematic terms which contribute to the observed orbital period derivative.
The accuracy of the distance and velocity determination from this thesis will permit tests
of GR using the double pulsar system  to proceed to the 0.01\% level. Furthermore, the 
confirmation of the system's low transverse velocity supports formation scenarios in which
neither pulsar received a large kick at birth.
Finally, the revised distance strongly supports a magnetospheric origin for the majority of the x--rays
observed in the \pfour\ system.
\item \pseven\ is shown to be even closer than its $DM$--derived distance of 180 pc, confirming
that it is the least luminous pulsar known at 1400 MHz.  The revision to its period derivative based
on the measurement of distance and proper motion places it even further past the traditionally
assumed pulsar ``death line".
\end{itemize}

The future of VLBI pulsar astrometry is bright, with continual improvement in instrumentation
bringing more and more pulsars within reach of the technique.  With this work having shown that
the measurement of pulsar parallaxes with the LBA is possible out to and beyond two kpc, a large
number of southern pulsars are accessible.  Several LBA proposals for pulsar astrometry are 
currently active, and the number of VLBI parallaxes in the south should continue to grow.
A proposal has also been submitted for the continuation of astrometry on \ptwo, which could
lead to the most accurate VLBI parallax measurement of any object to date, and allow
a limit on the time rate of change of $G$\ which surpasses that presently made by Lunar
Laser Ranging.

Beyond the LBA, the sensitivity upgrade of the VLBA which is presently underway should deliver
4 Gbps recording capabilities during 2009.  This upgrade will deliver instantaneous 
sensitivity surpassing that of the LBA and approaching that of the EVN, but with a much greater
field of view.  When combined with the full--time operation and concurrent systematic control 
available with the VLBA, the upgrade should allow the deepest VLBI--resolution imaging ever 
made.  This will enable almost any known pulsar to be targeted for VLBI astrometry, and 
allow larger and more complete astrometric surveys to be undertaken.  Further in the future,
the SKA will integrate VLBI into connected--element observing and make astrometry of pulsars 
almost routine; it is likely to revolutionise our knowledge of 
their spatial, velocity and luminosity distributions.  The trickle of measured pulsar parallaxes is likely
to become a flood over the next decade, and may help astronomers finally gain a solid
understanding of the birth, life and death of pulsars. 	

\newpage
\addcontentsline{toc}{chapter}{Bibliography}
\bibliographystyle{adamthesis}
\bibliography{deller_thesis}

\begin{thebibliography}{234}
\providecommand{\natexlab}[1]{#1}
\providecommand{\url}[1]{\texttt{#1}}
\expandafter\ifx\csname urlstyle\endcsname\relax
  \providecommand{\doi}[1]{doi: #1}\else
  \providecommand{\doi}{doi: \begingroup \urlstyle{rm}\Url}\fi

\bibitem[{Ables}(1974)]{ables74a}
{Ables}, J.~G.
\newblock \emph{{Maximum Entropy Spectral Analysis}}.
\newblock \aaps, 15:\penalty0 383, June 1974.

\bibitem[{Alford} et~al.(2005){Alford}, {Jotwani}, {Kouvaris}, {Kundu}, and
  {Rajagopal}]{alford05a}
{Alford}, M., {Jotwani}, P., {Kouvaris}, C., {Kundu}, J., \& {Rajagopal}, K.
\newblock \emph{{Astrophysical implications of gapless color-flavor locked
  quark matter: A hot water bottle for aging neutron stars}}.
\newblock \prd, 71\penalty0 (11):\penalty0 114011, June 2005.

\bibitem[{Alpar} et~al.(1982){Alpar}, {Cheng}, {Ruderman}, and
  {Shaham}]{alpar82a}
{Alpar}, M.~A., {Cheng}, A.~F., {Ruderman}, M.~A., \& {Shaham}, J.
\newblock \emph{{A new class of radio pulsars}}.
\newblock \nat, 300:\penalty0 728--730, December 1982.

\bibitem[{Avruch} et~al.(2006){Avruch}, {Pogrebenko}, and {Gurvits}]{avruch06a}
{Avruch}, I., {Pogrebenko}, S.~V., \& {Gurvits}, L.~I.
\newblock \emph{{VLBI observations of spacecraft}}.
\newblock In \emph{Proceedings of the 8th European VLBI Network Symposium},
  2006.

\bibitem[{Baade} and {Zwicky}(1934)]{baade34b}
{Baade}, W. \& {Zwicky}, F.
\newblock \emph{{Cosmic Rays from Super-novae}}.
\newblock Proceedings of the National Academy of Science, 20:\penalty0
  259--263, 1934.

\bibitem[{Backer}(1970)]{backer70a}
{Backer}, D.~C.
\newblock \emph{{Pulsar Nulling Phenomena}}.
\newblock \nat, 228:\penalty0 42, October 1970.

\bibitem[{Bailes}(1989)]{bailes89a}
{Bailes}, M.
\newblock \emph{{The origin of pulsar velocities and the velocity-magnetic
  moment correlation}}.
\newblock \apj, 342:\penalty0 917--927, July 1989.

\bibitem[{Bailes} et~al.(1990){Bailes}, {Manchester}, {Kesteven}, {Norris}, and
  {Reynolds}]{bailes90a}
{Bailes}, M., {Manchester}, R.~N., {Kesteven}, M.~J., {Norris}, R.~P., \&
  {Reynolds}, J.~E.
\newblock \emph{{The parallax and proper motion of PSR1451-68}}.
\newblock \nat, 343:\penalty0 240, January 1990.

\bibitem[{Bailes} et~al.(1994){Bailes}, {Harrison}, {Lorimer}, {Johnston},
  {Lyne}, {Manchester}, {D'Amico}, {Nicastro}, {Tauris}, and
  {Robinson}]{bailes94a}
{Bailes}, M., {Harrison}, P.~A., {Lorimer}, D.~R., {Johnston}, S., {Lyne},
  A.~G., {Manchester}, R.~N., {D'Amico}, N., {Nicastro}, L., {Tauris}, T.~M.,
  \& {Robinson}, C.
\newblock \emph{{Discovery of three binary millisecond pulsars}}.
\newblock \apjl, 425:\penalty0 L41--L44, April 1994.

\bibitem[{Bare} et~al.(1967){Bare}, {Clark}, {Kellermann}, {Cohen}, and
  {Jauncey}]{bare67a}
{Bare}, C., {Clark}, B.~G., {Kellermann}, K.~I., {Cohen}, M.~H., \& {Jauncey},
  D.~L.
\newblock \emph{{Interferometer Experiment with Independent Local
  Oscillators}}.
\newblock Science, 157:\penalty0 189--191, July 1967.

\bibitem[{Bartel} et~al.(1996){Bartel}, {Chandler}, {Ratner}, {Shapiro}, {Pan},
  and {Cappallo}]{bartel96a}
{Bartel}, N., {Chandler}, J.~F., {Ratner}, M.~I., {Shapiro}, I.~L., {Pan}, R.,
  \& {Cappallo}, R.~J.
\newblock \emph{{Toward a Frame TI via Millisecond Pulsar VLBI}}.
\newblock \aj, 112:\penalty0 1690, October 1996.

\bibitem[{Becker}(2000)]{becker00a}
{Becker}, W.
\newblock \emph{{X-ray Emission Characteristics of Pulsars}}.
\newblock In \emph{{Highly Energetic Physical Processes and Mechanisms for
  Emission from Astrophysical Plasmas}}, editors {Martens}, P.~C.~H.,
  {Tsuruta}, S., \& {Weber}, M.~A., volume 195 of \emph{IAU Symposium},
  page~49, May 2000.

\bibitem[{Becker} and {Tr{\"u}mper}(1999)]{becker99a}
{Becker}, W. \& {Tr{\"u}mper}, J.
\newblock \emph{{The X-ray emission properties of millisecond pulsars}}.
\newblock \aap, 341:\penalty0 803--817, January 1999.

\bibitem[{Becker} et~al.(2005){Becker}, {Jessner}, {Kramer}, {Testa}, and
  {Howaldt}]{becker05a}
{Becker}, W., {Jessner}, A., {Kramer}, M., {Testa}, V., \& {Howaldt}, C.
\newblock \emph{{A Multiwavelength Study of PSR B0628-28: The First
  Overluminous Rotation-powered Pulsar?}}
\newblock \apj, 633:\penalty0 367--376, November 2005.

\bibitem[{Belczynski} et~al.(2002){Belczynski}, {Kalogera}, and
  {Bulik}]{belczynski02a}
{Belczynski}, K., {Kalogera}, V., \& {Bulik}, T.
\newblock \emph{A Comprehensive Study of Binary Compact Objects as
  Gravitational Wave Sources: Evolutionary Channels, Rates, and Physical
  Properties}.
\newblock \apj, 572\penalty0 (1):\penalty0 407--431, 2002.

\bibitem[{Bell} and {Bailes}(1996)]{bell96a}
{Bell}, J.~F. \& {Bailes}, M.
\newblock \emph{{New Method for Obtaining Binary Pulsar Distances and Its
  Implications for Tests of General Relativity}}.
\newblock \apjl, 456:\penalty0 L33+, January 1996.

\bibitem[{Bell} et~al.(1993){Bell}, {Bailes}, and {Bessell}]{bell93a}
{Bell}, J.~F., {Bailes}, M., \& {Bessell}, M.~S.
\newblock \emph{{Optical detection of the companion of the millisecond pulsar
  J0437 - 4715}}.
\newblock \nat, 364:\penalty0 603--605, August 1993.

\bibitem[{Bell} et~al.(1995){Bell}, {Kulkarni}, {Bailes}, {Leitch}, and
  {Lyne}]{bell95a}
{Bell}, J.~F., {Kulkarni}, S.~R., {Bailes}, M., {Leitch}, E.~M., \& {Lyne},
  A.~G.
\newblock \emph{{Optical Observations of the Binary Millisecond Pulsars
  J2145-0750 and J0034-0534}}.
\newblock \apjl, 452:\penalty0 L121, October 1995.

\bibitem[{Bhat} et~al.(1999){Bhat}, {Rao}, and {Gupta}]{bhat99a}
{Bhat}, N.~D.~R., {Rao}, A.~P., \& {Gupta}, Y.
\newblock \emph{{Long-Term Scintillation Studies of Pulsars. I. Observations
  and Basic Results}}.
\newblock \apjs, 121:\penalty0 483--513, April 1999.

\bibitem[{Bhatnagar} et~al.(2008){Bhatnagar}, {Cornwell}, {Golap}, and
  {Uson}]{bhatnagar08a}
{Bhatnagar}, S., {Cornwell}, T.~J., {Golap}, K., \& {Uson}, J.~M.
\newblock \emph{{Correcting direction-dependent gains in the deconvolution of
  radio interferometric images}}.
\newblock ArXiv e-prints, 0805.0834, May 2008.

\bibitem[{Bhattacharya}(2002)]{bhattacharya02a}
{Bhattacharya}, D.
\newblock \emph{{Evolution of Neutron Star Magnetic Fields}}.
\newblock Journal of Astrophysics and Astronomy, 23:\penalty0 67, June 2002.

\bibitem[{Bhattacharya} and {van den Heuvel}(1991)]{bhattacharya91a}
{Bhattacharya}, D. \& {van den Heuvel}, E.~P.~J.
\newblock \emph{{Formation and evolution of binary and millisecond radio
  pulsars.}}
\newblock \physrep, 203:\penalty0 1--124, 1991.

\bibitem[{Bhattacharya} et~al.(1992){Bhattacharya}, {Wijers}, {Hartman}, and
  {Verbunt}]{bhattacharya92a}
{Bhattacharya}, D., {Wijers}, R.~A.~M.~J., {Hartman}, J.~W., \& {Verbunt}, F.
\newblock \emph{{On the decay of the magnetic fields of single radio pulsars}}.
\newblock \aap, 254:\penalty0 198--212, February 1992.

\bibitem[{Blaauw}(1961)]{blaauw61a}
{Blaauw}, A.
\newblock \emph{{On the origin of the O- and B-type stars with high velocities
  (the ''run-away'' stars), and some related problems}}.
\newblock \bain, 15:\penalty0 265, May 1961.

\bibitem[{Brisken}(2005)]{brisken05a}
{Brisken}, W.
\newblock \emph{{Pulsar Astrometry with the VLBA}}.
\newblock In \emph{Future Directions in High Resolution Astronomy}, editors
  {Romney}, J. \& {Reid}, M., volume 340 of \emph{Astronomical Society of the
  Pacific Conference Series}, page 489, December 2005.

\bibitem[{Brisken} et~al.(2002){Brisken}, {Benson}, {Goss}, and
  {Thorsett}]{brisken02a}
{Brisken}, W.~F., {Benson}, J.~M., {Goss}, W.~M., \& {Thorsett}, S.~E.
\newblock \emph{{Very Long Baseline Array Measurement of Nine Pulsar
  Parallaxes}}.
\newblock \apj, 571:\penalty0 906--917, June 2002.

\bibitem[{Brisken} et~al.(2003){Brisken}, {Fruchter}, {Goss}, {Herrnstein}, and
  {Thorsett}]{brisken03b}
{Brisken}, W.~F., {Fruchter}, A.~S., {Goss}, W.~M., {Herrnstein}, R.~M., \&
  {Thorsett}, S.~E.
\newblock \emph{{Proper-Motion Measurements with the VLA. II. Observations of
  28 Pulsars}}.
\newblock \aj, 126:\penalty0 3090--3098, December 2003.

\bibitem[{Burgay} et~al.(2003){Burgay}, {D'Amico}, {Possenti}, {Manchester},
  {Lyne}, {Joshi}, {McLaughlin}, {Kramer}, {Sarkissian}, {Camilo}, {Kalogera},
  {Kim}, and {Lorimer}]{burgay03a}
{Burgay}, M., {D'Amico}, N., {Possenti}, A., {Manchester}, R.~N., {Lyne},
  A.~G., {Joshi}, B.~C., {McLaughlin}, M.~A., {Kramer}, M., {Sarkissian},
  J.~M., {Camilo}, F., {Kalogera}, V., {Kim}, C., \& {Lorimer}, D.~R.
\newblock \emph{{An increased estimate of the merger rate of double neutron
  stars from observations of a highly relativistic system}}.
\newblock \nat, 426:\penalty0 531--533, December 2003.

\bibitem[{Burgay} et~al.(2006){Burgay}, {Joshi}, {D'Amico}, {Possenti}, {Lyne},
  {Manchester}, {McLaughlin}, {Kramer}, {Camilo}, and {Freire}]{burgay06a}
{Burgay}, M., {Joshi}, B.~C., {D'Amico}, N., {Possenti}, A., {Lyne}, A.~G.,
  {Manchester}, R.~N., {McLaughlin}, M.~A., {Kramer}, M., {Camilo}, F., \&
  {Freire}, P.~C.~C.
\newblock \emph{{The Parkes High-Latitude pulsar survey}}.
\newblock \mnras, 368:\penalty0 283--292, May 2006.

\bibitem[{Carlson} et~al.(1999){Carlson}, {Dewdney}, {Burgess}, {Casorso},
  {Petrachenko}, and {Cannon}]{carlson99a}
{Carlson}, B.~R., {Dewdney}, P.~E., {Burgess}, T.~A., {Casorso}, R.~V.,
  {Petrachenko}, W.~T., \& {Cannon}, W.~H.
\newblock \emph{{The S2 VLBI Correlator: A Correlator for Space VLBI and
  Geodetic Signal Processing}}.
\newblock \pasp, 111:\penalty0 1025--1047, August 1999.

\bibitem[{Casse}(1999)]{casse99a}
{Casse}, J.~L.
\newblock \emph{{The European VLBI Network MkIV Data Processor}}.
\newblock New Astronomy Review, 43:\penalty0 503--508, November 1999.

\bibitem[{Champion} et~al.(2008){Champion}, {Ransom}, {Lazarus}, {Camilo},
  {Bassa}, {Kaspi}, {Nice}, {Freire}, {Stairs}, {van Leeuwen}, {Stappers},
  {Cordes}, {Hessels}, {Lorimer}, {Arzoumanian}, {Backer}, {Bhat},
  {Chatterjee}, {Cognard}, {Deneva}, {Faucher-Gigu{\`e}re}, {Gaensler}, {Han},
  {Jenet}, {Kasian}, {Kondratiev}, {Kramer}, {Lazio}, {McLaughlin},
  {Venkataraman}, and {Vlemmings}]{champion08a}
{Champion}, D.~J., {Ransom}, S.~M., {Lazarus}, P., {Camilo}, F., {Bassa}, C.,
  {Kaspi}, V.~M., {Nice}, D.~J., {Freire}, P.~C.~C., {Stairs}, I.~H., {van
  Leeuwen}, J., {Stappers}, B.~W., {Cordes}, J.~M., {Hessels}, J.~W.~T.,
  {Lorimer}, D.~R., {Arzoumanian}, Z., {Backer}, D.~C., {Bhat}, N.~D.~R.,
  {Chatterjee}, S., {Cognard}, I., {Deneva}, J.~S., {Faucher-Gigu{\`e}re},
  C.-A., {Gaensler}, B.~M., {Han}, J., {Jenet}, F.~A., {Kasian}, L.,
  {Kondratiev}, V.~I., {Kramer}, M., {Lazio}, J., {McLaughlin}, M.~A.,
  {Venkataraman}, A., \& {Vlemmings}, W.
\newblock \emph{{An Eccentric Binary Millisecond Pulsar in the Galactic
  Plane}}.
\newblock Science, 320:\penalty0 1309, June 2008.

\bibitem[{Chandrasekhar}(1931)]{chandrasekhar31a}
{Chandrasekhar}, S.
\newblock \emph{{The Maximum Mass of Ideal White Dwarfs}}.
\newblock \apj, 74:\penalty0 81, July 1931.

\bibitem[{Chatterjee} et~al.(2001){Chatterjee}, {Cordes}, {Lazio}, {Goss},
  {Fomalont}, and {Benson}]{chatterjee01a}
{Chatterjee}, S., {Cordes}, J.~M., {Lazio}, T.~J.~W., {Goss}, W.~M.,
  {Fomalont}, E.~B., \& {Benson}, J.~M.
\newblock \emph{{Parallax and Kinematics of PSR B0919+06 from VLBA Astrometry
  and Interstellar Scintillometry}}.
\newblock \apj, 550:\penalty0 287--296, March 2001.

\bibitem[{Chatterjee} et~al.(2004){Chatterjee}, {Cordes}, {Vlemmings},
  {Arzoumanian}, {Goss}, and {Lazio}]{chatterjee04a}
{Chatterjee}, S., {Cordes}, J.~M., {Vlemmings}, W.~H.~T., {Arzoumanian}, Z.,
  {Goss}, W.~M., \& {Lazio}, T.~J.~W.
\newblock \emph{{Pulsar Parallaxes at 5 GHz with the Very Long Baseline
  Array}}.
\newblock \apj, 604:\penalty0 339--345, March 2004.

\bibitem[{Chatterjee} et~al.(2005){Chatterjee}, {Vlemmings}, {Brisken},
  {Lazio}, {Cordes}, {Goss}, {Thorsett}, {Fomalont}, {Lyne}, and
  {Kramer}]{chatterjee05a}
{Chatterjee}, S., {Vlemmings}, W.~H.~T., {Brisken}, W.~F., {Lazio}, T.~J.~W.,
  {Cordes}, J.~M., {Goss}, W.~M., {Thorsett}, S.~E., {Fomalont}, E.~B., {Lyne},
  A.~G., \& {Kramer}, M.
\newblock \emph{{Getting Its Kicks: A VLBA Parallax for the Hyperfast Pulsar
  B1508+55}}.
\newblock \apjl, 630:\penalty0 L61--L64, September 2005.

\bibitem[{Chatterjee} et~al.(2007){Chatterjee}, {Gaensler}, {Melatos},
  {Brisken}, and {Stappers}]{chatterjee07a}
{Chatterjee}, S., {Gaensler}, B.~M., {Melatos}, A., {Brisken}, W.~F., \&
  {Stappers}, B.~W.
\newblock \emph{{Pulsed X-Ray Emission from Pulsar A in the Double Pulsar
  System J0737-3039}}.
\newblock \apj, 670:\penalty0 1301--1306, December 2007.

\bibitem[{Chen} and {Ruderman}(1993)]{chen93a}
{Chen}, K. \& {Ruderman}, M.
\newblock \emph{{Pulsar death lines and death valley}}.
\newblock \apj, 402:\penalty0 264--270, January 1993.

\bibitem[{Cheng} et~al.(1986){Cheng}, {Ho}, and {Ruderman}]{cheng86a}
{Cheng}, K.~S., {Ho}, C., \& {Ruderman}, M.
\newblock \emph{{Energetic radiation from rapidly spinning pulsars. I - Outer
  magnetosphere gaps. II - VELA and Crab}}.
\newblock \apj, 300:\penalty0 500--539, January 1986.

\bibitem[{Chiang} and {Romani}(1994)]{chiang94a}
{Chiang}, J. \& {Romani}, R.~W.
\newblock \emph{{An outer gap model of high-energy emission from
  rotation-powered pulsars}}.
\newblock \apj, 436:\penalty0 754--761, December 1994.

\bibitem[{Chikada} et~al.(1987){Chikada}, {Ishiguro}, {Hirabayashi},
  {Morimoto}, {Morita}, {Kanzawa}, {Iwashita}, {Nakazima}, {Ishikawa},
  {Takahashi}, {Handa}, {Kasuga}, {Okumura}, {Miyazawa}, {Nakazuru}, {Miura},
  and {Nagasawa}]{chikada87a}
{Chikada}, Y., {Ishiguro}, M., {Hirabayashi}, H., {Morimoto}, M., {Morita},
  K.-I., {Kanzawa}, T., {Iwashita}, H., {Nakazima}, K., {Ishikawa}, S.-I.,
  {Takahashi}, T., {Handa}, K., {Kasuga}, T., {Okumura}, S., {Miyazawa}, T.,
  {Nakazuru}, T., {Miura}, K., \& {Nagasawa}, S.
\newblock \emph{{A 6$\times$320-MHz 1024-channel FFT cross-spectrum analyzer
  for radio astronomy.}}
\newblock IEEE Proceedings, 75:\penalty0 1203--1210, 1987.

\bibitem[{Clark} et~al.(1967){Clark}, {Cohen}, and {Jauncey}]{clark67a}
{Clark}, B.~G., {Cohen}, M.~H., \& {Jauncey}, D.~L.
\newblock \emph{{Angular Size of 3C 273B}}.
\newblock \apjl, 149:\penalty0 L151, September 1967.

\bibitem[{Cocke} et~al.(1969){Cocke}, {Disney}, and {Taylor}]{cocke69a}
{Cocke}, W.~J., {Disney}, M.~J., \& {Taylor}, D.~J.
\newblock \emph{{Discovery of Optical Signals from Pulsar NP 0532}}.
\newblock \nat, 221:\penalty0 525, 1969.

\bibitem[{Coles} et~al.(2005){Coles}, {McLaughlin}, {Rickett}, {Lyne}, and
  {Bhat}]{coles05a}
{Coles}, W.~A., {McLaughlin}, M.~A., {Rickett}, B.~J., {Lyne}, A.~G., \&
  {Bhat}, N.~D.~R.
\newblock \emph{{Probing the Eclipse of J0737-3039A with Scintillation}}.
\newblock \apj, 623:\penalty0 392--397, April 2005.

\bibitem[Cooley and Tukey(1965)]{cooley65a}
Cooley, J.~W. \& Tukey, J.~W.
\newblock \emph{An Algorithm for the Machine Calculation of Complex Fourier
  Series}.
\newblock Mathematics of Computation, 19\penalty0 (90):\penalty0 297--301,
  1965.

\bibitem[{Cooper}(1970)]{cooper70a}
{Cooper}, B.~F.~C.
\newblock \emph{{Correlators with two-bit quantization}}.
\newblock Australian Journal of Physics, 23:\penalty0 521, August 1970.

\bibitem[{Cordes}(1986)]{cordes86b}
{Cordes}, J.~M.
\newblock \emph{{Space velocities of radio pulsars from interstellar
  scintillations}}.
\newblock \apj, 311:\penalty0 183--196, December 1986.

\bibitem[{Cordes} and {Chernoff}(1998)]{cordes98a}
{Cordes}, J.~M. \& {Chernoff}, D.~F.
\newblock \emph{{Neutron Star Population Dynamics. II. Three-dimensional Space
  Velocities of Young Pulsars}}.
\newblock \apj, 505:\penalty0 315--338, September 1998.

\bibitem[{Cordes} and {Lazio}(2002)]{cordes02a}
{Cordes}, J.~M. \& {Lazio}, T.~J.~W.
\newblock \emph{{NE2001.I. A New Model for the Galactic Distribution of Free
  Electrons and its Fluctuations}}.
\newblock ArXiv e-prints, 0207156, July 2002.

\bibitem[{Cordes} and {Rickett}(1998)]{cordes98b}
{Cordes}, J.~M. \& {Rickett}, B.~J.
\newblock \emph{{Diffractive Interstellar Scintillation Timescales and
  Velocities}}.
\newblock \apj, 507:\penalty0 846--860, November 1998.

\bibitem[{Cordes} et~al.(1986){Cordes}, {Pidwerbetsky}, and
  {Lovelace}]{cordes86a}
{Cordes}, J.~M., {Pidwerbetsky}, A., \& {Lovelace}, R.~V.~E.
\newblock \emph{{Refractive and diffractive scattering in the interstellar
  medium}}.
\newblock \apj, 310:\penalty0 737--767, November 1986.

\bibitem[{Cordes} et~al.(1993){Cordes}, {Romani}, and {Lundgren}]{cordes93a}
{Cordes}, J.~M., {Romani}, R.~W., \& {Lundgren}, S.~C.
\newblock \emph{{The Guitar nebula - A bow shock from a slow-spin,
  high-velocity neutron star}}.
\newblock \nat, 362:\penalty0 133--135, March 1993.

\bibitem[{Cordes} et~al.(2006){Cordes}, {Rickett}, {Stinebring}, and
  {Coles}]{cordes06a}
{Cordes}, J.~M., {Rickett}, B.~J., {Stinebring}, D.~R., \& {Coles}, W.~A.
\newblock \emph{{Theory of Parabolic Arcs in Interstellar Scintillation
  Spectra}}.
\newblock \apj, 637:\penalty0 346--365, January 2006.

\bibitem[{Cornwell} and {Fomalont}(1999)]{cornwell99a}
{Cornwell}, T. \& {Fomalont}, E.~B.
\newblock \emph{{Self-Calibration}}.
\newblock In \emph{Synthesis Imaging in Radio Astronomy II}, editors {Taylor},
  G.~B., {Carilli}, C.~L., \& {Perley}, R.~A., volume 180 of \emph{Astronomical
  Society of the Pacific Conference Series}, page 187, 1999.

\bibitem[{Damour} and {Taylor}(1991)]{damour91a}
{Damour}, T. \& {Taylor}, J.~H.
\newblock \emph{{On the orbital period change of the binary pulsar PSR 1913 +
  16}}.
\newblock \apj, 366:\penalty0 501--511, January 1991.

\bibitem[{Daugherty} and {Harding}(1982)]{daugherty82a}
{Daugherty}, J.~K. \& {Harding}, A.~K.
\newblock \emph{{Electromagnetic cascades in pulsars}}.
\newblock \apj, 252:\penalty0 337--347, January 1982.

\bibitem[{Deller} et~al.(2007){Deller}, {Tingay}, {Bailes}, and
  {West}]{deller07a}
{Deller}, A.~T., {Tingay}, S.~J., {Bailes}, M., \& {West}, C.
\newblock \emph{{DiFX: A Software Correlator for Very Long Baseline
  Interferometry Using Multiprocessor Computing Environments}}.
\newblock \pasp, 119:\penalty0 318--336, March 2007.

\bibitem[{Deller} et~al.(2008){Deller}, {Verbiest}, {Tingay}, and
  {Bailes}]{deller08b}
{Deller}, A.~T., {Verbiest}, J.~P.~W., {Tingay}, S.~J., \& {Bailes}, M.
\newblock \emph{{Extremely High Precision VLBI Astrometry of PSR J0437-4715 and
  Implications for Theories of Gravity}}.
\newblock \apjl, 685:\penalty0 L67--L70, September 2008.

\bibitem[{Deller} et~al.(2009{\natexlab{a}}){Deller}, {Bailes}, and
  {Tingay}]{deller08c}
{Deller}, A.~T., {Bailes}, M., \& {Tingay}, S.~J.
\newblock \emph{{Implications of a VLBI distance to the double pulsar
  J0737-3039A/B}}.
\newblock Science, (10.1126/science.1167969), February 2009{\natexlab{a}}.

\bibitem[{Deller} et~al.(2009{\natexlab{b}}){Deller}, {Tingay}, , and
  {Brisken}]{deller08a}
{Deller}, A.~T., {Tingay}, S.~J., , \& {Brisken}, W.
\newblock \emph{{Precision Southern Hemisphere pulsar VLBI astrometry:
  Techniques and results for PSR J1559-4438}}.
\newblock \apj, 690\penalty0 (1):\penalty0 198--209, 2009{\natexlab{b}}.

\bibitem[{Dewey} and {Cordes}(1987)]{dewey87a}
{Dewey}, R.~J. \& {Cordes}, J.~M.
\newblock \emph{{Monte Carlo simulations of radio pulsars and their
  progenitors}}.
\newblock \apj, 321:\penalty0 780--798, October 1987.

\bibitem[{Dias} and {L{\'e}pine}(2005)]{dias05a}
{Dias}, W.~S. \& {L{\'e}pine}, J.~R.~D.
\newblock \emph{{Direct Determination of the Spiral Pattern Rotation Speed of
  the Galaxy}}.
\newblock \apj, 629:\penalty0 825--831, August 2005.

\bibitem[{Dodson} et~al.(2003){Dodson}, {Legge}, {Reynolds}, and
  {McCulloch}]{dodson03a}
{Dodson}, R., {Legge}, D., {Reynolds}, J.~E., \& {McCulloch}, P.~M.
\newblock \emph{{The Vela Pulsar's Proper Motion and Parallax Derived from VLBI
  Observations}}.
\newblock \apj, 596:\penalty0 1137--1141, October 2003.

\bibitem[{Ekers}(1999)]{ekers99a}
{Ekers}, R.~D.
\newblock \emph{{Error Recognition}}.
\newblock In \emph{Synthesis Imaging in Radio Astronomy II}, editors {Taylor},
  G.~B., {Carilli}, C.~L., \& {Perley}, R.~A., volume 180 of \emph{Astronomical
  Society of the Pacific Conference Series}, page 321, 1999.

\bibitem[{Faucher-Gigu{\`e}re} and {Kaspi}(2006)]{faucher-giguere06a}
{Faucher-Gigu{\`e}re}, C.-A. \& {Kaspi}, V.~M.
\newblock \emph{{Birth and Evolution of Isolated Radio Pulsars}}.
\newblock \apj, 643:\penalty0 332--355, May 2006.

\bibitem[{Fazio} et~al.(1972){Fazio}, {Helmken}, {O'Mongain}, and
  {Weekes}]{fazio72a}
{Fazio}, G.~G., {Helmken}, H.~F., {O'Mongain}, E., \& {Weekes}, T.~C.
\newblock \emph{{Detection of High-Energy Gamma Rays from the Crab Nebula}}.
\newblock \apjl, 175:\penalty0 L117, August 1972.

\bibitem[{Feast} and {Shuttleworth}(1965)]{feast65a}
{Feast}, M.~W. \& {Shuttleworth}, M.
\newblock \emph{{The kinematics of B stars, cepheids, galactic clusters and
  interstellar gas in the Galaxy}}.
\newblock \mnras, 130:\penalty0 245, 1965.

\bibitem[{Fey} et~al.(2004){Fey}, {Ma}, {Arias}, {Charlot}, {Feissel-Vernier},
  {Gontier}, {Jacobs}, {Li}, and {MacMillan}]{fey04a}
{Fey}, A.~L., {Ma}, C., {Arias}, E.~F., {Charlot}, P., {Feissel-Vernier}, M.,
  {Gontier}, A.-M., {Jacobs}, C.~S., {Li}, J., \& {MacMillan}, D.~S.
\newblock \emph{{The Second Extension of the International Celestial Reference
  Frame: ICRF-EXT.1}}.
\newblock \aj, 127:\penalty0 3587--3608, June 2004.

\bibitem[{Fomalont} and {Reid}(2004)]{fomalont04a}
{Fomalont}, E. \& {Reid}, M.
\newblock \emph{{Microarcsecond astrometry using the SKA}}.
\newblock New Astronomy Review, 48:\penalty0 1473--1482, December 2004.

\bibitem[{Fomalont}(2005)]{fomalont05a}
{Fomalont}, E.~B.
\newblock \emph{{From reference frames to relativistic experiments: Absolute
  and relative radio astrometry}}.
\newblock In \emph{EAS Publications Series}, editors {Gurvits}, L.~I., {Frey},
  S., \& {Rawlings}, S., volume~15 of \emph{EAS Publications Series}, pages
  131--155, 2005.

\bibitem[{Fomalont}(1999)]{fomalont99a}
{Fomalont}, E.~B.
\newblock \emph{{Astrometry and Geodesy}}.
\newblock In \emph{Synthesis Imaging in Radio Astronomy II}, editors {Taylor},
  G.~B., {Carilli}, C.~L., \& {Perley}, R.~A., volume 180 of \emph{Astronomical
  Society of the Pacific Conference Series}, page 463, 1999.

\bibitem[{Fomalont} et~al.(1997){Fomalont}, {Goss}, {Manchester}, and
  {Lyne}]{fomalont97a}
{Fomalont}, E.~B., {Goss}, W.~M., {Manchester}, R.~N., \& {Lyne}, A.~G.
\newblock \emph{{Improved proper motions for pulsars from VLA observations}}.
\newblock \mnras, 286:\penalty0 81--84, March 1997.

\bibitem[{Foster} and {Cordes}(1990)]{foster90a}
{Foster}, R.~S. \& {Cordes}, J.~M.
\newblock \emph{{Interstellar propagation effects and the precision of pulsar
  timing}}.
\newblock \apj, 364:\penalty0 123--135, November 1990.

\bibitem[{Fricke} et~al.(1988){Fricke}, {Schwan}, {Lederle}, {Bastian}, {Bien},
  {Burkhardt}, {Du Mont}, {Hering}, {J{\"a}hrling}, {Jahrei{\ss}}, {R{\"o}ser},
  {Schwerdtfeger}, and {Walter}]{fricke88a}
{Fricke}, W., {Schwan}, H., {Lederle}, T., {Bastian}, U., {Bien}, R.,
  {Burkhardt}, G., {Du Mont}, B., {Hering}, R., {J{\"a}hrling}, R.,
  {Jahrei{\ss}}, H., {R{\"o}ser}, S., {Schwerdtfeger}, H.-M., \& {Walter},
  H.~G.
\newblock \emph{{Fifth fundamental catalogue (FK5). Part 1: The basic
  fundamental stars}}.
\newblock Veroeffentlichungen des Astronomischen Rechen-Instituts Heidelberg,
  32:\penalty0 1--106, 1988.

\bibitem[{Fritz} et~al.(1969){Fritz}, {Henry}, {Meekins}, {Chubb}, and
  {Friedman}]{fritz69a}
{Fritz}, G., {Henry}, R.~C., {Meekins}, J.~F., {Chubb}, T.~A., \& {Friedman},
  H.
\newblock \emph{{X-ray Pulsar in the Crab Nebula}}.
\newblock Science, 164:\penalty0 709--712, May 1969.

\bibitem[{Fryer}(2004)]{fryer04a}
{Fryer}, C.~L.
\newblock \emph{{Neutron Star Kicks from Asymmetric Collapse}}.
\newblock \apjl, 601:\penalty0 L175--L178, February 2004.

\bibitem[{Fryer} and {Warren}(2002)]{fryer02a}
{Fryer}, C.~L. \& {Warren}, M.~S.
\newblock \emph{{Modeling Core-Collapse Supernovae in Three Dimensions}}.
\newblock \apjl, 574:\penalty0 L65--L68, July 2002.

\bibitem[{Gaensler} et~al.(2002){Gaensler}, {Arons}, {Kaspi}, {Pivovaroff},
  {Kawai}, and {Tamura}]{gaensler02a}
{Gaensler}, B.~M., {Arons}, J., {Kaspi}, V.~M., {Pivovaroff}, M.~J., {Kawai},
  N., \& {Tamura}, K.
\newblock \emph{{Chandra Imaging of the X-Ray Nebula Powered by Pulsar
  B1509-58}}.
\newblock \apj, 569:\penalty0 878--893, April 2002.

\bibitem[{Ghez} et~al.(2008){Ghez}, {Salim}, {Weinberg}, {Lu}, {Do}, {Dunn},
  {Matthews}, {Morris}, {Yelda}, {Becklin}, {Kremenek}, {Milosavljevic}, and
  {Naiman}]{ghez08a}
{Ghez}, A.~M., {Salim}, S., {Weinberg}, N.~N., {Lu}, J.~R., {Do}, T., {Dunn},
  J.~K., {Matthews}, K., {Morris}, M.~R., {Yelda}, S., {Becklin}, E.~E.,
  {Kremenek}, T., {Milosavljevic}, M., \& {Naiman}, J.
\newblock \emph{{Measuring Distance and Properties of the Milky Way's Central
  Supermassive Black Hole with Stellar Orbits}}.
\newblock \apj, 689:\penalty0 1044--1062, December 2008.

\bibitem[{Gold}(1968)]{gold68a}
{Gold}, T.
\newblock \emph{{Rotating Neutron Stars as the Origin of the Pulsating Radio
  Sources}}.
\newblock \nat, 218:\penalty0 731--+, May 1968.

\bibitem[{Goldreich} and {Julian}(1969)]{goldreich69a}
{Goldreich}, P. \& {Julian}, W.~H.
\newblock \emph{{Pulsar Electrodynamics}}.
\newblock \apj, 157:\penalty0 869, August 1969.

\bibitem[{Goldreich} and {Reisenegger}(1992)]{goldreich92a}
{Goldreich}, P. \& {Reisenegger}, A.
\newblock \emph{{Magnetic field decay in isolated neutron stars}}.
\newblock \apj, 395:\penalty0 250--258, August 1992.

\bibitem[{Gontier} et~al.(2001){Gontier}, {Le Bail}, {Feissel}, and
  {Eubanks}]{gontier01a}
{Gontier}, A.-M., {Le Bail}, K., {Feissel}, M., \& {Eubanks}, T.~M.
\newblock \emph{{Stability of the extragalactic VLBI reference frame}}.
\newblock \aap, 375:\penalty0 661--669, August 2001.

\bibitem[{Gothoskar} and {Gupta}(2000)]{gothoskar00a}
{Gothoskar}, P. \& {Gupta}, Y.
\newblock \emph{{Scintillation Velocities of Five Millisecond Pulsars}}.
\newblock \apj, 531:\penalty0 345--349, March 2000.

\bibitem[{Grindlay} et~al.(2002){Grindlay}, {Camilo}, {Heinke}, {Edmonds},
  {Cohn}, and {Lugger}]{grindlay02a}
{Grindlay}, J.~E., {Camilo}, F., {Heinke}, C.~O., {Edmonds}, P.~D., {Cohn}, H.,
  \& {Lugger}, P.
\newblock \emph{{Chandra Study of a Complete Sample of Millisecond Pulsars in
  47 Tucanae and NGC 6397}}.
\newblock \apj, 581:\penalty0 470--484, December 2002.

\bibitem[{Gupta}(1995)]{gupta95a}
{Gupta}, Y.
\newblock \emph{{On the Correlation between Proper Motion Velocities and
  Scintillation Velocities of Radio Pulsars}}.
\newblock \apj, 451:\penalty0 717, October 1995.

\bibitem[{Gupta} et~al.(1994){Gupta}, {Rickett}, and {Lyne}]{gupta94a}
{Gupta}, Y., {Rickett}, B.~J., \& {Lyne}, A.~G.
\newblock \emph{{Refractive Interstellar Scintillation in Pulsar Dynamic
  Spectra}}.
\newblock \mnras, 269:\penalty0 1035, August 1994.

\bibitem[{Hankins} and {Rickett}(1975)]{hankins75a}
{Hankins}, T.~H. \& {Rickett}, B.~J.
\newblock \emph{{Pulsar signal processing.}}
\newblock Methods in Computational Physics, 14:\penalty0 55--129, 1975.

\bibitem[{Harrison} et~al.(1993){Harrison}, {Lyne}, and
  {Anderson}]{harrison93a}
{Harrison}, P.~A., {Lyne}, A.~G., \& {Anderson}, B.
\newblock \emph{{New determinations of the proper motions of 44 pulsars}}.
\newblock \mnras, 261:\penalty0 113--124, March 1993.

\bibitem[{Helfand} et~al.(2001){Helfand}, {Gotthelf}, and
  {Halpern}]{helfand01a}
{Helfand}, D.~J., {Gotthelf}, E.~V., \& {Halpern}, J.~P.
\newblock \emph{{Vela Pulsar and Its Synchrotron Nebula}}.
\newblock \apj, 556:\penalty0 380--391, July 2001.

\bibitem[{Hessels} et~al.(2006){Hessels}, {Ransom}, {Stairs}, {Freire},
  {Kaspi}, and {Camilo}]{hessels06a}
{Hessels}, J.~W.~T., {Ransom}, S.~M., {Stairs}, I.~H., {Freire}, P.~C.~C.,
  {Kaspi}, V.~M., \& {Camilo}, F.
\newblock \emph{{A Radio Pulsar Spinning at 716 Hz}}.
\newblock Science, 311:\penalty0 1901--1904, March 2006.

\bibitem[{Hewish} et~al.(1968){Hewish}, {Bell}, {Pilkington}, {Scott}, and
  {Collins}]{hewish68a}
{Hewish}, A., {Bell}, S.~J., {Pilkington}, J.~D., {Scott}, P.~F., \& {Collins},
  R.~A.
\newblock \emph{{Observation of a Rapidly Pulsating Radio Source}}.
\newblock \nat, 217:\penalty0 709, February 1968.

\bibitem[{Hobbs} et~al.(2004){Hobbs}, {Lyne}, {Kramer}, {Martin}, and
  {Jordan}]{hobbs04a}
{Hobbs}, G., {Lyne}, A.~G., {Kramer}, M., {Martin}, C.~E., \& {Jordan}, C.
\newblock \emph{{Long-term timing observations of 374 pulsars}}.
\newblock \mnras, 353:\penalty0 1311--1344, October 2004.

\bibitem[{Hobbs} et~al.(2005){Hobbs}, {Lorimer}, {Lyne}, and
  {Kramer}]{hobbs05a}
{Hobbs}, G., {Lorimer}, D.~R., {Lyne}, A.~G., \& {Kramer}, M.
\newblock \emph{{A statistical study of 233 pulsar proper motions}}.
\newblock \mnras, 360:\penalty0 974--992, July 2005.

\bibitem[{Hobbs} et~al.(2006){Hobbs}, {Edwards}, and {Manchester}]{hobbs06a}
{Hobbs}, G.~B., {Edwards}, R.~T., \& {Manchester}, R.~N.
\newblock \emph{{TEMPO2, a new pulsar-timing package - I. An overview}}.
\newblock \mnras, 369:\penalty0 655--672, June 2006.

\bibitem[{Hobbs} et~al.(2008){Hobbs}, {Jenet}, {Lee}, {Verbiest}, {Yardley},
  {Manchester}, {Lommen}, {Coles}, {Edwards}, and {Shettigara}]{hobbs08a}
{Hobbs}, G.~B., {Jenet}, F.~A., {Lee}, K.~J., {Verbiest}, J.~P.~W., {Yardley},
  D., {Manchester}, R.~N., {Lommen}, A., {Coles}, W.~A., {Edwards}, R.~T., \&
  {Shettigara}, C.
\newblock \emph{{TEMPO2, a new pulsar-timing package - III. Gravitational wave
  simulations}}.
\newblock MNRAS, submitted, 2008.

\bibitem[{H{\"o}gbom}(1974)]{hogbom74a}
{H{\"o}gbom}, J.~A.
\newblock \emph{{Aperture Synthesis with a Non-Regular Distribution of
  Interferometer Baselines}}.
\newblock \aaps, 15:\penalty0 417, June 1974.

\bibitem[{Holmberg} and {Flynn}(2004)]{holmberg04a}
{Holmberg}, J. \& {Flynn}, C.
\newblock \emph{{The local surface density of disc matter mapped by
  Hipparcos}}.
\newblock \mnras, 352:\penalty0 440--446, August 2004.

\bibitem[{Horiuchi} et~al.(2000){Horiuchi}, {Kameno}, {Nan}, {Shibata},
  {Inoue}, {Kobayashi}, {Murata}, {Fomalont}, and {Carlson}]{horiuchi00a}
{Horiuchi}, S., {Kameno}, S., {Nan}, R., {Shibata}, K., {Inoue}, M.,
  {Kobayashi}, H., {Murata}, Y., {Fomalont}, E., \& {Carlson}, B.
\newblock \emph{{Imaging Capability of the Mitaka VSOP Correlator}}.
\newblock Advances in Space Research, 26:\penalty0 625--628, 2000.

\bibitem[{Hotan} et~al.(2004){Hotan}, {Bailes}, and {Ord}]{hotan04a}
{Hotan}, A.~W., {Bailes}, M., \& {Ord}, S.~M.
\newblock \emph{{PSR J1022+1001: profile stability and precision timing}}.
\newblock \mnras, 355:\penalty0 941--949, December 2004.

\bibitem[{Hotan} et~al.(2005){Hotan}, {Bailes}, and {Ord}]{hotan05a}
{Hotan}, A.~W., {Bailes}, M., \& {Ord}, S.~M.
\newblock \emph{{Geodetic Precession in PSR J1141-6545}}.
\newblock \apj, 624:\penalty0 906--913, May 2005.

\bibitem[{Hotan} et~al.(2006){Hotan}, {Bailes}, and {Ord}]{hotan06a}
{Hotan}, A.~W., {Bailes}, M., \& {Ord}, S.~M.
\newblock \emph{{High-precision baseband timing of 15 millisecond pulsars}}.
\newblock \mnras, 369:\penalty0 1502--1520, July 2006.

\bibitem[{Iguchi} et~al.(2005){Iguchi}, {Kkurayama}, {Kawaguchi}, and
  {Kawakami}]{iguchi05a}
{Iguchi}, S., {Kkurayama}, T., {Kawaguchi}, N., \& {Kawakami}, K.
\newblock \emph{{Gigabit Digital Filter Bank: Digital Backend Subsystem in the
  VERA Data-Acquisition System}}.
\newblock \pasj, 57:\penalty0 259--271, February 2005.

\bibitem[{Iwamoto} et~al.(1998){Iwamoto}, {Mazzali}, {Nomoto}, {Umeda},
  {Nakamura}, {Patat}, {Danziger}, {Young}, {Suzuki}, {Shigeyama},
  {Augusteijn}, {Doublier}, {Gonzalez}, {Boehnhardt}, {Brewer}, {Hainaut},
  {Lidman}, {Leibundgut}, {Cappellaro}, {Turatto}, {Galama}, {Vreeswijk},
  {Kouveliotou}, {van Paradijs}, {Pian}, {Palazzi}, and {Frontera}]{iwamoto98a}
{Iwamoto}, K., {Mazzali}, P.~A., {Nomoto}, K., {Umeda}, H., {Nakamura}, T.,
  {Patat}, F., {Danziger}, I.~J., {Young}, T.~R., {Suzuki}, T., {Shigeyama},
  T., {Augusteijn}, T., {Doublier}, V., {Gonzalez}, J.-F., {Boehnhardt}, H.,
  {Brewer}, J., {Hainaut}, O.~R., {Lidman}, C., {Leibundgut}, B., {Cappellaro},
  E., {Turatto}, M., {Galama}, T.~J., {Vreeswijk}, P.~M., {Kouveliotou}, C.,
  {van Paradijs}, J., {Pian}, E., {Palazzi}, E., \& {Frontera}, F.
\newblock \emph{{A hypernova model for the supernova associated with the
  {$\gamma$}-ray burst of 25 April 1998}}.
\newblock \nat, 395:\penalty0 672--674, October 1998.

\bibitem[{Jansky}(1933)]{jansky33a}
{Jansky}, K.~G.
\newblock \emph{{Electrical disturbances apparently of extraterrestrial
  origin}}.
\newblock Proceedings of the IRE, 21\penalty0 (10):\penalty0 1387--1396,
  October 1933.

\bibitem[{Jenet} et~al.(2005){Jenet}, {Hobbs}, {Lee}, and
  {Manchester}]{jenet05a}
{Jenet}, F.~A., {Hobbs}, G.~B., {Lee}, K.~J., \& {Manchester}, R.~N.
\newblock \emph{{Detecting the Stochastic Gravitational Wave Background Using
  Pulsar Timing}}.
\newblock \apjl, 625:\penalty0 L123--L126, June 2005.

\bibitem[{Jenet} et~al.(2006){Jenet}, {Hobbs}, {van Straten}, {Manchester},
  {Bailes}, {Verbiest}, {Edwards}, {Hotan}, {Sarkissian}, and {Ord}]{jenet06a}
{Jenet}, F.~A., {Hobbs}, G.~B., {van Straten}, W., {Manchester}, R.~N.,
  {Bailes}, M., {Verbiest}, J.~P.~W., {Edwards}, R.~T., {Hotan}, A.~W.,
  {Sarkissian}, J.~M., \& {Ord}, S.~M.
\newblock \emph{{Upper Bounds on the Low-Frequency Stochastic Gravitational
  Wave Background from Pulsar Timing Observations: Current Limits and Future
  Prospects}}.
\newblock \apj, 653:\penalty0 1571--1576, December 2006.

\bibitem[{Johnston} et~al.(1993){Johnston}, {Lorimer}, {Harrison}, {Bailes},
  {Lyne}, {Bell}, {Kaspi}, {Manchester}, {D'Amico}, and
  {Nicastro}]{johnston93a}
{Johnston}, S., {Lorimer}, D.~R., {Harrison}, P.~A., {Bailes}, M., {Lyne},
  A.~G., {Bell}, J.~F., {Kaspi}, V.~M., {Manchester}, R.~N., {D'Amico}, N., \&
  {Nicastro}, L.
\newblock \emph{{Discovery of a very bright, nearby binary millisecond
  pulsar}}.
\newblock \nat, 361:\penalty0 613--615, February 1993.

\bibitem[{Johnston} et~al.(1994){Johnston}, {Manchester}, {Lyne}, {Nicastro},
  and {Spyromilio}]{johnston94a}
{Johnston}, S., {Manchester}, R.~N., {Lyne}, A.~G., {Nicastro}, L., \&
  {Spyromilio}, J.
\newblock \emph{{Radio and Optical Observations of the PSR:B1259-63 / SS:2883
  Be-Star Binary System}}.
\newblock \mnras, 268:\penalty0 430, May 1994.

\bibitem[{Johnston} et~al.(1998){Johnston}, {Nicastro}, and
  {Koribalski}]{johnston98a}
{Johnston}, S., {Nicastro}, L., \& {Koribalski}, B.
\newblock \emph{{Scintillation parameters for 49 pulsars}}.
\newblock \mnras, 297:\penalty0 108--116, June 1998.

\bibitem[{Johnston} et~al.(1999){Johnston}, {Manchester}, {McConnell}, and
  {Campbell-Wilson}]{johnston99a}
{Johnston}, S., {Manchester}, R.~N., {McConnell}, D., \& {Campbell-Wilson}, D.
\newblock \emph{{Transient radio emission from the PSR B1259-63 system near
  periastron}}.
\newblock \mnras, 302:\penalty0 277--287, January 1999.

\bibitem[{Johnston} et~al.(2005){Johnston}, {Hobbs}, {Vigeland}, {Kramer},
  {Weisberg}, and {Lyne}]{johnston05a}
{Johnston}, S., {Hobbs}, G., {Vigeland}, S., {Kramer}, M., {Weisberg}, J.~M.,
  \& {Lyne}, A.~G.
\newblock \emph{{Evidence for alignment of the rotation and velocity vectors in
  pulsars}}.
\newblock \mnras, 364:\penalty0 1397--1412, December 2005.

\bibitem[{Johnston} et~al.(2007){Johnston}, {Kramer}, {Karastergiou}, {Hobbs},
  {Ord}, and {Wallman}]{johnston07a}
{Johnston}, S., {Kramer}, M., {Karastergiou}, A., {Hobbs}, G., {Ord}, S., \&
  {Wallman}, J.
\newblock \emph{{Evidence for alignment of the rotation and velocity vectors in
  pulsars - II. Further data and emission heights}}.
\newblock \mnras, 381:\penalty0 1625--1637, November 2007.

\bibitem[{Kalogera} et~al.(2004){Kalogera}, {Kim}, {Lorimer}, {Burgay},
  {D'Amico}, {Possenti}, {Manchester}, {Lyne}, {Joshi}, {McLaughlin}, {Kramer},
  {Sarkissian}, and {Camilo}]{kalogera04a}
{Kalogera}, V., {Kim}, C., {Lorimer}, D.~R., {Burgay}, M., {D'Amico}, N.,
  {Possenti}, A., {Manchester}, R.~N., {Lyne}, A.~G., {Joshi}, B.~C.,
  {McLaughlin}, M.~A., {Kramer}, M., {Sarkissian}, J.~M., \& {Camilo}, F.
\newblock \emph{{The Cosmic Coalescence Rates for Double Neutron Star
  Binaries}}.
\newblock \apjl, 601:\penalty0 L179--L182, February 2004.

\bibitem[{Kargaltsev} et~al.(2004){Kargaltsev}, {Pavlov}, and
  {Romani}]{kargaltsev04a}
{Kargaltsev}, O., {Pavlov}, G.~G., \& {Romani}, R.~W.
\newblock \emph{{Ultraviolet Emission from the Millisecond Pulsar J0437-4715}}.
\newblock \apj, 602:\penalty0 327--335, February 2004.

\bibitem[{Kargaltsev} et~al.(2006){Kargaltsev}, {Pavlov}, and
  {Garmire}]{kargaltsev06a}
{Kargaltsev}, O., {Pavlov}, G.~G., \& {Garmire}, G.~P.
\newblock \emph{{X-Ray Emission from the Nearby PSR B1133+16 and Other Old
  Pulsars}}.
\newblock \apj, 636:\penalty0 406--410, January 2006.

\bibitem[{Kellermann} and {Cohen}(1988)]{kellerman88a}
{Kellermann}, K.~I. \& {Cohen}, M.~H.
\newblock \emph{{The origin and evolution of the N.R.A.O.-Cornell VLBI
  system}}.
\newblock \jrasc, 82:\penalty0 248--265, October 1988.

\bibitem[{Kellermann} and {Moran}(2001)]{kellermann01a}
{Kellermann}, K.~I. \& {Moran}, J.~M.
\newblock \emph{{The Development of High-Resolution Imaging in Radio
  Astronomy}}.
\newblock \araa, 39:\penalty0 457--509, 2001.

\bibitem[{Kettenis} et~al.(2006){Kettenis}, {van Langevelde}, {Reynolds}, and
  {Cotton}]{kettenis06a}
{Kettenis}, M., {van Langevelde}, H.~J., {Reynolds}, C., \& {Cotton}, B.
\newblock \emph{{ParselTongue: AIPS Talking Python}}.
\newblock In \emph{Astronomical Data Analysis Software and Systems XV}, editors
  {Gabriel}, C., {Arviset}, C., {Ponz}, D., \& {Enrique}, S., volume 351 of
  \emph{Astronomical Society of the Pacific Conference Series}, page 497, July
  2006.

\bibitem[{Kondo} et~al.(2004){Kondo}, {Kimura}, {Koyama}, and
  {Osaki}]{kondo04a}
{Kondo}, T., {Kimura}, M., {Koyama}, Y., \& {Osaki}, H.
\newblock \emph{{Current Status of Software Correlators Developed at Kashima
  Space Research Center}}.
\newblock In \emph{{International VLBI Service for Geodesy and Astrometry 2004
  General Meeting Proceedings}}, editors {Vandenberg}, N.~R. \& {Baver}, K.~D.,
  page~36, June 2004.

\bibitem[{Koribalski} et~al.(1995){Koribalski}, {Johnston}, {Weisberg}, and
  {Wilson}]{koribalski95a}
{Koribalski}, B., {Johnston}, S., {Weisberg}, J.~M., \& {Wilson}, W.
\newblock \emph{{H I line measurements of eight southern pulsars}}.
\newblock \apj, 441:\penalty0 756--764, March 1995.

\bibitem[{Kovalev} et~al.(2007){Kovalev}, {Petrov}, {Fomalont}, and
  {Gordon}]{kovalev07a}
{Kovalev}, Y.~Y., {Petrov}, L., {Fomalont}, E.~B., \& {Gordon}, D.
\newblock \emph{{The Fifth VLBA Calibrator Survey: VCS5}}.
\newblock \aj, 133:\penalty0 1236--1242, April 2007.

\bibitem[{Koyama} et~al.(2004){Koyama}, {Kondo}, {Osaki}, {Hirabaru},
  {Takashima}, {Sorai}, {Takaba}, {Fujisawa}, {Lapsley}, {Dudevoir}, and
  {Whitney}]{koyama04a}
{Koyama}, Y., {Kondo}, T., {Osaki}, H., {Hirabaru}, M., {Takashima}, K.,
  {Sorai}, K., {Takaba}, H., {Fujisawa}, K., {Lapsley}, D., {Dudevoir}, K., \&
  {Whitney}, A.
\newblock \emph{{Geodetic VLBI Experiments with the K5 System}}.
\newblock In \emph{{International VLBI Service for Geodesy and Astrometry 2004
  General Meeting Proceedings}}, editors {Vandenberg}, N.~R. \& {Baver}, K.~D.,
  page~43, June 2004.

\bibitem[{Kramer} et~al.(1998){Kramer}, {Xilouris}, {Lorimer}, {Doroshenko},
  {Jessner}, {Wielebinski}, {Wolszczan}, and {Camilo}]{kramer98a}
{Kramer}, M., {Xilouris}, K.~M., {Lorimer}, D.~R., {Doroshenko}, O., {Jessner},
  A., {Wielebinski}, R., {Wolszczan}, A., \& {Camilo}, F.
\newblock \emph{{The Characteristics of Millisecond Pulsar Emission. I.
  Spectra, Pulse Shapes, and the Beaming Fraction}}.
\newblock \apj, 501:\penalty0 270, July 1998.

\bibitem[{Kramer} et~al.(2003){Kramer}, {Lyne}, {Hobbs}, {L{\"o}hmer}, {Carr},
  {Jordan}, and {Wolszczan}]{kramer03a}
{Kramer}, M., {Lyne}, A.~G., {Hobbs}, G., {L{\"o}hmer}, O., {Carr}, P.,
  {Jordan}, C., \& {Wolszczan}, A.
\newblock \emph{{The Proper Motion, Age, and Initial Spin Period of PSR
  J0538+2817 in S147}}.
\newblock \apjl, 593:\penalty0 L31--L34, August 2003.

\bibitem[{Kramer} et~al.(2006){Kramer}, {Stairs}, {Manchester}, {McLaughlin},
  {Lyne}, {Ferdman}, {Burgay}, {Lorimer}, {Possenti}, {D'Amico}, {Sarkissian},
  {Hobbs}, {Reynolds}, {Freire}, and {Camilo}]{kramer06a}
{Kramer}, M., {Stairs}, I.~H., {Manchester}, R.~N., {McLaughlin}, M.~A.,
  {Lyne}, A.~G., {Ferdman}, R.~D., {Burgay}, M., {Lorimer}, D.~R., {Possenti},
  A., {D'Amico}, N., {Sarkissian}, J.~M., {Hobbs}, G.~B., {Reynolds}, J.~E.,
  {Freire}, P.~C.~C., \& {Camilo}, F.
\newblock \emph{{Tests of General Relativity from Timing the Double Pulsar}}.
\newblock Science, 314:\penalty0 97--102, October 2006.

\bibitem[{Lai} et~al.(2001){Lai}, {Chernoff}, and {Cordes}]{lai01a}
{Lai}, D., {Chernoff}, D.~F., \& {Cordes}, J.~M.
\newblock \emph{{Pulsar Jets: Implications for Neutron Star Kicks and Initial
  Spins}}.
\newblock \apj, 549:\penalty0 1111--1118, March 2001.

\bibitem[{Large} et~al.(1969){Large}, {Vaughan}, and {Wielebinski}]{large69a}
{Large}, M.~I., {Vaughan}, A.~E., \& {Wielebinski}, R.
\newblock \emph{{Some Further Pulsar Observations at the Molonglo Radio
  Observatory}}.
\newblock \aplett, 3:\penalty0 123, 1969.

\bibitem[{Larson} and {Link}(2002)]{larson02a}
{Larson}, M.~B. \& {Link}, B.
\newblock \emph{{Simulations of glitches in isolated pulsars}}.
\newblock \mnras, 333:\penalty0 613--622, July 2002.

\bibitem[{Lattimer} and {Prakash}(2007)]{lattimer07a}
{Lattimer}, J.~M. \& {Prakash}, M.
\newblock \emph{{Neutron star observations: Prognosis for equation of state
  constraints}}.
\newblock \physrep, 442:\penalty0 109--165, April 2007.

\bibitem[{Lee} and {Jokipii}(1975)]{lee75a}
{Lee}, L.~C. \& {Jokipii}, J.~R.
\newblock \emph{{Strong scintillations in astrophysics. II - A theory of
  temporal broadening of pulses}}.
\newblock \apj, 201:\penalty0 532--543, October 1975.

\bibitem[{Legge}(2002)]{legge02a}
{Legge}, D.
\newblock \emph{Accurate Astrometry of Southern Pulsars}.
\newblock PhD thesis, University of Tasmania, 2002.

\bibitem[{Lindegren} et~al.(2008){Lindegren}, {Babusiaux}, {Bailer-Jones},
  {Bastian}, {Brown}, {Cropper}, {H{\o}g}, {Jordi}, {Katz}, {van Leeuwen},
  {Luri}, {Mignard}, {de Bruijne}, and {Prusti}]{lindegren08a}
{Lindegren}, L., {Babusiaux}, C., {Bailer-Jones}, C., {Bastian}, U., {Brown},
  A.~G.~A., {Cropper}, M., {H{\o}g}, E., {Jordi}, C., {Katz}, D., {van
  Leeuwen}, F., {Luri}, X., {Mignard}, F., {de Bruijne}, J.~H.~J., \& {Prusti},
  T.
\newblock \emph{{The Gaia mission: science, organization and present status}}.
\newblock In \emph{IAU Symposium}, volume 248 of \emph{IAU Symposium}, pages
  217--223, 2008.

\bibitem[{Liu} et~al.(2006){Liu}, {Wang}, {Urama}, and {Manchester}]{liu06a}
{Liu}, Z.~Y., {Wang}, N., {Urama}, J.~O., \& {Manchester}, R.~N.
\newblock \emph{{Monitoring of Pulse Intensity and Mode Changing for PSR
  B0329+54}}.
\newblock Chinese Journal of Astronomy and Astrophysics Supplement, 6\penalty0
  (2):\penalty0 020000--67, December 2006.

\bibitem[{L{\"o}hmer} et~al.(2004){L{\"o}hmer}, {Kramer}, {Driebe}, {Jessner},
  {Mitra}, and {Lyne}]{lohmer04a}
{L{\"o}hmer}, O., {Kramer}, M., {Driebe}, T., {Jessner}, A., {Mitra}, D., \&
  {Lyne}, A.~G.
\newblock \emph{{The parallax, mass and age of the PSR J2145-0750 binary
  system}}.
\newblock \aap, 426:\penalty0 631--640, November 2004.

\bibitem[{Loinard} et~al.(2007){Loinard}, {Torres}, {Mioduszewski},
  {Rodr{\'{\i}}guez}, {Gonz{\'a}lez-L{\'o}pezlira}, {Lachaume}, {V{\'a}zquez},
  and {Gonz{\'a}lez}]{loinard07a}
{Loinard}, L., {Torres}, R.~M., {Mioduszewski}, A.~J., {Rodr{\'{\i}}guez},
  L.~F., {Gonz{\'a}lez-L{\'o}pezlira}, R.~A., {Lachaume}, R., {V{\'a}zquez},
  V., \& {Gonz{\'a}lez}, E.
\newblock \emph{{VLBA Determination of the Distance to Nearby Star-forming
  Regions. I. The Distance to T Tauri with 0.4\% Accuracy}}.
\newblock \apj, 671:\penalty0 546--554, December 2007.

\bibitem[{Lorimer}(2008)]{lorimer08a}
{Lorimer}, D.~R.
\newblock \emph{{Binary and Millisecond Pulsars}}.
\newblock Living Reviews in Relativity, 11:\penalty0 8, November 2008.

\bibitem[Lorimer and Kramer(2005)]{lorimer05a}
Lorimer, D. \& Kramer, M.
\newblock \emph{Handbook of Pulsar Astronomy}, volume~4 of \emph{Cambridge
  Observing Handbooks for Research Astronomers}.
\newblock Cambridge University Press, Cambridge, U.K.; New York, U.S.A., 2005.

\bibitem[{Lundgren} et~al.(1996){Lundgren}, {Cordes}, {Foster}, {Wolszczan},
  and {Camilo}]{lundgren96a}
{Lundgren}, S.~C., {Cordes}, J.~M., {Foster}, R.~S., {Wolszczan}, A., \&
  {Camilo}, F.
\newblock \emph{{Optical Studies of Millisecond Pulsar Companions}}.
\newblock \apjl, 458:\penalty0 L33, February 1996.

\bibitem[{Lyne}(1971)]{lyne71a}
{Lyne}, A.~G.
\newblock \emph{{Mode changing in pulsar radiation}}.
\newblock \mnras, 153:\penalty0 27P, 1971.

\bibitem[{Lyne} et~al.(1985){Lyne}, {Manchester}, and {Taylor}]{lyne85a}
{Lyne}, A.~G., {Manchester}, R.~N., \& {Taylor}, J.~H.
\newblock \emph{{The galactic population of pulsars}}.
\newblock \mnras, 213:\penalty0 613--639, April 1985.

\bibitem[{Lyne} et~al.(1998){Lyne}, {Manchester}, {Lorimer}, {Bailes},
  {D'Amico}, {Tauris}, {Johnston}, {Bell}, and {Nicastro}]{lyne98a}
{Lyne}, A.~G., {Manchester}, R.~N., {Lorimer}, D.~R., {Bailes}, M., {D'Amico},
  N., {Tauris}, T.~M., {Johnston}, S., {Bell}, J.~F., \& {Nicastro}, L.
\newblock \emph{{The Parkes Southern Pulsar Survey - II. Final results and
  population analysis}}.
\newblock \mnras, 295:\penalty0 743--755, April 1998.

\bibitem[{Lyne} et~al.(2004){Lyne}, {Burgay}, {Kramer}, {Possenti},
  {Manchester}, {Camilo}, {McLaughlin}, {Lorimer}, {D'Amico}, {Joshi},
  {Reynolds}, and {Freire}]{lyne04a}
{Lyne}, A.~G., {Burgay}, M., {Kramer}, M., {Possenti}, A., {Manchester}, R.~N.,
  {Camilo}, F., {McLaughlin}, M.~A., {Lorimer}, D.~R., {D'Amico}, N., {Joshi},
  B.~C., {Reynolds}, J., \& {Freire}, P.~C.~C.
\newblock \emph{{A Double-Pulsar System: A Rare Laboratory for Relativistic
  Gravity and Plasma Physics}}.
\newblock Science, 303:\penalty0 1153--1157, February 2004.

\bibitem[{Lyutikov}(2004)]{lyutikov04a}
{Lyutikov}, M.
\newblock \emph{{On the nature of eclipses in binary pulsar J0737-3039}}.
\newblock \mnras, 353:\penalty0 1095--1106, October 2004.

\bibitem[{Lyutikov} et~al.(1999){Lyutikov}, {Blandford}, and
  {Machabeli}]{lyutikov99a}
{Lyutikov}, M., {Blandford}, R.~D., \& {Machabeli}, G.
\newblock \emph{{On the nature of pulsar radio emission}}.
\newblock \mnras, 305:\penalty0 338--352, April 1999.

\bibitem[{Ma} et~al.(1998){Ma}, {Arias}, {Eubanks}, {Fey}, {Gontier}, {Jacobs},
  {Sovers}, {Archinal}, and {Charlot}]{ma98a}
{Ma}, C., {Arias}, E.~F., {Eubanks}, T.~M., {Fey}, A.~L., {Gontier}, A.-M.,
  {Jacobs}, C.~S., {Sovers}, O.~J., {Archinal}, B.~A., \& {Charlot}, P.
\newblock \emph{{The International Celestial Reference Frame as Realized by
  Very Long Baseline Interferometry}}.
\newblock \aj, 116:\penalty0 516--546, July 1998.

\bibitem[{Malov} and {Malov}(2006)]{malov06a}
{Malov}, I.~F. \& {Malov}, O.~I.
\newblock \emph{{Integrated radio luminosities of pulsars}}.
\newblock Astronomy Reports, 50:\penalty0 483--495, June 2006.

\bibitem[{Manchester} and {Taylor}(1981)]{manchester81a}
{Manchester}, R.~N. \& {Taylor}, J.~H.
\newblock \emph{{Observed and derived parameters for 330 pulsars}}.
\newblock \aj, 86:\penalty0 1953--1973, December 1981.

\bibitem[{Manchester} et~al.(1996){Manchester}, {Lyne}, {D'Amico}, {Bailes},
  {Johnston}, {Lorimer}, {Harrison}, {Nicastro}, and {Bell}]{manchester96a}
{Manchester}, R.~N., {Lyne}, A.~G., {D'Amico}, N., {Bailes}, M., {Johnston},
  S., {Lorimer}, D.~R., {Harrison}, P.~A., {Nicastro}, L., \& {Bell}, J.~F.
\newblock \emph{{The Parkes Southern Pulsar Survey. I. Observing and data
  analysis systems and initial results.}}
\newblock \mnras, 279:\penalty0 1235--1250, April 1996.

\bibitem[{Manchester} et~al.(2005){Manchester}, {Hobbs}, {Teoh}, and
  {Hobbs}]{manchester05a}
{Manchester}, R.~N., {Hobbs}, G.~B., {Teoh}, A., \& {Hobbs}, M.
\newblock \emph{{The Australia Telescope National Facility Pulsar Catalogue}}.
\newblock \aj, 129:\penalty0 1993--2006, April 2005.

\bibitem[{Maron} et~al.(2000){Maron}, {Kijak}, {Kramer}, and
  {Wielebinski}]{maron00a}
{Maron}, O., {Kijak}, J., {Kramer}, M., \& {Wielebinski}, R.
\newblock \emph{{Pulsar spectra of radio emission}}.
\newblock \aaps, 147:\penalty0 195--203, December 2000.

\bibitem[{McLaughlin} et~al.(2004){McLaughlin}, {Camilo}, {Burgay}, {D'Amico},
  {Joshi}, {Kramer}, {Lorimer}, {Lyne}, {Manchester}, and
  {Possenti}]{mclaughlin04a}
{McLaughlin}, M.~A., {Camilo}, F., {Burgay}, M., {D'Amico}, N., {Joshi}, B.~C.,
  {Kramer}, M., {Lorimer}, D.~R., {Lyne}, A.~G., {Manchester}, R.~N., \&
  {Possenti}, A.
\newblock \emph{{X-Ray Emission from the Double Pulsar System J0737-3039}}.
\newblock \apjl, 605:\penalty0 L41--L44, April 2004.

\bibitem[{McLaughlin} et~al.(2006){McLaughlin}, {Lyne}, {Lorimer}, {Kramer},
  {Faulkner}, {Manchester}, {Cordes}, {Camilo}, {Possenti}, {Stairs}, {Hobbs},
  {D'Amico}, {Burgay}, and {O'Brien}]{mclaughlin06a}
{McLaughlin}, M.~A., {Lyne}, A.~G., {Lorimer}, D.~R., {Kramer}, M., {Faulkner},
  A.~J., {Manchester}, R.~N., {Cordes}, J.~M., {Camilo}, F., {Possenti}, A.,
  {Stairs}, I.~H., {Hobbs}, G., {D'Amico}, N., {Burgay}, M., \& {O'Brien},
  J.~T.
\newblock \emph{{Transient radio bursts from rotating neutron stars}}.
\newblock \nat, 439:\penalty0 817--820, February 2006.

\bibitem[{Michel}(1974)]{michel74a}
{Michel}, F.~C.
\newblock \emph{{Comment on ''Self-consistent solution for an axisymmetric
  pulsar model'' [Phys. Rev. Lett., Vol. 32, p. 1019 - 1022].}}
\newblock Physical Review Letters, 33:\penalty0 1521--1523, 1974.

\bibitem[{Middelberg}(2006)]{middelberg06a}
{Middelberg}, E.
\newblock \emph{{Automated Editing of Radio Interferometer Data with pieflag}}.
\newblock Publications of the Astronomical Society of Australia, 23:\penalty0
  64--68, May 2006.

\bibitem[{Migliazzo} et~al.(2002){Migliazzo}, {Gaensler}, {Backer}, {Stappers},
  {van der Swaluw}, and {Strom}]{migliazzo02a}
{Migliazzo}, J.~M., {Gaensler}, B.~M., {Backer}, D.~C., {Stappers}, B.~W., {van
  der Swaluw}, E., \& {Strom}, R.~G.
\newblock \emph{{Proper-Motion Measurements of Pulsar B1951+32 in the Supernova
  Remnant CTB 80}}.
\newblock \apjl, 567:\penalty0 L141--L144, March 2002.

\bibitem[{Mignani} et~al.(2003){Mignani}, {Manchester}, and
  {Pavlov}]{mignani03a}
{Mignani}, R.~P., {Manchester}, R.~N., \& {Pavlov}, G.~G.
\newblock \emph{{Search for the Optical Counterpart of the Nearby Pulsar PSR
  J0108-1431}}.
\newblock \apj, 582:\penalty0 978--983, January 2003.

\bibitem[{Mignani} et~al.(2008){Mignani}, {Pavlov}, and
  {Kargaltsev}]{mignani08a}
{Mignani}, R.~P., {Pavlov}, G.~G., \& {Kargaltsev}, O.
\newblock \emph{{A possible optical counterpart to the old nearby pulsar
  J0108-1431}}.
\newblock ArXiv e-prints, 0805.2586, May 2008.

\bibitem[{Mignard}(2000)]{mignard00a}
{Mignard}, F.
\newblock \emph{{Local galactic kinematics from Hipparcos proper motions}}.
\newblock \aap, 354:\penalty0 522--536, February 2000.

\bibitem[{Moran} et~al.(1967){Moran}, {Crowther}, {Burke}, {Barrett}, {Rogers},
  {Ball}, {Carter}, and {Bare}]{moran67a}
{Moran}, J.~M., {Crowther}, P.~P., {Burke}, B.~F., {Barrett}, A.~H., {Rogers},
  A.~E.~E., {Ball}, J.~A., {Carter}, J.~C., \& {Bare}, C.
\newblock \emph{{Spectral Line Interferometry with Independent Time Standards
  at Stations Separated by 845 Kilometers}}.
\newblock Science, 157:\penalty0 676--677, August 1967.

\bibitem[{Napier} et~al.(1994){Napier}, {Bagri}, {Clark}, {Rogers}, {Romney},
  {Thompson}, and {Walker}]{napier94a}
{Napier}, P.~J., {Bagri}, D.~S., {Clark}, B.~G., {Rogers}, A.~E.~E., {Romney},
  J.~D., {Thompson}, A.~R., \& {Walker}, R.~C.
\newblock \emph{{The Very Long Baseline Array.}}
\newblock IEEE Proceedings, 82:\penalty0 658--672, May 1994.

\bibitem[{Nicastro} and {Johnston}(1995)]{nicastro95a}
{Nicastro}, L. \& {Johnston}, S.
\newblock \emph{{Scintillation velocities for four millisecond pulsars}}.
\newblock \mnras, 273:\penalty0 122--128, March 1995.

\bibitem[{Niell} et~al.(2005){Niell}, {Whitney}, {Petrachenko}, {Schl{\"u}ter},
  {Vandenberg}, {Hase}, {Koyama}, {Ma}, {Schuh}, and {Tuccari}]{niell05a}
{Niell}, A., {Whitney}, A.~R., {Petrachenko}, W.~T., {Schl{\"u}ter}, W.,
  {Vandenberg}, N.~R., {Hase}, H., {Koyama}, Y., {Ma}, C., {Schuh}, H., \&
  {Tuccari}, G.
\newblock \emph{{VLBI2010: Current and Future Requirements for Geodetic VLBI
  Systems}}.
\newblock In \emph{{International VLBI Service for Geodesy and Astrometry 2005
  Annual Report}}, editors {Behrend}, D. \& {Baver}, K., June 2005.

\bibitem[{Ott} et~al.(2006){Ott}, {Burrows}, {Thompson}, {Livne}, and
  {Walder}]{ott06a}
{Ott}, C.~D., {Burrows}, A., {Thompson}, T.~A., {Livne}, E., \& {Walder}, R.
\newblock \emph{{The Spin Periods and Rotational Profiles of Neutron Stars at
  Birth}}.
\newblock \apjs, 164:\penalty0 130--155, May 2006.

\bibitem[{Pacini}(1968)]{pacini68a}
{Pacini}, F.
\newblock \emph{{Rotating Neutron Stars, Pulsars and Supernova Remnants}}.
\newblock \nat, 219:\penalty0 145, July 1968.

\bibitem[{Paragi} et~al.(2007){Paragi}, {Kouveliotou}, {Garrett},
  {Ramirez-Ruiz}, {van Langevelde}, {Szomoru}, and {Argo}]{paragi07a}
{Paragi}, Z., {Kouveliotou}, C., {Garrett}, M.~A., {Ramirez-Ruiz}, E., {van
  Langevelde}, H.~J., {Szomoru}, A., \& {Argo}, M.
\newblock \emph{{e-VLBI detection of SN2007gr}}.
\newblock The Astronomer's Telegram, 1215:\penalty0 1, September 2007.

\bibitem[{Pavlov} et~al.(2008){Pavlov}, {Kargaltsev}, {Wong}, and
  {Garmire}]{pavlov08a}
{Pavlov}, G.~G., {Kargaltsev}, O., {Wong}, J.~A., \& {Garmire}, G.~P.
\newblock \emph{{Detection of X-ray Emission from the Very Old Pulsar
  J0108-1431}}.
\newblock ArXiv e-prints, 0803.0761, March 2008.

\bibitem[{Pawsey} et~al.(1946){Pawsey}, {Payne-Scott}, and
  {McCready}]{pawsey46a}
{Pawsey}, J.~L., {Payne-Scott}, R., \& {McCready}, L.~L.
\newblock \emph{{Radio-Frequency Energy from the Sun}}.
\newblock \nat, 157\penalty0 (3980):\penalty0 158--159, February 1946.

\bibitem[{Pearson}(1999)]{pearson99a}
{Pearson}, T.~J.
\newblock \emph{{Non-Imaging Data Analysis}}.
\newblock In \emph{Synthesis Imaging in Radio Astronomy II}, editors {Taylor},
  G.~B., {Carilli}, C.~L., \& {Perley}, R.~A., volume 180 of \emph{Astronomical
  Society of the Pacific Conference Series}, page 335, 1999.

\bibitem[{Pearson} and {Readhead}(1984)]{pearson84a}
{Pearson}, T.~J. \& {Readhead}, A.~C.~S.
\newblock \emph{{Image Formation by Self-Calibration in Radio Astronomy}}.
\newblock \araa, 22:\penalty0 97--130, 1984.

\bibitem[{Pellizzoni} et~al.(2008){Pellizzoni}, {Tiengo}, {De Luca},
  {Esposito}, and {Mereghetti}]{pellizzoni08a}
{Pellizzoni}, A., {Tiengo}, A., {De Luca}, A., {Esposito}, P., \& {Mereghetti},
  S.
\newblock \emph{{PSR J0737-3039: Interacting Pulsars in X-Rays}}.
\newblock \apj, 679:\penalty0 664--674, May 2008.

\bibitem[{Perley} et~al.(2004){Perley}, {Napier}, and {Butler}]{perley04a}
{Perley}, R.~A., {Napier}, P.~J., \& {Butler}, B.~J.
\newblock \emph{{The Expanded Very Large Array: goals, progress, and plans}}.
\newblock In \emph{{Ground-based Telescopes}}, editor {Oschmann}, Jr., J.~M.,
  volume 5489 of \emph{Proceedings of the SPIE}, pages 784--795, October 2004.

\bibitem[{Petrov} et~al.(2008){Petrov}, {Gordon}, {Gipson}, {MacMillan}, {Ma},
  {Fomalont}, {Walker}, and {Carabajal}]{petrov08a}
{Petrov}, L., {Gordon}, D., {Gipson}, J., {MacMillan}, D., {Ma}, C.,
  {Fomalont}, E., {Walker}, R.~C., \& {Carabajal}, C.
\newblock \emph{{Precise geodesy with the Very Long Baseline Array}}.
\newblock ArXiv e-prints, 0806.0167, June 2008.

\bibitem[{Phillips} et~al.(2007){Phillips}, {Deller}, {Amy}, {Tingay},
  {Tzioumis}, {Reynolds}, {Jauncey}, {Stevens}, {Ellingsen}, {Dickey},
  {Fender}, {Tudose}, and {Nicolson}]{phillips07a}
{Phillips}, C.~J., {Deller}, A., {Amy}, S.~W., {Tingay}, S.~J., {Tzioumis},
  A.~K., {Reynolds}, J.~E., {Jauncey}, D.~L., {Stevens}, J., {Ellingsen},
  S.~P., {Dickey}, J., {Fender}, R.~P., {Tudose}, V., \& {Nicolson}, G.~D.
\newblock \emph{{Detection of compact radio emission from Circinus X-1 with the
  first Southern hemisphere e-VLBI experiment}}.
\newblock \mnras, 380:\penalty0 L11--L14, September 2007.

\bibitem[{Piran} and {Shaviv}(2005)]{piran05a}
{Piran}, T. \& {Shaviv}, N.~J.
\newblock \emph{{Origin of the Binary Pulsar J0737-3039B}}.
\newblock Physical Review Letters, 94\penalty0 (5):\penalty0 051102, February
  2005.

\bibitem[{Podsiadlowski} et~al.(2005){Podsiadlowski}, {Pfahl}, and
  {Rappaport}]{podsiadlowski05a}
{Podsiadlowski}, P., {Pfahl}, E., \& {Rappaport}, S.
\newblock \emph{{Neutron-Star Birth Kicks}}.
\newblock In \emph{Binary Radio Pulsars}, editors {Rasio}, F.~A. \& {Stairs},
  I.~H., volume 328 of \emph{Astronomical Society of the Pacific Conference
  Series}, page 327, July 2005.

\bibitem[{Possenti} et~al.(2008){Possenti}, {Rea}, {McLaughlin}, {Camilo},
  {Kramer}, {Burgay}, {Joshi}, and {Lyne}]{possenti08a}
{Possenti}, A., {Rea}, N., {McLaughlin}, M.~A., {Camilo}, F., {Kramer}, M.,
  {Burgay}, M., {Joshi}, B.~C., \& {Lyne}, A.~G.
\newblock \emph{{The Very Soft X-Ray Spectrum of the Double Pulsar System
  J0737-3039}}.
\newblock \apj, 680:\penalty0 654--663, June 2008.

\bibitem[{Pradel} et~al.(2006){Pradel}, {Charlot}, and {Lestrade}]{pradel06a}
{Pradel}, N., {Charlot}, P., \& {Lestrade}, J.-F.
\newblock \emph{{Astrometric accuracy of phase-referenced observations with the
  VLBA and EVN}}.
\newblock \aap, 452:\penalty0 1099--1106, June 2006.

\bibitem[{Prakash}(2007)]{prakash07a}
{Prakash}, M.
\newblock \emph{{Quark matter and the astrophysics of neutron stars}}.
\newblock Journal of Physics G Nuclear Physics, 34:\penalty0 253, August 2007.

\bibitem[Press et~al.(2002)Press, Vetterling, Teukolsky, and
  Flannery]{press02a}
Press, W.~H., Vetterling, W.~T., Teukolsky, S.~A., \& Flannery, B.~P.
\newblock \emph{Numerical Recipes in C++: the art of scientific computing}.
\newblock Cambridge University Press, 2002.
\newblock ISBN 0-521-75033-4.

\bibitem[{Rankin}(2007)]{rankin07a}
{Rankin}, J.~M.
\newblock \emph{{Further Evidence for Alignment of the Rotation and Velocity
  Vectors in Pulsars}}.
\newblock \apj, 664:\penalty0 443--447, July 2007.

\bibitem[{Ransom} et~al.(2004){Ransom}, {Kaspi}, {Ramachandran}, {Demorest},
  {Backer}, {Pfahl}, {Ghigo}, and {Kaplan}]{ransom04a}
{Ransom}, S.~M., {Kaspi}, V.~M., {Ramachandran}, R., {Demorest}, P., {Backer},
  D.~C., {Pfahl}, E.~D., {Ghigo}, F.~D., \& {Kaplan}, D.~L.
\newblock \emph{{Green Bank Telescope Measurement of the Systemic Velocity of
  the Double Pulsar Binary J0737-3039 and Implications for Its Formation}}.
\newblock \apjl, 609:\penalty0 L71--L74, July 2004.

\bibitem[{Rickett}(1990)]{rickett90a}
{Rickett}, B.~J.
\newblock \emph{{Radio propagation through the turbulent interstellar plasma}}.
\newblock \araa, 28:\penalty0 561--605, 1990.

\bibitem[{Roberts}(1997)]{roberts97a}
{Roberts}, P.~P.
\newblock \emph{{Calculating quantization correction formulae for digital
  correlators with digital fringe rotation}}.
\newblock \aaps, 126:\penalty0 379--383, December 1997.

\bibitem[{Romney}(1999)]{romney99a}
{Romney}, J.~D.
\newblock \emph{{Cross Correlators}}.
\newblock In \emph{Synthesis Imaging in Radio Astronomy II}, editors {Taylor},
  G.~B., {Carilli}, C.~L., \& {Perley}, R.~A., volume 180 of \emph{Astronomical
  Society of the Pacific Conference Series}, page~57, 1999.

\bibitem[{Roy} et~al.(2006){Roy}, {Rottmann}, {Teuber}, and {Keller}]{roy06a}
{Roy}, A., {Rottmann}, H., {Teuber}, U., \& {Keller}, R.
\newblock \emph{{Phase correction of VLBI with water vapour radiometry}}.
\newblock In \emph{Proceedings of the 8th European VLBI Network Symposium},
  2006.

\bibitem[{Ruderman} and {Sutherland}(1975)]{ruderman75a}
{Ruderman}, M.~A. \& {Sutherland}, P.~G.
\newblock \emph{{Theory of pulsars - Polar caps, sparks, and coherent microwave
  radiation}}.
\newblock \apj, 196:\penalty0 51--72, February 1975.

\bibitem[{Ryle}(1952)]{ryle52a}
{Ryle}, M.
\newblock \emph{{A New Radio Interferometer and Its Application to the
  Observation of Weak Radio Stars}}.
\newblock Royal Society of London Proceedings Series A, 211:\penalty0 351--375,
  March 1952.

\bibitem[{Ryle} and {Vonberg}(1946)]{ryle46a}
{Ryle}, M. \& {Vonberg}, D.~D.
\newblock \emph{{Solar Radiation on 175 Mc./s}}.
\newblock \nat, 158\penalty0 (4010):\penalty0 339--340, September 1946.

\bibitem[{Sandhu} et~al.(1997){Sandhu}, {Bailes}, {Manchester}, {Navarro},
  {Kulkarni}, and {Anderson}]{sandhu97a}
{Sandhu}, J.~S., {Bailes}, M., {Manchester}, R.~N., {Navarro}, J., {Kulkarni},
  S.~R., \& {Anderson}, S.~B.
\newblock \emph{{The Proper Motion and Parallax of PSR J0437-4715}}.
\newblock \apjl, 478:\penalty0 L95, April 1997.

\bibitem[{Sandulescu} et~al.(2004){Sandulescu}, {van Giai}, and
  {Liotta}]{sandulescu04a}
{Sandulescu}, N., {van Giai}, N., \& {Liotta}, R.~J.
\newblock \emph{{Superfluid properties of the inner crust of neutron stars}}.
\newblock \prc, 69\penalty0 (4):\penalty0 045802, April 2004.

\bibitem[{Sault} et~al.(1995){Sault}, {Teuben}, and {Wright}]{sault95a}
{Sault}, R.~J., {Teuben}, P.~J., \& {Wright}, M.~C.~H.
\newblock \emph{{A Retrospective View of MIRIAD}}.
\newblock In \emph{Astronomical Data Analysis Software and Systems IV}, editors
  {Shaw}, R.~A., {Payne}, H.~E., \& {Hayes}, J.~J.~E., volume~77 of
  \emph{Astronomical Society of the Pacific Conference Series}, page 433, 1995.

\bibitem[{Schaab} et~al.(1999){Schaab}, {Sedrakian}, {Weber}, and
  {Weigel}]{schaab99a}
{Schaab}, C., {Sedrakian}, A., {Weber}, F., \& {Weigel}, M.~K.
\newblock \emph{{Impact of internal heating on the thermal evolution of neutron
  stars}}.
\newblock \aap, 346:\penalty0 465--480, June 1999.

\bibitem[{Sekido} et~al.(2004){Sekido}, {Kondo}, {Kawai}, and
  {Imae}]{sekido04a}
{Sekido}, M., {Kondo}, T., {Kawai}, E., \& {Imae}, M.
\newblock \emph{{Evaluation of Global Ionosphere TEC by Comparison with VLBI
  Data}}.
\newblock In \emph{International VLBI Service for Geodesy and Astrometry 2004
  General Meeting Proceedings}, editors {Vandenberg}, N.~R. \& {Baver}, K.~D.,
  page~86, June 2004.

\bibitem[{Shapiro}(1964)]{shapiro64a}
{Shapiro}, I.~I.
\newblock \emph{{Fourth Test of General Relativity}}.
\newblock Physical Review Letters, 13:\penalty0 789--791, December 1964.

\bibitem[{Shepherd}(1997)]{shepherd97a}
{Shepherd}, M.~C.
\newblock \emph{{Difmap: an Interactive Program for Synthesis Imaging}}.
\newblock In \emph{Astronomical Data Analysis Software and Systems VI}, editors
  {Hunt}, G. \& {Payne}, H., volume 125 of \emph{Astronomical Society of the
  Pacific Conference Series}, page~77, 1997.

\bibitem[{Shklovskii}(1970)]{shklovskii70a}
{Shklovskii}, I.~S.
\newblock \emph{{Possible Causes of the Secular Increase in Pulsar Periods.}}
\newblock Soviet Astronomy, 13:\penalty0 562, February 1970.

\bibitem[{Siegman} et~al.(1993){Siegman}, {Manchester}, and
  {Durdin}]{siegman93a}
{Siegman}, B.~C., {Manchester}, R.~N., \& {Durdin}, J.~M.
\newblock \emph{{Timing parameters for 59 pulsars}}.
\newblock \mnras, 262:\penalty0 449--455, May 1993.

\bibitem[{Stairs} et~al.(2006){Stairs}, {Thorsett}, {Dewey}, {Kramer}, and
  {McPhee}]{stairs06a}
{Stairs}, I.~H., {Thorsett}, S.~E., {Dewey}, R.~J., {Kramer}, M., \& {McPhee},
  C.~A.
\newblock \emph{{The formation of the double pulsar PSR J0737-3039A/B}}.
\newblock \mnras, 373:\penalty0 L50--L54, November 2006.

\bibitem[{Tauris} et~al.(1994){Tauris}, {Nicastro}, {Johnston}, {Manchester},
  {Bailes}, {Lyne}, {Glowacki}, {Lorimer}, and {D'Amico}]{tauris94a}
{Tauris}, T.~M., {Nicastro}, L., {Johnston}, S., {Manchester}, R.~N., {Bailes},
  M., {Lyne}, A.~G., {Glowacki}, J., {Lorimer}, D.~R., \& {D'Amico}, N.
\newblock \emph{{Discovery of PSR J0108-1431: The closest known neutron star?}}
\newblock \apjl, 428:\penalty0 L53--L55, June 1994.

\bibitem[{Taylor} and {Cordes}(1993)]{taylor93a}
{Taylor}, J.~H. \& {Cordes}, J.~M.
\newblock \emph{{Pulsar distances and the galactic distribution of free
  electrons}}.
\newblock \apj, 411:\penalty0 674--684, July 1993.

\bibitem[{Taylor} and {Weisberg}(1989)]{taylor89a}
{Taylor}, J.~H. \& {Weisberg}, J.~M.
\newblock \emph{{Further experimental tests of relativistic gravity using the
  binary pulsar PSR 1913 + 16}}.
\newblock \apj, 345:\penalty0 434--450, October 1989.

\bibitem[{Thompson}(1999)]{thompson99a}
{Thompson}, A.~R.
\newblock \emph{{Fundamentals of Radio Interferometry}}.
\newblock In \emph{Synthesis Imaging in Radio Astronomy II}, editors {Taylor},
  G.~B., {Carilli}, C.~L., \& {Perley}, R.~A., volume 180 of \emph{Astronomical
  Society of the Pacific Conference Series}, page~11, 1999.

\bibitem[{Thompson} et~al.(1994){Thompson}, {Moran}, and
  {Swenson}]{thompson94a}
{Thompson}, A.~R., {Moran}, J.~M., \& {Swenson}, Jr., G.~W.
\newblock \emph{{Interferometry and Synthesis in Radio Astronomy}}.
\newblock Wiley, New York, 1994.

\bibitem[{Thompson} and {Duncan}(1996)]{thompson96a}
{Thompson}, C. \& {Duncan}, R.~C.
\newblock \emph{{The Soft Gamma Repeaters as Very Strongly Magnetized Neutron
  Stars. II. Quiescent Neutrino, X-Ray, and Alfven Wave Emission}}.
\newblock \apj, 473:\penalty0 322, December 1996.

\bibitem[{Tingay}(2008)]{tingay08b}
{Tingay}, S.~J.
\newblock \emph{{e-VLBI observations of SN1987A with the LBA}}.
\newblock \apjl, submitted, 2008.

\bibitem[{Tingay} et~al.(2008){Tingay}, {Alef}, {Graham}, and
  {Deller}]{tingay08a}
{Tingay}, S.~J., {Alef}, W., {Graham}, D., \& {Deller}, A.~T.
\newblock \emph{{Geodetic VLBI correlation in software I. Feasibility of using
  the DiFX software correlator for geodetic VLBI}}.
\newblock Journal of Geodesy, submitted, 2008.

\bibitem[{Titov} et~al.(2004){Titov}, {Tesmer}, and {Boehm}]{titov04a}
{Titov}, O., {Tesmer}, V., \& {Boehm}, J.
\newblock \emph{{OCCAM v.6.0 Software for VLBI Data Analysis}}.
\newblock In \emph{International VLBI Service for Geodesy and Astrometry 2004
  General Meeting Proceedings, Ottawa, Canada.}, editors {Vandenberg}, N.~R. \&
  {Baver}, K.~D., page~53, June 2004.

\bibitem[{Titov}(2007)]{titov07a}
{Titov}, O.~A.
\newblock \emph{{Apparent proper motions of radio sources from geodetic VLBI
  data}}.
\newblock Astronomy Letters, 33:\penalty0 481--487, July 2007.

\bibitem[{Torii} et~al.(1999){Torii}, {Tsunemi}, {Dotani}, {Mitsuda}, {Kawai},
  {Kinugasa}, {Saito}, and {Shibata}]{torii99a}
{Torii}, K., {Tsunemi}, H., {Dotani}, T., {Mitsuda}, K., {Kawai}, N.,
  {Kinugasa}, K., {Saito}, Y., \& {Shibata}, S.
\newblock \emph{{Spin-Down of the 65 Millisecond X-Ray Pulsar in the Supernova
  Remnant G11.2-0.3}}.
\newblock \apjl, 523:\penalty0 L69--L72, September 1999.

\bibitem[{van den Heuvel}(1975)]{van-den-heuvel75a}
{van den Heuvel}, E.~P.~J.
\newblock \emph{{Modes of mass transfer and classes of binary X-ray sources}}.
\newblock \apjl, 198:\penalty0 L109--L112, June 1975.

\bibitem[{van den Heuvel} and {Bonsema}(1984)]{van-den-heuvel84a}
{van den Heuvel}, E.~P.~J. \& {Bonsema}, P.~T.~J.
\newblock \emph{{Formation of a single millisecond pulsar by the coalescence of
  a neutron star and a massive white dwarf}}.
\newblock \aap, 139:\penalty0 L16--L18, October 1984.

\bibitem[{van Straten} et~al.(2001){van Straten}, {Bailes}, {Britton},
  {Kulkarni}, {Anderson}, {Manchester}, and {Sarkissian}]{van-straten01a}
{van Straten}, W., {Bailes}, M., {Britton}, M., {Kulkarni}, S.~R., {Anderson},
  S.~B., {Manchester}, R.~N., \& {Sarkissian}, J.
\newblock \emph{{A test of general relativity from the three-dimensional
  orbital geometry of a binary pulsar}}.
\newblock \nat, 412:\penalty0 158--160, July 2001.

\bibitem[{Verbiest} et~al.(2008){Verbiest}, {Bailes}, {van Straten}, {Hobbs},
  {Edwards}, {Manchester}, {Bhat}, {Sarkissian}, {Jacoby}, and
  {Kulkarni}]{verbiest08a}
{Verbiest}, J.~P.~W., {Bailes}, M., {van Straten}, W., {Hobbs}, G.~B.,
  {Edwards}, R.~T., {Manchester}, R.~N., {Bhat}, N.~D.~R., {Sarkissian}, J.~M.,
  {Jacoby}, B.~A., \& {Kulkarni}, S.~R.
\newblock \emph{{Precision Timing of PSR J0437-4715: An Accurate Pulsar
  Distance, a High Pulsar Mass, and a Limit on the Variation of Newton's
  Gravitational Constant}}.
\newblock \apj, 679:\penalty0 675--680, May 2008.

\bibitem[{Vo{\^u}te} et~al.(2002){Vo{\^u}te}, {Kouwenhoven}, {van Haren},
  {Langerak}, {Stappers}, {Driesens}, {Ramachandran}, and {Beijaard}]{voute02a}
{Vo{\^u}te}, J.~L.~L., {Kouwenhoven}, M.~L.~A., {van Haren}, P.~C., {Langerak},
  J.~J., {Stappers}, B.~W., {Driesens}, D., {Ramachandran}, R., \& {Beijaard},
  T.~D.
\newblock \emph{{PuMa, a digital Pulsar Machine}}.
\newblock \aap, 385:\penalty0 733--742, April 2002.

\bibitem[{Walker}(1998)]{walker98a}
{Walker}, M.~A.
\newblock \emph{{Interstellar scintillation of compact extragalactic radio
  sources}}.
\newblock \mnras, 294:\penalty0 307, February 1998.

\bibitem[{Walker}(1999)]{walker99a}
{Walker}, R.~C.
\newblock \emph{{Very Long Baseline Interferometry}}.
\newblock In \emph{Synthesis Imaging in Radio Astronomy II}, editors {Taylor},
  G.~B., {Carilli}, C.~L., \& {Perley}, R.~A., volume 180 of \emph{Astronomical
  Society of the Pacific Conference Series}, page 433, 1999.

\bibitem[{Wang} et~al.(2007){Wang}, {Manchester}, and {Johnston}]{wang07a}
{Wang}, N., {Manchester}, R.~N., \& {Johnston}, S.
\newblock \emph{{Pulsar nulling and mode changing}}.
\newblock \mnras, 377:\penalty0 1383--1392, May 2007.

\bibitem[{Weisberg} et~al.(1989){Weisberg}, {Romani}, and
  {Taylor}]{weisberg89a}
{Weisberg}, J.~M., {Romani}, R.~W., \& {Taylor}, J.~H.
\newblock \emph{{Evidence for geodetic spin precession in the binary pulsar
  1913 + 16}}.
\newblock \apj, 347:\penalty0 1030--1033, December 1989.

\bibitem[West(2004)]{west04a}
West, C.
\newblock \emph{Development of disk-based baseband recorders and software
  correlators for radio astronomy}.
\newblock Master's thesis, Swinburne University of Technology, 2004.

\bibitem[{Whitney}(2003)]{whitney03a}
{Whitney}, A.~R.
\newblock \emph{{Mark 5 Disk-Based Gbps VLBI Data System}}.
\newblock In \emph{Astronomical Society of the Pacific Conference Series},
  editor {Minh}, Y.~C., volume 306 of \emph{Astronomical Society of the Pacific
  Conference Series}, page 123, 2003.

\bibitem[{Whitney}(1993)]{whitney93a}
{Whitney}, A.~R.
\newblock \emph{{The Mark IV VLBI Data-Acquisition and Correlation System}}.
\newblock In \emph{Developments in Astrometry and their Impact on Astrophysics
  and Geodynamics}, editors {Mueller}, I.~I. \& {Kolaczek}, B., volume 156 of
  \emph{IAU Symposium}, pages 151--+, 1993.

\bibitem[{Whitney} et~al.(1971){Whitney}, {Shapiro}, {Rogers}, {Robertson},
  {Knight}, {Clark}, {Goldstein}, {Marandino}, and {Vandenberg}]{whitney71a}
{Whitney}, A.~R., {Shapiro}, I.~I., {Rogers}, A.~E.~E., {Robertson}, D.~S.,
  {Knight}, C.~A., {Clark}, T.~A., {Goldstein}, R.~M., {Marandino}, G.~E., \&
  {Vandenberg}, N.~R.
\newblock \emph{{Quasars Revisited: Rapid Time Variations Observed Via
  Very-Long-Baseline Interferometry}}.
\newblock Science, 173:\penalty0 225--230, July 1971.

\bibitem[{Willems} et~al.(2006){Willems}, {Kaplan}, {Fragos}, {Kalogera}, and
  {Belczynski}]{willems06a}
{Willems}, B., {Kaplan}, J., {Fragos}, T., {Kalogera}, V., \& {Belczynski}, K.
\newblock \emph{{Formation and progenitor of PSR J0737-3039: New constraints on
  the supernova explosion forming pulsar B}}.
\newblock \prd, 74\penalty0 (4):\penalty0 043003, August 2006.

\bibitem[{Williams} et~al.(2004){Williams}, {Turyshev}, and
  {Boggs}]{williams04a}
{Williams}, J.~G., {Turyshev}, S.~G., \& {Boggs}, D.~H.
\newblock \emph{{Progress in Lunar Laser Ranging Tests of Relativistic
  Gravity}}.
\newblock Physical Review Letters, 93\penalty0 (26):\penalty0 261101, December
  2004.

\bibitem[{Wilson} et~al.(1996){Wilson}, {Roberts}, and {Davis}]{wilson96a}
{Wilson}, W., {Roberts}, P., \& {Davis}, E.
\newblock \emph{The ATNF S-2 VLBI Correlator}.
\newblock In \emph{Proceedings of the 4th APT Workshop}, editor {King}, E.,
  page~16, 1996.

\bibitem[{You} et~al.(2007){You}, {Hobbs}, {Coles}, {Manchester}, {Edwards},
  {Bailes}, {Sarkissian}, {Verbiest}, {van Straten}, {Hotan}, {Ord}, {Jenet},
  {Bhat}, and {Teoh}]{you07a}
{You}, X.~P., {Hobbs}, G., {Coles}, W.~A., {Manchester}, R.~N., {Edwards}, R.,
  {Bailes}, M., {Sarkissian}, J., {Verbiest}, J.~P.~W., {van Straten}, W.,
  {Hotan}, A., {Ord}, S., {Jenet}, F., {Bhat}, N.~D.~R., \& {Teoh}, A.
\newblock \emph{{Dispersion measure variations and their effect on precision
  pulsar timing}}.
\newblock \mnras, 378:\penalty0 493--506, June 2007.

\bibitem[{Young} et~al.(1999){Young}, {Manchester}, and {Johnston}]{young99a}
{Young}, M.~D., {Manchester}, R.~N., \& {Johnston}, S.
\newblock \emph{{A radio pulsar with an 8.5-second period that challenges
  emission models}}.
\newblock \nat, 400:\penalty0 848--849, August 1999.

\bibitem[{Yue} et~al.(2007){Yue}, {Xu}, and {Zhu}]{yue07a}
{Yue}, Y.~L., {Xu}, R.~X., \& {Zhu}, W.~W.
\newblock \emph{{What can the braking indices tell us about the nature of
  pulsars?}}
\newblock Advances in Space Research, 40:\penalty0 1491--1497, 2007.

\bibitem[{Zakamska} and {Tremaine}(2005)]{zakamska05a}
{Zakamska}, N.~L. \& {Tremaine}, S.
\newblock \emph{{Constraints on the Acceleration of the Solar System from
  High-Precision Timing}}.
\newblock \aj, 130:\penalty0 1939--1950, October 2005.

\bibitem[{Zavlin}(2006)]{zavlin06a}
{Zavlin}, V.~E.
\newblock \emph{{XMM-Newton Observations of Four Millisecond Pulsars}}.
\newblock \apj, 638:\penalty0 951--962, February 2006.

\bibitem[{Zavlin} et~al.(2002){Zavlin}, {Pavlov}, {Sanwal}, {Manchester},
  {Tr{\"u}mper}, {Halpern}, and {Becker}]{zavlin02a}
{Zavlin}, V.~E., {Pavlov}, G.~G., {Sanwal}, D., {Manchester}, R.~N.,
  {Tr{\"u}mper}, J., {Halpern}, J.~P., \& {Becker}, W.
\newblock \emph{{X-Radiation from the Millisecond Pulsar J0437-4715}}.
\newblock \apj, 569:\penalty0 894--902, April 2002.

\bibitem[{Zhang} et~al.(2000){Zhang}, {Harding}, and {Muslimov}]{zhang00a}
{Zhang}, B., {Harding}, A.~K., \& {Muslimov}, A.~G.
\newblock \emph{{Radio Pulsar Death Line Revisited: Is PSR J2144-3933
  Anomalous?}}
\newblock \apjl, 531:\penalty0 L135--L138, March 2000.

\bibitem[{Zharikov} et~al.(2004){Zharikov}, {Shibanov}, {Mennickent},
  {Komarova}, {Koptsevich}, and {Tovmassian}]{zharikov04a}
{Zharikov}, S.~V., {Shibanov}, Y.~A., {Mennickent}, R.~E., {Komarova}, V.~N.,
  {Koptsevich}, A.~B., \& {Tovmassian}, G.~H.
\newblock \emph{{Multiband optical observations of the old PSR B0950+08}}.
\newblock \aap, 417:\penalty0 1017--1030, April 2004.

\end{thebibliography}
\newpage



\clearpage
\addcontentsline{toc}{chapter}{Publications}
{\bf Publications linked to this thesis}

\begin{enumerate}

\item {\bf Deller, A. T.}, Bailes, M. \& Tingay, S. J.

{\it Implications of a VLBI distance to the double pulsar J0737--3039A/B}

2009, Science (10.1126/science.1167969)

\item {\bf Deller, A. T.}, Tingay, S. J., \& Brisken, W. F.

{\it Precision Southern Hemisphere pulsar VLBI astrometry: 
techniques and results for PSR J1559--4438}

2009, ApJ, 690, 198

\item {\bf Deller, A. T.}, Verbiest, J. P. W., Tingay, S. J., \& Bailes, M

{\it Extremely high precision 
VLBI astrometry of PSR J0437--4715 and implications for theories of gravity}

2008, ApJL, 685, L67

\item Tingay, S. J., Alef, W., Graham, D., \& {\bf Deller, A. T.}

{\it Geodetic VLBI correlation in software I. 
Feasibility of using the DiFX software correlator for geodetic VLBI}

2008, Journal of Geodesy, submitted

\item Johnston, S. et al.,

{\it Science with the ASKAP: The Australian square--kilometre--array pathfinder}

2008, Experimental Astronomy, 22, 151

\item Johnston, S. et al.,

{\it Science with the Australian SKA Pathfinder}

2007, PASA, 24, 124

\pagebreak

\item Phillips, C. J., {\bf Deller, A.}, Amy, S. W., Tingay, S. J., Tzioumis, A. K., Reynolds, J. E., Jauncey, D. L., Stevens, J., Ellingsen, S. P., Dickey, J., Fender, R. P., Tudose, V. \& Nicolson, G. D.

{\it Detection of compact radio emission from Circinus X-1 with the first Southern hemisphere e-VLBI experiment}

2007, MNRAS, 380, L11

\item Norris, R. P., Tingay, S. J., Phillips, C. J., Middelberg, E., {\bf Deller, A. T.} \& Appleton, P. N.

{\it Very long baseline interferometry detection of an Infrared-Faint Radio Source}

2007, MNRAS, 378, 1434

\item {\bf Deller, A. T.}; Tingay, S. J.; Bailes, M. \& West, C.

{\it DiFX: A Software Correlator for Very Long Baseline Interferometry Using Multiprocessor \ Environments}

2007, PASP, 119, 318

\end{enumerate}

\end{document}